\documentclass[12pt,letterpaper]{article}
\usepackage[usenames, dvipsnames]{xcolor}
\usepackage{jcapmod}

\usepackage{tocloft}
\usepackage[]{todonotes}
\usepackage{verbatim}
\usepackage{mathrsfs}
\usepackage{cleveref}
\usepackage[shortlabels]{enumitem}
\usepackage{pgf}
\usepackage{tikz-cd}
\usetikzlibrary{shapes,arrows,decorations.markings}
\tikzset{
  big arrow/.style={
    decoration={markings,mark=at position 1 with {\arrow[scale=2,#1]{>}}},
    postaction={decorate},
    shorten >=0.4pt},
  big arrow/.default=black}
\usetikzlibrary{calc}
\usepackage{pgfplots}
\pgfplotsset{compat=1.12}
\usepackage{tikz-3dplot}

\usepackage{amsthm}
\usepackage{array}

\usepackage{tikz}
\usepackage{setspace,caption}
\usepackage{lipsum}
\usetikzlibrary{matrix,arrows,calc}
\makeatletter
\newsavebox\myboxA
\newsavebox\myboxB
\newlength\mylenA
\usepackage[margin=2.85cm]{geometry}

\definecolor{cornellRed}{HTML}{B31B1B}
\definecolor{cornellBlue}{HTML}{0068AC}
\definecolor{cornellGreen}{HTML}{6EB43F}

\usepackage{bbm} 					%%% Blackboard Bold math symbols
\usepackage{slashed} 				%%% Feynman slash
\usepackage{graphicx}				%%% Graphics
\usepackage{subcaption}			%%% Subcaptions for subfigures
\usepackage{psfrag}				%%% Editing EPS files in TeX
\usepackage{tensor}				%%% The \indices command
\usepackage{fouridx}				%%% Arbitrary subscripts/superscripts
\usepackage{bm}					%%% Bold Math symbols
\usepackage{mdframed}				%%% Box for \aside
\usepackage{multirow}				%%% More flexible tables
\usepackage{soul}					%%% Strike-through text
\usepackage{bbold}				%%% Blackboard Bold
\usepackage{multicol}				%%% Pages with multiple columns
\usepackage{tikz-cd}
\usepackage{rotating}

\usetikzlibrary{arrows}

\tikzset{
commutative diagrams/.cd,
arrow style=tikz,
diagrams={>=latex}}

\usepackage{amsmath}
\usepackage{amssymb}
\usepackage{amsfonts}
\usepackage{mathtools}

\usepackage{feynmf}
\usepackage{marvosym}

\usepackage{import}

\newcommand*\xoverline[2][0.75]{%
    \sbox{\myboxA}{$\m@th#2$}%
    \setbox\myboxB\null% Phantom box
    \ht\myboxB=\ht\myboxA%
    \dp\myboxB=\dp\myboxA%
    \wd\myboxB=#1\wd\myboxA% Scale phantom
    \sbox\myboxB{$\m@th\overline{\copy\myboxB}$}%  Overlined phantom
    \setlength\mylenA{\the\wd\myboxA}%   calc width diff
    \addtolength\mylenA{-\the\wd\myboxB}%
    \ifdim\wd\myboxB<\wd\myboxA%
       \rlap{\hskip 0.5\mylenA\usebox\myboxB}{\usebox\myboxA}%
    \else
        \hskip -0.5\mylenA\rlap{\usebox\myboxA}{\hskip 0.5\mylenA\usebox\myboxB}%
    \fi}
\makeatother

\definecolor{cobalt}{RGB}{44, 98, 120}
\definecolor{celadon}{rgb}{0.67, 0.88, 0.69}
\definecolor{dm}{cmyk}{.20, 0, .30, 0}
\definecolor{burgundy}{rgb}{0.5, 0.0, 0.13}
\definecolor{plotBlue}{RGB}{94, 130, 181}

\hypersetup{
  colorlinks,
  citecolor=Violet,
  linkcolor=cobalt,
  urlcolor=Blue}

\DeclareSymbolFontAlphabet{\mathbb}{AMSb}

\newif\iffastcompile

\fastcompilefalse
\iffastcompile
\newcommand{\mk}[1]{}
\newcommand{\lm}[1]{}
\else
\newcommand{\mk}[1]{\todo[color=burgundy!30, size=\scriptsize, bordercolor=burgundy!30]{MK: #1}}
\newcommand{\lm}[1]{\todo[color=dm!90, size=\scriptsize, bordercolor=dm!90]{LM: #1}}

\fi

\ProvideTextCommandDefault{\Dbar}{%
\leavevmode\lower.5ex\rlap{\hskip-.07em\accent"16}D%
}

\usepackage{environ}
\usepackage{changepage}

\begin{document}
	\newcommand{\main}{.}
\begin{titlepage}

\setcounter{page}{1} \baselineskip=15.5pt \thispagestyle{empty}
\setcounter{tocdepth}{2}

\bigskip\
{\hfill \small MIT-CTP/5470}
\vspace{0.5cm}
\vspace{1cm}
\begin{center}
{\fontsize{18}{24} \bfseries On the intermediate Jacobian of M5-branes}
\end{center}

\vspace{0.55cm}

\begin{center}
\scalebox{0.95}[0.95]{{\fontsize{14}{30}\selectfont Patrick Jefferson$^{a}$ and Manki Kim$^{a}$\vspace{0.25cm}}}

\end{center}

\begin{center}

\vspace{0.15 cm}
{\fontsize{11}{30}
\textsl{$^{a}$Center for Theoretical Physics, Department of Physics, Massachusetts Institute of Technology, Cambridge, MA 02139}}\\
\vspace{1cm}
	\normalsize{\bf Abstract} \\
%\vskip .5cm
\end{center}

%\vspace{0.8cm}
\noindent 
 
We study Euclidean M5-branes wrapping vertical divisors in elliptic Calabi-Yau fourfold compactifications of M/F-theory that admit a Sen limit. We construct these Calabi-Yau fourfolds as elliptic fibrations over coordinate flip O3/O7 orientifolds of toric hypersurface Calabi-Yau threefolds. We devise a method to analyze the Hodge structure (and hence the dimension of the intermediate Jacobian) of vertical divisors in these fourfolds, using only the data available from a type IIB compactification on the O3/O7 Calabi-Yau orientifold. Our method utilizes simple combinatorial formulae (that we prove) for the equivariant Hodge numbers of the Calabi-Yau orientifolds and their prime toric divisors, along with a formula for the Euler characteristic of vertical divisors in the corresponding elliptic Calabi-Yau fourfold. Our formula for the Euler characteristic includes a conjectured correction term that accounts for the contributions of pointlike terminal $\Bbb{Z}_2$ singularities corresponding to perturbative O3-planes. We check our conjecture in a number of explicit examples and find perfect agreement with the results of direct computations.

\begin{center}
	\begin{minipage}[h]{15.0cm}

	\end{minipage}
\end{center}

\end{titlepage}
\tableofcontents

\section{Introduction}

One of the main ambitions of string phenomenology is to identify stable cosmological solutions with realistic features such as positive cosmological constant, as in the case of de Sitter spacetime, and separation of characteristic length scales between the external 4D spacetime and the internal compactification space. Despite the fact that the community as a whole has put significant effort into constructing semi-realistic cosmological solutions of string theory over the past few decades, a fully explicit and well-controlled semi-realistic cosmological solution of string theory has not been constructed so far.\footnote{Two of the most well studied proposals of this sort are the KKLT proposal \cite{Kachru:2003aw} and the large volume scenario \cite{Balasubramanian:2005zx}. For the recent work on explicit constructions of string vacua with scale separation, see \cite{Demirtas:2021ote,Demirtas:2021nlu}.} Partly, this is due to the geometric intricacies that must be dealt with to ensure dynamical stability, which requires precise computational control over the superpotential of the corresponding low energy effective 4D action---in particular, control over the non-perturbative contributions to the superpotential. The purpose of this paper is to provide techniques for computing the non-perturbative contributions to the 4D $\mathcal N=1$ superpotential in an effort to further develop the mathematical tools necessary to identify stable cosmological string vacua of the type described above. 

Much of the string phenomenology literature focused on type IIB/F-theory compactifications has been produced under the assumption that most or all complex structure moduli in generic setups can be made sufficiently heavy (relative to other moduli) to justify essentially ignoring the details of Euclidean D3-instanton contributions to the non-perturbative superpotential. However, recent work on flux vacua with small flux superpotential \cite{Dudas:2019pls,Demirtas:2019sip,Demirtas:2020ffz,Alvarez-Garcia:2020pxd}\footnote{See also \cite{Bena:2018fqc,Bena:2019sxm,Lust:2022xoq}, which showcase the lightness of conifold moduli in generic KKLT-like constructions.} strongly suggests that light complex structure moduli are unavoidable in generic KKLT-like constructions, and hence that the moduli-dependence of the one-loop pfaffian of Euclidean D3-instanton contributions cannot safely be ignored in realistic vacuum solutions.\footnote{For recent studies of flux vacua with small flux superpotential, see \cite{Honma:2021klo,Marchesano:2021gyv,Broeckel:2021uty,Bastian:2021hpc,Grimm:2021ckh,Carta:2021kpk,Carta:2022oex,Cicoli:2022vny}.} 

Let us review some of what is known about Euclidean D3-instantons in this context. In \cite{Witten:1996bn}, it was shown that for 3D M-theory vacua defined by compactification on a Calabi-Yau fourfold $Y_4$, a non-perturbative superpotential can be generated by Euclidean M5-instantons wrapping certain divisors $\overline D \subset Y_4$.\footnote{ A sufficient condition for the generation of the superpotential is $h^{\bullet}(\overline{D},\mathcal{O}_{\overline{D}})=(1,0,0,0)$ for a smooth $\overline{D}.$ An important open problem is to understand when a non-rigid divisor $\overline{D}$ can contribute to the superpotential by means of rigidification by spacetime filling D-branes or fluxes \cite{Bianchi:2011qh,Palti:2020qlc}. We expect that these conditions continue to hold even in the presence of O3-plane singularities. For recent work on divisors with singularities, see \cite{Gendler:2022qof}.} When $Y_4$ is furthermore elliptically-fibered over a threefold base $B_3$ and $\overline D$ is a vertical divisor (i.e. $\overline D$ is an irreducible component of the pullback of a divisor $\widehat D$ in $B_3$), one finds that a Euclidean M5-instanton wrapping $\overline D$ is mapped via M/F-theory duality to either a Euclidean D3-instanton wrapping $\widehat D$ \cite{Witten:1996bn} or gaugino condensate due to a seven-brane stack on $\widehat{D}$ \cite{Katz:1996fh,Katz:1996th,Diaconescu:1998ua,Denef:2008wq}, and hence M5-instanton contributions to the non-perturbative 3D superpotential lift to Euclidean D3-instanton or gaugino condensate contributions to the 4D superpotential. M5-instantons have therefore become an essential tool for analyzing the non-perturbative superpotential in F-theory compactifications.

The contribution of a single M5-instanton to the non-perturbative superpotential takes the form
	\begin{align}
		W_{\text{n.p.}} = \mathcal A_{\overline D} e^{\text{vol}(\overline D) + i \int_{\overline D} C_6}
	\end{align}
where $C_6$ is the Hodge dual of the M-theory 3-form and $\text{vol}(\overline{D})$ is the calibrated volume of the divisor $\overline{D}.$\footnote{For multi-covering effects on the non-perturbative superpotential, see \cite{Witten:1999eg,Garcia-Etxebarria:2007fvo,Blumenhagen:2008ji,Blumenhagen:2012kz,Alexandrov:2022mmy}.} Complete characterization of the non-perturbative superpotential including the moduli-dependent one-loop Pfaffian $\mathcal{A}_{\overline{D}}$ is an important open problem \cite{Ganor:1996pe,Witten:1999eg,Harvey:1999as,Buchbinder:2002ic,Buchbinder:2002pr,Beasley:2003fx,Berg:2004ek,Kallosh:2005gs,Lust:2005cu,Ibanez:2006da,Akerblom:2006hx,Blumenhagen:2006xt,Baumann:2006th,Braun:2007xh,Braun:2007vy,Koerber:2007xk,Blumenhagen:2007zk,Beasley:2008dc,Blumenhagen:2010ja,Baumann:2010sx,Donagi:2010pd,Bianchi:2011qh,Marsano:2011xfj,Grimm:2011dj,Kerstan:2012cy,Cvetic:2012ts,Anderson:2015yzz,Hamada:2018qef,Kim:2018vgz,Kachru:2019dvo,Bena:2019mte,Hamada:2019ack,Gautason:2019jwq,Carta:2019rhx,Alexandrov:2022mmy,Kim:2022uni}. In the absence of spacetime filling M2/D3-branes, the moduli-dependent\footnote{The one-loop determinant $\mathcal A_{\overline D}$ could in principle depend on the complex structure moduli of $X_4$, the $h^{2,1} + h^{1,2}$ moduli that descend from the M-theory 3-form $C_3$, and moduli corresponding to the positions of spacetime filling M2 branes.} one-loop determinant $\mathcal A_{\overline D}$ was shown \cite{Witten:1996hc,Belov:2006jd}  to be a holomorphic section of a line bundle whose first Chern class is the principal polarization of the intermediate Jacobian
	\begin{align}
		\mathcal J({\overline D}) := H^{3}(\overline D, \mathbb R)/ H^{3}(\overline D,\mathbb Z).
	\end{align}
Much of the analysis of the non-perturbative superpotential can be reduced to computing various properties of $\mathcal J(\overline D)$, such as its dimension and dependence on various moduli. In this paper we focus on the task of computing the dimension of $\mathcal J(\overline D)$. While it is in principle possible to compute the dimension of $\mathcal J(\overline D)$ with the aid of various (co)homological exact sequences \cite{Denef:2005mm}, sophisticated computations of this sort can be cumbersome and quickly become intractable for Calabi-Yau manifolds with large Hodge numbers, and thus it would be desirable to have some sort of combinatorial formula that can be used to compute the dimension of $\mathcal J(\overline D)$ and other relevant data in terms of characteristic numbers of the underlying Calabi-Yau fourfold.

To address this problem, in this paper we continue the work of \cite{Kim:2021lwo} and devise an algorithm to compute the Hodge numbers of toric vertical divisors $\overline D$ in Calabi-Yau fourfolds $Y_4$ that are elliptic fibrations over coordinate flip O3/O7-orientifolds of toric hypersurface Calabi-Yau threefolds $B_3 = Z_3/\Bbb{Z}_2$.\footnote{In F-theory compactifications, we refer to a Sen limit \cite{Sen:1996vd,Sen:1997gv} such that the string coupling is finite and all D7-brane stacks carry $SO(8)$ gauge bundle as the ``global Sen limit''. The Calabi-Yau fourfolds that we construct explicitly in this paper, which are elliptically-fibered over a base $B_3 = Z_3/\mathbb Z_2$ and only contain $SO(8)$ D7-brane stacks, are examples of Calabi-Yau fourfolds that admit a global Sen limit. For recent study of O3/O7 orientifolds, see \cite{Gao:2013pra,Carta:2020ohw,Altman:2021pyc,Crino:2022zjk,Gao:2021xbs}.} Conceptually, our algorithm consists of two steps. The first step is to explicitly construct an orientifold of a toric hypersurface Calabi-Yau threefold $B_3$, and to use aspects of this construction to prove simple combinatorial formulae for the equivariant Hodge numbers $h_{\pm{}}^{p,q}(Z_3)$ and the Hodge numbers $h^{p,q}(\widehat D)$ for its prime toric divisors, $\widehat D \subset B_3$. This procedure is not quite sufficient to compute all of the Hodge numbers of the prime toric divisors of $Y_4$. The second step, the purpose of which is to capture the remaining Hodge number $h^{2,1}(\widehat D)$, proceeds as follows: we use the existence of a Calabi-Yau fourfold defining an F-theory uplift to identify a local fibration $\overline D$ over $\widehat D$. Then, following the methods of \cite{Esole:2017kyr}, we compute the pushforward of the Euler characteristic of $\overline D$ to $B_3$ to recover the missing Hodge number.\footnote{In this work, we assume that moduli are at a generic point in the moduli space to ensure that the dimension of the intermediate Jacobian we compute captures all possible moduli dependence.}

The algorithm presented in this paper provides a method to characterize the vertical prime toric divisors in an elliptic Calabi-Yau fourfold $Y_4$ without the need to explicitly construct and analyze $Y_4$. Our reason for devising such a method is that despite the fact that it is desirable to be able to perform a direct analysis of an explicitly constructed Calabi-Yau fourfold and its prime toric divisors, this is not necessarily the easiest computation to perform. There are three major technical challenges responsible for making direct analysis rather complicated. First, toric resolutions of the toric ambient variety in which we embed the elliptic Calabi-Yau fourfold need not be described by fine regular star triangulations, but rather may involve \emph{vex} triangulations whose assoicated toric fan is not convex\footnote{We thank W. Taylor for explaining this to us.} \cite{Berglund:2016yqo,Berglund:2016nvh,vex}, thereby going beyond the standard framework studied by Batyrev and Borisov \cite{Batyrev:1994ju,Batyrev:1994pg}. Second, the toric ambient varieties need not be constructed from reflexive polytopes, as we demonstrate by way of example in \S\ref{sec:quintic}. This also forces us to go beyond the standard framework. Third, typically the elliptic Calabi-Yau fourfolds that admit a global Sen limit are not simple hypersurfaces but rather are complete intersections, which further complicates the direct analysis \cite{Batyrev:1994pg,Batyrev:1995ca,Kreuzer:2001fu}. We will present a direct analysis of prime toric divisors in complete intersection Calabi-Yau manifolds in \cite{nefpartition}.

Although our algorithm is restricted to elliptic Calabi-Yau fourfolds that admit a global Sen limit, this represents an extremely interesting class of Calabi-Yau fourfolds to study for various reasons. The most obvious of these reasons is motivated by string phenomenology: By construction, F-theory compactified on a Calabi-Yau fourfold $Y_4$ defines a weakly-coupled type IIB string vacuum solution in the global Sen limit; moreover, since many of the F-theory bases $B_3$ admit a $\mathbb P^1$ fibration\footnote{The $\mathbb P^1$ fibration structure of $B_3$ descends from the elliptic fibration structure of $Z_3$.}, this construction also admits a dual heterotic/type I string theory description. Thus, elliptic Calabi-Yau fourfold that admit a global Sen limit represent one of the most accessible subsets of candidate string vacua amenable to detailed analysis in multiple string duality frames, and the algorithm we describe in this paper has numerous applications for the study of these vacua. 

Another reason that elliptic Calabi-Yau fourfolds that admit a global Sen limit are interesting, which is to some extent independent of their connection to string phenomenology, is that they often contain pointlike terminal $\Bbb{Z}_2$ orbifold singularities.\footnote{From a mathematical perspective, the fact that terminal $\mathbb Z_2$ orbifold singularities can appear in elliptic Calabi-Yau fourfolds should come as no great surprise, as Calabi-Yau varieties of complex dimension greater than three are not guaranteed to have smooth phases \cite{batyrev1993dual}. Although the physics of terminal singularities is still not well-understood, fortunately, it has been shown that the ``worst'' type of singularity that can occur in toric hypersurface Calabi-Yau varieties with maximal projective crepant partial (MPCP) desingularizations are orbifold singularities, and thus if we can gain a complete understanding of the physics of Calabi-Yau manifolds with orbifold singularities, then we will be adequately equipped to deal with the full range of singularity types that appear in a rather large and interesting class of candidate string vacua.} Although the physics associated with terminal singularities in F-theory vacua is still being explored, it is at least known that terminal cyclic fourfold $\mathbb Z_k$ singularities \cite{morrison1984terminal} indicate the presence of O3-planes, with $k=2$ corresponding to perturbative O3-planes \cite{Collinucci:2008zs,Collinucci:2009uh} and the cases $k=3,4,6$ corresponding to their non-perturbative generalizations \cite{Garcia-Etxebarria:2015wns}. The limited literature that exists on this subject given the richness of the physics associated with these singularities indicates that additional mathematical techniques need to be developed in order to fully study their properties. 

It is tempting to regard terminal orbifold singularities as exotic features of F-theory vacua. However, as will be shown in a forthcoming paper \cite{stats}, there is evidence suggesting that a significant fraction of elliptic Calabi-Yau fourfolds that admit a global Sen limit include O3-planes, and hence that terminal $\mathbb Z_2$ singularities seem to be a rather unavoidable feature of such constructions. An important feature of our algorithm, which accounts for the apparent prevalence of O3-planes, is a conjectured formula for the Euler characteristic of vertical divisors $\overline D$, which depends combinatorially on the number of pointlike terminal $\mathbb Z_2$ orbifold singularities intersecting $\widehat D \subset B_3$: 
	\begin{align}
		\boxed{\chi_n(\overline D):=\int_{\overline{D}}c_3(\overline{D}) = \sum_{p,q} (-1)^{p+q} h^{p,q}(\overline D) - 2 n_{\text{O3}}(\widehat D)\,.}
	\end{align}
To our knowledge, this is the first time such a formula has appeared in the string theory literature. We believe a reasonably straightforward mathematical proof of the above formula should be possible if one indeed does not exist at the time of writing. Using the above conjectural formula, we find that we are able to use our algorithm to recover the Hodge numbers of all prime toric divisors $\widehat D$ appearing in orientifolds of elliptic Calabi-Yau threefolds. We have checked the validity of this formula in a large number of examples. 

The remainder of this paper is structured as follows. In \S\ref{sec:toric review}, we review various tools from toric geometry relevant for the computations of this paper. In \S\ref{sec:toric CYs}, we review Batyrev's construction of toric hypersurface Calabi-Yau threefolds. In \S\ref{sec:toric CY orientifolds}, we construct coordinate flip O3/O7 orientifolds $B_3$ as hypersurfaces in toric fourfolds $\widehat{V}_4$. We furthermore prove combinatorial formulae for the Hodge numbers $h^{p,q}_\pm(Z_3)$ and $h^{p,q}(\widehat{D})\in B_3.$ In \S\ref{sec:fourfold analysis}, we study the F-theory uplifts of type IIB compactifications on the O3/O7 orientifolds $B_3$, and study the topology of vertical prime toric divisors in the resulting F-theory geometry. In particular, in \S\ref{sec:fourfold analysis}, we classify the possible vertical divisors that can appear in Calabi-Yau fourfolds, in terms of the Iitaka dimension of the twisting line bundle of the elliptic fibration. We show that for each case, a subset of the Hodge numbers of the vertical divisors can be computed using the combinatoric formulae developed in \S\ref{sec:toric CY orientifolds}. In \S\ref{sec:push forward and local}, we prove the pushforward formula for the integral of the top Chern class $\int_{\overline{D}}c_3(\overline{D}),$ and study local models of divisors in the presence of O3-planes, as well as elliptic fibrations with twisting line bundle of Iitaka dimension 1---the tools developed in this section give us the means to compute the remaining Hodge numbers that were not captured by classification scheme described previously, in \S\ref{sec:fourfold analysis}. In \S\ref{Examples}, as an illustration, we compute the Hodge numbers of vertical divisors in the context of four explicit examples. In \S\ref{sec:discussion}, we conclude and discuss possible future directions. In \S\ref{app:mirrorexample}, we record some numerical results relevant to the final example in \S\ref{Examples}, and a compendium of commonly-used notation is given in \S\ref{app:notation}.

\section{Review of toric geometry}\label{sec:toric review}

\subsection{Constructing toric varieties from fans}
Recall that a $d$-dimensional toric variety $V_d$ is a complex algebraic variety that contains an algebraic torus $T = (\mathbb C^*)^d$ as a dense open set, together with an action $T$ on $V_d$ such that the restriction of this action to the algebraic torus $T \subset V_d$ itself is multiplicative. In this paper, we restrict our attention to normal projective toric varieties, which can be described both by means of a fan and a lattice polytope. Below, we review aspects of both constructions relevant for the analysis of this paper. 

A fan $\Sigma$ is a set of strongly convex rational polyhedral cones, such that the face of any cone is a cone and the intersection of any two cones is a cone. It is convenient to regard each cone as being spanned by a finite number of rays with primitive generators $\vec v$, which are themselves elements of a lattice $N$. Strong convexity guarantees that no non-trivial subspace of the ambient vector space in which $\Sigma$ is embedded lies inside any cone of $\Sigma$. Consider a fan $\Sigma$ that consists of cones whose rays are given by the non-negative span of the generators 
	\begin{align}
		 \vec v \in N. 
	\end{align}
We associate to $\Sigma$ a (normal projective) toric variety $\mathbb P_\Sigma$ defined as follows:
	\begin{align}
	\label{PSigma}
		\mathbb P_\Sigma = \frac{\mathbb C^r \backslash {SRI}}{G}	
	\end{align}
where the Stanley-Reisner ideal ${SRI}$ consists of all subsets of $I \subset \{ \vec v\}$ for which the one-dimensional cones generated by $\vec v$ do not share a common higher-dimensional cone. By definition, each such subset in $\text{SRI}$ corresponds to a subspace of $\mathbb C^r$ given by the zero locus $\{ x_{\vec v} = 0 \, |\, \vec v \in  I \}$ for some choice of correspondence between the coordinates of $\mathbb C^r$ and the vertices $\vec v \in N$:	
	\begin{align}
		 x_{\vec v} \leftrightarrow \vec v .	
	\end{align}
The abelian group $G$ is the kernel of the morphism
	\begin{align}
	\label{tk}
		x_{\vec v} \rightarrow t_k = \prod_{\vec v} x_{\vec v}^{ \vec e_k \cdot \vec v}
	\end{align}
where $\vec e_k$ form a basis of the dual lattice $M = \text{Hom}(N,\mathbb Z)$.\footnote{We abuse notation and use the usual vector notation $\vec m$ to denote elements of $M$. Moreover, we use the dot product $\vec m \cdot \vec v$ to denote the action of an element $\vec m \in M$ on an element $\vec v \in N$.} To every linear dependence relation
	\begin{align}
	\label{linrel}
		\sum_{\vec v} a_{\vec v} \vec v = 0,~~~~ a_{\vec v} \in \mathbb Z
	\end{align}
there corresponds a one-parameter subgroup $g \subset G$ whose action on $\mathbb C^r$ in (\ref{PSigma}) is given by 
	\begin{align}
	\label{morphism}
		g: x_{\vec v} \rightarrow \lambda^{a_{\vec v}} x_{\vec v},~~~~\lambda \in \mathbb C^*. 
	\end{align}
Note that while in many relevant examples the abelian group $G$ can be specified completely by its subgroups of the above form, there are some cases in which $G$ contains discrete subgroups, and consequently $G$ cannot be completely determined in terms its continuous subgroups.

\subsection{Stratifications}\label{sec:stratification}

When $x_{\vec v} \ne 0$ (for all $\vec v$ generating the cones of $\Sigma$), the coordinates $t_k$ in (\ref{morphism}) parametrize the algebraic torus $(\mathbb C^*)^d$ as a dense open set in $\mathbb C^d$. The full toric variety $\mathbb P_\Sigma$ can be viewed as a disjoint union of algebraic tori $(\mathbb C^*)^k$, $0 \leq k \leq d$ of various dimensions. These collections of algebraic tori, called \emph{strata}, are associated with various cones $\sigma$ in the fan $\Sigma$ in the following manner. The torus $(\mathbb C^*)^d$, to which we refer as the \emph{prime stratum}, is associated with the unique zero-dimensional cone $\sigma^{(0)}$, i.e. the origin of $\Sigma$. More generally, each $n$-dimensional cone $\sigma^{(n)}$ is naturally associated with the subset of homogeneous coordinates $x_{\vec v}$ corresponding to the rays of $\sigma^{(n)}$ generated by $\vec v$. Given a cone $\sigma$, if we restrict the basis of dual vectors $\vec e_k  \in M$ in (\ref{tk}) to span the subset of lattice vectors contained in the dual cone $\sigma^\vee$, defined as
	\begin{align}
		\label{dualcone}
			\sigma^\vee := \{ \vec m \in M | \vec m \cdot \vec v \geq 0, \, \forall \vec v \in \sigma\},
		\end{align}
then it is possible to take a limit of (\ref{tk}) in which $x_{\vec v}=0$ for all $\vec v$ generating $\sigma$; since $\sigma^{(n)}$ contains $n$ generators $\vec v$ by definition, this shows that the $n$-dimensional cones $\sigma^{(n)}$ correspond to $(d-n)$-dimensional algebraic tori 
	\begin{align}
		T_{\sigma^{(n)}} \cong (\mathbb C^*)^{d-n}. 
	\end{align}
Note that the simplicial structure of $\Sigma$ guarantees that the strata are glued together in a consistent manner. It therefore follows that we can use the data of the fan $\Sigma$ to stratify any toric variety $\mathbb P_\Sigma$ as follows:
	\begin{align}
	\label{toricstrat}
		\mathbb P_\Sigma &= \coprod_n \coprod_{\sigma^{(n)} \in \Sigma(n) } T_{\sigma^{(n)}},~~~~ T_{\sigma^{(n)}} \cong (\mathbb C^*)^{d-n}.
	\end{align}
Note that the symbol $\Sigma(n)$ in the above expression denotes the set of $n$-dimensional cones in the fan $\Sigma$. In terms of  coordinates, the stratum $T_{\sigma}$ associated with each cone is parametrized by 
	\begin{align}
		T_\sigma = \{ x_{\vec v} \, | \, x_{\vec v} \ne 0 \text{ if } \vec v  \notin \sigma \text{ and } x_{\vec v}= 0 \text{ if }\vec  v \in \sigma \}.
	\end{align}

We have described above how every projective normal toric variety $\mathbb P_\Sigma$ admits a stratification of the form (\ref{toricstrat}). There is a complementary viewpoint whereby one may regard any toric variety $\mathbb P_\Sigma$ as a compactification of some choice of prime stratum (note that the algebraic torus $(\mathbb C^*)^d$ is non-compact). We now explore this viewpoint. Let us begin by describing the simplest example of such a compactification, which will provide intuition for more complicated examples. The simplest case is $\Bbb{C}^*$. The open set $\Bbb{C}^*$ can be understood as a copy of the complex plane with the origin excised. The toric fan for $\mathbb C^*$ consists of two toric rays, which we denote by (resp.) $\vec v_1 = (1)$ and $\vec v_2 = (-1),$ each of which corresponds to a homogeneous coordinate. Without loss of generality, we use $x_1$ to denote the homogeneous coordinate associated to $\vec v_1$ and $x_2$ to denote the homogeneous coordinate associated to $-\vec v_{2}$. When $x_1x_2\neq0,$ we can parametrize the one-dimensional algebraic torus $\mathbb C^*$ by a complex variable $t= x_1/x_2\in \Bbb{C}^*.$ Because there are two missing points, namely the origin and infinity, we see clearly that $\Bbb{C}^*$ is non-compact. To compactify $\Bbb{C}^*,$ we can glue two points $t=0$ and $t=\infty$ to $\Bbb{C}^*.$ As the result, we have compactified $\Bbb{C}^*$ into $\Bbb{P}^1$, where we keep in mind that $t=0$ is equivalent to $x_1=0,$ and similarly $t=\infty$ is given by $x_2=0.$ 

This procedure generalizes to higher-dimensional toric varieties. To compactify an $d$-dimensional algebraic torus $(\mathbb C^*)^d$, one can glue in $(d-1)$-dimensional algebraic tori, each of which is identified with the vanishing locus of a homogeneous coordinate parametrizing the higher dimensional toric variety. Analogous to the case of $\Bbb{P}^1,$ the vanishing locus of a homogeneous coordinate $x_{\vec{v}}$ is represented by a toric ray generated by the vertex $\vec v$. Let us suppose that there are $d+1$ homogeneous coordinates $x_{\vec v}$. The vanishing locus $T_{\sigma^{(0)}}$ is then parametrized by $d$ independent toric coordinates $t$, so we conclude that there is an isomorphism $T_{\sigma^{(0)}}\cong (\Bbb{C}^*)^{d}.$ Similar reasoning then leads to the conclusion that an $n$-dimensional toric cone $\sigma^{(n)}$ corresponds to a $(d-n)$-dimensional algebraic torus $T_{\sigma^{(n)}}\cong (\Bbb{C}^*)^{(d-n)}.$ This identification now then completes the compactification of the prime stratum, and yields a compact toric variety $\Bbb{P}_{\Sigma}$ (compare to (\ref{toricstrat}))
\begin{equation}\label{eqn:compactification}
\Bbb{P}_{\Sigma}=\coprod_n \coprod_{\sigma^{(n)} \in \Sigma(n) } T_{\sigma^{(n)}}\,.
\end{equation}

\subsection{Toric hypersurfaces}
\label{newtonnormal}
In this paper, we are primarily concerned with algebraic subvarieties of toric varieties $\mathbb P_\Sigma$. In well-behaved constructions, most or all of the geometric properties of a subvariety that are relevant to our analysis can be inferred from the geometry of the ambient toric variety $\mathbb P_\Sigma$. As an example, if all strata $T_\sigma \subset \mathbb P_\Sigma$ intersect the subvariety transversely, then the subvariety naturally inherits a stratification from $\mathbb P_\Sigma$. In the special case that the subvariety is a hypersurface $Z \subset \mathbb P_\Sigma$, one can associate to every $n$-dimensional cone $\sigma^{(n)} \in \Sigma$ a $(d-n-1)$-dimensional stratum
	\begin{align}
		Z_{\sigma^{(n)}} = Z \cap T_{\sigma^{(n)}}
	\end{align}
so that $Z$ admits the following stratification:
	\begin{align}
	\label{hypersurfacestrata}
		Z = \coprod_n \coprod_{\sigma^{(n)} \in \Sigma(n) } Z_{\sigma^{(n)}}. 
	\end{align}
In the following subsections, we review how to determine the geometry of the strata $Z_{\sigma^{(n)}}$ of a toric hypersurface $Z \subset \mathbb P_\Sigma$ appearing in (\ref{hypersurfacestrata}). To this end, we will find it convenient to first review some useful facts about Newton polytopes and normal fans. 

We assume that the data specifying the topology of $Z $ comprises the fan $\Sigma$ and a choice of line bundle $\mathcal O_{\mathbb P_\Sigma} (Z) $ such that $Z$ is the zero locus of a section of $ \mathcal O_{\mathbb P_\Sigma} (Z) $. It follows that the class\footnote{Whenever the distinction is clear from the context, we abuse notation and use the same symbol $D$ to denote both a divisor and its representative in the appropriate Chow ring. In this case, we use $Z$ to denote the class of the divisor $Z \subset \mathbb P_{\Sigma}$.} 
	\begin{align}
		c_1(\mathcal O_{\mathbb P_\Sigma}(Z)) =Z
	\end{align}
of the divisor can be expanded in a basis of classes of toric divisors $D_{\vec v} \subset \mathbb P_\Sigma$:
		\begin{align}
			Z = \mathcal \sum_{\vec v} a_{\vec v} D_{\vec v}. 
		\end{align}
	The group of holomorphic sections of $ \mathcal O_{\mathbb P_\Sigma} (Z) $ can be encoded in the data of a polytope called the \emph{Newton polytope}:
		\begin{align}
			\Delta = \{ \vec m \in M \, | \,  \vec m \cdot \vec v \geq - a_{\vec v} \,, \forall \, \vec v \in\Sigma \}. 
		\end{align}
	It is convenient to use the monomials
		\begin{align}
			p(\vec m) &= \prod_{\vec v} x_{\vec v}^{\vec m \cdot \vec  v + a_{\vec v}}
		\end{align}
	as a basis for this group. One simple and useful application of the Newton polytope is the computation of the dimension of the zeroth cohomology group with coefficients in $\mathcal O_{\mathbb P_\Sigma}(Z)$, i.e.
		\begin{align}
			h^0(\mathbb P_\Sigma, \mathcal O_{\mathbb P_\Sigma}(Z)) = | \Delta \cap M|. 
		\end{align}
The Newton polytope $\Delta$ also determines a blowdown of the toric variety $\mathbb P_\Sigma$ in a manner that we now describe in more detail. To begin, first suppose that $\Delta \subset M$ is any polytope. Then the normal fan $\Sigma(\Delta)$ of $\Delta \subset M$ defines a toric variety $\mathbb P_{\Sigma(\Delta)}$ along with a divisor $D_\Delta$. The normal fan $\Sigma(\Delta)$ is the set of cones
	\begin{align}
	\label{normalfan}
		\Sigma(\Delta) = \{ \sigma(\Theta) \},
	\end{align}
where in the above expression $\Theta$ is a face of $\Delta$ and the cone $\sigma(\Theta)$ associated to $\Theta$ is defined as follows: Given a $k$-dimensional face of $\Delta$, $\Theta^{(k)}$, the cone $\sigma(\Theta^{(k)})$ is the dual\footnote{Recall that the dual $\sigma^\vee$ of a cone $\sigma$ is defined in (\ref{dualcone}).} of the cone	
	\begin{align}
	\label{normalcone}
			\sigma^\vee(\Theta^{(k)}) = \cup_{r \geq 0} \{ r (\vec p_\Delta - \vec p_{\Theta^{(k)}} ) \}\,,
		\end{align}
	where $p_\Delta\in\Delta$ and $p_{\Theta^{(k)}} \in\Theta^{(k)}$. In this construction, the $k$-dimensional faces $\Theta^{(k)}$ of $\Delta$ are associated with $(d-k)$-dimensional cones $\sigma(\Theta^{(k)})$---in particular, the vertices $\vec m = \Theta^{(0)}$ of $\Delta$ are associated with cones $\sigma(\vec m) = \sigma(\Theta^{(0)}) $ of maximal dimension. The polytope $\Delta$ also determines a line bundle over (or equivalently, a Cartier divisor $D_\Delta$ in) $\mathbb P_{\Sigma(\Delta)}$, via a strongly convex support function $\Psi_\Delta$, which we now describe: The restriction of $\Psi_{\Delta}$ to each cone of maximal dimension in $\Sigma(\Delta)$ can be described by means of the dual vertex $\vec m $ by setting 
		\begin{align}
			\left. \Psi_{\Delta} \right|_{\sigma(\vec m)}(\vec v) \equiv \vec m\cdot \vec v,~~~~\forall \, \vec v \in \sigma(\vec m). 
		\end{align}
	The above assignment can be straightforwardly extended to define the restriction of $\Psi_\Delta$ to all cones of lower dimension. Then, the class of the divisor $D_\Delta$ is given by
		\begin{align}
		\label{Newtonpolytopedivisor}
			D_\Delta = \sum_{\vec v \in \Sigma(\Delta)(1)} a_{\vec v} D_{\vec v},~~~~ a_{\vec v} = - \left. \Psi_{\Delta}\right|_{\sigma(\vec m)}(\vec v),~~~~ \forall \, \vec v \in \sigma(\vec m).
		\end{align}
Following the above analysis, one can associate a cone-wise support function $\Psi_\Delta$ to any divisor $D_\Delta$, where $\Psi_\Delta$ is strongly convex if the line bundle $\mathcal O_{\mathbb P_\Sigma}(D_\Delta)$ is ample. Importantly, a toric variety is projective if and only if its fan is the normal fan of some lattice polytope.

\subsection{Resolution of singularities}
\label{sec:resolution}

As described in the previous subsection, we find it useful to construct certain toric hypersurfaces $Z$ from the data of a lattice polytope $\Delta$. However, before describing how to determine the topology of the strata of such a hypersurface $Z_\Delta$, there is an additional issue that we must address, namely that the projective variety $\mathbb P_{\Sigma(\Delta)}$ associated to the normal fan $\Sigma(\Delta)$ is not in general smooth. Thus, before proceeding further, we need to understand how to resolve the singularities of $\mathbb P_{\Sigma(\Delta)}$, as these singularities restrict to singularities of $Z_\Delta$. 

One way in which a fan $\Sigma$ can correspond to a singular projective variety $\mathbb P_\Sigma$ is if $\Sigma$ is not simplicial. (A fan $\Sigma$ is simplicial provided that each of its $d$-dimensional cones is generated by $d$ rays.) However, although $\Sigma$ being simplicial is a necessary condition for $\mathbb P_\Sigma$ to be smooth, this condition not sufficient to guarantee smoothness, as $\mathbb P_\Sigma$ may still contain orbifold (ie. $\mathbb Q$-factorial Gorenstein) singularities corresponding to simplicial cones with volume greater the one \cite{reid1983decomposition}.\footnote{A more complete discussion of possible types of singularities of toric varieties can be found in, e.g. \cite{reid1983decomposition,batyrev1993dual,cox2011toric}. Note that non-simplicial fans correspond to toric varieties with more general classes of singularities than orbifold singularities.} There are many examples of toric varieties $\mathbb P_{\Sigma}$ with such singularities, and in many instances it is desirable to resolve these singularities by means of a choice of refinement $\Sigma \rightarrow \Sigma (\Delta)$.\footnote{A refinement $\Sigma'$ of a fan $\Sigma$ is a fan such that every cone of $\Sigma'$ is contained in a cone of $\Sigma$, and such that the union of cones of $\Sigma'$ is the same as that of $\Sigma$.} 

\begin{figure}
\begin{center}
$
\begin{tikzpicture}
\node[] (A) at (-3.5,0) {$
	\begin{tikzpicture}[scale=1.4]
		\node[label={135:$\vec v_1$}] (v1) at (-1,1) {};
		\node[label={225:$\vec v_2$}] (v1) at (-1,-1) {};
		\node[label={right:$\vec v_3$}] (v3) at (1,0) {};
		\draw[big arrow] (0,0) --  (1,0){};
		\draw[big arrow] (0,0) -- (-1,1){};
		\draw[big arrow] (0,0) -- (-1,-1){};
	\end{tikzpicture}$};
\node[] (B) at (3.5,0) {$
	\begin{tikzpicture}[scale=1.4]
		\node[label={135:$\vec v_1$}] (v1) at (-1,1) {};
		\node[label={225:$\vec v_2$}] (v1) at (-1,-1) {};
		\node[label={right:$\vec v_3$}] (v3) at (1,0) {};
		\node[label={left:$\vec v_E$}] (vE) at (-1,0) {};
		\draw[big arrow] (0,0) --  (1,0){};
		\draw[big arrow] (0,0) -- (-1,1){};
		\draw[big arrow] (0,0) -- (-1,-1){};
		\draw[big arrow] (0,0) -- (-1,0){};
	\end{tikzpicture}$};
\end{tikzpicture}
$
\end{center}
	\caption{Left: toric fan for $\mathbb P_{[1,1,2]}$. The vectors are $\vec v_1 = (-1,1), \vec v_2 = (-1,-1), \vec v_3 = (1,0)$. Right: toric fan for the Hirzebruch surface $\mathbb F_2$, which is the blowup of the singular point of $\mathbb P_{[1,1,2]}$. The additional vector corresponding to the exception divisor is $\vec v_E = (-1,0)$. The toric fan for $\mathbb F_2$ is a refinement of the toric fan for $\mathbb P_{[1,1,2]}$.}
	\label{P112}
\end{figure}
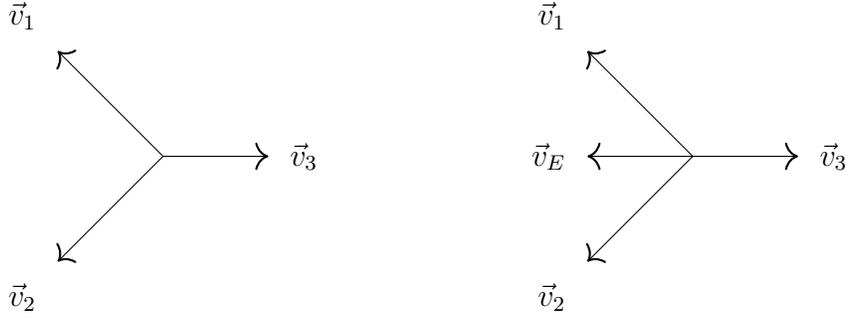

To see how this works in more detail, we again study a very simple example, namely the case $\mathbb P_{\Sigma} =\Bbb{P}_{[1,1,2]}$. The toric fan of $\Bbb{P}_{[1,1,2]}$ can be seen in the left-hand diagram in Figure \ref{P112}. 
 As one can see, the cone spanned by $\vec v_1:=(-1,1)$ and $\vec v_2:=(-1,-1)$ has volume equal to 2, indicating an orbifold $\Bbb{Z}_2$ singularity at $x_1=x_2=0.$ The stratification of the singular $\Bbb{P}_{[1,1,2]}$ is given by
\begin{equation}
\Bbb{P}_{[1,1,2]}= (\Bbb{C}^*)^2\coprod_{i=1}^3 T_{\sigma_{\vec v_i}}\coprod_{1\leq i<j\leq3} T_{\sigma_{\vec v_i\cup \vec v_j}}\,.
\end{equation}
To resolve the singularity, we can blow up $x_1=x_2=0$, which introduces an exceptional divisor $E$ corresponding to the toric ray generated by $\vec v_{E} = (-1,0).$ Introduction of this toric ray now forbids a non-trivial intersection between $x_1$ and $x_2$ in the toric variety, and thus removes the $\Bbb{Z}_2$ singularity. As the result, we obtain the new stratification
\begin{equation}
\hat{\Bbb{P}}_{[1,1,2]}= (\Bbb{C}^*)^2\coprod_{i=1}^3 T_{\sigma_{\vec v_i}}\coprod T_{\sigma_{\vec v_1\cup \vec v_2}} \times  E_{\sigma_{\vec v_1\cup \vec v_2}}\coprod T_{\sigma_{\vec v_1\cup \vec v_3}}\coprod T_{\sigma_{\vec v_2\cup \vec v_3}}\,,
\end{equation}
where $E_{\sigma_{\vec v_1\cup \vec v_2}}=\Bbb{C}^*\coprod (\Bbb{C}^*)^0\coprod (\Bbb{C}^*)^0.$ Note that the $\Bbb{C}^*$ is due to the toric ray $\sigma_{\vec v_E}$ and each $(\Bbb{C}^*)^0$ factor in $E_{\sigma_{\vec v_1\cup \vec v_2}}$ is given by the cone $\sigma_{\vec v_i\cup \vec v_E}$ for $i=1,2.$ The refinement $\Sigma'$ is the new fan defined by the addition of the ray $\sigma_{\vec v_E}$---see the right-hand diagram in Figure \ref{P112}.

Unlike in two-dimensional toric varieties like the example of $\mathbb P_{[1,1,2]}$ described above, in higher-dimensional toric varieties, just adding toric rays is not enough to fully specify a resolution because it is also necessary to specify which divisors can have non-trivial intersections. This data is provided by a refinement of the normal fan $\Sigma,$ or equivalently a triangulation $\mathcal T$ of $\Delta^\circ$ if $\Delta$ is reflexive, where $\Delta^\circ$ is the polar dual of $\Delta.$  In this paper, we choose a particular class of triangulations called fine-regular-star-triangulations (FRST) $\mathcal{T}$ of $\Delta^\circ$. A triangulation $\mathcal{T}$ is called \emph{fine} if all the integral points in $\Delta^\circ$ are used in the triangulation. This means that the resolution involves a sequence of blowups that introduces the maximal number of exceptional prime toric divisors. We call $\mathcal{T}$ \emph{regular} if it can be obtained as a projection of the lower dimensional faces of the convex hull of a point set in one higher dimension. Regularity ensures that the corresponding toric variety is projective. A triangulation $\mathcal{T}$ is called \emph{star}, if all of its maximal dimensional simplices have the origin as the vertex of the cones. The star property guarantees that each simplex of the triangulation defines a refined toric cone.  

Given an FRST $\mathcal{T},$ we can construct the associated refined toric fan $\Sigma(\mathcal{T})$, which is simply defined to be the toric fan of the triangulation $\mathcal T$. From the toric fan $\Sigma(\mathcal T)$, one can then define the toric variety $\Bbb{P}_{\Sigma(\mathcal{T})}$ whose stratification is 
\begin{equation}
\Bbb{P}_{\Sigma(\mathcal{T})}=T_{\sigma^{(0)}}\coprod T_{\sigma^{(1)}}\coprod_{n\geq2} T_{\sigma^{(n)}} \times E_{\sigma^{(n)}}\,.
\end{equation} 
When the FRST $\mathcal T$ defines a refinement $\Sigma(\mathcal T) \rightarrow \Sigma(\Delta)$, we denote the set of vertices $\vec v$ whose non-negative span is the FRST $\Sigma(\mathcal T)$ by $\mathcal{V}_\Delta$. Each ray $\vec{v}\in\Sigma(\mathcal{T})$ corresponds to a homogeneous coordinate $x_{\vec{v}}$, which together generate the Cox ring of the toric variety $\Bbb{P}_{\Sigma(\mathcal{T})}$. We call the vanishing locus of a homogeneous coordinate $x_{\vec{v}}$ a prime toric divisor $\hat{D}_{\vec{v}}$.

\subsection{Strata of resolved toric hypersurfaces}

Let us now return to our discussion of the stratification of a resolution of the possibly singular hypersurface $Z_\Delta$. Assume that the corresponding divisor $Z_\Delta$ transversely intersects the toric strata of $\mathbb P_{\Sigma}$. We regard the singular hypersurface $Z_\Delta$ as a divisor of the blowdown $\mathbb P_{\Sigma(\Delta)}$. Since there is a one-to-one correspondence between faces of $\Delta$ and cones of $\Sigma(\Delta)$, one may write the stratification of such a divisor $Z_\Delta$ in terms of faces of $\Delta$ rather than cones of $\Sigma(\Delta)$: 
	\begin{align}
		Z_\Delta = \coprod_k \coprod_{\Theta^{(k)}} Z_{\Theta^{(k)}}.
	\end{align}
Since faces of dimension $k$ correspond to cones of dimension $d-k$, the dimension of the strata $Z_{\Theta^{(k)}}$ is $d-(d-k)-1= k-1$. 

In what follows, we assume that there exists a refinement of the fan $\Sigma(\Delta)$ corresponding to an FRST $\mathcal T$ of the polytope $\Delta$,
	\begin{align}
		\pi: \Sigma(\mathcal T) \rightarrow \Sigma(\Delta),	
	\end{align} 
which resolves the singular hypersurface $Z_\Delta$:
	\begin{align}
		Z \rightarrow Z_\Delta.	
	\end{align}
The face stratification of the resolution $Z$ is 
	\begin{align}
		Z = Z_\Delta \coprod_{\Theta^{(d-1)}} Z_{\Theta^{(d-1)}} \coprod_{k \geq 2} \coprod_{\Theta^{(d-k)}} E_{\Theta^{(d-k)}} \times Z_{\Theta^{(d-k)}}.
	\end{align}
where in the above expression $Z_\Delta$ is the blowdown of the resolution $Z$ and $E_{\Theta^{(d-k)}}$ is the exceptional set of the refinement of the cone in $\Sigma(\Delta)$ associated with the face $\Theta^{(d-k)}$, i.e.
	\begin{align}
		E_{\Theta^{(d-k)}} &= \coprod_{i=0}^{d-k-1} (\mathbb C^*)^{i}.
	\end{align}
Note that to every $l$-dimensional cone in $\pi^{-1}(\sigma(\Theta^{(d-k)}))$ 
%(where $\sigma \in \Sigma(\Delta)$ is associated with the face $\Theta^{(d-k)}$)
, there corresponds a stratum $(\mathbb C^*)^{k-l}$ in $E_{\Theta^{(d-k)}}$. 

\subsection{Hodge-Deligne numbers}
\label{sec:HD numbers}
In the previous sections, we studied how to decompose toric varieties and their hypersurfaces into a collection of algebraic tori and hypersurfaces therein. The fact that a stratum of a hypersurface in a toric variety is a hypersurface of an algebraic torus makes it desirable to have a method to compute the Hodge numbers of the toric hypersurface by combining certain ``building blocks'' associated to each stratum, where each building block is a sort of character. However, we are not only interested in the dimensions of cohomology groups of a toric hypersurface; we would also like to be able to determine their Hodge structures. For this reason, we shall make a careful choice when deciding what character best suits our purposes.

In order to make a suitable choice of character, there are a few problems to consider. First, we observe that by definition, each stratum is non-compact. Hence, we need a cohomology theory that is well-behaved for non-compact varieties. Furthermore, we wish to recover the Hodge structure of the toric hypersurface from some structure associated to the hypersurface strata that is analogous to Hodge structure. This implies that the character we desire must have a structure that parallels Hodge structure, and also that the gluing of strata should preserve this data. 

Although a given solution satisfying this set of constraints may not be unique, we just need one solution. An elegant solution was provided by Danilov and Khovanski in their pioneering paper \cite{Danilov_1987}. The starting point of \cite{Danilov_1987} is the cohomology with compact support $H^k_c(X)$ associated to an algebraic variety $X.$ This choice is very natural for several reasons. First, cohomology with compact support has a natural Hodge structure, albeit mixed. Second, a proper morphism between algebraic varieties respects the Hodge structures of the cohomology with compact support. Third, the mixed Hodge structure satisfies $h^{p,q}(H_c^k(X))=0$ for $p+q>k.$ Fourth, for a closed subvariety $Y\subset X$ the short exact sequence
\begin{equation}\label{eqn:short 1}
0\rightarrow X\backslash Y\rightarrow X\rightarrow Y\rightarrow 0
\end{equation}
respects the Hodge structure, in the sense that the associated long exact sequence
\begin{equation}\label{eqn:short 2}
\cdots\rightarrow  H_c^k(X\backslash Y)\rightarrow H_c^k(X)\rightarrow H_c^k(Y)\rightarrow H_c^{k+1}(X\backslash Y)\rightarrow \cdots
\end{equation}
is a sequence of the Hodge structures. Finally, and perhaps most importantly, for a compact quasi-smooth algebraic variety, the Hodge structure of the cohomology with compact support is the same as of the usual cohomology theories. As a result, it seems quite plausible that the characters associated with the cohomology with compact support provide us with the building blocks we seek. 

To construct the character in question from the Hodge structure of the cohomology with compact support, we stress a few important ideas. As described in \S\ref{sec:stratification}, a toric variety can be understood as a compactification of an algebraic torus. This implies that one can sequentially apply the exact sequences \eqref{eqn:short 1} and \eqref{eqn:short 2} to study the Hodge structure of the cohomology with compact support of a toric variety and its hypersurfaces. Because of \eqref{eqn:short 2}, we remark that the character $e$
\begin{equation}
\label{echar}
e(X):= \sum_k (-1)^k \dim \left(H_c^k(X)\right)\,, 
\end{equation}
 of the cohomology with compact support satisfies a simple addition rule
\begin{equation}
e(X)= e(X\backslash Y)+e(Y)\,,
\end{equation}
which will allow us to compute the cohomology of the total space from the cohomology of the building blocks. In fact, we can take this a step further. Because \eqref{eqn:short 2} is a sequence of the Hodge structures, it is possible to define a more refined character
\begin{equation}\label{eqn:def HD}
e^{p,q}(X):=\sum_k (-1)^k h^{p,q} \left(H_c^k(X)\right)\,,
\end{equation}
which again satisfies a simple addition rule
\begin{equation}\label{eqn:addition HD}
e^{p,q}(X)= e^{p,q}(X\backslash Y)+e^{p,q}(Y)\,.
\end{equation}
We call this character $e^{p,q}$ the Hodge-Deligne number, and the Hodge-Deligne number will be heavily used to understand the Hodge structures in this paper.\footnote{For recent applications of the Hodge-Deligne numbers, see \cite{Braun:2014xka,Braun:2017nhi,Kim:2021lwo}.} It is important to note that according to the definition \eqref{eqn:def HD}, if $X$ is compact and quasi-smooth, we have an identity
\begin{equation}
e^{p,q}(X)=(-1)^{p+q} h^{p,q}(X)\,.
\end{equation} 

For later convenience, we also introduce the characteristic polynomial 
\begin{equation}
\label{HDcharpoly}
e(X;x,\bar{x}) = \sum_{p,q} e^{p,q}(X) x^p \overline{x}^q\,.
\end{equation}
This characteristic polynomial has many nice properties. The character $e(X)$ is obviously related to $e(X;x,\bar{x})$ by
\begin{equation}
e(X)=e(X;1,1)\,.
\end{equation}
As was studied in \cite{Danilov_1987}, one can use \eqref{eqn:addition HD} to prove the following equalities
	\begin{align}
		e(X_1 \amalg X_2;x,\bar{x}) &= e(X_1;x,\bar{x}) +e(X_2;x,\bar{x})\,,\\
		e(X_1 \times X_2;x,\bar{x}) &=e(X_1;x,\bar{x}) \times e(X_2;x,\bar{x}). \label{eqn:HD identities}
	\end{align}
In fact, the equality \eqref{eqn:HD identities} can be generalized for fiber products. If $f:X\rightarrow Y$ is a fiber bundle with fiber $F,$ which is locally trivial in Zariski topology, then one can apply the equality \eqref{eqn:HD identities} fiberwise to obtain
\begin{equation}
e (X;x,\bar{x})=e(F;x,\bar{x})\times e(Y;x,\bar{x})\,.
\end{equation}

%The Hodge numbers of toric hypersurfaces and their strata can be computed in a systematic fashion using the Hodge-Deligne numbers
%	\begin{align}
%		e^{p,q}(X) = \sum_k (-1)^k h^{p,q}(H^k(X,\mathbb C)).
%	\end{align}
%These numbers agree (up to sign) with the usual Hodge numbers when the underlying topological space $X$ exhibits a pure Hodge structure. Furthermore, they behave in an analogous fashion to the Euler characteristic with regards to disjoint unions and Cartesian products of pairs of topological spaces:
%	\begin{align}
%	\begin{split}
%		e^{p,q}(X_1 \amalg X_2) &= e^{p,q}(X_1) + e^{p,q}(X_2)\\
%		e^{p,q}(X_1 \times X_2) &= \sum_{\substack{p_1 + p_2 = p \\ q_1 + q_2 = q}} e^{p_1,q_1}(X_1) \times e^{p_2,q_2}(X_2). 
%	\end{split}
%	\end{align}
%The above properties imply that the Hodge-Deligne numbers of the strata of a toric hypersurface are sufficient to compute the Hodge numbers of the full hypersurface. We now describe an algorithm for computing the Hodge-Deligne numbers, following the results of  \cite{Danilov_1987}.

As a warm up, we first explain how to compute the Hodge-Deligne numbers of a toric variety. Let us begin with the very simplest example, namely a point. Because a point is compact and zero-dimensional, we easily conclude
\begin{equation}
e(pt;x,\bar{x})=1\,.
\end{equation}
Next, consider a two-sphere $\Bbb{P}^1.$ Because $\Bbb{P}^1$ is compact and smooth, we have $h^{p,q}(\Bbb{P}^1)=(-1)^{p+q}e^{p,q}(\Bbb{P}^1).$ As a result, we conclude
\begin{equation}
e(\Bbb{P}^1;x,\bar{x})=1+x\bar{x}\,.
\end{equation}
Now let us move on to non-compact spaces with the goal of understanding algebraic tori. $\Bbb{P}^1$ can be obtained as a one-point compactification of $\Bbb{C}.$ This means that the following equality holds
\begin{equation}
e(\Bbb{C};x,\bar{x})=e(\Bbb{P}^1;x,\bar{x})-e(pt;x,\bar{x})=x\bar{x}\,.
\end{equation}
Similarly, $\Bbb{C}^*$ can be thought of as a punctured $\Bbb{C}.$ Following the same logic as above, we obtain
\begin{equation}\label{eqn:torus e}
e(\Bbb{C}^*;x,\bar{x})=e(\Bbb{C};x,\bar{x})-e(pt;x,\bar{x})=x\bar{x}-1\,.
\end{equation}
The result \eqref{eqn:torus e} can be used to compute the Hodge-Deligne numbers of an arbitrary algebraic torus: by using the identity
\begin{equation}
e((\Bbb{C}^*)^n;x,\bar{x})= \left(e (\Bbb{C}^*;x,\bar{x})\right)^n\,,
\end{equation}
we obtain
\begin{align}
		e^{p,q}((\mathbb C^*)^{n}) &= \delta_{p,q}(-1)^{n+p} \begin{pmatrix} n \\ p \end{pmatrix}. 
\end{align}
Note that the above equation implies that for a quasi-smooth toric variety, $h^{p,0} = 0$ for $p > 0$. Combining the above formula with the stratification of a toric variety encoded in its fan, we can reproduce a well-known formula \cite{Fulton} for the non-trivial Hodge numbers of a smooth toric $n$-fold:
	\begin{align}
		h^{p,p}(\mathbb P_\Sigma) &= \sum_{k=0}^n (-1)^{p+k} \begin{pmatrix} k \\ p \end{pmatrix} | \Sigma(n-k)|.
	\end{align}
	
\subsubsection{Hodge-Deligne numbers of a toric hypersurface}
	
We now compute the Hodge-Deligne numbers of a hypersurface in a toric variety. Most of the time, we follow the discussion of \cite{Danilov_1987}. However, we do not require the Newton polytope to be integral; rather, we only require that the hypersurface is free of a non-trivial base locus. We wish to make this generalization because in general the Newton polytope for the orientifold need not be integral. Furthermore, we do not construct the ambient toric variety via the normal fan construction, although we still assume that the corresponding line bundle is big without loss of generality.\footnote{Because orientifolding generically forces the moduli of a Calabi-Yau toric hypersurface to be tuned, one may expect that there can be a non-trivial base locus. This is a possibility that we cannot exclude. In the presence of such loci, the analysis is subtler, and it is outside the scope of this paper. For the purposes of this paper, we assume that there is no base locus. We wish to revisit the issue of a non-trivial base locus in future work.}

Our starting point is the compactification of the $d$-dimensional prime stratum given in $(\Bbb{C}^*)^n$ \eqref{eqn:compactification}, namely
\begin{equation}
\Bbb{P}_{\Sigma}=\coprod_n \coprod_{\sigma^{(n)} \in \Sigma(n) } T_{\sigma^{(n)}}\,.
\end{equation}
We assume that every cone $\sigma^{(n)} \in \Sigma$ is simplicial, and hence that the toric variety $\Bbb{P}_{\Sigma}$ is quasi-smooth. Although we can try to resolve the residual orbifold singularities by refining the toric fan $\Sigma,$ we shall not do so. The reason for not doing this is quite simple. For the correct identification of the Hodge numbers of strata, it suffices to study quasi-smooth varieties. Furthermore, we wish to study a hypersurface 
	\begin{align}
		Z\subset \Bbb{P}_{\Sigma}\,,
	\end{align}
such that in every stratum $T_{\sigma^{(n)}}$ the hypersurface stratum
\begin{equation}
Z_{\sigma^{(n)}}=Z\cap T_{\sigma^{(n)}}\,,
\end{equation} 
is a generic codimension-one hypersurface in $T_{\sigma^{(n)}}$. It should be noted that if there are non-trivial exceptional divisors, in general the hypersurface restricted to a stratum $T_{\sigma^{(n)}}\times E_{\sigma^{(n)}}$ is not the most generic hypersurface, but rather $Z_{\sigma^{(n)}}\times E_{\sigma^{(n)}}$. Because this fact unnecessarily complicates the analysis, we avoid further resolving the singularities.
	
Before we start computing the Hodge-Deligne numbers of hypersurfaces of algebraic tori, we need to introduce some more machinery. In this section, we denote the complement of the prime stratum in $\Bbb{P}_{\Sigma}$ by $D$, i.e. 
\begin{equation}
D:=\Bbb{P}_{\Sigma}\backslash T_{\sigma^{(0)}}\,.
\end{equation}
Note that $D$ is a normal crossing divisor in $\Bbb{P}_{\Sigma}.$ We then define $\Omega_{(\Bbb{P}_\Sigma,D)}^p$ to be the kernel of the restriction
\begin{equation}
\Omega^p_{(\Bbb{P}_\Sigma,D)}:=\ker \left(\Omega^p_{\Bbb{P}_\Sigma}\rightarrow \oplus_{D_i} \Omega_{D_i}^p \right)\,,
\end{equation}
where $D_i$ is an irreducible component of $D,$ and $\Omega_{V}$ is the sheaf of differential forms on the toric variety $V.$\footnote{For a careful treatment of differential forms on toric varieties, see Section 4 of \cite{danilov1978geometry}.} The sheaf $\Omega_{(\Bbb{P}_\Sigma,D)}^p$ can also be thought of as a sheaf of meromorphic $p$-forms with logarithmic poles along $D.$ The sheaf $\Omega_{(\Bbb{P}_\Sigma,D)}$ is particularly useful for us due to the following property: Let $X$ be an (affine) algebraic variety, and let $\overline{X}$ be its compactification such that $D=\overline{X}\backslash X$ is a normal crossing divisor in $\overline{X}.$ Then, we have
\begin{equation}
\label{primestratumechar}
e^p(X):=\sum_q e^{p,q}(X)=(-1)^p \chi(\overline{X},\Omega_{(\overline{X},D)}^p)\,.
\end{equation}
So, to summarize, the sheaf of differential forms with logarithmic poles $\Omega_{(\overline{X},D)}$ can be used to compute the character $e$ of the prime stratum of $\overline{X}$! We note that there is an isomorphism between sheaves
\begin{equation}\label{eqn:iso between sheaves1}
\Lambda^p(M)\otimes \mathcal{O}(-D)\equiv \Omega_{(\Bbb{P}_{\Sigma},D)}^p\,,
\end{equation}
 which is induced by the explicit form of a local section 
\begin{equation}
f \frac{d x^{\vec m_1}}{x^{\vec m_1}}\wedge \dots\wedge \frac{d x^{\vec m_p}}{x^{\vec m_p}}\,,
\end{equation}
where $f$ is a local section of $\mathcal{O}(-D)$, and $\wedge^p_i dx^{\vec m_i} /x^{\vec m_i}$ is an element of the space $\Lambda^p(M)$ of $p$-forms on the algebraic torus corresponding to the spectrum of the ring $\mathbb C[M]$.\footnote{We regard the commutative ring of functions on the algebraic torus $(\mathbb C^*)^{d} = \text{Spec}\,\mathbb C[M]$ as being spanned by linear combinations of monomials $x^{\vec m} := \prod_{k} x_k^{\vec m \cdot \vec e_k}$, where $\vec e_k$ are a basis of lattice vectors for $M$, and $x_k$ is a coordinate on the $k$th factor $\mathbb C^*$ of the algebraic torus. Similarly, the monomials $x^{\vec m}$ provide a convenient basis for $p$-forms on the algebraic torus.} 

We are now ready to introduce the basic building blocks of the computation. Let $\vec v_i \in \Sigma(1)$ be the generators of toric rays, and let $D_i \subset \mathbb P_{\Sigma}$ be the corresponding divisors. We define a (non-negative) line bundle $\mathcal{L}$ whose first Chern class $L = c_1(\mathcal L)$ is given by
\begin{equation}
L=\sum_i a_i D_i\,,
\end{equation}
where $a_i$ is a non-negative integer. The corresponding Newton polytope is then given by 
\begin{equation}
\Delta_{L}= \{\vec{m} \in M | \vec{m}\cdot \vec{v_i}\geq -a_i, \forall v_i\}\,.
\end{equation}
Each point $\vec{m}\in\Delta_L$ that is integral with respect to $M$ defines a global section. As was proven in \cite{khovanskii1977newton}, the combinatorial properties of $\Delta_L$ are related to the line bundle cohomology groups
\begin{align}
H^0(\Bbb{P}_\Sigma,\mathcal{L})&=\Gamma_{\Delta_L}(\mathcal L)\,\\
H^i(\Bbb{P}_\Sigma,\mathcal{L})&=0\,.
\end{align}
In the above equations, $\Gamma_{L}(\mathcal L)$ is the space of all globally-defined sections of $\mathcal{L}$ with support in $\Delta_L$.\footnote{Let $s = \sum_{\vec m \in M} s_{\vec m} x^{\vec m}$ (where again we adopt the shorthand notation $x^{\vec m} := \prod_{k} x_k^{\vec m \cdot \vec e_k}$ with $\vec e_k$ being a basis of lattice vectors for $M$ and $x_k$ being a coordinate on the $k$th factor of the algebraic torus $(\mathbb C^*)^d$). Every section $s$ can be uniquely expanded as finite linear combination of the monomials $x^{\vec m}$. The \emph{support} of $s$ is defined to be the set of $\vec m \in M$ such that $s_{\vec m} \ne 0$.} It should be noted that the Newton polytope $\Delta_L$ not only encodes the dimensions of the line bundle cohomology groups, but also their group structure. Because of this, we can similarly study the cohomology of the sheaf 
	\begin{align}
		\Omega_{(\Bbb{P}_\Sigma,D)}^p( \mathcal{L}) := \Omega_{(\Bbb{P}_\Sigma,D)}^p\otimes \mathcal{L}
	\end{align} 
rather straightforwardly. The isomorphism \eqref{eqn:iso between sheaves1} implies that 
\begin{equation}\label{eqn:cohom2}
H^0(\Bbb{P}_\Sigma,\Omega_{(\Bbb{P}_\Sigma,D)}^p(\mathcal{L}))= \Lambda^p(M) \otimes \Gamma_{\Delta^*_L}(\mathcal L)\,,
\end{equation}
and
\begin{equation}\label{eqn:cohom3}
H^i(\Bbb{P}_\Sigma,\Omega_{(\Bbb{P}_\Sigma,D)}^p(\mathcal{L}))= 0\,,
\end{equation}
for $i>0.$ In (\ref{eqn:cohom2}), we note that the Newton polytope $\Delta^*_L$ is defined to be the strict interior of $\Delta_L$. To arrive at \eqref{eqn:cohom2}, we have used the fact that the Newton polytope of a line bundle $\mathcal{L}\otimes \mathcal K_{\Bbb{P}_{\Sigma}}$ is equivalent to the strict interior of the Newton polytope of the line bundle $\mathcal{L}$, namely\footnote{This can be easily seen from the fact that the Newton polytope $\Delta_{L+K_{\Bbb{P}_\Sigma}}$ is defined to be $\Delta_{L+K_{\Bbb{P}_\Sigma}}:=\{\vec{m} \in M | \vec{m}\cdot \vec{v_i}\geq -a_i+1, \forall v_i\}.$} 
\begin{align}
	\Delta_{L + K_{\mathbb P_\Sigma}} \equiv \Delta^*_{L}
\end{align}	
and hence
\begin{equation}
\Gamma_{L+K_{\Bbb{P}_\Sigma}}(\mathcal L)=\Gamma_{\Delta^*_{L}}(\mathcal L)\,.
\end{equation}

We are finally ready to explain an algorithm to compute the Hodge-Deligne numbers of the strata of a toric hypersurface $Z \subset \mathbb P_{\Sigma}$. Because of the assumption that we are considering a generic hypersurface $Z$, \emph{before} the blow up, we can treat the hypersurface $Z$ as a nondegenerate hypersurface. Therefore, the Gysin homomorphism
\begin{equation}\label{eqn:gysin 1}
H^i(Z,\Bbb{C})\rightarrow H^{i+2}(\Bbb{P}_\Sigma,\Bbb{C})
\end{equation}
is an isomorphism for $i>d-1$ and is a surjection for $i=d-1$ \cite{Danilov_1987}. Similarly, we have the Gysin isomorphism
\begin{equation}\label{eqn:gysin 2}
H^i_c(Z_{\sigma^{(0)}},\Bbb{C})\rightarrow H_c^{i+2}(T_{\sigma^{(0)}},\Bbb{C})
\end{equation}
for $i>d-1.$ Note that \eqref{eqn:gysin 2} can be generalized to arbitrary toric cones $\sigma^{(n)}.$ The Gysin isomorphisms \eqref{eqn:gysin 1} and \eqref{eqn:gysin 2} greatly simplify the computation of the Hodge-Deligne numbers. Because $Z$ is  quasi-smooth, Poincar\'e duality holds, hence
\begin{equation}
e^{p,q}(Z)=e^{d-1-p,d-1-q}(Z)\,.
\end{equation}
This basically reduces the computation of the Hodge-Deligne numbers of $Z$ to a computation for the middle-dimensional cohomology groups $p+q=d-1.$ 

To complete the computation of the Hodge structure of $Z,$ we recall two short exact sequences
\begin{equation}\label{eqn:short1 HD}
0\rightarrow \Omega_{({Z},D_Z)}^p\otimes_{\mathcal{O}_{\Bbb{P}_\Sigma}}\mathcal{O}_{\Bbb{P}_\Sigma}(-{Z})\rightarrow \Omega_{(\Bbb{P}_\Sigma,D)}^{p+1}\otimes_{\mathcal{O}_{\Bbb{P}_\Sigma}}\mathcal{O}_{\Bbb{P}_{\Sigma}}({Z})\rightarrow \Omega_{({Z},D_Z)}^{p+1}\rightarrow0\,,
\end{equation}
and
\begin{equation}\label{eqn:short2 HD}
0\rightarrow \mathcal{O}_{\Bbb{P}_\Sigma}(-L)\rightarrow \mathcal{O}_{\Bbb{P}_\Sigma}\rightarrow \mathcal{O}_{{Z}}\rightarrow0\,.
\end{equation}
where in the above equation we have introduced the notation 
\begin{align}
	D_Z := D \cap Z =  Z \backslash Z_{\sigma^{(0)}},~~~~ D = \mathbb P_\Sigma \backslash T_{\sigma^{(0)}}.
\end{align}
By computing the character of the long exact sequence built from \eqref{eqn:short1 HD} tesnored with 
	\begin{align}
		\mathcal L =\mathcal{O}(Z)\,,
	\end{align}
we obtain
\begin{equation}
\chi({Z},\Omega^p_{({Z},D_Z)})=\chi(\Bbb{P}_\Sigma,\Omega_{(\Bbb{P}_\Sigma,D)}^{p+1}(\mathcal{L})\otimes_{\mathcal{O}_{\Bbb{P}_\Sigma}}\mathcal{O}_{{Z}})-\chi({Z},\Omega_{({Z},D_Z)}^{p+1}(\mathcal{L}))\,.
\end{equation}
To eliminate the term $\chi({Z},\Omega_{({Z},D_Z)}^{p+1}(\mathcal{L})),$ we can then compute the character of the long exact sequence built from \eqref{eqn:short1 HD} tensored with $\Omega_{({Z},D_Z)}(\mathcal{L}).$ We can reiterate this procedure, to arrive at
\begin{equation}\label{eqn:char 1}
\chi({Z},\Omega^p_{({Z},D_Z)})=\sum_k (-1)^k\chi(\Bbb{P}_\Sigma,\Omega_{(\Bbb{P}_\Sigma,D)}^{p+k+1}((k+1)\mathcal{L})\otimes_{\mathcal{O}_{\Bbb{P}_\Sigma}}\mathcal{O}_{{Z}})\,.
\end{equation}
The equation \eqref{eqn:char 1} can be further simplified, by using the exact sequence \eqref{eqn:short2 HD}:
\begin{equation}\label{eqn:char 2}
\chi({Z},\Omega^p_{({Z},D_Z)})=\sum_{k\geq0} (-1)^k \left[\chi(\Bbb{P}_\Sigma,\Omega_{(\Bbb{P}_\Sigma,D)}^{p+k+1}(\mathcal{L}^{\otimes k+1}))-\chi(\Bbb{P}_\Sigma,\Omega_{(\Bbb{P}_\Sigma,D)}^{p+k+1}(\mathcal{L}^{\otimes k}))\right]\,.
\end{equation}
Since \eqref{eqn:cohom2} and \eqref{eqn:cohom3} provide us with a thorough understanding of the relevant cohomonology groups, we therefore have enough information to determine the Hodge structure of the prime stratum of ${Z}$ using (\ref{primestratumechar}). 

To compute the Hodge structure of the strata of $Z$, we can then use
\begin{equation}
e({Z};x,\bar{x})=\sum_n\sum_{\sigma^{(n)}} e(Z_{\sigma^{(n)}};x,\bar x)\,.
\end{equation}
We record the Hodge-Deligne numbers of hypersurface strata of dimension up to three. From now on, we slightly change our conventions and denote by $Z_{\Theta^{(k)}}$ a generic hypersurface in $(\Bbb{C}^*)^k$, whose corresponding Newton polytope  specified by the defining equation for the hypersurface $Z_{\Theta^{(k)}}$, is given by $\Theta^{(k)}$, i.e.
	\begin{equation}
		\Delta_{Z_{\Theta^{(k)}}} \leftrightarrow \Theta^{(k)}. 	
	\end{equation} 
For $p > 0$, we have 
	\begin{align}
		e^{p,0}( Z_{\Theta^{(k)}}) = (-1)^{k-1} \sum_{\Theta^{(p+1)} \leq \Theta^{(k)}} l^*(\Theta^{(p+1)})
	\end{align}
where $l^*(\Theta^{(p+1)})$ counts the numbers of lattice points in the strict interior of $\Theta^{(p+1)}$ and the sum is taken over all $(p+1)$-dimensional faces $\Theta^{(p+1)}$ contained in $\Theta^{(k)}$. The remaining Hodge-Deligne numbers satisfy the following identity:	
	\begin{align}
		(-1)^{k-1} \sum_{p} e^{p,q}(Z_{\Theta^{(k)}} ) = (-1)^{p} \begin{pmatrix} k \\ p+1 \end{pmatrix} + \varphi_{k-p}(\Theta^{(k)}).
	\end{align}
The function $\varphi_n(\Theta^{(k)})$ in the above equation is defined as follows:
	\begin{align}
	\label{phifn}
		\varphi_n(\Theta^{(k)}) = \sum_{j=1}^n (-1)^{n+j} \begin{pmatrix} k+1 \\ n-j \end{pmatrix} l^*(j \Theta^{(k)}),
	\end{align}
where $j\Theta$ denotes the polytope the results from scaling the face $\Theta$ by the factor $j \in \mathbb Z_{>0}$. Given a face $\Theta^{(k)}$ of dimension $k \geq 4$, we have the simple formula
	\begin{align}
		e^{k-2,1}(Z_{\Theta^{(k)}}) = (-1)^{k-1} \biggl( \varphi_2(\Theta^{(k)}) - \sum_{\Theta^{(k-1)} \leq \Theta^{(k)}} \varphi_1(\Theta^{(k-1)}) \biggr). 
	\end{align}
Finally, for higher Hodge numbers $p+q \geq n$ of $n$-dimensional strata $Z_{\Theta^{(n)}}$, we have 
	\begin{align}
		e^{p,q}(Z_{\Theta^{(n)}}) = \delta_{p,q}(-1)^{n + p + 1} \begin{pmatrix} n \\ p+1 \end{pmatrix}. 
	\end{align}
	
One can use the above formulae to derive the Hodge-Deligne numbers for strata corresponding to $k$-faces $\Theta^{(k)}$ for arbitrary $k$. We illustrate the general method in some simple, schematic examples. Let us use $l^n(\Theta)$ to denote the number of points contained in the $n$-skeleton of $\Theta$. The Hodge-Deligne numbers of strata $Z_{\Theta^{(k)}}$ for $k \leq 4$:
\begin{equation}
e^{0,0}(Z_{\Theta^{(1)}})=(l^1(\Theta^{(1)})-1)\,,
\end{equation}
\begin{equation}
e^{i,j}(\Theta^{(2)})=
\begin{tabular}{|c|c|}
\hline
$-l^*(\Theta^{(2)})$&1\\
\hline
$1-l^1(\Theta^{(2)})$&$-l^*(\Theta^{(2)})$\\
\hline
\end{tabular}\,,
\end{equation}
\begin{equation}
e^{i,j}(\Theta^{(3)})=\begin{tabular}{|c|c|c|}
\hline
$l^*(\Theta^{(3)})$&0&1\\
\hline
$l^2(\Theta^{(3)})-l^1(\Theta^{(3)})$&$ e^{1,1}(\Theta^{(3)})$&0\\
\hline
$l^1(\Theta^{(3)})-1$& $l^2(\Theta^{(3)})-l^1(\Theta^{(3)})$ & $l^*(\Theta^{(3)})$\\
\hline
\end{tabular}\,,
\end{equation}
where $e^{1,1}(\Theta^{(3)})=\varphi_2(\Theta^{(3)})-l^2(\Theta^{(3)})+l^1(\Theta^{(3)})-3,$ and
\begin{equation}
\begin{tabular}{|c|c|c|c|}
\hline
$-l^*(\Theta^{(4)})$&0&0&1\\
\hline
$-l^3(\Theta^{(4)})+l^2(\Theta^{(4)})$&$ e^{2,1}(\Theta^{(4)})$&-4&0\\
\hline
$-l^2(\Theta^{(4)})+l^1(\Theta^{(4)})$&$ e^{1,1}(\Theta^{(4)})$&$e^{2,1}(\Theta^{(4)})$&0\\
\hline
$1-l^1(\Theta^{(4)}) $& $-l^2(\Theta^{(4)})+l^1(\Theta^{(4)})$&$-l^3(\Theta^{(4)})+l^2(\Theta^{(4)})$&$-l^*(\Theta^{(4)})$\\
\hline
\end{tabular}
\end{equation}
where $e^{2,1}(\Theta^{(4)})=l^3(\Theta^{(4)})-l^2(\Theta^{(4)})-\varphi_2(\Theta^{(4)})$ and $e^{1,1}=-\varphi_3(\Theta^{(4)})+\varphi_2(\Theta^{(4)})-l^3(\Theta^{(4)})+2l^2(\Theta^{(4)})-l^1(\Theta^{(4)})+6.$

\section{Calabi-Yau toric hypersurfaces and reflexive pairs}
\label{sec:toric CYs}

In this section, following the methods of \cite{batyrev1993dual}, we now explain how to construct a $(d-1)$-dimensional Calabi-Yau (CY) toric hypersurface and its stratifications from a pair of $d$-dimensional reflexive lattice polytopes $\Delta,\Delta^\circ$ where $\Delta \subset M$ is the usual Newton polytope and its polar dual $\Delta^{\circ} \subset N$ is defined below. \cite{batyrev1993dual}.

\subsection{Preliminaries}

To begin, we note that CY manifolds are defined to be manifolds with vanishing first Chern class. A hypersurface $Z'$ of a projective variety $\mathbb P_\Sigma$ satisfies the CY condition if and only if $Z'$ is the zero locus of a section of the anti-canonical line bundle $-K_{\mathbb P_\Sigma}$, as one can easily check using the adjunction formula. The divisor class of such a (toric) hypersurface is given by
	\begin{align}
	\label{anticanonicalclass}
		 Z'= -K_{\mathbb P_\Sigma} = c_1(\mathbb P_\Sigma) = \sum_{\vec v \in \Sigma(1)} D_{\vec v}.
 \end{align}
Since the coefficients of the divisors in the expression (\ref{anticanonicalclass}) are all equal to one, it follows that sections of $-K_{\mathbb P_\Sigma}$ can be identified with the Newton polytope
	\begin{align}
		\Delta = \{ \vec m \in M \, | \, \vec m \cdot \vec v \geq 1, \, \forall \vec v \in \Sigma(1) \}. 
	\end{align}
A general section of $-K_{\mathbb P_\Sigma}$ can be expressed as 
	\begin{align}
		\sum_{\vec m} \alpha_{\vec m } p(\vec m) = \sum_{\vec m \in \Delta } \prod_{\vec v \in \Sigma(1)} \alpha_{\vec m} x^{\vec m},~~~~ x^{\vec m} :=x_{\vec v}^{\vec m \cdot \vec v + 1 } 
	\end{align}
where $\alpha_{\vec m}$ are complex constants. 
%	
%The toric fan $\Sigma(\Delta^\circ)$ is constructed as follows. First, to the origin of the reflexive polytope $\Delta^\circ,$ we assign a zero-dimensional cone $\sigma^{(0)}.$ Similarly, to each $n$-dimensional face $\Theta^{\circ(n)},$ we construct an $(n+1)$-dimensional cone $\sigma_{\Theta^{\circ(n)}}$ defined by
%\begin{equation}
%\sigma_{\Theta^{\circ(n)}}:= \{ (r,r \vec{v})\, |\,r\in \Bbb{R}_{\geq0},~ \vec{v}\in\Theta^{\circ(n)}\}\,.
%\end{equation}
%Oftentimes, we use the shorthand notation $\sigma^{(n+1)}$ for $\sigma_{\Theta^{\circ(n)}},$ when the explicit choic of $\Theta^{(n)}$ is not very important. From the collection of the cones over faces, we now construct the toric fan $\Sigma(\Delta^\circ)$
%\begin{equation}
%\Sigma(\Delta^\circ):=\cup_n \,\sigma^{(n)}\,.
%\end{equation}
%This toric fan now defines the corresponding toric variety $\Bbb{P}_{\Sigma(\Delta^\circ)}$ as follows. Each one-dimensional toric ray $\sigma_{\Theta^{\circ(0)}}$ corresponds to the span of a vertex $\vec v \in \Delta^\circ$; moreover, to each vertex $\vec v:=\Theta^{\circ(0)}$, there corresponds a homogeneous coordinate $x_{\vec{v}}$ that generates the Cox ring associated to the toric variety. When none of the homogeneous coordinates vanish, the coordinates $t_{\vec{m}}$ defined by (cf. (\ref{tk})) 
%\begin{equation}
%t_{\vec{m}}:=\prod_{v} x_{\vec{v}}^{\vec{v}\cdot \vec{m}}
%\end{equation}
%parametrize the prime stratum $(\Bbb{C}^*)^d$ of the toric variety. 

The vertices of $\Delta$ may not in general be lattice points in $M$, nor is it guaranteed that a generic section of $-K_{\mathbb P_\Sigma}$ defines a smooth CY hypersurface. In cases when all vertices of $\Delta$ are lattice points in $M$ (and hence $\Delta$ is referred to as a \emph{lattice} polytope), it follows that the vertices $\vec v \in \Sigma(1)$ belong to a lattice polytope $\Delta^\circ$ defined by the property
	\begin{align}
	\label{polardual}
		\Delta \cdot \Delta^\circ \geq -1.
	\end{align}
The lattice polytope $\Delta^\circ$ is in general called the \emph{polar dual} of $\Delta$; when both $\Delta$ and $\Delta^\circ$ are lattice polytopes, the pair $\Delta, \Delta^\circ$ is referred to as a \emph{reflexive pair}.\footnote{A lattice polytope is reflexive only if its unique interior point is the origin.} Given a reflexive pair, one may construct the normal fan $\Sigma(\Delta)$, which is equivalent to the fan over the faces of $\Delta^\circ$; by abuse of notation, we denote the fan over the faces of $\Delta^\circ$ by $\Sigma(\Delta^\circ)$. In the forthcoming discussion, we find it convenient to regard the ambient toric variety (of which $Z'$ is a hypersurface) as corresponding to the fan $\Sigma(\Delta^\circ)$ over the faces of the polar dual $\Delta^\circ$. 

As we have discussed, the fan $\Sigma(\Delta^\circ)$ may not necessarily define a smooth toric variety $\mathbb P_{\Sigma(\Delta^\circ)}$. We assume that there exists an FRST $\mathcal T$ of $\Delta^\circ$ that resolves $\mathbb P_{\Sigma(\Delta^\circ)}$. While an FRST is not sufficient to remove all singularities of a $d$-dimensional toric variety $\mathbb P_{\Sigma(\Delta^\circ)}$, in the case $d=4$ (i.e. CY threefolds), an FRST is sufficient to resolve all singularities of the toric fourfold $\mathbb P_{\Sigma(\Delta^\circ)}$ up to pointlike orbifold singularities, and hence a generic CY threefold hypersurface of $\mathbb P_{\Sigma(\mathcal T)}$. Thus, FRSTs are sufficient for the purpose of construction smooth CY threefold toric hypersurfaces.

%Consider a CY
%
%Although our goal is to study the Calabi-Yau hypersurface in the resolved toric variety $\Bbb{P}_{\Sigma(\mathcal{T})},$ to illustrate a few important points, let us first study the Calabi-Yau hypersurface embedded in the toric variety $\Bbb{P}_{\Sigma(\Delta^\circ)}.$ The anti-canonical class of a toric variety constructed from a toric fan $\Sigma$ is given by
%\begin{equation}
%-K_{\Bbb{P}_\Sigma(\Delta^\circ)}=\sum [D_{\Theta^{(0)}}]\,.
%\end{equation}
%The corresponding Newton polytope $\Delta_{-K_{\Bbb{P}_{\Sigma(\Delta^\circ)}}}$ is defined to be the region bound by the following inequalities
%\begin{equation}
%\Delta_{-K_{\Bbb{P}_{\Sigma(\Delta^\circ)}}}:=\{\vec{m}\in M| \vec{m}\cdot \vec{v}\geq -1,~ \forall \vec{v}\in\mathcal{V}_{\Delta^\circ}  \}\,,
%\end{equation}
%where we define $\mathcal{V}_{\Delta^\circ}$ be the set of the vertices of $\Delta^\circ.$ Each point $\vec{m}\in\Delta$ represents a monomial of the section of the anti-canonical class which we denote by a shorthand notation
%\begin{equation}
%x^{\vec{m}}:=\prod_{\vec{v}\in\mathcal{V}_{\Delta^\circ}} x^{\vec{v}\cdot \vec{m}+1}\,.
%\end{equation}
%The vanishing locus of a section of $\Delta_{-K_{\Bbb{P}_{\Sigma(\Delta^\circ)}}}$ defines the Calabi-Yau hypersurface $\widehat{X}\in \Bbb{P}_{\Sigma(\Delta^\circ)}.$ 

\subsection{Strata of CY hypersurfaces}
To study the stratification of $Z'$, we decompose $Z'$ into hypersurfaces in the strata of the toric variety $\mathbb P_{\Sigma(\Delta^\circ)}$. First, let us study the hypersurface in the prime stratum $T_{\sigma^{(0)}},$ in which no homogeneous coordinate vanishes. Within the prime stratum, no section of the anti-canonical line bundle vanishes, and therefore $\Delta$ defines the Newton polytope for the hypersurface $Z'_{\Delta}$ within the prime stratum $T_{\sigma^{(0)}}.$ As a lower dimensional stratum $T_{\sigma^{(n)}}$ is given as a complete intersection of a set $I$ of $n$ prime toric divisors,
\begin{equation}
T_{\sigma^{(n)}}= \cap_{\vec v_i \in I} D_{\vec v_i}\,,
\end{equation} 
not all monomials in the section of the anti-canonical line bundle will survive the restriction to $T_{\sigma^{(n)}}.$ As the collection of the monomials $\{x^{\vec{m}}\}$ that do not vanish in $T_{\sigma^{(n)}}$ should all satisfy
\begin{equation}
\vec{v}\cdot \vec{m}+1=0
\end{equation}
for all $\vec{v} \in I,$ we find that the corresponding Newton polytope for the hypersurface in $T_{\sigma^{(n)}}$ is therefore defined to be
\begin{equation}
\Delta^{(n-1)}_I:=\{\vec{m}\in M| \vec{m}\cdot \vec{v}\geq -1 ~ \forall \vec{v}\in\mathcal{V}_{\Delta^\circ}\backslash I  \, \text{ and } \, \vec{m}\cdot \vec{v}= -1 ~\forall \vec{v}\in I\}\,.
\end{equation}
Correspondingly, we denote the hypersurface in $T_{\sigma^{(n)}}$ by $Z'_{\Delta^{(n-1)}}$.\footnote{Unless explicitly required for clarity, we omit the subcript $I$.} As a result, we now arrive at the stratification of the Calabi-Yau hypersurface $Z'$:
\begin{equation}
Z'=Z'_{\Delta}\coprod_{\Delta^{(3)}} Z'_{\Delta^{(3)}}\coprod_{\Delta^{(2)}}  Z'_{\Delta^{(2)}}\coprod_{\Delta^{(1)}}  Z'_{\Delta^{(1)}}\coprod_{\Delta^{(0)}}  Z'_{\Delta^{(0)}}\,.
\end{equation}
We note that because $Z'_{\Delta^{(0)}}$ is a hypersurface in a point, unless there is a reason that all the sections in $\Delta^{(0)}$ should vanish (for instane, at a special point in the complex structure moduli space $Z'$), $Z'_{\Delta^{(0)}}$ is the zero set at a generic point in the moduli space. Therefore, we omit $Z'_{\Delta^{(0)}}$. 

We next turn our attention to the study of CY hypersurfaces in the resolved toric variety $\Bbb{P}_{\Sigma(\mathcal{T})}.$ The Newton polytope $\Delta_{-K_{\Bbb{P}_{\Sigma(\mathcal{T})}}}$ for the anti-canonical class is defined by the following inequalities
\begin{equation}
\Delta_{-K_{\Bbb{P}_{\Sigma(\mathcal{T})}}}:=\{\vec{m}\in M| \vec{m}\cdot \vec{v}\geq -1,~ \forall \vec{v}\in\Delta^{\circ}\}\,.
\end{equation} 
Because the $\Delta^\circ$ is a convex hull of vertices $\vec v\in \mathcal{V}_{\Delta^\circ},$ we find that as integral polytopes $\Delta_{-K_{\Bbb{P}_{\Sigma(\mathcal{T})}}}$ is isomorphic to $\Delta_{-K_{\Bbb{P}_{\Sigma(\Delta^\circ)}}}.$ We will therefore, unless needed,denote the Newton polytope for the anti-canonical class by $\Delta.$ This is a roundabout way to say that the resolution we performed $\pi:\Bbb{P}_{\Sigma(\mathcal{T})}\rightarrow \Bbb{P}_{\Sigma(\Delta^\circ)}$ is crepant. Similar to earlier discussions, each point $\vec{m}\in\Delta$ represents a monomial of a section of the anti-canonical class. We denote these momomials by the shorthand notation
\begin{equation}
x^{\vec{m}}: = \prod_{\vec{v}\in \partial\Delta^\circ} x^{\vec{v}\cdot \vec{m}+1}\,.
\end{equation}
The stratification of the CY embedded in the resolved toric variety is now then given as
\begin{equation}
Z=Z_{\Delta}\coprod Z_{\Delta^{(3)}}\coprod Z_{\Delta^{(2)}} \times E_{\sigma^{(2)}}\coprod Z_{\Delta^{(1)}} \times E_{\sigma^{(3)}}\,.
\end{equation}
We note that as was shown in \cite{batyrev1993dual}, choosing an FRST of $\Delta^\circ$ corresponds to a maximal projective crepant partial desingularization (MPCP desingularization). An MPCP desingularization can completely resolve singularities of toric CY $n$-folds for $n\leq 3,$ but need not for $n>3.$ 

\subsection{Hodge numbers of CY hypersurfaces}

We are now ready to compute the Hodge numbers of the CY manifold $Z.$ Although this is a well known result, we present the computation explicitly as a warmup for the computation of the Hodge numbers of an orientifold of a CY threefold. Let us start with the computation of the structure sheaf cohomology 
\begin{equation}
h^\bullet(Z,\mathcal{O}_{Z})=h^{\bullet,0}(Z)\,.
\end{equation}
We first compute
\begin{equation}
h^{3,0}(Z)=-e^{3,0}(Z_\Delta)=l^*(\Delta)\,.
\end{equation}
By the assumption that the Newton polytope is reflexive, we obtain 
\begin{equation}
h^{3,0}(Z)=1\,.
\end{equation}
Next, we compute $h^{2,0}(Z)$:
\begin{align}
\begin{split}
h^{2,0}(Z)=&~e^{2,0}(Z)\,,\\
=&~e^{2,0}(Z_{\Delta})\sum_{\Delta^{(3)}}e^{2,0}(Z_{\Delta^{(3)}})\,,\\
=&~(-l^3(\Delta)+l^2(\Delta))+\sum_{\Delta^{(3)}}(l^3(\Delta^{(3)})-l^2(\Delta^{(3)}))\,,\\
=&~0\,.
\end{split}
\end{align}
Similarly, we compute
\begin{equation}
h^{1,0}(Z)=0\,,
\end{equation}
and
\begin{equation}
h^{0,0}(Z)=1\,.
\end{equation}

To compute $h^{1,1}(Z)$, we use the fact that $h^{1,1}(Z)=h^{2,2}(Z)$, and compute $h^{2,2}(Z)$ using the Hodge-Deligne number of the strata:
\begin{align}
\begin{split}
h^{1,1}(Z)=&~e^{2,2}(Z_{\Delta})+ \sum_{\Delta^{(3)}} e^{2,2}(Z_{\Delta^{(3)}})+\sum_{\Delta^{(2)}}e^{1,1}(Z_{\Delta^{(2)}})e^{1,1}(E_{\sigma^{(2)}})\\
&+\sum_{\Delta^{(3)}}e^{0,0}(Z_{\Delta^{(1)}}) e^{2,2}(E_{\sigma^{(3)}})\,, \\
=&~l(\Delta^\circ)-5 -\sum_{\Theta^{\circ(3)}}l^*(\Theta^{\circ(3)})+\sum_{\Theta^{\circ(2)}} l^*(\Theta^{\circ(2)})l^*(\Theta^{(1)})\,.
\end{split}
\end{align}
The above formula begs further explanation. Naively, one may conclude that $h^{1,1}$ of the CY $Z$ is equal to the number of toric rays modulo the linear relations among the prime toric divisors, namely $l(\Delta^\circ)-5.$ This is not quite correct for two reasons: First, as we explained earlier, $Z_{\Delta^{(0)}}$ is a zero set at a generic point in the moduli space. This means that all the would-be divisors that have the topology
\begin{equation}
Z_{\Delta^{(0)}}\times E_{\sigma^{(4)}}
\end{equation}
should be also considered as zero sets in the $Z$. To account for this fact, we need to subtract the number of prime toric divisors due to points interior to three-dimensional faces of $\Delta^\circ.$ By doing so, we obtain the first correction term
\begin{equation}
-\sum_{\Theta^{\circ(3)}}l^*(\Theta^{\circ(3)})\,.
\end{equation}
The second correction term arises due to the fact that $Z_{\Delta^{(1)}}$ is a vanishing locus of the degree $l^*(\Theta^{(2)})+1$ hypersurface in a one-dimensional algebraic torus. Since at a generic point in the moduli space the zero set of the degree $n$ hypersurface is a collection of $n$ points, we conclude that each of the prime toric divisors due to points interior to two-dimensional faces of $\Theta^{\circ(2)}$ is $l^*(\Theta^{(2)})+1$ copies of an irreducible divisor; this leads to the second correction term
\begin{equation}
\sum_{\Theta^{\circ(2)}}l^*(\Theta^{\circ(2)})l^*(\Theta^{(2)})\,.
\end{equation}

The computation of $h^{2,1}(Z)$ proceeds similarly:
\begin{align}\label{eqn:h21 CY3}
-h^{2,1}(Z)=e^{2,1}(Z_{\Delta})+\sum_{\Delta^{(2)}}e^{1,0}(Z_{\Delta^{(2)}})e^{1,1}(E_{\sigma^{(2)}}).
\end{align}
By applying the identities
\begin{align}
-e^{2,1}(Z_{\Delta})&=l^*(2\Delta)-5 l^*(\Delta)-\sum_{\Delta^{(3)}}l^*(\Delta^{(3)})\\
l^*(2\Delta)&=l(\Delta)\,
\end{align}
to a polytope with only one interior point, we rewrite \eqref{eqn:h21 CY3} as
\begin{equation}\label{eqn:h21 CY3 2}
h^{2,1}(Z)=l(\Delta)-5-\sum_{\Delta^{(3)}}l^*(\Delta^{(3)})+\sum_{\Delta^{(2)}}l^*(\Delta^{(2)})l^*(\Theta^{\circ(1)})\,.
\end{equation}
The equation \eqref{eqn:h21 CY3 2} is intuitive to understand. Due to the fact that
\begin{equation}
h^1(Z,TZ)=h^{2,1}(Z)\,,
\end{equation}
we see that $h^{2,1}(Z)$ counts the number of inequivalent complex structure deformations by monomials to the defining equation of the CY manifold $Z$.  The total number of monomial deformations are counted by $l(\Delta),$ but it should be noted that the number of equivalence relations must also be taken into account. Since there are four toric equivalence relations and furthermore the overall scale of the defining equation is redundant, one must subtract 5 from the number of monomials. This is not the end, however, because one must also take the root automorphism group actions into account \cite{cox1995homogeneous}, which are counted by $\sum_{\Delta^{(3)}}l^*(\Delta^{(3)})$. We are still not done: some monomial deformations can actually correspond to more than one complex structure deformation, a subtle feature that cannot be easily determined from the toric description. This subtlety is counted by the last term in \eqref{eqn:h21 CY3 2}.

\subsection{Topology of prime toric divisors of CY hypersurfaces}\label{sec:divisors in CY3}
\label{sec:CY3hyptopology}
In this section, we study the topological properties of prime toric divisors in CY manifolds. This section is a warmup for the study of the topological properties of prime toric divisors in the CY orientifolds. In this section, we make the exceptional parts of the stratification very explicit in order to make the computation more tractable. Unless explicitly noted otherwise, we denote a vertex of $\Delta^\circ$ by $\vec v.$ Similarly, by $\vec v_e,~\vec v_f$ we denote points interior to an edge $e$ and a 2-face $f$ of $\Delta^\circ$, respectively. By $t_i$ we denote an $i$-dimensional simplex in $\mathcal{T}\cap \partial \Delta^\circ$. For a vector $\vec{v}\in N,$ we define $v^\circ$ to be
\begin{equation}
v^\circ:= \{ \vec{m}\in \Delta| \vec{m}\cdot \vec{v}=-1 \}\,.
\end{equation}
Hence, $v^\circ$ lives in $M$ lattice. Similarly, for a set $S\in\Delta^\circ,$ we define $S^\circ$ to be
\begin{equation}
v^\circ:= \{ \vec{m}\in \Delta| \vec{m}\cdot \vec{v}=-1 \,\forall \vec{v}\in S\}\,.
\end{equation}

\subsubsection{Vertex divisors}
We first study the topology of prime toric divisors due to vertices of $\Delta^\circ.$ The stratification of the prime toric divisor $D_{\vec v}$ is 
\begin{equation}
D_{\vec v}= Z_{\vec v} \coprod_{e\supset  \vec v} Z_{e}\coprod_{f\supset \vec v} Z_{f} \left(\coprod_{f\supset t_1\supset \vec v} \Bbb{C}^*\coprod_ {f\supset t_2 \supset \vec v}(pt)\right)\,.
\end{equation}

To compute $h^{0,0}(D_{\vec v})=h^{2,2}(D_{\vec v})$, we can simply note that the only contribution to $e^{2,2}(D_{\vec v})$ comes from $e^{2,2}(Z_{\vec v})=1.$ As a result, we determine $h^{2,2}(D_{\vec v})=1.$ Similarly, $h^{2,0}(D_{\vec v})=h^{0,2}(D_{\vec v})$ receives contribution only from $e^{2,0}(Z_{\vec v})=l^*(v^\circ)$. Now, we compute $h^{1,1}(D_{\vec v})$:
\begin{align}
\begin{split}
\label{eqn:h11 vertex}
h^{1,1}(D_{\vec v})=&~e^{1,1}(D_{\vec v})\\
=&~ e^{1,1}(Z_{\vec v})+\sum_{e\supset \vec v} e^{1,1}(Z_{e})+\sum_{f\supset \vec v} e^{0,0}(Z_{f^\circ})\sum_{f\supset t_1\supset \vec v} e^{1,1}(\Bbb{C}^*)\\
=&~ l^*(2v^\circ)-4l^*( v^\circ)-\sum_{e\supset \vec v} l^*(e^\circ)-3+\sum_{e\supset \vec v}1+\sum_{f\supset \vec v}\sum_{f\supset t_1\supset \vec v} (l^*(f^\circ)+1)\,.
\end{split}
\end{align}
Finally, we determine $h^{1,0}(D_{\vec v})$:
\begin{align}
\begin{split}
h^{1,0}(D_{\vec v})=&~-e^{1,0}(D_{\vec v})\\
=&~-(e^{1,0}(Z_{\vec v})+\sum_{e\supset \vec v} e^{1,0}(Z_{e}))\\
=&~0\,.
\end{split}
\end{align}
We note that the vertex divisors have the topology of blow ups at points of a generic hypersurface in a toric threefold.

\subsubsection{Edge divisors}
In this section, we study topology of prime toric divisors due to a point $\vec v_e$ interior to an edge $e$ of $\Delta^\circ.$ The stratification of the prime toric divisor $D_{\vec v_e}$ is 
\begin{equation}
D_{\vec v_e}=Z_{e^{\circ}}\times (\Bbb{C}^*\coprod (pt) \coprod (pt) )\coprod_{f\supset e} Z_{f^\circ}\left(\coprod_{f\supset t_1\supset \vec v_{e}}\Bbb{C}^*\coprod_{f\supset t_2\supset \vec v_e} (pt) \right)\,.
\end{equation}

We again start by computing $h^{0,0}(D_{\vec v_e})=h^{2,2}(D_{\vec v_e}).$ The only contribution to $h^{2,2}(D_{\vec v_e})$ comes from $e^{1,1}(Z_{e}) e^{1,1}(\Bbb{C}^*)=1,$ which leads to $h^{2,2}(D_{\vec v_e})=1.$ There is no stratum that yields non-trivial $e^{2,0}(D_{\vec v_e})$, hence we conclude $h^{2,0}(D_{\vec v_e})=0.$ Next, we compute $h^{1,1}(D_{\vec v})$:
\begin{align}
h^{1,1}(D_{\vec v_e})=e^{1,1}(Z_{e})+e^{0,0}(Z_{e})+\sum_{f\supset e}\sum_{f\supset t_1 \supset \vec v_e}e^{0,0}(Z_{f^\circ})\,.
\end{align}
By using the identity
\begin{equation}
e^{0,0}(Z_{e})=1-l^1(e^\circ)=1-\sum_{f\supset e}(1+l^*(f^\circ))\,,
\end{equation}
we thus obtain
\begin{equation}
h^{1,1}(D_{\vec v_e})=2 +\sum_{f\supset e}(-1+\sum_{f\supset t_1\supset \vec v_e} 1)(1+l^*(f^\circ))\,.
\end{equation}
Finally, we compute $h^{1,0}(D_{\vec v_e})$
\begin{align}
\begin{split}
h^{1,0}(D_{\vec v_e})=&~-e^{1,0}(D_{\vec v_e})\\
=&~l^*(e^\circ)\,.
\end{split}
\end{align}
We conclude that the topology of an edge divisor is blow ups at points of a $\Bbb{P}^1$ bundle over a genus $h^{1,0}(D_{\vec v_e})$ curve.

\subsubsection{Face divisors}
Finally, we come to the last type of prime toric divisor, namely face divisors due to a point $\vec v_f$ interior to a face $f$ of $\Delta^\circ.$ The stratification of the face divisors $D_{\vec v_f}$ is 
\begin{equation}
D_{\vec v_f}=Z_{f^\circ}\times \left((\Bbb{C}^*)^2\coprod_{f\supset t_1\supset \vec v_f}\Bbb{C}^*\coprod_{f\supset t_2\supset \vec v_f} (pt)\right)\,.
\end{equation} 
We first compute $h^{2,2}(D_{\vec v_f})$:
\begin{align}
\begin{split}
h^{2,2}(D_{\vec v_f})=& ~e^{0,0}(Z_{f^\circ})e^{2,2}((\Bbb{C}^*)^2)\\
=&~(1+l^*(f^\circ))\,.
\end{split}
\end{align}
Next, we compute $h^{1,1}(D_{\vec v_f})$:
\begin{align}
\begin{split}
h^{1,1}(D_{\vec v_f})=& ~e^{0,0}(Z_{f^\circ}) \left( e^{1,1}((\Bbb{C}^*)^2)+\sum_{f\supset t_1\supset \vec v_f} e^{1,1}(\Bbb{C}^*)\right)\\
=&~\left(-2+\sum_{f\supset t_1\supset \vec v_f} 1\right)(1+l^*(f^\circ))\,.
\end{split}
\end{align}
Note that $h^{2,0}(D_{\vec v_f})$ is simply 0, because there is no stratum that contributes non-trivial $e^{2,0}(D_{\vec v_f})$. For the same reason, we find that $h^{1,0}(D_{\vec v_f})=0.$ As a result, we conclude that face divisors $D_{\vec v_f}$ have the topology of $h^{0,0}(D_{\vec v_f})$ copies of toric twofolds.

\section{Construction of O3/O7 orientifolds}\label{sec:toric CY orientifolds}
In this section, we study how to embed an O3/O7 orientifold into a toric variety $\widehat{V}_4$ as a hypersurface $B_3\subset \widehat{V}_4.$ 

In \S\ref{sec:orientifoldinvolution} we study the action of the orientifold involution on the ambient toric fourfold $V_4$ in which we embed a CY threefold, $Z_3 \subset V_4.$ In \S\ref{sec:refinementmap}, we define a toric morphism called a \emph{refinement map}, which we use to construct an orbifold of $V_4$, namely a new variety that we denote by $\widehat{V}_4\equiv V_4/\Bbb{Z}_2$, which is invariant under the orientifold involution. Constructing $\widehat{V}_4$ provides us with an appropriate in which we may embed the orientifold of the CY hypersurface $Z_3$ as a hypersurface $B_3\subset \widehat{V}_4$; this is the subject of \S\ref{sec:constructCY3orientifold}. In \S\ref{sec:divisors in CY3orientifolds}, we study the topology of prime toric divisors in $B_3$. 

Throughout the paper, we focus on orientifold involutions given by coordinate flips of homogeneous coordinates of a toric variety, which we call \emph{coordinate flip orientifolding}.\footnote{For more general orientifolds, see \cite{Gao:2013pra,Carta:2020ohw,Altman:2021pyc,orientifoldingks}.} 

\subsection{Orientifold involution}
\label{sec:orientifoldinvolution}

Let $V_4$ be a toric fourfold, and let the basis of divisor $D_{i}$ denote the vanishing loci $x_i =0$ corresponding to rays $\vec v_i$ in a toric fan, hence points in $\partial \Delta^\circ.$ Let us consider an involution $\mathcal{I}_{\vec p}:V_4\rightarrow V_4,$ such that
\begin{equation}
\label{pinvolution}
\mathcal{I}_{\vec p}: x_{\vec p}\mapsto -x_{\vec p}\,,
\end{equation}
where $x_{\vec p}$ is a homogeneous coordinate in the Cox ring corresponding to a point $\vec p\in \partial\Delta^\circ.$ Then the vanishing locus $x_{\vec p}=0$ is a fixed locus of the involution $\mathcal{I}_{\vec p}.$ To find the fixed loci under $\mathcal{I}_{\vec{p}},$ we will use an efficient algorithm developed in \cite{orientifoldingks}. We will review the algorithm here. Because of the toric rescaling relations (recall (\ref{linrel}) and (\ref{morphism}))
\begin{equation}
g:x_i\rightarrow \lambda^{a_i}x_i\,, ~\lambda\in \Bbb{C}^*\,,
\end{equation}
if there is a linear relation between points $\vec v_i\in\partial \Delta^\circ$
\begin{equation}\label{eqn:linear relation}
\sum_i a_i \vec v_i=0\,,~a_i\in\Bbb{Z}\,,
\end{equation}
such that $a_{\vec p}\equiv 1 \mod 2,$ then the involution $\mathcal I_{\vec p}$ can be conjugated into an equivalent representation
\begin{equation}\label{eqn:orientifolding 2}
g\cdot \mathcal{I}_{\vec p} \cdot g^{-1}: x_i \mapsto (-1)^{a_i} x_i,~~  i\neq p\,.
\end{equation}
The presentation of the involution (\ref{pinvolution}) given in \eqref{eqn:orientifolding 2} makes locating all the fixed loci very easy. There are two cases of interest, corresponding (respectively) to O7-planes and O3-planes. The first case is when all but one $a_i$ satisfy $a_i\equiv0\mod2,$, while the second case occurs when all but three or four $a_i$ satisfy $a_i\equiv0\mod2.$\footnote{Codimension four fixed loci in $V_4$ can still be O3-planes, if the Calabi-Yau hypersurface $Z_3$ at the $\Bbb{Z}_2$ symmetric locus contains the fixed loci. We will see this more explicitly in \S\ref{sec:quintic}.} In the first case, the fixed locus of the involution $g\cdot \mathcal{I}_{\vec p} \cdot g^{-1}$ is simply given by $x_{j}=0,$ for $a_{j}\not\equiv0\mod2.$ To find such point $\vec v_j \in \partial\Delta^\circ,$ we can massage the linear relation \eqref{eqn:linear relation} into 
\begin{equation}\label{eqn:other o7}
\sum_i a_i \vec v_i\equiv \vec 0\mod2 \,.
\end{equation}
Using our assumptions about the first case, we then conclude that \eqref{eqn:other o7} is equivalent to
\begin{equation}
\vec v_j+\vec p\equiv \vec 0\mod2\,.
\end{equation}
As a result, we learn that to find another fixed locus $x_j=0,$ we can simply search for $\vec v_j\in \partial\Delta^\circ$ such that $\vec v_j+ \vec p\equiv \vec 0\mod2.$ Similarly, one can reach a conclusion that in the second case scenario, there is a codimension three fixed locus $x_{j_1}=x_{j_2}=x_{j_3}=0,$ where $\vec v_{j_1} + \vec v_{j_2}+\vec v_{j_3}+\vec p\equiv \vec 0\mod2.$ For the later convenience, we define a set of divisors that are fixed under the $g \cdot \mathcal I_{\vec p} \cdot g^{-1}$:
\begin{equation}
\label{hatI}
\widehat {I}_{\vec p}:=\{ x_i | g \cdot \mathcal{I}_{\vec p}\cdot g^{-1} (x_i)=-x_i ~ \text{for some} ~ g\in G\}\,.
\end{equation}
We oftentimes abuse notation and denote the set of indices $i$ corresponding to points $\vec v_i \in\partial\Delta^\circ$ that satisfy $\vec v_{i} +\vec p\equiv \vec 0\mod2$ by $\widehat{I}_{\vec p}$ if the meaning is clear from the context.

\subsection{Refinement map}
\label{sec:refinementmap}

Now as we promised, we shall construct the new toric variety $\widehat{V}_4.$ We will denote the homogeneous coordinates of $\widehat{V}_4$ by $\widehat{x}_i$ and the corresponding prime toric divisor by $\widehat{D}_{i}.$ To construct $\widehat{V}_4,$ we consider a map between the homogeneous coordinates of $V_4$ and $\widehat{V}_4$,
\begin{equation}
\label{refinementmap}
\varphi_{\mathcal I_{\vec p}}:\widehat{V}_4\rightarrow V_4\,,
\end{equation}
such that
\begin{equation}
\varphi_{\mathcal I_{\vec p}}:\widehat{x}_i\mapsto x_i^2\,, ~ \forall i\in \widehat{I}_{\vec p}\,,
\end{equation}
and
\begin{equation}
\varphi_{\mathcal I_{\vec p}}:\widehat{x}_i\mapsto x_i\,,~\forall i\not\in \widehat{I}_{\vec p} ~\text{ and } ~\vec v_i\in\partial\Delta^\circ\,.
\end{equation}
Therefore, the refinement map $\varphi_{\mathcal I_{\vec p}}$ can be used to define the prime toric divisor $\widehat{D}_i$ corresponding to any point $\vec v_i\in\partial\Delta^\circ.$ This map $\varphi_{\mathcal I_{\vec p}}$ can be understood in two different ways. In terms of the gauged linear sigma model\footnote{For readers who are not familiar with gauged linear sigma models, see \cite{Witten:1993yc,hori2003mirror}.} (GLSM), the map $\varphi_{\mathcal I_{\vec p}}$ simply doubles the GLSM charges of the chiral multiplet, whose scalar components' vanishing loci define a fixed locus of $\varphi_{\mathcal I_{\vec p}}.$ Alternatively, one can understand $\varphi_{\mathcal I_{\vec p}}$ as a refinement of a toric fan in the following sense: Let us recall the FRST $\mathcal{T}$ of $\Delta^\circ,$ where to be more explicit we now denote $\mathcal{T}$ as a collection of simplices indexed by a set ${I}_{\mathcal{T}}$ 
\begin{equation}
\mathcal{T}:= \left\{ \{0,\vec v_{i},\vec v_{j},\vec v_{k},\vec v_{l}\}| \vec v_{m = i,j,k,l}\in\partial\Delta^\circ\,,~\{i,j,k,l\}\in{I}_{\mathcal{T}}  \right\} \,.
\end{equation}
We choose a gauge in which the point $\vec v_p$ that defines the orientifold involution $\mathcal I_{\vec p}$ is simply 
\begin{equation}
\vec p=(1,0,0,0)\,.
\end{equation}
Then, noting that the lift to the orientifold covering space acts by doubling the volume of divisors (and hence sending vertices $\vec v$ to $2\vec v$) we define the action of $\varphi_{\mathcal I_{\vec p}}^{-1}$ on a point $\vec v\in \partial\Delta^\circ$ as follows
\begin{equation}
\varphi_{\mathcal I_{\vec p}}^{-1}: \vec v=(v_1,v_2,v_3,v_4)\mapsto\vec{\widehat{v}}= 
\begin{cases}
(2v_1,v_2,v_3,v_4)& \text{ if } \vec p+ \vec v\not\equiv \vec 0\mod 2\\
\frac{1}{2}(2v_1,v_2,v_3,v_4)& \text{ if } \vec p+\vec v\equiv \vec 0\mod2
\end{cases}\,.
\end{equation}
Then the refined toric fan under $\varphi_{\mathcal{I}_{\vec p}}^{-1}$ is given as
\begin{equation}
\mathcal{T}_{\widehat{V}_4}:= \varphi_{\mathcal I_{\vec p}}^{-1}(\mathcal{T})\,.
\end{equation}

\subsection{Orientifolds of CY threefold hypersurfaces}
\label{sec:constructCY3orientifold}

We are now ready to apply the orientifold involution on toric varieties to orientifolding of CY threefolds. Let us recall that the CY hypersurface $Z_3$ is defined to be the vanishing locus of a section of the anti-canonical class
\begin{equation}
-K_{V_4}=\sum_{\vec v\in\partial\Delta^\circ} [D_{\vec v}]\,,
\end{equation}
where the corresponding Newton polytope is defined to be
\begin{equation}
\Delta_{-K_{V_4}}:=\{\vec{m}\in M| \vec{m}\cdot \vec{v}\geq -1,~ \forall \vec{v}\in\partial\Delta^{\circ}\}\,.
\end{equation}
As explained earlier, each point $\vec{m}\in\Delta_{-K_{V_4}}$ defines a monomial 
\begin{equation}
x^{\vec{m}}:=\prod_{\vec{v}\in\partial\Delta^\circ} x_{\vec v}^{\vec{m}\cdot \vec{v}+1}\,.
\end{equation}
The orientifold involution $\mathcal{I}_{\vec p}$ acts on these monomials as follows:
\begin{equation}
\mathcal{I}_{\vec p} : x^{\vec{m}}\mapsto \pm x^{\vec{m}}\,.
\end{equation}
The collection of monomials $\Delta_-$ that are not invariant under the involution $\mathcal{I}_{\vec p}$, i.e. the monomials for which
\begin{equation}
\mathcal{I}_{\vec p} (x^{\vec{m}})=-x^{\vec{m}}\,,
\end{equation}
are projected out and cannot be used to define a $\Bbb{Z}_2$-symmetric CY threefold $Z_3$. On the other hand, the monomial deformations that are invariant under the involution respect the desired $\Bbb{Z}_2$-symmetry and can therefore be used to define a $\Bbb{Z}_2$-symmetric CY threefold. We define a collection of these invariant monomials by 
	\begin{equation}
		\Delta_+:=\{ x^{\vec{m}}|\vec{m}\in\Delta,~\mathcal{I}_{\vec p}(x^{\vec{m}})=x^{\vec{m}}\}.
	\end{equation}
 We are mainly interested in involutions $\mathcal I_{\vec p}$ such that there is at least one distinct monomial in $\Delta_+.$

There are two types of the fixed loci in $Z_3$: O7-planes and O3-planes. An O7-plane is a divisor in $Z_3$ that is fixed under the $\Bbb{Z}_2$ involution. Similarly, an O3-plane is a point in $Z_3$ that is fixed under the involution. Other types of fixed loci cannot occur, as they would fail to satisfy the calibration conditions necessary to preserve supersymmetry \cite{Gimon:1996rq}.

To define the orientifold $B_3:=Z_3/\Bbb{Z}_2,$ we wish to embed the $\Bbb{Z}_2$-invariant Calabi-Yau threefold $Z_3$ into $\widehat{V}_4.$ To do so, we shall study how the refinement map $\varphi_{\mathcal I_{\vec p}}$ acts on the monomials in $\Delta.$ Under the refinement map $\varphi_{\mathcal I_{\vec p}},$ the anti-canonical class of $V_4$ maps to a line bundle with first Chern class
\begin{equation}\label{eqn:refined CY3 defining equation}
-K_{\widehat{V}_4}-L:= \varphi_{\mathcal I_{\vec p}}^{-1}(-K_{V_4})\,,
\end{equation}
where \eqref{eqn:refined CY3 defining equation} can be understood as a definition for $\mathcal{L}$, 
\begin{align}
\begin{split}
L:= &\sum_{i\in\widehat{I}_{\vec p}} \varphi_{\mathcal I_{\vec p}}^{-1}\left(D_i\right)\\
=&\frac{1}{2}\sum_{i\in\widehat{I}_{\vec p}}\widehat{D}_i\,.
\end{split}
\end{align}
Existence of the global section in $\Delta_+$ guarantees that there exists a relation
\begin{equation}\label{eqn:orientifold class}
-K_{\widehat{V}_4}-L=\sum b_i \widehat{D}_i\,,
\end{equation}
where $b_i\in\Bbb{Z}_{\geq0}$ as required for $- K_{\widehat{V}_4} -  L$ to be a non-trivial line bundle. We can therefore define the Newton polytope for the orientifold to be
\begin{equation}
\label{orientifoldNewtonpolytope}
\widehat{\Delta}:=\{ \vec{m}\in M|\vec{m}\cdot \vec{\widehat{v}}_i\geq -b_i \,, ~\forall \vec{v}_i\in\partial\Delta^\circ \}\,.
\end{equation}
Note that $\widehat{\Delta}$ is not necessarily a convex hull of integral points. Because $\widehat{\Delta}$ can be understood as $GL(4,\Bbb{Z})$ transformation and translation of $\Delta,$ we find that there is a non-ambiguous one-to-one map between $\Delta^{(n)}$ and $\widehat{\Delta}^{(n)}.$ We therefore oftentimes denote $\widehat{\Delta}^{(n)}$ by $\varphi_{\mathcal I_{\vec p}}^{-1}(\Delta^{(n)}).$ As in the case of a CY hypersurface, to each integral point $\vec{m}$ in $\widehat{\Delta}$ we associate a monomial
\begin{equation}
\widehat x^{\vec{m}}=\prod_{i} \widehat x_i^{\vec{m}\cdot\vec{\widehat{v}}_i+b_i}\,.
\end{equation}
Next, we define the orientifold $B_3,$ which is topologically equivalent to $Z_3/\Bbb{Z}_2,$ to be the vanishing locus of a section of the line bundle $\mathcal O(-K_{\widehat{V}_4}-L).$ Because the first Chern class of $B_3$ is
\begin{equation}
c_1(B_3)=L\,,
\end{equation}
$B_3$ cannot be a CY threefold. Rather, $L$ is non-negative, and we find that there is a relation
\begin{equation}
L=\sum c_i \widehat{D}_i\,,~~~~ c_i \in \mathbb Z_{\geq 0}.
\end{equation}
In special cases where there exists a prime toric divisor 
\begin{equation}
\widehat{D}_i=-K_{\widehat{V}_4}-L\,,
\end{equation}
$B_3$ can be understood as a toric threefold. Generically, however, the F-theory uplift of a type IIB compactification on $B_3$ should be understood as an F-theory compactification on an elliptic fibration over a non-toric threefold $B_3$. 

To compute the intersection numbers, one can certainly follow the standard method to compute the intersection numbers given the toric fan $\varphi_{\mathcal I_{\vec p}}^{-1}(\mathcal{T})$ and the hypersurface class $[B_3]$.\footnote{For the intersection theory of toric varieties, see \cite{Fulton}.} However, given that we already have the CY construction at hand, it turns out that there is a simpler way to compute the intersection numbers
\begin{equation}\label{eqn:ori intersection}
\widehat{D}_i \cdot \widehat{D}_j \cdot \widehat{D}_k=\frac{1}{2} \varphi_{\mathcal I_{\vec p}}(\widehat{D}_i)\cdot \varphi_{\mathcal I_{\vec p}}(\widehat{D}_j) \cdot \varphi_{\mathcal I_{\vec p}}(\widehat{D}_k)\,.
\end{equation}
This formula \eqref{eqn:ori intersection} is intuitive to understand as follows. Intersection numbers between three divisors, if positive, denote the number of points in the $Z_3$ where these three divisors meet simultaneously. Intersection in $B_3$ should count half of the intersections points in $Z_3$ if none of the three divisors hosts an O7-plane. If one or more divisors involved in the above intersection host O7-planes, then we should account for the fact that an O7-plane class in the $B_3$ is double the class of the prime toric divisor that hosts corresponding O7-plane in $Z_3$. This subtlety is automatically taken care of by the refinement map.

As the last step towards a full characterization of $B_3$, we compute the Hodge numbers. Let us first recall that the stratification of $V_4$ is given as
\begin{equation}
V_4=T_{\sigma^{(0)}}\coprod T_{\sigma^{(1)}}\coprod_{n\geq 2} T_{\sigma^{(n)}}\times E_{\sigma^{(n)}}\,,
\end{equation}
where $\sigma^{(0)}$ is the origin and $\sigma^{(n)}$ for $n\geq1$ is a cone over $n-1$ dimensional face of $\Delta^\circ.$ Similarly, the stratification of the hypersurface $Z_3$ in $V_4$ is 
\begin{equation}
Z_3=Z_{\Delta}\coprod Z_{\Delta^{(3)}}\coprod Z_{\Delta^{(2)}}\times E_{\sigma^{(2)}}\coprod Z_{\Delta^{(1)}}\times E_{\sigma^{(3)}}\coprod Z_{\Delta^{(0)}}\times E_{\sigma^{(4)}}\,.
\end{equation}
Note that we made the factor $Z_{\Delta^{(0)}}$ explicit for reasons we explain shortly. Very conveniently, the stratifications of $\widehat{V}_4$ and $B_3$ are inherited from the stratifications of $V_4$ and $Z_3$, respectively. Introducing the shorthand notation $\widehat{\sigma}^{(n)}=\varphi_{\mathcal I_{\vec p}}^{-1}(\sigma^{(n)}),$ we may write the stratification of $\widehat{V}_4$ and $B_3$ as
\begin{equation}
\widehat{V}_4=T_{\widehat{\sigma}^{(0)}}\coprod T_{\widehat{\sigma}^{(1)}}\coprod_{n\geq2} T_{\widehat{\sigma}^{(n)}}\times E_{\widehat{\sigma}^{(n)}}\,,
\end{equation}
\begin{equation}
B_3=Z_{\widehat{\Delta}}\coprod Z_{\widehat{\Delta}^{(3)}}\coprod Z_{\widehat{\Delta}^{(2)}}\times E_{\widehat{\sigma}^{(2)}}\coprod Z_{\widehat{\Delta}^{(1)}}\times E_{\widehat{\sigma}^{(3)}}\coprod Z_{\widehat{\Delta}^{(0)}}\times E_{\widehat{\sigma}^{(4)}}\,.
\end{equation}
If some of the vertices of $\widehat{\Delta}$ are not integral points, namely $l(\widehat{\Delta}^{(0)})=0,$ some of the prime toric divisors in $V_4$ that missed $Z_3$ at a generic point in the moduli space can now intersect the $\Bbb{Z}_2$-symmetric CY threefold. This happens because forcing $Z_3$ to be $\Bbb{Z}_2$ symmetric generically tunes the complex structure moduli of $Z_3$. We have to pay extra attention to these extra divisors in order to compute the Hodge numbers. 

To avoid overwhelming the discussion, let us begin by studying an orientifold for which the Newton polytope is an integral polytope.  We start by computing the Hodge vector $h^\bullet(B_3,\mathcal{O}_{B_3})=h^{\bullet,0}(B_3)$,
\begin{equation}
h^{3,0}(B_3)=-e^{3,0}(Z_{\widehat{\Delta}})=l^*(\widehat{\Delta})\,.
\end{equation}
\begin{align}
\begin{split}
h^{2,0}(B_3)=&~e^{2,0}(Z_{\widehat{\Delta}})+\sum_{\widehat{\Delta}^{(3)}} e^{2,0}(Z_{\widehat{\Delta}^{(3)}})\\
=&~\left(-\sum_{\widehat{\Delta}^{(3)}}l^*(\widehat{\Delta}^{(3)})\right)+\sum_{\widehat{\Delta}^{(3)}}l^*(\widehat{\Delta}^{(3)})\\
=&~0\,.
\end{split}
\end{align}
Similarly, we compute
\begin{align}
h^{1,0}(B_3)=&~e^{1,0}(Z_{\widehat{\Delta}})+\sum_{\widehat{\Delta}^{(3)}}e^{1,0}(Z_{\widehat{\Delta}}^{(3)})+\sum_{\widehat{\Delta}^{(2)}} e^{1,0}(Z_{\widehat{\Delta}^{(2)}})\,.
\end{align}
By using the fact that
\begin{equation}
l^2(\widehat{\Delta})-l^1(\widehat{\Delta})=\left(\sum_{\widehat{\Delta}^{(3)}}l^2(\widehat{\Delta}^{(3)})-\sum_{\widehat{\Delta}^{(3)}}l^1(\widehat{\Delta}^{(3)})\right)-\left(\sum_{\widehat{\Delta}^{(2)}}l^2(\widehat{\Delta}^{(2)})-\sum_{\widehat{\Delta}^{(2)}}l^1(\widehat{\Delta}^{(2)})\right)\,,
\end{equation}
we conclude 
\begin{equation}
h^{1,0}(B_3)=0\,.
\end{equation}
And similarly, we compute
\begin{align}
h^{0,0}(B_3)=&\sum_{k=4}^1 \sum_{\widehat{\Delta}^{(k)}}(-1)^{k-1} (1-l^1(\widehat{\Delta}^{(k)})).
\end{align}
By using the fact that Euler characteristic of the boundary of a 4-dimensional polytope is 0, and using the relation\footnote{The following relation \eqref{eqn:combinatoric trick} comes from the fact that the number of intergral points in 1 simplices of $\Delta$ can be counted as follows. First, one can collect the number of integral points in 1 simplices of all the codimension 1 faces in $\Delta.$ But, this is an overcounting because the number of integral points in 1 simplices of codimension 2 faces in $\Delta$ are counted twice. Subracting the number of points in 1 simplices of codimension 2 faces now leads to undercounting because now the points in 1 simplices of codim 3 faces are not counted. By keep doing this procedure, we obtain the formula \eqref{eqn:combinatoric trick}.}
\begin{equation}\label{eqn:combinatoric trick}
\sum_{k=4}^1 \sum_{\widehat{\Delta}^{(k)}}(-1)^{k}l^1(\widehat{\Delta}^{(k)})+\sum_{\widehat{\Delta}^{(0)}} 1=0\,,
\end{equation}
we obtain
\begin{equation}
h^{0,0}(B_3)=1\,.
\end{equation} 
As a result, we find that $B_3$ is a good candidate for a base of an elliptic CY fourfold if and only if $l^*(\widehat{\Delta})=0.$ In the following, we therefore assume $l^*(\widehat{\Delta})=0.$

We next compute $h^{2,1}(B_3)=h^{2,1}_+(Z_3)$:
\begin{align}
\begin{split}
\label{eqn:h21p}
h^{2,1}(B_3)=&~-e^{2,1}(Z_{\widehat{\Delta}})-\sum_{\widehat{\Delta}^{(2)}}e^{1,0}(Z_{\widehat{\Delta}^{(2)}})e^{1,1} (E_{\widehat{\sigma}^{(2)}})\\
=&~l^*(2\widehat{\Delta})-\sum_{\widehat{\Delta}^{(3)}}l^*(\widehat{\Delta}^{(3)})+\sum_{\widehat{\Delta}^{(2)}}l^*(\widehat{\Delta}^{(2)})l^*(\Theta^{\circ(1)})\,.
\end{split}
\end{align}
Note that in the last term we have $l^*(\Theta^{\circ(1)})$, which is not a quantity defined in the refined toric fan. This is merely due to the fact that (since the quotient map $\varphi_{\mathcal I_{\vec p}}$ does not change the number of toric rays)
\begin{equation}
e^{1,1}(E_{\widehat{\sigma}^{(2)}})=e^{1,1}(E_{\sigma^{(2)}})\,.
\end{equation}
To understand what each term in \eqref{eqn:h21p} is counting, let us first recall \eqref{eqn:h21 CY3},
\begin{equation}\label{eqn:h21 CY32}
h^{2,1}(Z_3)=l(\Delta)-5-\sum_{\Delta^{(3)}}l^*(\Delta^{(3)})+\sum_{(\Delta^{(2)})^\circ=\Theta^{\circ(1)}}l^*(\Delta^{(2)})l^*(\Theta^{\circ(1)})\,.
\end{equation}
The first three terms in \eqref{eqn:h21 CY32}, namely
\begin{equation}\label{eqn:toric deformations}
l(\Delta)-5-\sum_{\Delta^{(3)}}l^*(\Delta^{(3)})
\end{equation} 
count the number of inequivalent monomial deformations of the complex structure of $Z_3$. On the other hand, the origin of the last term in \eqref{eqn:h21 CY32} 
\begin{equation}
\sum_{\Delta^{(2)}} l^*(\Delta^{(2)})l^*(\Theta^{\circ(1)})
\end{equation}
is subtler. Let us take a close look at the stratum
\begin{equation}\label{eqn:stratum 2-face}
Z_{\Delta^{(2)}}\times E_{\sigma^{(2)}}\,.
\end{equation}
Since $E_{\sigma^{(2)}}$ has complex dimension one, the stratum \eqref{eqn:stratum 2-face} by construction is a multiple $\Bbb{P}^1$ fibration over a punctured Riemann surface 
\begin{equation}
Z_{\Delta^{(2)}}\subset T_{\sigma^{(2)}}\,.
\end{equation}
The natural compactification of $Z_{\Delta^{(2)}}$ is thus a closed Riemann surface whose stratification is
\begin{equation}
\overline{Z}_{\Delta^{(2)}}=Z_{\Delta^{(2)}}\coprod_{\Delta^{(1)}\subset \Delta^{(2)}} Z_{\Delta^{(1)}}\,.
\end{equation}
Now, the genus of the Riemann surface $\overline{Z}_{\Delta^{(2)}}$ is
\begin{equation}
g(\Delta^{(2)})=l^*(\Delta^{(2)})\,.
\end{equation}
Quite importantly, all of the complex structure moduli of the Riemann surface $\overline{Z}_{\Delta^{(2)}}$ are captured by the monomials in $\Delta^{(2)}$. The complex structure moduli of $\overline{Z}_{\Delta^{(2)}}$ are already counted in \eqref{eqn:toric deformations}. However, some care is required in deducing the number of complex structure moduli of $Z_3$ from the complex structure moduli of $\overline{Z}_{\Delta^{(2)}}$: If the stratum \eqref{eqn:stratum 2-face} has more than one $\Bbb{P}^1$ fibration, it turns out that the total space \eqref{eqn:stratum 2-face} should have $1+l^*(\Delta^{(2)})l^*(\Theta^{\circ(1)})$ complex structure moduli, despite the fact that only $(1+g)$ of the linear combinations can be captured by the monomial deformations in $\Delta^{(2)}$; these extra complex structure moduli are precisely counted by the term
\begin{equation}
\sum_{\Delta^{(2)}} l^*(\Delta^{(2)}) l^*(\Theta^{\circ(1)})\,.
\end{equation}
Now we are ready to understand the formula \eqref{eqn:h21p}. The first two terms in \eqref{eqn:h21p}, namely
\begin{equation}
l^*(2\widehat{\Delta})-\sum_{\widehat{\Delta}^{(3)}}l^*(\widehat{\Delta}^{(3)})\,,
\end{equation}
count the number of monomial deformations that are projected out by the $\mathbb Z_2$ quotient corresponding to orientifolding. The last term in \eqref{eqn:h21p} can be understood as follows. Under the orientifolding, the Riemann surface $\overline{Z}_{\Delta^{(2)}}$ transforms into a new Riemann surface $\overline{Z}_{\widehat{\Delta}^{(2)}}.$ In particular, the orientifolding can reduce the genus of the Riemann surface. The change in the genus also then leads to the change in the number of complex structure moduli, which is precisly counted by
\begin{equation}
\sum_{\widehat{\Delta}^{(2)}}l^*(\widehat{\Delta}^{(2)})l^*(\Theta^{\circ(1)})\,.
\end{equation}

Similarly, for $h^{1,1}(B_3)=h^{1,1}_+(Z_3),$ we compute
\begin{align}
\begin{split}
h^{1,1}(B_3)=&~e^{2,2}(Z_{\widehat{\Delta}})+ \sum_{\widehat{\Delta}^{(3)}} e^{2,2}(Z_{\widehat{\Delta}^{(3)}})+\sum_{\widehat{\Delta}^{(2)}}e^{1,1}(Z_{\widehat{\Delta}^{(2)}})e^{1,1}(E_{\widehat{\sigma}^{(2)}})\\
&+\sum_{\widehat{\Delta}^{(3)}}e^{0,0}(Z_{\widehat{\Delta}^{(1)}}) e^{2,2}(E_{\widehat{\sigma}^{(3)}})\,, \end{split} \\ 
\begin{split}
=&~l(\Delta^\circ)-5 -\sum_{\Theta^{\circ(3)}}l^*(\Theta^{\circ(3)})+\sum_{\Theta^{\circ(2)}} l^*(\Theta^{\circ(2)})l^*(\widehat{\Delta}^{(1)})\,.\end{split}
\end{align}
This implies that
\begin{equation}\label{eqn:h11n}
h^{1,1}_-(Z_3)=\sum_{\Theta^{\circ(2)}}l^*(\Theta^{\circ(2)})\left(l^*(\Delta^{(1)})-l^*(\widehat{\Delta}^{(1)})\right)\,.
\end{equation}
A comment on \eqref{eqn:h11n} is in order. The orientifold involution $\mathcal I_{\vec p}$ does not change the number of toric rays of $V_4.$ This means that $h^{1,1}(V_4)=h^{1,1}(\widehat{V}_4).$ But, this does not imply that $h^{1,1}(Z_3)=h^{1,1}(B_3).$ To understand the counting of $h^{1,1}(B_3)$ better, recall the fact that a prime toric divisor $D_{\vec v_f}$ corresponding to a point $\vec v_f$ interior to a two-dimensional face $\Theta^{\circ(2)} \subset \partial\Delta^\circ$ is a (generically) reducible toric two-fold with $1+l^*(\Delta^{(1)})$ components, where $\Delta^{(1)}$ is polar dual to $\Theta^{\circ(2)}.$ The important point to note here is that the orientifold involution can identify some of the irreducible components of a two-face divisor $D_{\vec v_f}$. If such a non-trivial identification occurs, only $\Bbb{Z}_2$-symmetric combinations of the irreducible components can contribute to $h^{1,1}(B_3)=h^{1,1}_+(Z_3).$ The formula \eqref{eqn:h11n} counts the number of asymmetric combinations under the orientifold involution. 

Next, we turn to the subtler case where the Newton polytope $\widehat{\Delta}$ is not integral. A non-integral $\widehat{\Delta}$ can exhibit many different characteristics. For the purposes of this paper, however, we focus on the simplest case where only vertices of $\widehat{\Delta}$ are allowed to not have integral points. The only Hodge numbers that are affected by the non-integrality of some of the vertices of $\widehat{\Delta}$ are $h^{n,n}(B_3)$ for $n=0,\dots,3.$ If there is a $\widehat{\Delta}^{(0)}$ such that $e^{n,n}(E_{\widehat{\sigma}^{(4)}})$ is non-trivial and $l(\widehat{\Delta}^{(0)}),$ then the Hodge number $h^{n,n}(B_3)$ receives a correction term
\begin{equation}
e^{n,n}(E_{\widehat{\sigma}}^{(4)})\,,
\end{equation}
for $n\neq0,$ and
\begin{equation}
-1+e^{0,0}(E_{\widehat{\sigma}^{(4)}})\,,
\end{equation}
for $n=0.$ As a result, non-vanishing
\begin{equation}
-1+e^{0,0}(E_{\widehat{\sigma}^{(4)}})
\end{equation}
indicates the existence of a base locus, which makes the CY orientifold reducible. As the meaning of string compactifications on such manifolds is rather ambiguous, we restrict the discussion to triangulations with vanishing $-1+e^{0,0}(E_{\widehat{\sigma}^{(4)}}).$

\subsection{Topology of prime toric divisors in orientifolds of CY hypersurfaces}\label{sec:divisors in CY3orientifolds}
In this section, we continue studying the topology of prime toric divisors in the orientifold $B_3.$ This section parallels much of the discussion in section \S\ref{sec:divisors in CY3}. Because the stratification of the CY orientifold $B_3$ inherits the stratification of $Z_3,$ the stratification of the prime toric divisors in $B_3$ is also inherited from the prime toric divisors in $Z_3.$ Therefore, even if there is no reason that a vertex divisor in $Z_3$ should (after orientifolding) still describe a divisor in $B_3$ corresponding to the vertex of a polytope, we nonetheless abuse terminology and call $\widehat D_{\vec{\widehat{v}}}$ a vertex divisor if $\widehat D_{\vec{\widehat{v}}} =\varphi_{\mathcal I_{\vec p}}^{-1}(D_{\vec v})$ where $D_{\vec v}$ is a vertex divisor in $Z_3$. We analogously abuse the terminology ``edge divisor'' and ``face divisor''. Unless otherwise noted, we retain the conventions of \S\ref{sec:divisors in CY3}. For example, let $\vec q\in\partial\Delta^\circ$ be a vertex. We then denote the subpolytope of the Newton polytope $\widehat{\Delta}$ for which no corresponding monomial has a factor of $\widehat{x}_{\vec{\widehat{q}}}$ by $\widehat{\Delta}^{(3)}_{\vec{\widehat{q}}}$, whose equivalent definition is
\begin{equation}
\label{qhat}
\widehat{\Delta}^{(3)}_{\vec{\widehat{q}}}:=\{ \vec{m}\in M| \vec{m}\cdot \vec{\widehat{q}} =-b_{\vec{\widehat{q}}}\, ,~\vec{m}\cdot \widehat{v}_i\geq -b_i~\forall \vec v_i\in\partial\Delta^{\circ},~\text{ and } \vec v_i\neq \vec q\}\,.
\end{equation} 
As the notation $\widehat{\Delta}^{(3)}_{\vec{\widehat{q}}}$ is rather cumbersome, we instead use the shorthand notation 
	\begin{equation}
		\widehat{q}^\circ:=\widehat{\Delta}^{(3)}_{\vec{\widehat{q}}}.
	\end{equation}
 Likewise, a codimension-one subpolytope $\widehat{\Delta}^{(2)}$ in $\widehat{q}$ can be understood as the preimage of the refinement map acting on an edge $e$ containing $\vec q.$ Hence, we denote such a codimension-one subpolytope by 
 	\begin{equation}
	\label{ehat}
		\widehat{e}^\circ:= \varphi_{\mathcal I_{\vec p}}^{-1}(e).
	\end{equation}
Likewise, we follow an analogous convention for $\widehat{f}^\circ.$ 

\subsubsection{Vertex divisor}
Let $\vec v$ be a vertex of $\partial\Delta^\circ.$ The stratification of the vertex divisor $\widehat D_{\vec{\widehat{v}}}$ is
\begin{align}
\widehat D_{\vec{\widehat{v}}}=Z_{\widehat{v}}\coprod_{\vec v \subset e} Z_{\widehat{e}}\coprod_{\vec v \subset f} Z_{\widehat{f}}\left(\coprod_{f\supset t_1\supset \vec v} (\Bbb{C}^*)\coprod_{f\supset t_2 \supset \vec v}(pt)\right) \,.
\end{align}
It is then straightforward to read off the Hodge numbers $h^{p,q}(\widehat D_{\vec{\widehat{v}}})=h^{p,q}_+( D_{\vec v}).$ We compute
\begin{align}
\begin{split}
h^{2,0}(\widehat{D}_{\vec{\widehat{v}}})=&~h^{2,0}_+(\varphi_{\mathcal I_{\vec p}}({D}_{\vec v}))\\
=&~l^*(\widehat{v}^\circ)\,,
\end{split}
\end{align}
and for $\varphi_{\mathcal I_{\vec p}}(\widehat{D}_{\vec{\widehat{v}}})=D_{\vec v}$ we have
\begin{equation}
h^{2,0}_-(D_{\vec v})=l^*( v^\circ)-l^*(\widehat{v}^\circ)\,.
\end{equation}
Similarly, we compute
\begin{align}
\begin{split}
\label{eqn:h11p vertex}
h^{1,1}(\widehat{D}_{\vec{\widehat{v}}})=&~h^{1,1}_+(\varphi_{\mathcal I_{\vec p}}(\widehat{D}_{\vec{\widehat{v}}}))\\
=&~l^*(2\widehat{v}^\circ)-4l^*(\widehat{v}^\circ)-\sum_{e\supset \vec v} l^*(\widehat{e}^\circ)-3+\sum_{e\supset \vec v}1+\sum_{f\supset \vec v}\sum_{f\supset t_1\supset \vec v} (l^*(\widehat{f}^\circ)+1)\,,
\end{split}
\end{align}
and for $\varphi_{\mathcal I_{\vec p}}(\widehat{D}_{\vec{\widehat{v}}})=D_{\vec v}$ we have
\begin{align}\label{eqn:h11n vertex}
h^{1,1}_-(D_{\vec v})=&\left(l^*(2 v^\circ)-4l^*( v^\circ)-\sum_{e\supset v} l^*(e^\circ)\right)-\left(l^*(2\widehat{v}^\circ)-4l^*(\widehat{v}^\circ)-\sum_{e\supset v} l^*(\widehat{e}^\circ)\right)\nonumber\\
&\quad +\sum_{f\supset \vec v}\sum_{f\supset t_1\supset \vec v} (l^*(f^\circ)-l^*(\widehat{f}^\circ))\,.
\end{align}
Lastly, we compute 
\begin{align}
h^{1,0}(\widehat{D}_{\vec{\widehat{v}}})=0\,,
\end{align}
and
\begin{align}
h^{0,0}(\widehat{D}_{\vec{\widehat{v}}})=1\,,
\end{align}
assuming that $B_3$ is not reducible.

In order to understand the formulae \eqref{eqn:h11p vertex} and \eqref{eqn:h11n vertex}, we start by first understanding how the computation of $h^{1,1}$ works for $\varphi_{\mathcal I_{\vec p}}(\widehat{D}_{\vec{\widehat{v}}})=D_{\vec v}$ in $Z_3.$ Let us recall \eqref{eqn:h11 vertex}:
\begin{equation}
h^{1,1}(D_{\vec v})= l^*(2 v^\circ)-4l^*( v^\circ)-\sum_{e\supset \vec v} l^*(e^\circ)-3+\sum_{e\supset \vec v}1+\sum_{f\supset  \vec v}\sum_{f\supset t_1\supset \vec v} (l^*(f^\circ)+1)\,.
\end{equation}
Much like the case of K3 manifold, the quantity $h^{1,1}(D_{\vec v})$ consists of two components: the Picard rank, and the number of monomial deformations of the defining equation for $Z_{\vec v}$:
	\begin{equation}
		h^{1,1}(D_{\vec v}) =h^{1,1}_{Pic}(D_{\vec v}) + h_{mon}^{1,1}(D_{\vec v}).
	\end{equation}
If $D_{\vec{v}}$ is not rational, the Picard rank of $D_{\vec v}$ at a generic point in the moduli space is given by:
\begin{equation}
h^{1,1}_{Pic}(D_{\vec v}):=-3+\sum_{e\supset \vec v}1+\sum_{f\supset \vec v}\sum_{f\supset t_1\supset \vec v} (l^*(f^\circ)+1)\,,
\end{equation}
where $\sum_{f\supset \vec v}\sum_{f\supset t_1\supset \vec v}l^*(f)$ is the correction term due to the reducible divisor in $D_{\vec v}.$ The rest of the terms
\begin{equation}
 h^{1,1}_{mon}(D_{\vec v}):=l^*(2 v^\circ)-4l^*( v^\circ)-\sum_{e\supset \vec v} l^*(e^\circ)
\end{equation}
therefore count the number of complex structure deformations of the defining equation for $D_{\vec v}$. Note that if $D_{\vec{v}}$ is rational, then $h^{1,1}_{Pic}+h^{1,1}_{mon}$ shall be understood as the correct Picard rank.

The orientifold action modifies the computation of $h^{1,1}(\widehat{D}_{\vec{\widehat{v}}})$ in two major ways. First, the orientifold action either projects in or out the monomials counted by $h^{1,1}_{mon}(D_{\vec v}).$ This is reflected by the contribution
\begin{equation}
\left(l^*(2 v^\circ)-4l^*( v^\circ)-\sum_{e\supset \vec v} l^*(e^\circ)\right)-\left(l^*(2\widehat{v}^\circ)-4l^*(\widehat{v}^\circ)-\sum_{e\supset \vec v} l^*(\widehat{e}^\circ)\right)
\end{equation}
in the expression for $h^{1,1}_-(D_{\vec v})$ above. Second, the orientifold action can identify points in $Z_{f^\circ}.$ If such things happen, some of the reducible divisors within $D_{\vec v}$ are also identified, leading to the second contribution to $h^{1,1}_-(D_{\vec v})$, namely
\begin{equation}
\sum_{f\supset \vec v}\sum_{f\supset t_1\supset \vec v} (l^*(f^\circ)-l^*(\widehat{f}^\circ))\,.
\end{equation}

\subsubsection{Edge divisors}
Let $\vec v_e$ be a point interior to an edge $e \subset \partial\Delta^\circ.$ Then the stratification of the prime toric divisor $\widehat{D}_{\vec{\widehat{v}}_e}$ is
\begin{equation}\label{eqn:edge div in b stratification}
\widehat{D}_{\vec{\widehat{v}}_e}=Z_{\widehat{e}^{\circ}}\times (\Bbb{C}^*\coprod (2pts))\coprod_{f\supset e} Z_{\widehat{f}}\left(\coprod_{f\supset t_1\supset \vec v_{e}}\Bbb{C}^*\coprod_{f\supset t_2\supset \vec v_e} (pt) \right)\,.
\end{equation}

It is easy to check $h^{2,2}(\widehat{D}_{\vec{\widehat{v}}_e})=h^{0,0}(\widehat{D}_{\vec{\widehat{v}}_e})=1,$ again assuming the absence of a base locus. Like in the CY case, there is no stratum that yields non-trivial $e^{2,0}(\widehat{D}_{\vec{\widehat{v}}_e}),$ hence we conclude $h^{2,0}(D_{\vec v_e})=0.$ We now compute $h^{1,1}(\widehat{D}_{\vec{\widehat{v}}_e})$:
\begin{align}
h^{1,1}(\widehat{D}_{\vec{\widehat{v}}_e})=e^{1,1}(Z_{\widehat{e}})+e^{0,0}(Z_{\widehat{e}})+\sum_{f\supset  e}\sum_{f\supset t_1 \supset \vec v_e}e^{0,0}(Z_{\widehat{f}})\,.
\end{align}
We again use an identity, namely
\begin{equation}
e^{0,0}(Z_{\widehat{e}})=1-l^1(\widehat{e}^\circ)=1-\sum_{f\supset e}(1+l^*(\widehat{f}^\circ))\,,
\end{equation}
to obtain
\begin{equation}
h^{1,1}(\widehat{D}_{\vec{\widehat{v}}_e})=h^{1,1}_+(\varphi_{\mathcal I_{\vec p}}(\widehat{D}_{\vec{\widehat{v}}_e}))=2 +\sum_{f\supset  e}(-1+\sum_{f\supset t_1\supset \vec v_e} 1)(1+l^*(\widehat{f}^\circ))\,,
\end{equation}
and for $\varphi_{\mathcal I_{\vec p}}(\widehat{D}_{\vec{\widehat{v}}_e})=D_{\vec v_e}$ we have
\begin{equation}\label{eqn:h11n edge}
h^{1,1}_-(D_{\vec v_e})=\sum_{f\supset e}(-1+\sum_{f\supset t_1\supset v_e} 1)(l^*(f^\circ)-l^*(\widehat{f}^\circ))\,.
\end{equation}
The origin of \eqref{eqn:h11n edge} is simple to understand. If some of the reducible divisors in $D_{\vec v_e}$ that are described by the strata
\begin{equation}
\coprod_{f\supset e} Z_{f^\circ}\left(\coprod_{f\supset t_1\supset v_{e}}\Bbb{C}^*\coprod_{f\supset t_2\supset v_e} (pt) \right)
\end{equation}
get identified under the orientifolding, only combinations of reducible divisors in $D_{\vec v_e}$ even under the orientifold involution can contribute to $h^{1,1}_+(D_{\vec v_e})$---the number of such combinations is precisely what is counted by \eqref{eqn:h11n edge}. Finally, we compute $h^{1,0}(\widehat{D}_{\vec{\widehat{v}}_e})$
\begin{align}
h^{1,0}(\widehat{D}_{\vec{\widehat{v}}_e})=&~l^*(\widehat{e}^\circ)\,.
\end{align}

\subsubsection{Face divisors}
We finally study the face divisors due to a point $\vec v_f$ interior to a face $f$ of $\Delta^\circ.$ The stratification of the face divisors $\widehat{D}_{\vec{\widehat{v}}_f}$ is given by
\begin{equation}\label{eqn:face div in b stratification}
\widehat{D}_{v_f}=Z_{\widehat{f}}\times \left((\Bbb{C}^*)^2\coprod_{f\supset t_1\supset \vec v_f}\Bbb{C}^*\coprod_{f\supset t_2\supset \vec v_f} (pt)\right)\,.
\end{equation} 
We first compute $h^{2,2}(\widehat{D}_{\vec{\widehat{v}}_f})$:
\begin{align}
\begin{split}
h^{2,2}(\widehat{D}_{\vec{\widehat{v}}_f})=& ~e^{0,0}(Z_{\widehat{f}})e^{2,2}((\Bbb{C}^*)^2)\\
=&~(1+l^*(\widehat{f}^\circ))\,.
\end{split}
\end{align}
Next, we compute $h^{1,1}(\widehat{D}_{\vec{\widehat{v}}_f})$:
\begin{align}
\begin{split}
h^{1,1}(\widehat{D}_{\vec{\widehat{v}}_f})=&~ e^{0,0}(Z_{\widehat{f}}) \left( e^{1,1}((\Bbb{C}^*)^2)+\sum_{f\supset t_1\supset \vec v_f} e^{1,1}(\Bbb{C}^*)\right)\\
=&~(-2+\sum_{f\supset t_1\supset \vec v_f} 1)(1+l^*(\widehat{f}^\circ))\,.
\end{split}
\end{align}
$h^{2,0}(\widehat{D}_{\vec{\widehat{v}}_f})$ is simply equal to zero, because there is no stratum that contributes non-trivial $e^{2,0}.$ For the same reason, we obtain $h^{1,0}(\widehat{D}_{\vec{\widehat{v}}_f})=0.$ 

%\subsubsection{Index formulae}

\section{Topology of vertical divisors in elliptic Calabi-Yau fourfolds}\label{sec:fourfold analysis}
\subsection{Comments on the F-theory uplift}
\label{sec:Comments on the F-theory uplift}
In this section, we comment on the construction of an elliptic CY fourfold $Y_4$ 
\begin{equation}
E \hookrightarrow Y_4\overset{\pi_E}{\rightarrow} B_3\,.
\end{equation}
A more complete exposition on the construction of the F-theory uplift will be presented in elsewhere \cite{nefpartition}. As was studied in previous sections, the orientifold $B_3$ is embedded into $\widehat{V}_4$ as a hypersurface belonging to the class $-K_{\widehat{V}_4}-\mathcal{L}.$ The first Chern class of the base threefold
\begin{equation}
c_1(B_3)=L\,,
\end{equation}
is the same as half the O7-plane divisor class. We embed the elliptic fibration $\pi_E$ as the anti-canonical hypersurface over a $\Bbb{P}_{[2,3,1]}$ fibration over $B_3$. We choose the $\mathbb P_{[2,3,1]}$ fibration such that the homogeneous coordinates $X,~Y,$ and $Z$ of $\Bbb{P}_{[2,3,1]}$ are sections of $\mathcal{O}(1)^{\otimes 2} \otimes \mathcal{L}^{\otimes 2}$, $\mathcal{O}(1)^{\otimes 3}\otimes \mathcal{L}^{\otimes 3}$, and $\mathcal{O}(1)$, respectively. Because $B_3$ is generically non-toric, the elliptic fourfold $Y_4$ is generically a complete intersection in a toric six-fold which is a $\Bbb{P}_{[2,3,1]}$ fibration over $\widehat{V}_4.$ 

We recall that the first Chern class $c_1(B_3)$ can be represented as 
\begin{equation}
L=\sum c_i \widehat D_i\,,
\end{equation}
where $c_i\in\Bbb{Z}_{\geq0}$ and $\widehat{D}_i$ is a basis of divisors for $B_3$. This implies that a $\Bbb{P}_{[2,3,1]}$ fibration twisted by appropriate tensor powers of the line bundle $\mathcal{L}$ can be constructed as follows. We define a map $\pi_{\Bbb{P}_{[2,3,1]}}:V_6\rightarrow \widehat{V}_4$ such that, given $\vec{\widehat{v}} \in \partial \Delta^{\circ}$, we have
\begin{equation}
\pi_{\Bbb{P}_{[2,3,1]}}^{-1}:(\widehat{v}_1,\widehat{v}_2,\widehat{v}_3,\widehat{v}_4)\mapsto (\widehat{v}_1,\widehat{v}_2,\widehat{v}_3,\widehat{v}_4,-2c_i,-3c_i)\,,
\end{equation}
and add the following points to the resulting toric rays:
\begin{equation}
\left(\begin{array}{cccccc}
0&0&0&0&0&0\\
0&0&0&0&0&0\\
0&0&0&0&0&0\\
0&0&0&0&0&0\\
1&0&-2&0&-1&-1\\
0&1&-3&-1&-1&-2
\end{array}
\right)\,.
\end{equation}
This procedure is a generalization of the standard stacking\footnote{For studies of elliptic fibration structures in toric hypersurface Calabi-Yau manifolds, including those realized by standard stacking, see, for example, \cite{Candelas:1996su,Kreuzer:1997zg,Candelas:1997eh,Perevalov:1997vw,Skarke:1998yk,Huang:2018gpl,Huang:2018vup}.} for complete intersection elliptic CY manifolds. The convex hull of the toric rays of $V_6$ can define a reflexive polytope, but not always. It should be noted that having the complete intersection alone will not define the elliptic phase of the CY fourfold $Y_4,$ as one has to find a phase that respects the toric morphism $\pi_E.$ We note that because $Y_4$ admits a global Sen limit, the only non-Higgsable\footnote{In F-theory compactifications on a singular elliptic CY defined by a Weierstrass model $y^2 = x^3 + f x +g $, a non-Higgsable gauge group factor appears over a codimension-one component of the discriminant locus $\Delta = 4 f^3 + 27 g^2 = 0$ in the base of the elliptic CY when the restriction of the sections $f,g$ to this component vanish to orders greater then (resp.) $1,2$. These gauge factors are referred to as ``non-Higgsable'' because they cannot be broken by charged matter in a manner compatible with supersymmetry---see, e.g., \cite{Morrison:2012np,Morrison:2012js,Morrison:2014lca,Halverson:2015jua} for further discussion.} clusters that can appear in $Y_4$ have $SO(8)$ gauge groups. Furthermore, these non-Higgsable clusters and O7-planes cannot intersect each other, because otherwise this would imply the existence of codimension-two fixed loci under the orientifold involution, which would break supersymmetry. We leave a detailed study of such elliptic phases for future work.

\subsection{Anatomy of vertical divisors}

\begin{figure}
\centering
\begin{tikzcd}[row sep=tiny, column sep=tiny]
&&&h^{0,0}&&&\\
&&h^{1,0}&&h^{0,1}&&\\
&h^{2,0}&&h^{1,1}&&h^{0,2}&\\
h^{3,0}&&h^{2,1}&&h^{1,2}&&h^{0,3}\\
&h^{3,1}&&h^{2,2}&&h^{1,3}&\\
&&h^{3,2}&&h^{2,3}&&\\
&&&h^{3,3}&&&
\end{tikzcd}
\caption{Hodge diamond of a divisor $\overline{D}\subset Y_4.$}
\end{figure}
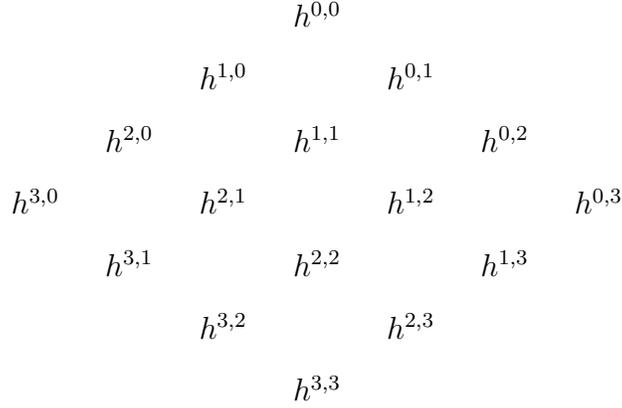

We are finally in a position to study the topology of vertical prime toric divisors in the CY fourfold $Y_4$. We call an irreducible divisor $\overline{D}\subset Y_4$ \emph{vertical} if there is a divisor $\widehat{D}\subset B_3$ such that $\overline{D}$ is an irreducible component of $\pi_E^{-1}(\widehat{D}).$ We have introduced the notion of irreducibility here because if $\widehat{D}$ is a non-Higgsable component of the discriminant locus of the elliptic fibration, then in the resolved CY fourfold $Y_4,$ $\pi_E^{-1}(\widehat{D})$ is not a torus fibration but rather, is a multiple $\Bbb{P}^1$ fibration over $\widehat{D}$ as a result of the resolution.

We list below all different possible classes of vertical prime toric divisors $\overline{D}$ :
\begin{itemize}
	\item $\overline{D}=\pi_E^{-1}(\widehat{D}),$ where $\dim \mathcal{L}|_{\widehat{D}}=2.$
	\item $\overline{D}=\pi_E^{-1}(\widehat{D}),$ where $\dim \mathcal{L}|_{\widehat{D}}=1.$
	\item $\overline{D}=\pi_E^{-1}(\widehat{D}),$ where $ \mathcal{L}|_{\widehat{D}}\neq\mathcal{O}_{\widehat{D}}$ but $\dim\mathcal{L}|_{\widehat{D}}=0.$
	\item $\overline{D}=\pi_E^{-1}(\widehat{D}),$ where $ \mathcal{L}|_{\widehat{D}}=\mathcal{O}_{\widehat{D}}.$ In this case, $\dim\mathcal{L}|_{\widehat{D}}=0.$
	\item An irreducible component of $\pi_{E}^{-1}(\widehat{D})$ where $\widehat{D}$ is a non-Higgsable $SO(8)$ seven-brane locus. In this case, $\dim\mathcal{L}|_{\widehat{D}}=-\infty.$
\end{itemize}
Let us explain the above classification in more detail. The reader may be confused by the above usage of ``dimension'' in connection with the line bundle $\mathcal{L}|_{\widehat{D}},$ as any line bundle is by definition a one-dimensional vector space. In this context, we use $\dim \, \mathcal L_{\widehat{D}}$ to denote the Iitaka dimension of the line bundle $\mathcal L|_{\widehat{D}}$, which is a mathematical quantity that characterizes the structure of the line bundle $\mathcal{L}|_{\widehat{D}}$ and of the corresponding elliptic fibration twisted by $\mathcal{L}|_{\widehat{D}}.$

To define the Iitaka dimension of a line bundle, we need to introduce some additional mathematical notions (see, e.g., \cite{lazarsfeld2017positivity} for more details). Let $\mathcal L$ be a line bundle defined on an algebraic variety $X$, and consider
%we define the graded ring 
%\begin{equation}
%C_{\mathcal L}:=\{ H^0(X,\mathcal L^{\otimes n})|0\leq n,~ n\in \Bbb{Z}\}\,.
%\end{equation}
%The graded ring $C_{\mathcal L}$ can be also understood as a cone or a semigroup. Note that projectivization of a graded subring defines a projective variety. %$S_L.$ T
%This notion is particularly useful when $X$ is a toric variety. In particular, we can use this fact to define a r
the rational map
\begin{equation}
\phi_n:= X\rightarrow \mathbb PH^0(X,\mathcal L^{\otimes n})\,
\end{equation}
associated to the complete linear system $|\mathcal L^{\otimes n}|$. We denote by $\Phi_n$ the image of $\phi_n$:
	\begin{equation}
		\Phi_n:=\phi_n(X)\subset \mathbb PH^0(X,\mathcal L^{\otimes n}).
	\end{equation} 
The Iitaka dimension of a line bundle $\mathcal L$ on $X$ is defined to be the maximal dimension of the image of the rational map $\phi_n$, 
 \begin{equation}
 	\dim \mathcal  L:=\max_{n} \{\dim \Phi_n\},
\end{equation}
where in the above equation $n \in \mathbb Z_{\geq 0}$ and is taken over all values such that $H^{0}(X,\mathcal L^{\otimes n}) \ne 0$. Note that when $H^{0}(X,\mathcal L^{\otimes n} ) =0$ for all $n >0$, by convention $\text{dim}\, \mathcal L = -\infty$. We then have 
	\begin{align}
	\label{Iitakadims}
		\dim\, \mathcal L \in \{ -\infty, 0,\dots, \dim(X) \}.
	\end{align}
 When $\widehat{D}$ is a toric two-fold, the Iitaka dimension of $\mathcal{L}|_{\widehat{D}}$ is given by the equation \cite{lazarsfeld2017positivity}
\begin{equation}
\dim \mathcal{L}|_{\widehat{D}}:=\dim \Delta_{\mathcal{L}|_{\widehat{D}}}\,,
\end{equation}
where $\Delta_{\mathcal{L}|_{\widehat{D}}}$ the Newton polytope for the global sections of $\mathcal{L}|_{\widehat{D}}.$ In the special case that $\mathcal L$ is the canonical bundle of $X$, the Iitaka dimension is equal to the Kodaira dimension of $X$.

The reason why we are interested in the dimension of the line bundle is quite simple. By design, the Iitaka dimension counts the number of directions that can change the relative sizes of the sections of $\mathcal L^{\otimes n}.$ Therefore, the Iitaka dimension of the twisting line bundle of the elliptic fibration counts the number of independent directions along which the discriminant (and more importantly, the $j$-invariant) can vary. Hence, at a generic point in moduli space, the elliptic fibration over $\widehat{D}$ can vary over a $(\dim \mathcal{L}|_{\widehat{D}})$-dimensional subspace of $\widehat{D}.$ Since the structure of the Hodge numbers of vertical divisors depends implicitly on the Iitaka dimension of $\mathcal{L}|_{\widehat{D}},$ we can use the Iitaka dimension to partition the possible types of vertical divisors into separate cases and study them one-by-one.

Before analyzing each type of irreducible vertical divisor that can appear in $Y_4$, we first explain our overall strategy for computing the Hodge numbers of $\overline{D}$. Other than exceptional cases where $\mathcal{L}|_{\widehat{D}}$ has Iitaka dimension less than or equal to 0, the Hodge numbers of $\widehat{D}$ we computed in the previous sections can be used to compute all Hodge numbers but $h^{2,1}(\overline{D})$ of the divisor $\overline{D} =\pi_E^{-1}(\widehat{D})$. Although we do not know of a simple combinatorial way to compute $h^{2,1}(\overline{D})$ at this point, because the only missing data in the Hodge diamond is $h^{2,1}(\overline{D})$, we can compute the Euler characteristic of $\overline{D}$ to indirectly compute $h^{2,1}(\overline{D})$, thus completing the computation of the Hodge numbers of $\overline{D}$. On the other hand, in the exceptional cases, we can compute all of the Hodge numbers directly without needing to compute Euler characteristic of $\overline{D}$ as an intermediate step.

\subsubsection{Computation of the Hodge vector for $\dim\mathcal{L}|_{\widehat{D}}\geq0$ and $\mathcal{L}|_{\widehat{D}}\neq\mathcal{O}_{\widehat{D}}.$}\label{sec:521}
Let us start by computing the Hodge vector for the cases where $\dim\mathcal{L}|_{\widehat{D}}\geq0$ and $\mathcal{L}|_{\widehat{D}}\neq\mathcal{O}_{\widehat{D}}.$ An important comment is in order: By construction, $Y_4$ is an elliptic fibration over an O3/O7-orientifold. This implies that the only allowed non-Higgsable clusters are SO(8) stacks, and there is always a global Sen limit. This weakly-coupled type IIB limit will be useful for us. Let $\cup_a U_a$ be a collection of local patches that cover $\widehat{D}$ such that in each $U_a,$ the elliptic fiber trvilalizes. For each local patch $U_a,$ we can define integral one-forms $dz_a$ and $d\bar{z}_a$ on the elliptic fiber. When moving from one patch to another, say from $U_a$ to $U_b,$ the integral one-forms undergo an $SL(2,\Bbb{Z})$ transformation. Now let us make a complete loop around an $SO(8)$ stack, or if there is no $SO(8)$ stack intersecting $\widehat{D}$  a complete loop around a system of branes where the total RR-charge is zero. The monodromy transformation of the one-forms $dz_a$ and $d\bar{z}_a$ along such a complete loop is $-1\in SL(2,\Bbb{Z})$ \cite{Sen:1996vd}. Unlike the space $H^{1,0} \oplus H^{0,1}$ of one-forms on the elliptic fiber, the group $H^{0,0}$ remains invariant---this can be seen from the fact that $dz_a\wedge d\bar{z}_a$ is invariant under $-1\in SL(2,\Bbb{Z}).$  This implies that a cycle in $B_3$ that is even (odd) under the orientifold involution can be combined with an even (odd) cycle in the fiber to form a non-trivial cycle in $\varphi_{\mathcal I_{\vec p}}(\widehat{D})$.  As a result, we obtain an inequality\footnote{More explicitly: since $\overline D$ is elliptically-fibered, we enumerate $(p,q)$ cycles of $\overline D$ by counting cycles of $(p',q')$ cycles of the $T^2$ fiber that pair with $(p-p',q-q')$ cycles of $D$. Since $h^{0,0}(T^2)= h^{1,0}(T^2) = h^{0,1}(T^2) = h^{1,1}(T^2)=1$, we must sum over all contributions from $H^{p-p',q-q'}(D)$ with $(p',q') \in \{ (0,0), (1,0), (0,1), (1,1)\}$.}
\begin{equation}\label{eqn:hpq inequality}
h^{p,q}(\overline{D})\geq h^{p,q}_+\left(\varphi_{\mathcal I_{\vec p}}(\widehat{D})\right)+h^{p-1,q}_-\left(\varphi_{\mathcal I_{\vec p}}(\widehat{D})\right)+h^{p,q-1}_-\left(\varphi_{\mathcal I_{\vec p}}(\widehat{D})\right)+h^{p-1,q-1}_+\left(\varphi_{\mathcal I_{\vec p}}(\widehat{D})\right)\,.
\end{equation}
Note that if the divisor $\widehat{D}$ is wrapped by an O7-plane\footnote{Because we are assuming that the Iitaka dimension of the twisting line bundle is not $-\infty,$ $\widehat{D}$ cannot be rigid}, then we have
\begin{equation}
\varphi_{\mathcal I_{\vec p}}(\widehat{D})=2D\,,
\end{equation}
otherwise we have
\begin{equation}
\varphi_{\mathcal I_{\vec p}}(\widehat{D})=D\,.
\end{equation}
It is worthwhile to stress that we have an inequality not an equality in \eqref{eqn:hpq inequality}, as can be inferred from numerous examples---see \S\ref{Examples}. The reason for having an inequality comes from the fact that there can be a non-trivial loop or even a chain in $\widehat{D}$ that can go between seven-branes that can generate a non-trivial $SL(2,\Bbb{Z})$ action to form a non-trivial cycle. Because this counting can be very subtle in general, we do not attempt to compute $h^{1,1}(\overline{D})$ and $h^{2,1}(\overline{D})$ using this method except in a special case that we describe in the next section.

We can now compute the Hodge vector $h^{\bullet,0}(\overline{D}).$ Despite the fact that counting of $h^{p,q}(\overline{D})$ by carefully examining the $SL(2,\Bbb{Z})$ monodromy transformations is very difficult in general, at least for $h^{\bullet,0}(\overline{D}),$ we can get an exact answer. A notable fact about $h^{\bullet,0}(\overline{D})$ is that it counts the number of uncharged zero modes of an Euclidean M5-branes wrapping $\overline{D}.$ Since, in the orientifold picture, the uncharged zero modes of a Euclidean D3-brane wrapping $\widehat{D}$ are counted by $h^{\bullet,0}_\pm(\varphi_{\mathcal I_{\vec p}}(\widehat{D})),$ on physical grounds, we expect that the following equation must hold true to match the total number of fermion zero modes:
\begin{equation}\label{eqn:zero mode sum}
\sum_{i=0}^3 h^{i,0}(\overline{D})=\sum_{i=0}^2h^{i,0}_\pm \left(\varphi_{\mathcal I_{\vec p}}(\widehat{D})\right)\,.
\end{equation}
Because none of the Hodge numbers can be negative, \eqref{eqn:zero mode sum} implies that the following equality is also true:
\begin{equation}\label{eqn:hodge vectors}
h^{\bullet,0}(\overline{D})=h^{\bullet-1,0}_-\left(\varphi_{\mathcal I_{\vec p}}(\widehat{D})\right)+h^{\bullet,0}_+\left(\varphi_{\mathcal I_{\vec p}}(\widehat{D})\right)\,.
\end{equation}
Note that \eqref{eqn:zero mode sum} was obtained under the assumption that the presence of O3-planes does not alter the total number of fermion zero modes.\footnote{For the proof of \eqref{eqn:hodge vectors} in the absence of singularities, see \cite{Grassi:1997mr}.} We expect that \eqref{eqn:zero mode sum} remains true even when $\widehat D$ contains termina $\mathbb Z_2$ singularities signaling the presence of O3-planes. Although we are not aware of a mathematical proof for why \eqref{eqn:zero mode sum} remains true when $\widehat D$ contains terminal $\mathbb Z_2$ singularities, we have checked that this indeed the case in numerous examples. On physical grounds, we remark that it is not very surprising that the formula \eqref{eqn:zero mode sum} holds even when there are O3-planes in the vicinity of a Euclidean D3-brane, since O3-planes can only project in or out the zero modes on a Eulcidean D3-brane in a CY and do not generate any additional uncharged zero modes beyond the ones that existed before the orientifolding \cite{Polchinski:1995mt,Gimon:1996rq,Polchinski:1998rq,Polchinski:1998rr}.

\subsubsection{Computation of the Hodge numbers for $\dim\mathcal{L}|_{\widehat{D}}=0$ and $\mathcal{L}|_{\widehat{D}}\neq\mathcal{O}_{\widehat{D}}.$}
In this case, the complex structure of the elliptic fiber is frozen and does not vary along points in $\widehat{D},$ because all seven brane stacks that intersect $\widehat{D}$ are non-Higgsable $SO(8)$ stacks. As a result, we can regard the singular vertical divisor $\overline{D}^{(0)}$ before the blow up\footnote{That is, $\overline{D}$ maps to $\overline{D}^{(0)}$ under the blowdown map contracting the exceptional divisors of $Y_4$.} as a $\Bbb{Z}_2$ orbifold
\begin{equation}
\overline{D}=\left(\widehat{D}\times T^2\right)/\Bbb{Z}_2\,,
\end{equation}
where the action $\overline{\mathcal{I}}_p$ of $\Bbb{Z}_2$ group is given by
\begin{equation}
\overline{\mathcal{I}}_{\vec p}: (x,z)\mapsto (\mathcal{I}_{\vec p}(x),-z)\,,
\end{equation}
where in the above equation $x\in  D,$ and $z\in T^2.$ As a result, we find that in the case where $\dim\mathcal{L}|_{\widehat{D}}=0$ and $\mathcal{L}|_{\widehat{D}}\neq\mathcal{O}_{\widehat{D}}$ the inequality \eqref{eqn:hpq inequality} is saturated for singular $\overline{D}^{(0)}$:
\begin{equation}
h^{p,q}\left(\overline{D}^{(0)}\right)= h^{p,q}_+(D)+h^{p-1,q}_-(D)+h^{p,q-1}_-(D)+h^{p-1,q-1}_+(D)\,.
\end{equation} 
Note that by assumption, $\widehat{D}$ cannot correspond to a non-Higgsable $SO(8)$ stack. After resolving the singularities in the elliptic fiber due to $SO(8)$ seven brane stacks, we obtain
\begin{equation}
h^{p,q}(\overline{D})= h^{p,q}_+(D)+h^{p-1,q}_-(D)+h^{p,q-1}_-(D)+h^{p-1,q-1}_+(D)\,,
\end{equation}
for $(p,q)\neq(1,1),~(2,2),$ and
\begin{equation}
h^{1,1}(\overline{D})=h^{2,2}(\overline{D})= h^{1,1}_+(D)+2h^{1,0}_-(D)+1+4n_{SO(8)}(\widehat{D})\,.
\end{equation}

\subsubsection{Computation of the Hodge numbers for  $\mathcal{L}|_{\widehat{D}}=\mathcal{O}_{\widehat{D}}.$}
In this case, the elliptic fibration is trivial, hence $\pi_E^{-1}(\widehat{D})$ has the topology of
\begin{equation}
\overline{D}=T^2\times \widehat{D}\,.
\end{equation}
Thus, by direct computation, we obtain the following Hodge numbers for $\overline{D}$:
\begin{equation}
h^{p,q}(\overline{D})=h^{p,q}(\widehat{D})+h^{p-1,q}(\widehat{D})+h^{p,q-1}(\widehat{D})+h^{p-1,q-1}(\widehat{D})\,.
\end{equation}

\subsubsection{Computation of the Hodge numbers for  $\dim\mathcal{L}|_{\widehat{D}}=-\infty.$}
In this case, the divisor $\widehat{D}$ is wrapped by a non-Higgsable $SO(8)$ stack. Therefore, an irreducible component of $\pi_E^{-1}(\widehat{D})$ has the topology of a $\mathbb P^1$ fibration, i.e. $\pi_E^{-1}(\widehat D)$ is birational to 
\begin{equation}
\Bbb{P}^1\times \widehat{D}\,.
\end{equation}
Using the fact that $h^{0,0}(\mathbb P^1) = h^{1,1}(\mathbb P^1) = 1,$  $h^{1,0}(\mathbb P^1) = h^{0,1}(\mathbb P^1)=0$, we find that the Hodge numbers of an irreducible component of $\pi_{E}^{-1}(\widehat D)$ are:\footnote{There is a conjecture, due to Batyrev and Dais \cite{Batyrev:1994ju}, which asserts that the Hodge numbers of a smooth crepant resolution of an algebraic variety defined over the complex numbers with at most Gorenstein canonical singularities are independent of the choice of resolution. Since the singular SO(8) model defined over a threefold base $B_3$ (where $\widehat D \subset B_3$ is the component of the discriminant locus over which the elliptic fibers develop an SO(8) Kodaira singularity) admits a crepant resolution in which the irreducible divisors comprising $\pi_{E}^{-1}(\widehat D)$ are $\mathbb P^1$ bundles over $\widehat D$ \cite{Esole:2017qeh}, we assume that the conjecture of Batyrev and Dais holds for $\pi_E^{-1}(\widehat D)$ and compute the Hodge numbers of $\pi_E^{-1}(\widehat D )$ by summing over the Hodge numbers of the irreducible components (birational to $\mathbb P^1 \times \widehat D$). See Section \ref{sec:push forward} for further discussion of the geometry of the SO(8) model.}
\begin{equation}
h^{p,q}(\mathbb P^1 \times \widehat{D})=h^{p,q}(\widehat{D})+h^{p-1,q-1}(\widehat{D})\,.
\end{equation} 

\subsubsection{Vertical divisors with $\dim\mathcal{L}|_{\widehat{D}}=2.$}
In this case, the lifted divisor $\overline{D}$ can be embedded into $\Bbb{P}_{[2,3,1]}$ fibration over $\widehat{D},$ call it  $V_4(\widehat{D})$, as a generic hypersurface,
\begin{equation}
\overline{D}\subset V_4(\widehat{D})\,,
\end{equation} 
where the hypersurface is defined by a section of a big line bundle in $V_4(\widehat{D}).$ As described in \S\ref{sec:521}, the only Hodge numbers we have not yet computed are $h^{1,1}(\overline{D})$ and $h^{2,1}(\overline{D}).$ 

To compute $h^{1,1}(\overline{D}),$ we apply the Shioda-Tate-Wazir theorem \cite{shioda1972elliptic,tate1965algebraic,tate1965conjectures,wazir2004arithmetic}, which states that the rank of Neron-Severi group of $\overline{D}$ is\footnote{As stated, (\ref{eqn:STW d2}) assumes that $\overline D$ only has a zero section and does not have any additional independent rational sections, which would correspond to non-trivial generators of the group of Mordell-Weil group (extended fiber-wise over the $\widehat D$). Physically, this means that the gauge group associated to the singular limit of $\overline D$ does not have any ``geometric'' U(1) factors (i.e., that the free abelian part of the gauge group associated with $\overline D$ is trivial) \cite{Weigand:2018rez}.}
\begin{equation}\label{eqn:STW d2}
\text{rank}( NS(\overline{D})) =\text{rank}( NS(\widehat{D}))+1+\sum_a\text{rank}(G_a)\,,
\end{equation}
where $G_a$ is the gauge group of the seven-branes that intersect $\widehat{D}.$ This theorem teaches us that the divisor class is generated by the divisors in the base $\widehat{D},$ the section of the elliptic fibration, and the exceptional divisors due to the blow up. Because all the non-Higgsable clusters carry $SO(8)$ gauge group, we can further simplify \eqref{eqn:STW d2} to
\begin{equation}
{rank}( NS(\overline{D})) ={rank}( NS(\widehat{D}))+1+4 n_{SO(8)}(\widehat{D})\,,
\end{equation}
where $n_{SO(8)}(\widehat{D})$ counts the number of non-Higgsable $SO(8)$ stacks seen in $\widehat{D}.$ Note that $n_{SO(8)}$ can be larger than the number of SO(8) stacks intersecting $\widehat{D}$.\footnote{Although there can be multiple seven brane loci in $B_3$ over which there is no non-abelian gauge symmetry enhancement in codimension one, it is nevertheless possible for there to be a gauge enhancement at their intersections (i.e., in codimension two). When these codimension-two loci lie along O7-planes, the enhanced gauge group must be SO(8). It is furthermore possible for $\widehat D$ to intersect these codimension-two SO(8) loci, which implies that the number of SO(8) singularities in $\widehat D$ can, strictly speaking, be larger than the number of SO(8) stacks over codimension-one loci in $B_3$---see Section \ref{Examples} for examples.} 

Next, we relate the Picard rank to $h^{1,1}(\overline{D})$ on a case-by-case basis. Let us start with the case that $\widehat{D}$ is a 2-face divisor. As a 2-face divisor is a toric two-fold, we have
\begin{equation}
\text{rank}( NS(\widehat{D}))=h^{1,1}(\widehat{D})\,.
\end{equation}
Furthermore, due to the fact that $h^2(\overline{D},\mathcal{O}_{\overline{D}})=0,$ $H^2(\overline{D},\Bbb{Z})$ is a lattice of divisors 
\begin{equation}
\text{rank}( NS(\overline{D}))=h^{1,1}(\overline{D})\,.
\end{equation}
As a result, we compute for a 2-face divisor
\begin{equation}
h^{1,1}(\overline{D})=h^{1,1}(\widehat{D})+1+4 n_{SO(8)}(\widehat{D})\,.
\end{equation}

Next we come to the case that $\widehat{D}$ is an edge divisor. To study this case, we first recall that the topology of an edge divisor is given as a set of blow ups at points in the base of $\Bbb{P}^1$ fibration over a Riemann surface $\Sigma_g$ of genus $g$. To simplify the analysis, we blow down the exceptional divisors. Let us denote the blow down of $D$ by $D^{(0)},$ and similarly the blow down of $\widehat{D}$ by $\widehat{D}^{(0)}.$ By construction, $\Sigma_g$ is a hypersurface in a toric twofold. Let us first establish our notational conventions for $D.$ We denote the hyperplane class of $\Sigma_g$ by $H,$ the section of the $\Bbb{P}^1$ by $S,$ and the twisting line bundle of the $\Bbb{P}^1$ by $T=t H.$ We denote the homogeneous coordinate (in the Cox ring) of the class $S$ by $y_1,$ and similarly the homogeneous coordinate of the class $S+T$ by $y_2.$ We will argue that the only possible action of the orientifolding on the divisors in $D$ such that the Iitaka dimension of $\mathcal{L}|_{\widehat{D}}$ is given by
\begin{equation}\label{eqn:dim2 edge orientifold action1}
\mathcal I_{\vec p}:y_1\mapsto -y_1\,,
\end{equation}
and
\begin{equation}\label{eqn:dim2 edge orientifold action2}
g\cdot \mathcal I_{\vec p} \cdot g^{-1}:y_2\mapsto -y_2\,.
\end{equation}
This restriction comes from the fact that a line bundle $n H$ for $n>0$ has Iitaka dimension 1, so for the Iitaka dimension of $\mathcal{L}|_{\widehat{D}}$ to be 2, the first Chern class of $\mathcal{L}|_{\widehat{D}}$ must satisfy
\begin{equation}
L=n H+m S\,,
\end{equation}
where $n,m>0.$ This implies that the orientifold action should flip the sign of at least one of $\{y_1,y_2\}.$ But then due to the toric rescaling, a map $y_1\mapsto -y_1$ is equivalent to $y_2\mapsto -y_2.$ So, the orientifolding must flip the sign of $y_1$ and $y_2,$ leading to our claims \eqref{eqn:dim2 edge orientifold action1} and \eqref{eqn:dim2 edge orientifold action2}. We next argue that $y_1=0$ and $y_2=0$ are the only O7-plane loci. This can be easily seen from the supersymmetry condition: To preserve supersymmetry in the presence of O3/O7-planes, there cannot be any non-trivial intersection between O7-plane loci. If an O7-plane also wraps a divisor class $nH,$ for $n>0,$ then this O7-plane inevitably intersects an O7-plane at $y_1=0$ or $y_2=0.$ Hence, there cannot be an O7-plane that wraps a divisor with class $nH.$ 

We next examine the implications of the orientifold actions \eqref{eqn:dim2 edge orientifold action1} and \eqref{eqn:dim2 edge orientifold action2} on the topology of $\widehat{D}$ and $\overline{D}.$ Because $y_1=0$ and $y_2=0$ are the O7-plane loci in $D^{(0)},$ under the refinement map, we have
\begin{equation}
\varphi_I(\widehat{y}_i)=y^2_i\,.
\end{equation}
This means that the topology of $\widehat{D}$ is again a $\Bbb{P}^1$ fibration over the same Riemann surface $\Sigma_g,$ but with the different twisting for the $\Bbb{P}^1$ fiber, given by
\begin{equation}
\widehat{T}=2t H\,.
\end{equation}
As a result, we obtain
\begin{equation}
h^{p,q}_+(D^{(0)})=h^{p,q}(D^{(0)})\,,
\end{equation}
and also
\begin{equation}
h^{p,q}_+(D)=h^{p,q}(D)\,.
\end{equation}
Hence, we have
\begin{equation}
h^{2,0}(\overline{D})=h^{2,0}(\widehat{D})=0\,,
\end{equation}
which allows us to conclude
\begin{equation}
\text{rank}( NS(\widehat{D}))=h^{1,1}(\widehat{D})\,,
\end{equation}
and
\begin{equation}
\text{rank}( NS(\overline{D}))=h^{1,1}(\overline{D})\,.
\end{equation}
For an edge divisor $\widehat{D}$, we thus compute
\begin{equation}
h^{1,1}(\overline{D})=h^{1,1}(\widehat{D})+1+4 n_{SO(8)}(\widehat{D})\,.
\end{equation}

We finally come to the case that $\widehat D$ is a vertex divisor. We have to be careful here, because in this case $h^{2,0}(\widehat{D})$ can be non-zero. If $h^{2,0}$ is non-zero, then not all contributions to $h^{1,1}$ correspond to holomorphic divisors. Let us fix the complex structure $z$ of $\widehat{D}.$ At fixed complex structure, there are $h^{2,0}(\widehat{D})$ independent 2-cycles $\gamma_i$ for $i=1,\dots,h^{2,0}(\widehat{D}).$ These 2-cycles $\gamma_i$ are not divisors. Now let us change the complex structure by $\epsilon.$ Then the 2-cycles $\gamma_i$ are now linear combinations of cycles in $H^{2,0}$ and $H^{1,1}.$ Again, the cycles $\gamma_i$ cannot be divisors because the K\"ahler form restricted to an arbitrary linear combination of $\gamma_i$ vanishes. We therefore deduce that $H^2$ splits into horizontal and vertical subgroups, where the vertical part is spanned by divisors. In general, we have
\begin{equation}
\text{rank}( NS(\widehat{D}))\leq h^{1,1}(\widehat{D})\,.
\end{equation}
To use the Shioda-Tate-Wazir theorem to compute $h^{1,1}(\overline{D}),$ we first argue that the following equation holds
\begin{equation}\label{eqn:horizontal}
h^{1,1}(\widehat{D})-\text{rank}( NS(\widehat{D}))=h^{1,1}(\overline{D})-\text{rank}( NS(\overline{D}))\,.
\end{equation}
The reason is quite simple. As was shown in the previous section,
\begin{equation}\label{eqn:cohom under projection0}
h^{2,0}(\overline{D})=h^{2,0}(\widehat{D})\,.
\end{equation}
In fact, we have an even stronger statement, where the projection map $\pi_E:\overline{D}\rightarrow \widehat{D}$ induces an isomorphism
\begin{equation}\label{eqn:cohom under projection}
\pi_{E*}\left(H^{2,0}(\overline{D})\right)=H^{2,0}(\widehat{D})\,.
\end{equation}
To study $H^{2,0}(\widehat{D})$ in more detail, let us treat $\widehat{D}$ as a hypersurface in a toric threefold $V_3,$ where the defining equation $f|_{\widehat{D}} = 0$ of $\widehat{D}$  is inherited from the defining equation $f$ of $B_3.$ At fixed complex structure of $\widehat{D}$ at $z_0,$ which only depends on the complex structure of $B_3,$ elements of $H^{2,0}(\widehat{D}_{z_0})$ are generated by
\begin{equation}\label{eqn:h20 vert}
\oint_{f_{\widehat{D}}=\epsilon} x^{\vec{m}} \frac{\omega_{V_3}}{f|_{\widehat{D}}} \,,
\end{equation}
where $\omega_{V_3}$ is the top differential form defined on $V_3,$ and $\vec{m}$ is in strict interior of $\widehat{v}^\circ.$ The crucial point here again is that the elements of $H^{2,0}$ only depend on the complex structure of $B_3.$ Now, we can study $\overline{D}$ as a complete intersection in $V_5,$ which is  $\Bbb{P}_{[2,3,1]}$ fibration over $V_3.$ The pullback of
\begin{equation}
\pi_{V}:V_5\rightarrow V_3
\end{equation}
has a natural restriction to the pullback of the projection map $\pi_E,$ and therefore we can conclude that $H^{2,0}(\overline{D})$ is spanned by
\begin{equation}\label{eqn:h20 vert2}
\oint_{\pi_E^*(f|_{\widehat{D}})=\epsilon} \pi_E^{*}(x^{\vec{m}}) \frac{\pi_E^{*}(\omega_{V_3})}{\pi_E^*(f|_{\widehat{D}})} \,.
\end{equation}
Now, the elements in $H^{1,1}(\widehat{D})$ that are orthogonal to the Neron-Severi group are then generated by the complex structure deformation $z_0\mapsto z_0+\epsilon$ of the forms \eqref{eqn:h20 vert}. And similarly, the elements in $H^{1,1}(\overline{D})$ that are orthogonal to the Neron-Severi group $NS(\overline{D})$ are generated by the complex structure deformation $z_0\mapsto z_0+\epsilon$ of the forms \eqref{eqn:h20 vert2}. Because changing coefficients of the Weierstrass form does not deform \eqref{eqn:h20 vert2}, we conclude that
\begin{equation}
h^{1,1}(\widehat{D})-\text{rank}( NS(\widehat{D}))=h^{1,1}(\overline{D})-\text{rank}( NS(\overline{D}))\,.
\end{equation}
This means that the horizontal component of $H^2(\widehat{D})$ is isomorphic to $H^2(\overline{D}).$ Then, assuming that $H^2$ is a direct sum of the horizontal cohomology groups and the Neron-Severi group, for vertex divisors we conclude
\begin{equation}
h^{1,1}(\overline{D})=h^{1,1}(\widehat{D})+1+4 n_{SO(8)}(\widehat{D})\,.
\end{equation}

%To compute $h^{1,1}(\overline{D}),$ let us first study the case where $\widehat{D}$ is simply connected, hence $h^{1,0}(\widehat{D})=0.$ In this case, the rank of the Neron-Severi group is equal to $h^{1,1}(\overline{D}),$ hence 
%\begin{equation}
%h^{1,1}(\overline{D})=h^{1,1}(\widehat{D})+1+4 n_{SO(8)}(\widehat{D})\,,~ \text{ if }~ \pi_1(\widehat{D})=0\,.
%\end{equation}
%Now let us move on to the more non-trivial case where $h^{1,0}(\widehat{D})\neq0.$ In this case, from the inequality \eqref{eqn:hpq inequality} we expect at least
%\begin{equation}\label{eqn:weak bound on h11}
%h^{1,1}(\overline{D})\geq h^{1,1}(\widehat{D})+1+2h^{1,0}_-(\varphi_{\mathcal I_{\vec p}}(\widehat{D}))\,.
%\end{equation}
%In fact, we can improve the inequality \eqref{eqn:weak bound on h11} by noting that blow up along the SO(8) singularities will increase $h^{1,1}(\overline{D}),$ which leads us to a more stronger bound
%\begin{equation}\label{eqn:strong bound on h11}
%h^{1,1}(\overline{D})\geq h^{1,1}(\widehat{D})+1+2h^{1,0}_-(\varphi_{\mathcal I_{\vec p}}(\widehat{D}))+4 n_{SO(8)}(\widehat{D})\,.
%\end{equation}
%It is coceivable that the equality for \eqref{eqn:strong bound on h11} may hold, but we will not atttempt to argue for it. 

Finally, we use the Euler characteristic $\chi(\overline{D})$ to complete the Hodge diamond. As we explain in \S\ref{sec:o3-planes}, in the presence of O3-planes that intersect $\widehat{D},$ the Euler characteristic we compute using the result of \S\ref{sec:push forward} is not the same as $\sum_{p,q} (-1)^{p,q}h^{p,q}(\overline{D}).$ The correct identification is
\begin{equation}
\chi_n(\overline{D})+2n_{O3}(\widehat{D})=\sum_{p,q} (-1)^{p,q}h^{p,q}(\overline{D})\,,
\end{equation} 
where $n_{O3}$ is the number of O3-planes that intersect $\widehat{D}.$ From this, we can compute $h^{2,1}(\overline{D})$
\begin{equation}\label{eqn:h21 dim2}
h^{2,1}(\overline{D})=-n_{O3}(\widehat{D})-\frac{1}{2} \chi_n(\overline{D})+\frac{1}{2}\sum_{(p,q)\neq(2,1)\&(1,2)} (-1)^{p,q}h^{p,q}(\overline{D})\,.
\end{equation}
Note that if $h^{1,0}(\widehat D) \ne 0$, we do not know of a simple combinatorial prescription that can be used to compute $h^{1,1}(\overline D)$, and hence the right-hand side of (\ref{eqn:h21 dim2}) must be computed by other means. On the other hand, if the divisor $\overline{D}$ is rigid (i.e. satisfies $h^{\bullet,0}(\overline{D})=(1,0,0,0)$, which is a case of prime interest because Euclidean M5-brane wrapping such divisors can generate a non-perturbative superpotential), we can simplify \eqref{eqn:h21 dim2} to
\begin{align}
h^{2,1}(\overline{D})=&~1+h^{1,1}(\overline{D})-n_{O3}(\widehat{D})-\frac{1}{2}\chi_n(\overline{D})\\
=&~2+h^{1,1}(\widehat{D})+4n_{SO(8)}(\widehat{D})-n_{O3}(\widehat{D})-\frac{1}{2}\chi_n(\overline{D})\,.
\end{align}
This completes the Hodge diamond.

\subsubsection{Vertical divisors with $\dim\mathcal{L}|_{\widehat{D}}=1.$}\label{sec:dim1 line bundle}
This is perhaps the most non-trivial case. We do not know of a way to complete the Hodge diamond for this case in full generality. We approach the various possibilities for $\widehat{D}$ on a case-by-case basis. 

Let us start with 2-face divisors $\widehat{D}$. Per the stratification \eqref{eqn:face div in b stratification}, we know that 2-face divisors are toric two folds. Because all toric twofolds are rational surfaces, upon blowing down all the exceptional curves, we either arrive at $\Bbb{P}^2$ or $\Bbb{F}_n$ with $n\neq1.$ But, in this section, we will proceed somewhat differently. In the case of $\Bbb{F}_1$ we shall not blow down the exceptional curve and treat $\Bbb{F}_1$ seprately from $\Bbb{P}^2.$ So, $\Bbb{P}^2$ will correspond to the case in which $\widehat{D}$ has a topology of $\Bbb{P}^2.$ Because adding exceptional divisors to a line bundle does not change the Iitaka dimension of the line bundle, without any loss of generality, we can analyze a minimal model of $\widehat{D}$, which we denote by $\widehat{D}_{m}.$\footnote{As a simple example, consider a blow up of $\Bbb{P}^2$ at a point. After the blow up, $\Bbb{P}^2$ becomes $\Bbb{F}_1.$ The hyperplane class $[H]$ in $\Bbb{P}^2$ is now mapped to a sum of the base and the fiber class. As a result, we see that the Iitaka dimension is 2 in both cases.} There is one subtlety. If the twisting line bundle defined on $\widehat{D},$ does not have a counterpart in the blow down of $\widehat{D},$ then such blow downs will destroy the crucial data we seek to understand. In this section, we will assume that the twisting line bundle is well defined under the blow downs to $\Bbb{F}_n.$ For more general cases, one can blow down the exceptional curves as far as one can, such that the twisting line bundle is well defined under such procedures, and understand the properties of the twisting line bundle in the blow down.

Now, we note that no line bundle of $\Bbb{P}^2$ has Iitaka dimension 1.\footnote{The first Chern class of any line bundle of $\Bbb{P}^2$ can be represented by $a H$, where $a$ is an integer and $H$ is the hyperplane class of $\Bbb{P}^2.$ When $a$ is negative, no tensor power of the line bundle will have global sections, which implies that the Iitaka dimension of the line bundle is $-\infty.$ When $a=0,$ the corresponding line bundle is the structure sheaf, which by definition has dimension $0.$ When $a>0,$ as the dimension of the Newton polytope is $2,$ the line bundle also has dimension $2.$} Therefore, the existence of a line bundle $\mathcal{L}|_{\widehat{D}}$ with dimension 1 indicates that the 2-face divisor must be $\Bbb{F}_{n\ne 1}$ or $\Bbb{F}_1.$ Furthermore, the line bundle $\mathcal{L}|_{\widehat{D}_m}$ should be a multiple of the hyperplane class $H$ in the base of the $\Bbb{P}^1$ fibration in $\widehat{D}_m$.\footnote{The reason for this is that any power of the line bundle whose first Chern class is the class of the section of the fibration can only have one global section. This implies that the section corresponds to a line bundle of Iitaka dimension 0. If the first Chern class of a line bundle $L=aH+b S$ for $a,b>0,$ where $H$ is the hyperplane class and $S$ is the section, then the corresponding Newton polytope again has dimension 2. Because the dimension of the Newton polytope for a line bundle $L=aH$ for any positive integer $a$ is always 1, the only possibility here is $L=aH.$ } To ensure that the class $c_1(\mathcal L|_{\widehat D_m})$ (which is one-half times the class of the O7-plane locus restricted to $\widehat D_m$) is an integer class, for convenience, we write
\begin{equation}
2c_1(\mathcal{L}|_{\widehat{D}_m})= a H\,,
\end{equation} 
where $a\in2 \Bbb{Z}_{>0}.$ Now, by acting on $\widehat{D}_m$ with a composition of blowups and blowdowns to go back to $\widehat{D},$ we learn that 
\begin{equation}
2c_1(\mathcal{L}|_{\widehat{D}})=a H+\sum_{i} b_i E_i\,,
\end{equation}
where $b_i$ is $0$ or $1,$ and $E_i$ is an exceptional divisor. Note that $a H$ class can be understood as the intersection between a non-rigid O7-plane class and $\widehat{D}.$ Similarly, for $b_i=1,$ $E_i$ class should be understood as the intersection between a rigid $SO(8)$ stack and $\widehat{D}.$ 

Because the elliptic fiber does not vary along the $\Bbb{P}^1$ fiber of $\widehat{D}_m,$ we can now understand the topology of $\overline{D}$ as a sequence of blowups of an algebraic threefold $\overline{D}_m$ with two fibers,
\begin{equation}
\begin{tikzcd}
&\Bbb{P}^1\arrow[d, "\pi_2"]&\Bbb{P}^1\arrow[d, "\pi_2"]\\
E\arrow[r,"\pi_1"]&\overline{D}_m \arrow[r] \arrow[d, ]& \widehat{D}_m \arrow[d,] \\
E\arrow[r,"\pi_1"]&S \arrow[r]&  \Bbb{P}^1
\end{tikzcd}\,,
\end{equation}
where $\pi_1$ is the elliptic fibration, and $\pi_2$ is the $\Bbb{P}^1$ fibration. To study the Hodge numbers of $\overline{D},$ we first study the topology of $S.$ Note that the Euler characteristic of $S$ is given by
\begin{equation}\label{eqn:euler S}
\chi(S)=12\int_{\Bbb{P}^1} c_1(\mathcal{L}|_{\Bbb{P}^1})=6a\,.
\end{equation}
Unlike higher-dimensional elliptic spaces, \eqref{eqn:euler S} cannot change due to a change of complex structure. Because the structure sheaf cohomology also does not change under the complex structure deformation, $h^{1,1}(S)$ is also an invariant. Although we know how to directly compute $h^{1,1}(S),$ this invariance allows us to read off $h^{1,1}(S)$ using an alternative method that we now describe. Note that 
\begin{equation}
\chi(S)=2-2h^{1,0}(S)+h^{1,1}(S)\,.
\end{equation}
Because we can easily compute $h^{1,0}(S)$ either using the Hirzebruch-Riemann-Roch theorem, or using the combinatorial structure of the Newton polytope, we can therefore compute $h^{1,1}(S)$ by computing $\chi(S).$ This result allows us to compute $h^{1,1}(\overline{D})$ as follows:
\begin{equation}
h^{1,1}(\overline{D})=1+h^{1,1}(S)\,.
\end{equation}
As a result, we can again compute $h^{2,1}(\overline{D})$ by using $\chi(\overline{D}).$  

We next consider the case that $\widehat{D}$ is a 1-face divisor or a vertex divisors. Using the stratification \eqref{eqn:edge div in b stratification}, we see that 1-face divisors are either blowups of ruled surfaces or rational surfaces. Although it is redundant, we can understand this from a slightly different perspective: As we have studied in \S\ref{sec:divisors in CY3} and \S\ref{sec:divisors in CY3orientifolds}, $h^{2,0}$ of 1-face divisors always vanishes. According to the adjunction formula, this implies that the normal bundle $N_{D/Z_3}$ is not effective (here, we view $D\subset Z_3$ as a divisor in a CY threefold $Z_3$). As a result, any positive power of $K_{D}$ is also not effective, which implies that the Kodaira dimension of $\mathcal O(D)$ is $-\infty$. Because $\widehat{D}$ can be understood as a $\Bbb{Z}_2$ orbifold of $D,$ it is expected that the Kodaira dimension of $\widehat{D}$ is also $-\infty.$ In fact, this can be argued in a very simple manner. Due to the adjunction formula, we have
\begin{equation}
K_{\widehat{D}}= -\mathcal{L}|_{\widehat{D}}+\widehat{D}|_{\widehat{D}}\,.
\end{equation}
As $\widehat{D}$ is rigid, we have that $\widehat{D}|_{\widehat{D}}$ is not effective. Because $c_1(\mathcal{L}_{\widehat{D}})$ is effective, $K_{\widehat{D}}$ must not be effective for $\widehat{D}|_{\widehat{D}}$ to be not effective. This leads to the conclusion that the Kodaira dimension of $\widehat{D}$ is $-\infty.$ From this, we reach the conclusion that, as claimed, 1-face divisors are either blowups of ruled surfaces or rational surfaces. Finally, we remark that by the assumption that $\dim\mathcal{L}|_{\widehat{D}}=1$, we again have that the 1-face divisor in question cannot be $\Bbb{P}^2.$ 

The argument we just gave above can be equally applied to a rigid vertex divisor. If a vertex divisor $D$ is rigid, the normal bundle to $D$ is not effective. Similarly, $K_{\widehat{D}}$ is also not effective. As a result, the Kodaira dimension of a rigid vertex divisor is again $-\infty.$ This implies again that the topology of a rigid vertex divisor is a blowup of either ruled surface or a rational surface. Likewise, because of the assumption that $\mathcal{L}_{\widehat{D}}$ is a line bundle of dimension 1, $\widehat{D}$ cannot be $\Bbb{P}^2.$

If $\widehat{D}$ is rigid, same as in the 2-face divisor case, we can again complete the Hodge diamond of $\overline{D},$ using the equation
\begin{equation}
h^{1,1}(\overline{D})=1+h^{1,1}(S)\,.
\end{equation}
When $\widehat{D}$ is not rigid, we do not know of a simple combinatorial method to compute $h^{1,1}(\overline{D})$, and hence we also do not know of a combinatorial formula for $h^{2,1}(\overline{D}).$ We would like to revisit the computation of the Hodge diamond for more general cases in the future.

\section{Euler characteristic of vertical divisors in elliptic Calabi-Yau fourfolds}\label{sec:push forward and local}

In \S\ref{sec:fourfold analysis}, we described in detail an algorithm for computing the Hodge structure of vertical divisors $\overline{D}  \subset Y_4$ where $Y_4$ is an elliptic CY fourfold admitting a global Sen limit and $\widehat{D} \subset B_3$ is a divisor in the orientifold CY threefold $B_3$. Our algorithm relies essentially on the classification of these vertical divisors' images under the canonical projection, $\pi_E(\overline{D}) = \widehat{D}$, in terms of the Iitaka dimension of the line bundle $\mathcal L|_{\widehat{D}}$. For certain cases, a key part of this algorithm involves using the Euler characteristic $\chi(\widehat{D})$ to compute the Hodge number $h^{2,1}(\widehat{D})$ in terms of the remaining ones. 

In this section, we present a conjectural combinatorial formula for the Euler characteristic those cases for which our algorithm does not explicitly determine $h^{2,1}(\widehat{D})$, namely
	\begin{align}
	\label{O3conjecture}
	\boxed{
		\chi(\overline{D}) = \chi_n(\overline{D}) + 2 n_{O3}(\widehat{D})}.
	\end{align}
We explain how to compute the ``naive'' Euler characteristic $\chi_n(\overline{D})$ in \S\ref{sec:push forward}; see (\ref{finalEuler}). Although the naive Euler characteristic agrees with the apparently correct Euler characteristic 
	\begin{align}
	\label{correctEuler}
		\chi(\overline{D})=\sum_{p,q} (-1)^{p+q} h^{p,q}(\overline{D}). 
	\end{align} 
when $\widehat{D}$ is smooth, in \S\ref{sec:o3-planes} we argue that when $\overline{D}$ contains terminal $\mathbb Z_2$ orbifold singularities signaling the presence of perturbative O3-planes, the correct Euler characteristic is obtained by adding the correction term $2n_{O3}(\widehat{D})$, where $n_{O3}(\widehat{D})$ is the number of O3-plane singularities. In \S\ref{sec:Iitakadimone}, we clarify some aspects of cases where the base $\widehat{D}$ of a vertical divisor $\overline{D}$ is characterized by the property $\dim\, \mathcal L|_{\widehat{D}}=1$ previously discussed in \S\ref{sec:dim1 line bundle}. 

Later, in \S\ref{Examples}, we verify that (\ref{O3conjecture}) holds in a number of examples, by directly computing the Hodge numbers of $\overline{D}$ in terms of Hodge-Deligne numbers, and substituting the Hodge numbers into the expression (\ref{correctEuler}).

\subsection{``Naive'' Euler characteristic via pushforwards}\label{sec:push forward}

The purpose of this section is to describe how to compute the naive Euler characteristic $\chi_n(\overline{D})$ of a smooth prime divisor $\overline D \subset Y_4 \rightarrow B_3$ where $ Y_4$ is a smooth elliptically-fibered CY fourfold that admits a global Sen limit and $B_3 = Z_3/\mathbb Z_2$ is an orientifold of a CY threefold. The CY fourfold $Y_4$ is in general characterized by some number of $SO(8)$ (Kodaira) singular fibers appearing over a collection of disjoint seven brane ``gauge divisors'' $\Sigma \subset B_3$, i.e. codimension-one components of the discriminant locus. Here, ``disjointness'' means that $\Sigma \Sigma' = 0$ for any two distinct gauge divisors $\Sigma, \Sigma'$. We assume that the divisor $\overline D$ is also elliptically fibered, 
	\begin{equation}
		\overline D \rightarrow \widehat D \subset B_3
	\end{equation}
and moreover that $\overline D$ generically intersects a non-trivial subset of the $SO(8)$ gauge divisors $\Sigma_a \subset B_3$. We use
	\begin{align}
	S := \Sigma \cap \widehat D
\end{align} 
to denote the restriction of the gauge divisors to $\widehat D$. We may regard $\widehat D$ as the pullback of $\widehat D$, which has the topology of an elliptically-fibered threefold (not CY) resolving some number of $SO(8)$ singularities over disjoint divisors $S \subset \widehat D$:\footnote{Note that the intersection of two prime toric divisors is necessarily primitive.} 
	\begin{equation}
		S\cap  S' = 0.
	\end{equation}
When $\widehat D$ is smooth, we may regard $\overline D$ is a complete resolution of a singular elliptic threefold. However, we are eventually interested in studying cases where $\widehat D$ contains $\mathbb Z_2$ singularities away from gauge divisors $S \subset \widehat D$, due to the presence of O3-planes. In such cases, $\overline D$ is a partial resolution, and the results of this subsection require a slight modification. As advertised, we discuss these modifications in detail in the following subsection.

Our starting point for analyzing the geometry of $\overline D$ is the singular $SO(8)$ Weierstrass model, which can be constructed as the zero locus of a section of a suitable choice of line bundle over the base $\widehat D$. For now we focus on a single $SO(8)$ Kodaira singularity, as the generalization to cases with multiple disjoint $SO(8)$ Kodaira singularities is completely straightforward. The defining data for this elliptic fibration consists of a choice of characteristic line bundle 
	\begin{align}
	\label{charlinebundle}
		\mathcal L \rightarrow \widehat D,
	\end{align}
where we write $L \equiv c_1(\mathcal L)$, along with a choice of divisor $S \subset \widehat D$ over which the $SO(8)$ singularity is imposed. With this choice of defining data, along with a local parametrization of the divisor $S \subset \widehat D$ as the zero locus $s=0$, we can construct the singular $SO(8)$ model explicitly as the zero section \cite{Esole:2017qeh}
	\begin{equation}
	\label{affineSO8model}
		 -y^2 + \prod_{n=1}^3 (x-x_n s) + s^2 r x^2  + s^3 q x + s^4 t = 0
	\end{equation}
where we assume $(x_1-x_2,x_2-x_3,x_3-x_1) \ne (0,0,0)$ and $(r,q,t) \ne (0,0,0)$. In (\ref{affineSO8model}), $x,y,q,r,t$ are sections of tensor powers of $\mathcal L$, where the specific powers are fixed by the constraint that the above Weierstrass polynomial is a section of the line bundle $\mathcal L^{\otimes 6}$. To produce a globally well-defined model, we take the projective closure\footnote{The local presentation of the Weierstrass model realizes the elliptic fibers as subvarieties of the affine space $\mathbb C^2$ and hence taking the projective closure of the Weierstrass model entails compactifying the ambient space $\mathbb C^2$ in a manner compatible with the elliptic fibration. Although the choice of projective closure is not unique, we stress that the results of this subsection are independent of this choice and hence we are free to take the projective closure to be $\mathbb P^2$ rather than, e.g., $\mathbb P_{[2,3,1]}$. The reason for this is that the topological data relevant to the results of this section are encoded in the singularities of the Weierstrass model, which can regarded as the set of singular elliptic fibers located over the discriminant locus $\Delta = 4 f^3 + 27 g^2 =0$ in the base $D$; one can show that the singularities of the elliptic fibers do not lie on the ``$\mathbb P^1$ at infinity" that is glued in when taking the projective closure of the ambient space $\mathbb C^2$. The choice $\mathbb P^2$ is a matter of convenience, as the pushforward formulae presented in this section are most simply stated for smooth elliptic fibrations that resolve singular (hypersurface) elliptic fibrations of smooth $\mathbb P^2$ bundles.} of the above equation in $\mathbb P^2$ and write
	\begin{equation}
		\overline D^{(0)} ~:~   -y^2z + \prod_{n=1}^3 (x-x_n sz) + s^2 r x^2 z + s^3 q x z^2 + s^4 tz^3 = 0.
	\end{equation} 
We regard the above zero locus, $\widehat D^{(0)}$, as a singular hypersurface of the ambient projective fourfold 
	\begin{align}
		V_4^{(0)} = \mathbb P(\mathcal L^{\otimes 2} \oplus \mathcal L^{\otimes 3} \oplus \mathcal O) \rightarrow \widehat D.
	\end{align}
We denote by $H  := c_1(\mathcal O(1)_{V^{(0)}_4})$ the hyperplane class of the $\mathbb P^2$ fibers. 

The next step is to resolve the singularities of $\overline D^{(0)}$. It was shown in \cite{Esole:2017qeh} that a resolution\footnote{The resolution we describe is smooth through codimension-two loci of the discriminant locus in $\widehat D$, assuming there are no additional singular fibers forced by the geometry described by the Weierstrass model.} of this model is given by the following sequence of blowups of complete intersections of hyperplanes, $g_{i1} = g_{i2} = \cdots = 0$ in the ambient space $V_4^{(0)}$:
	\begin{align}
	%	Y_4 \overset{(x-x_j s z,e_2|e_4)}{\longrightarrow} Y_3 \overset{(x-x_i s z,e_2|e_3)}{\longrightarrow} Y_2 \overset{(y, e_1|e_2)}{\longrightarrow} Y_1 \overset{(x,y,s|e_1)}{\longrightarrow} Y_0.
	\begin{array}{lllll}
	\label{blowups}
		(\vec g_1|e_1) &~=~& (x,y,s|e_1) &~:~& V_4^{(1)} \rightarrow V_{4}^{(0)}\\
		(\vec g_2|e_2) &~=~& (y,e_1|e_2) &~:~&V^{(2)}_4 \rightarrow V_4^{(1)}\\
		(\vec g_3|e_3) &~=~& (x-x_i s z,e_2|e_3) &~:~& V^{(3)}_4 \rightarrow V^{(2)}_4\\
		(\vec g_4|e_4) &~=~& (x-x_j s z,e_2|e_4) &~:~& V^{(4)}_4 \rightarrow V^{(3)}_4.
	\end{array}
	\end{align}
Let us explain the notation appearing in (\ref{blowups}). Each arrow in the above expression represents a blowdown map of the schematic form 
	\begin{align}
		(\vec g_i|e_i): V^{(i)}_4 \rightarrow V^{(i-1)}_4,
	\end{align}
 which contracts the exceptional divisor $e_{i}=0$ in $V^{(i)}_4$. The array $\vec g_i = (g_{i1},g_{i2},\dots)$ indicates that the $i$th blowup occurs along the complete intersection $g_{i1} = g_{i2} = \dots =0$ in $V^{(i)}_4$. Note that we abuse notation and retain the same variables for blowups along complete intersections of hyperplanes that lie in the $x=0$ or $y=0$ planes in the ambient space; for example, the first blowup entails the substitution 
 	\begin{align}
		x \rightarrow e_1 x, ~~~~y \rightarrow e_1 y,~~~~ s \rightarrow e_1 s.
	\end{align}
	
Each blowup of the ambient space described above introduces an exceptional divisor $e_i =0$ in $V_4^{(i)}$, which is contracted by the blowdown map $V_4^{(i)} \rightarrow V_4^{(i-1)}$. The proper transform, $\overline D^{(i)} \subset V^{(i)}_4$, of the elliptic threefold $\overline D^{(i-1)} \subset V^{(i-1)}_4$ under each blowup is obtained after factoring out a suitable number of copies of the exceptional divisor from total transform of $\overline D^{(i-1)}$, which is a reducible algebraic variety. We denote the resolved space by 
	\begin{align}
		\overline D := \overline D^{(4)},
	\end{align}
 where the superscript is a reminder that the resolution is comprised of a sequence of four blowups,
	\begin{equation}
	\label{resolution}
		\overline D \equiv \overline D^{(4)} \longrightarrow \overline D^{(3)} \longrightarrow \overline D^{(2)} \longrightarrow \overline  D^{(1)} \longrightarrow \overline D^{(0)}.
	\end{equation}	
Similarly, we write
	\begin{align}
		V_4 := V_4^{(4)}.
	\end{align}			
 Note again that $\widehat D$ may be a partial, and not a complete, resolution of $\overline D^{(0)}$, as the base $\widehat D$ may contain $\mathbb Z_2$ singularities due to O3-planes. However, as these $\mathbb Z_2$ singularities do not intersect the gauge divisor $S \subset \widehat D$, they do not affect the topology of the exceptional divisors introduced by the sequence of blowups described above, as the centers of the blowups have been chosen to lie along $S$.

For the purposes of this paper, the only data we require from the resolution of the $SO(8)$ model are the divisor classes of the blowup loci, $[g_{ij}]$, in the Chow ring of the ambient space $V^{(i-1)}_4$. The classes of each of these divisors can be expressed as linear combinations of the classes of a basis of divisors of $V^{(i-1)}_4$, namely 
	\begin{equation}
		L,S,H,[e_{j < i-1}]
	\end{equation} 
where $H$ is the (pullback to $V^{(i-1)}_4$ of the) hyperplane class of the $\mathbb P^2$ fibers of $V_4^{(0)}$, $[e_{j}]$ are linear combinations of the classes $E_j$ of the exceptional divisors $e_j =0$, $L$ is the pullback of $c_1(\mathcal L)$ (see (\ref{charlinebundle})), and $S$ is the pullback of the $SO(8)$ gauge divisor in $\widehat D$.\footnote{Note that we do not explicit indicate the pullbacks unless the Chow ring to which a given divisor class belongs is not clear from the context of the discussion.} In terms of this basis, we may write
	\begin{align}
	\begin{array}{lllll}
		([g_{11}] ,[g_{12}], [g_{13}]) &~=~&( [x],[y],[s] ) &~=~& (H + 2 L , H  +3L, S)\\
		([g_{21}] ,[g_{22}]) &~=~& ([y],[e_1]) &~=~& (H+3L - E_1,E_1) \\
		([g_{31}],[g_{32}]) &~=~&([x-x_i s z] ,[e_2]) &~=~& (H+2L , E_2)\\
		([g_{41}],[g_{42}]) &~=~& ([x-x_j s z],[e_2] ) &~=~& (H+2L , E_2-E_3).
	\end{array}
	\end{align}
\emph{When $\overline{D}$ is completely smooth}, the above divisor classes can be used to construct the total Chern class $c(\overline D)$, which is given by the following expression:
	\begin{equation}
	\label{totalChern}
		c(\overline D) = \left( \prod_{i=1}^{4} (1+ [e_i]) \cdot \prod_{j=1}^{n_i} \frac{1+ [g_{ij}] - [e_i]}{1+ [g_{ij}]} \right)\cdot \frac{c(\overline D_0)}{1 + [\overline D] } \cap [\overline D]
	\end{equation}
where 
	\begin{equation}
	c(\overline D_0) = (1 + H + 2L)\cdot ( 1+ H +3L )\cdot ( 1+H)\cdot c(\widehat D)
	\end{equation}
 with $c(\widehat D)$ being the total Chern class of the base $\widehat D$, and where
 	\begin{equation}
	\label{hypclass}
		\overline D = 3H + 6L - \sum_{i=1}^4 n_i [e_i].
	\end{equation}
(Note that $n_i$ is the codimension of the blowup locus $g_{i1} = g_{i2} =\cdots=0$, or equivalently the length of the $n_i$-tuple $\vec g_i$.) By rescaling each divisor class by a factor of $\varepsilon$ and expanding in powers of $\varepsilon$, the total Chern class can be expressed as the Chern polynomial
	\begin{equation}
	\label{Cpoly}
		c_{\varepsilon}(\overline D) =  1 + c_1(\overline D) \varepsilon + c_2(\overline D) \varepsilon^2 + \cdots. 
	\end{equation}
We define the Euler characteristic $\chi_n(\overline D)$ to be the degree of the top Chern class,
	\begin{equation}
		\chi_n(\overline D) = c_3(\overline D) = c_3(T\overline D) \cap \overline D	
	\end{equation} 
and can be recovered from the Chern polynomial by isolating the coefficient of the $O(\varepsilon^3)$ term. Note that third Chern class $c_3(\overline D)$ (i.e. the top Chern class for a threefold) can be expressed as a homogeneous quartic polynomial in the divisor classes of the Chow ring of the ambient fourfold $Y_4$. In particular, this implies we may write
	\begin{equation}
	\label{thirdChernclass}
		c_3(\overline D) = c_3(T\overline D) \cap \overline D
	\end{equation}
where $c_3(T \overline D)$ is a homogeneous cubic polynomial in the classes $L,S,H,[e_j]$.

In order to evaluate the above expression, it is necessary to supply the quadruple intersection numbers of the divisor classes $L,S,H,[e_j]$. A convenient way to compute the quadruple intersection numbers is to compute their pushforward to the Chow ring of $\widehat D$, with respect to the lift of the projection 
	\begin{align}
		\overline D \rightarrow \widehat D,
	\end{align}
to the projection of the blown-up ambient space, namely 
	\begin{align}
	\label{blowdownmap}
		\pi: V_4^{(4)} \rightarrow \widehat D.
	\end{align}
We follow the strategy of \cite{Esole:2017kyr}. This strategy enables us to express any formal analytic function of the classes $H, [e_j]$ in the Chow ring $Y_4$ as a formal analytic function of the classes $L,S$ in the Chow ring of $\widehat D$ (note that since the pushforward acts trivially on pullbacks of divisor classes in $\widehat D$, we do not bother to explicitly write the action of the pushforward on the classes $L,S$):
	\begin{equation}
			\pi_{E*}: F(H,[e_j]) \rightarrow \tilde F(L,S). 
	\end{equation}
There are various ways we can use the above pushforward map to compute the Euler characteristic. We describe two methods below:
	\begin{enumerate} 
		\item One method is to directly compute the pushforward of the total Chern class $c(\overline D)$, regarded as an analytic function of the classes $H,[e_j]$:
	\begin{equation}
		\pi_{E*}: c(\overline D)(H,[e_j])  \rightarrow  Q(L,S)\cdot c(\widehat D),~~~~ c(\widehat D) = c(T\widehat D) \cap \widehat D.
	\end{equation}
The resulting expression, $Q(L,S) c(\widehat D)$, which is an analytic function of the divisor classes $L,S$, can be regarded as generating function for top Chern classes; in particular, in the case of a twofold base $\widehat D$, the top Chern class can be extracted by introducing the formal rescaling $L \rightarrow \varepsilon L, S \rightarrow \varepsilon S$ and extracting the coefficient of the $ O(\varepsilon^2)$ term:
	\begin{align}
	\begin{split}
		Q(\varepsilon L, \varepsilon S) \cdot c(\widehat D) &=(Q_0 + \varepsilon Q_1 + \varepsilon^2 Q_2) \cdot(1 +\varepsilon c_1(\widehat D) + \varepsilon^2 c_2(\widehat D)) \\		
		 \chi_n(\overline D)&= Q_0 c_2(\widehat D) + Q_1 c_1(\widehat D) + Q_2.
	\end{split}
	\end{align} 
\item A second method, which seems in practice to be considerably less computationally expensive, is to first extract the triple intersection numbers of $\overline D$ (expressed as quadruple intersection numbers of $V_4$) and then substitute them into the expression for $c_3(\overline D)$ appearing in (\ref{Cpoly}). Following the methods of \cite{Jefferson:2022xft}, an efficient way to compute these intersection numbers of $V_4$ is to first encode them in a particularly simple analytic generating function, and then compute the pushforward of this generating function to $\widehat D$. Notice from (\ref{thirdChernclass}) that $c_3(T\widehat D)$ is a cubic polynomial in the classes $L, S, H, [e_j]$, the generating function of interest here is simply the product of the Chern characters associated with each divisor class of $V_4$:
	\begin{equation}
		Z_\alpha \equiv  \exp(\sum_m \alpha_m d_m+\frac{1}{3} \alpha^0 H+\alpha_1 [e_1]+\alpha_2 [e_2]+\alpha_3 [e_3]+\alpha_4 [e_4]) \cap \widehat D
	\end{equation}
where $d_m$ are pullbacks of the subset of divisors $d_m \in \widehat D$ that appear in the expansion
	\begin{align}
		S = S^m d_m,~~~~L = L^m d_m
	\end{align} 
and the expression for $\overline D$ was given in (\ref{hypclass}). The pushforward of the above function with respect to the blowdown map $\pi_E$ (see \ref{blowdownmap}) is 
	\begin{align}
		\pi_{E*}(Z_\alpha) &= \frac{e^{\alpha_m d_m - \alpha_0 L}}{L\cdot (4L-3S)\cdot (3L-2S)\cdot (2L-S)\cdot(L-S)} \cdot \sum_{j=0}^4 \mathcal Z_{j,4-j}S^{j} L^{4-j} ,
	\end{align}
with
	\begin{align}
		\begin{split}
			\mathcal Z_{0,4} &= 24 \left(e^{\alpha _0 L+\left(\alpha _1+\alpha _2+\alpha _3\right) S}-1\right)\\
			\mathcal Z_{1,3}&=-2(-35+6 e^{\left(\alpha _0+2 \alpha _1+\alpha _2\right) L}+23 e^{\alpha _0 L+\left(\alpha _1+\alpha _2+\alpha _3\right) S} \\
			&~~~~+3 e^{\alpha _0 L+\alpha _3 (2 L-S)+2 \alpha _4 (S-L)+\left(\alpha _1+\alpha _2\right) S}  \\
			&~~~~+2 e^{\alpha _0 L+\alpha _2 (3 L-S)+\left(\alpha _3+\alpha _4\right) (2 L-S)+\alpha _1 S} \\
			&~~~~+ e^{\alpha _0 L+\left(\alpha _3+\alpha _4\right) (2 L-S)+\left(\alpha _1+\alpha _2\right) S} )  \\
			\mathcal Z_{2,2}&=(-75 +29 e^{\left(\alpha _0+2 \alpha _1+\alpha _2\right) L} +29 e^{\alpha _0 L+\left(\alpha _1+\alpha _2+\alpha _3\right) S} \\
			&~~~~+7 e^{\alpha _0 L+\alpha _3 (2 L-S)+2 \alpha _4 (S-L)+\left(\alpha _1+\alpha _2\right) S}\\
			&~~~~+7 e^{\alpha _0 L+\alpha _2 (3 L-S)+\left(\alpha _3+\alpha _4\right) (2 L-S)+\alpha _1 S}\\
			&~~~~+3 e^{\alpha _0 L+\left(\alpha _3+\alpha _4\right) (2 L-S)+\left(\alpha _1+\alpha _2\right) S})\\
			\mathcal Z_{3,1} &= -(-35 + 23 e^{\left(\alpha _0+2 \alpha _1+\alpha _2\right) L}+6 e^{\alpha _0 L+\left(\alpha _1+\alpha _2+\alpha _3\right) S}\\
			&~~~~+2 e^{\alpha _0 L+\alpha _3 (2 L-S)+2 \alpha _4 (S-L)+\left(\alpha _1+\alpha _2\right) S}\\
			&~~~~+3 e^{\alpha _0 L+\alpha _2 (3 L-S)+\left(\alpha _3+\alpha _4\right) (2 L-S)+\alpha _1 S}\\
			&~~~~+e^{\alpha _0 L+\left(\alpha _3+\alpha _4\right) (2 L-S)+\left(\alpha _1+\alpha _2\right) S})\\
			\mathcal Z_{4,0} &=6 \left(e^{\left(\alpha _0+2 \alpha _1+\alpha _2\right) L}-1\right)
		\end{split}	
	\end{align}
and can be used to compute intersection numbers by taking derivatives, e.g.,
	\begin{align}
		[e_1][e_2][e_3] &= \left. \frac{\partial}{\partial \alpha_1}  \frac{\partial}{\partial \alpha_2}  \frac{\partial}{\partial \alpha_3} \pi_{E*}(Z_\alpha)\right|_{\alpha_a = \alpha_0 = \alpha_j = 0} = - S^2.
	\end{align}
\end{enumerate}
Our strategy for computing the Euler characteristic is as follows: we first use the first method outlined above to compute the Euler characteristic $\chi_n(\overline D)$ pushed down to the Chow ring of $\widehat D$. Then, we use the structure of the expression for $\chi_n(\overline D)$, along with some general considerations, to deduce what the Euler characteristic must look like for a smooth threefold resolving an arbitrary number of $SO(8)$ singularities tuned on disjoint divisors $S_a$. Later, as a consistency check, we use the second method outlined above to directly compute the Euler characteristic for the more complicated case of a smooth threefold resolving three $SO(8)$ singularities, and compare the resulting expression to our proposal for the general answer. We find that the two answers agree, which is a strong check that our proposal for the general answer (i.e. the generalization to an arbitrary number of $SO(8)$ singularities), is correct.

Before writing down the final answer, we first explicitly state the action of the pushforward map $\pi_{E*}$ needed to convert $c(\overline D)$ into a formal analytic expression in the Chow ring of $\widehat D$. The key point is that the lift $\pi_E$ of the projection $\overline D \rightarrow \widehat D$ to the ambient space $V_4$ can be expressed as a composition of blowdown maps $f_i : V^{(i)}_4 \rightarrow V^{(i-1)}_4$ with the canonical projection $\pi_V : V_4^{(0)} \rightarrow \widehat D$. Thus, it follows that we can express the action of the pushforward map as the composition 
	\begin{equation}
		\pi_{V*} \circ f_{1*} \circ f_{2*} \circ f_{3*} \circ f_{4*}.
	\end{equation}
This enables us to compute pushforward explicitly in terms of the actions of each individual map appearing in the above expression. The pushforward maps $\pi_*, f_{i*}$ were worked out in a series of papers \cite{Esole:2017kyr,Esole:2017qeh,Esole:2018tuz,Esole:2018bmf,Esole:2019asj,Esole:2019hgr,Esole:2019ocl,Esole:2020alo}; we now briefly summarize the key results of these papers. First, given a blowup $f_i : V^{(i)}_4 \rightarrow V^{(i-1)}_4$ of a smooth projective variety $V^{(i-1)}_4$ along a complete intersection of hypersurfaces, the pushforward of an analytic function $F([e_i])$ to the Chow ring of $V^{(i-1)}_4$ is
	\begin{equation}
		f_{i*} (F([e_i]) )= \sum_{k=1}^{n_i} F([g_{ik}]) \prod_{\substack{m=1\\m\ne k}}^{n_i} \frac{[g_{im}]}{[g_{im}]-[g_{ik}]}.
	\end{equation} 
Next, given a smooth rank two projective bundle $V_4^{(0)} = \mathbb P(\oplus_{i=1}^3 \mathcal L_i)$ equipped with canonical projection $\pi_V : V^{(0)}_4 \rightarrow \widehat D$, the pushforward of an analytic function $F(H)$ (where again $H = c_1(\mathcal O(1)_{V^{(0)}_4})$ is the hyperplane class of the fibers) to the Chow ring of $\widehat D$ is 
	\begin{equation}
		\pi_{V*}(F(H)) = \sum_{a=1}^3 \frac{F(-L_i)}{\prod_{j\ne i} (L_j - L_i) },~~~~ L_k \equiv c_1(\mathcal L_k). 
	\end{equation}
Applying the composition of the above pushforward maps to the analytic expression $c(\widehat D)$, it was shown in \cite{Esole:2017kyr} that the generating function for the Euler characteristic of the $SO(8)$ model is given by
	\begin{equation}
		\pi_{V*}\circ f_{1*} \circ f_{2*} \circ f_{3*} \circ f_{4*} (c(\overline D) ) = 12 \frac{L + 3 S\cdot L - 2 S^2}{(1+S)\cdot (1+6L - 4S)} \cdot c(\widehat D). 
	\end{equation}
Extracting the degree two term from the above formal power series gives the Euler characteristic for the elliptically-fibered threefold $\overline D$:
	\begin{equation}
	\label{SO8Euler}
		\chi_n(\overline D) = 12 ( c_1 \cdot L - 6L^2 + 6 L\cdot S - 2S^2)
	\end{equation}
where $c_1$ is the first Chern class of $\widehat D$. 

So far, we have only described the case of a single $SO(8)$ singularity. To generalize this result to resolutions of a singular threefold $\overline D^{(0)}$ containing an arbitrary number of $SO(8)$ singularities tuned on disjoint divisors 
	\begin{align}
		S_{I=1,2,\dots} \subset \widehat D,
	\end{align} 
we simply iterate the above process for each $SO(8)$ singularity separately. In the absence of any preferential distinctions between the divisors $S_a$, and to be consistent with the fact that the trivialization $S_b \rightarrow 0$ for all $b \ne a$ must reduce to the answer in (\ref{SO8Euler}), the pushforward of the final answer must depend on the classes $S_a$ in a symmetric fashion. We obtain the answer for multiple SO(8) tunings by making the replacement
	\begin{equation}
		S \rightarrow \sum_a S_a
	\end{equation}
in (\ref{SO8Euler}) and iteratively imposing the ``freshman's dream'' constraint $(\sum_a S_a)^p = \sum_a S_a^p$:
	\begin{equation}
	\label{finalEuler}
		\chi_n(\overline D) = 12 (c_1\cdot L - 6 L^2 + \sum_a (6 L\cdot S_a - 2 S_a^2 ) )
	\end{equation}
	
As a consistency check of (\ref{finalEuler}), we compute the Euler characteristic for a resolution $\widehat D^{(3)}$ of the ${SO(8)} \times {SO(8)} \times {SO(8)}$ model, where the three gauge groups are tuned on divisors $S_1,S_2,S_3 \subset D$. The resolution we study is simply the composition in (\ref{blowups}) applied separately to each $SO(8)$ singularity, in order, and consists of twelve blowups in total. In order to compute the Euler characteristic, we first use the \emph{Mathematica} package of \cite{Jefferson:2022xft} to compute the pushforward of the generating function of intersection numbers, namely the function
	\begin{equation}
		Z = \exp\biggl( \sum_m \alpha_m d_m + \frac{1}{3} \alpha_0 H + \sum_{a=1}^{3} \sum_{j_a=1}^{4} \alpha_{j_a} [e_{j_a}] \biggr)\cap \widehat D
	\end{equation} 
where 
	\begin{equation}
		\widehat D = 3H + 6L -\sum_{a=1}^3 ([e_{1_a}]- \sum_{j=1}^{4} [e_{j_a}]).
	\end{equation}
The pushforward of $Z$ is a horrendously long and unilluminating expression, so we do not include it here; however, the interested reader can easily compute this expression using \emph{Mathematica}. We then substitute the intersection numbers into the expression for $c_3(\overline D)$, which can be extracted from the appropriate generalization of the total Chern class (\ref{totalChern}) to this case. The final answer is:
	\begin{align}
	\begin{split}
		\chi_n(\overline D)& =12 c_1(\widehat D) \cdot L-72 L^2+72 L\cdot S_1+72 L\cdot S_2+72 L \cdot S_3\\
		&~~-24 S_1^2-24 S_2^2-24 S_3^2-32 S_1\cdot S_2-32 S_1 \cdot S_3-32 S_2 \cdot S_3.
	\end{split}
	\end{align}
Setting $S_1 \cdot S_2 = S_2 \cdot S_3 =S_3\cdot S_1 =0$, we find that the above expression clearly matches (\ref{finalEuler}).

\subsection{Correcting for O3-planes}\label{sec:o3-planes}
In the previous section, we computed the Euler characteristic $\chi_n(\overline{D})$ of a divisor $\overline{D}\subset Y_4$ using the top Chern class of $\overline{D}.$ It should be noted that if $\overline{D}$ is smooth, the integral of the top Chern class $\chi_n(\overline{D})$ 
\begin{equation}
\chi_n(\overline{D}):=\int_{\overline{D}}c_3(\overline{D})\,,
\end{equation}
is related to the Betti numbers as follows
\begin{equation}\label{eqn:betti euler}
\chi_n(\overline{D})=\sum_i (-1)^i b_i\,.
\end{equation}
We define the usual topological Euler characteristic $\chi_{}(\overline{D})$ to be
\begin{equation}
\label{chidef}
\chi_{}(\overline{D}):=\sum_i (-1)^i b_i = \sum_{p,q} (-1)^{p+q} h^{p,q},.
\end{equation}
Because $\chi_n(\overline{D})=\chi_{}(\overline{D})$ for smooth $\overline{D},$ when $\overline{D}$ is smooth we denote the Euler characteristic by $\chi(\overline{D})$ as there is no ambiguity. Note, however, that this relation \eqref{eqn:betti euler} need not hold true for a singular $\overline{D}.$ This poses an important problem for us, as a generic orientifold $B_3$ contains many O3-planes. Because of this, we need to understand how to relate the Euler characteristic to the Hodge numbers of $\widehat{D}$ in the presence of point-like $\Bbb{Z}_2$ orbifold singularities. Because O3-planes are terminal singularities, we unfortunately have the luxury of blowing up the singularities in a matter compatible with supersymmetry in order to carry out the analysis. 

In this section, we present two distinct local analyses. First, we consider a family of elliptic fibrations $\overline{D}$ parametrized by an integer $a$ with a point-like orbifold $\Bbb{Z}_2$ singularity. We will find that when $a$ is even, there exists a crepant resolution that resolves the $\Bbb{Z}_2$ singularity, and confirm that \eqref{eqn:betti euler} holds. On the other hand, when $a$ is odd, we find that there is no crepant resolution, and hence the $\Bbb{Z}_2$ singularity should be interpreted as an O3-plane. In such cases, we find that the equation \eqref{eqn:betti euler} is modified to $\chi_n(\overline{D})=\chi_{}(\overline{D})-2.$ 

In the second local analysis, we consider local affine orbifolds and their resolutions. We find that the resolution of the affine orbifold is not unique, and in F-theory perspective, distinct resolutions are related by a flip \cite{Denef:2005mm}. By carefully analyzing the flip, we arrive at the conclusion that in the presence of an O3-plane the equation \eqref{eqn:betti euler} is modified to  $\chi_n(\overline{D})=\chi_{}(\overline{D})-2.$ Supported by these local analyses we present in this section, as well as by numerous examples for which we have computed $h^{2,1}$ ``manually'' (i.e. by computing Hodge-Deligne numbers and using them to compute the Hodge numbers of $\overline{D}$), we are led to conjecture
\begin{equation}\label{eqn:o3 conjecture}
\boxed{\chi_n(\overline{D})=\chi(\overline{D})-2n_{O3}(\widehat{D})\,}
\end{equation}
where we emphasize that we have adopted the definition (\ref{chidef}).

Let us begin by analyzing the aformentioned one-parameter family of the local models. As a simple model of a two-dimensional surface with a point-like orbifold $\Bbb{Z}_2$ singularity, let us consider the singular projective variety $\Bbb{P}_{[1,1,2]},$\footnote{This local geometry can actually be found in an orientifold of a CY threefold described in \S\ref{Examples}.} whose toric rays are given by 
\begin{equation}
\left(
\begin{array}{ccc}
\vec v_1&\vec v_2&\vec v_3\\
1&-1&-1\\
0&1&-1
\end{array}
\right)\,.
\end{equation}
The corresponding GLSM is
\begin{equation}
\begin{array}{c|c|c}
x_1&x_2&x_3\\\hline
2&1&1
\end{array}
\end{equation}
For simplicity, we denote the class $[x_2]=[x_3]$ by $H.$ The cone spanned by $\vec v_2$ and $\vec v_3$ has volume two, indicating a $\Bbb{Z}_2$ singularity. This orbifold singularity at $x_2=x_3=0$ will serve as our local model for an O3-plane locus in $\Bbb{P}_{[1,1,2]}.$ 

Let us construct an elliptic fibration over $\Bbb{P}_{[1,1,2]}$ in terms of a Weierstrass model $\overline{D}$ with twisting line bundle $\mathcal{L}$ satisfying $L=aH.$ The Weierstrass model is embedded as a hypersurface in a $\Bbb{P}_{[2,3,1]}$ fibration over $\Bbb{P}_{[1,1,2]}.$ The corresponding GLSM is 
\begin{equation}\label{eqn:glsm E over P112}
\begin{array}{c|c|c|c|c|c}
x_1&x_2&x_3&X&Y&Z\\\hline
2&1&1&2a&3a&0\\\hline
0&0&0&2&3&1
\end{array}
\end{equation}
where the Weierstrass model is given by the vanishing locus of a section of a line bundle $2[Y]=[X]+[Y]+[Z]+\mathcal{L}.$ The toric rays for the corresponding GLSM are given as
\begin{equation}\label{eqn:rays E over P112}
\left(
\begin{array}{cccccc}
\vec v_1&\vec v_2&\vec v_3&\vec v_X&\vec v_Y&\vec v_Z\\
1&-1&-1&0&0&0\\
0&1&-1&0&0&0\\
0&0&-2a&1&0&-2\\
0&0&-3a&0&1&-3

\end{array}
\right)\,.
\end{equation}
For $0\leq a\leq 3,$ $\overline{D}$ is a threefold with positive first Chern class; for $a=4,$ $\overline{D}$ is an elliptic CY threefold.\footnote{When $a=0,$ because the line bundle $\mathcal{L}|_{\widehat{D}}$ is not big, we cannot use the formulae derived for line bundles with Iitaka dimension equal to 2. Nevertheless, the analysis is still quite simple in this case because when $n=0,$ the topology of $\overline{D}$ is $\overline{D}=T^2\times \Bbb{P}_{[1,1,2]}.$} 

We first compute the Hodge numbers of $\overline{D}$ using its stratification.\footnote{Note that the Hodge-Deligne number method \'a la Danilov-Khovanski (see \S\ref{sec:HD numbers}), computes the Hodge structure of cohomology with compact support. As was proved in \cite{Danilov_1987}, as long as the algebraic variety in question is quasi-smooth, the Hodge numbers of the cohomology groups with compact support coincide with the usual Hodge numbers. Furthermore, cohomology with compact support is a natural structure to study in relation to the physics of Euclidean M5-branes, as the Euclidean M5-brane partition function is computed in terms of integrals of the M-theory three-form $C_3$ over three-cycles in $\overline{D}.$ Therefore, it is in fact the numbers of distinct \emph{homology} classes that are relevant for understanding the physics of the Euclidean M5-brane partition function. Because homology is Poincar\'e dual to cohomology with compact support, we conclude that the Hodge-Deligne numbers, which determine for us the dimensions of the cohomology groups with compact support, are the relevant characteristic numbers needed to analyze the Euclidean M5-brane partition function.} We define the Newton polytope for the Weierstrass model as follows:
\begin{equation}
\Delta:=\{\vec{m}\in M| \vec{m}\cdot \vec v_1\geq0,~\vec{m}\cdot \vec v_2\geq-a,~\vec{m}\cdot \vec v_3\geq0,~\vec{m}\cdot \vec v_X\geq-1,~\vec{m}\cdot \vec v_Y\geq-1,~\vec{m}\cdot \vec v_Z\geq-1\}\,.
\end{equation}
One can easily check that $\Delta$ is a three-dimensional Newton polytope for $a>0.$ The combinatorial properties of $\Delta$ determine the Hodge numbers of $\overline{D}.$ Notably, we have
\begin{equation}
h^{3,0}(\overline{D})=l^*(\Delta)\,.
\end{equation}
The relative interior of $\Delta$ is parametrized as
\begin{equation}
\Delta^*:=\{\vec{m}\in M| \vec{m}\cdot \vec v_1>0,~\vec{m}\cdot \vec v_2>-a,~\vec{m}\cdot \vec v_3>0,~\vec{m}\cdot \vec v_X>-1,~\vec{m}\cdot \vec v_Y>-1,~\vec{m}\cdot \vec v_Z>-1\}\,.
\end{equation}
Equivalently, we can rewrite $\Delta^*$ as
\begin{equation}
\Delta^*=\{ \vec{m}\in M| m_1>0,~-m_1+m_2>-a,~-m_1-m_2>0,~m_3=m_4=0 \}\,.
\end{equation}
That is, the number of integral points in $\Delta^*$ is the same as the number of integral points in the Newton polytope for the line bundle $\mathcal O({L}+K_{\Bbb{P}_{[1,1,2]}})$
\begin{equation}
\Delta_{\mathcal{L}+K_{\Bbb{P}_{[1,1,2]}}}=\{ \vec{m}\in M| m_1\geq1,~-m_1+m_2\geq-a+1,~-m_1-m_2\geq1,~m_3=m_4=0 \}\,.
\end{equation}
As a result, we prove the relations
\begin{align}
\begin{split}
h^3(\overline{D},\mathcal{O}_{\overline{D}})=&~ h^0(\widehat{D},\mathcal{L}_{\widehat{D}}+K_{\Bbb{P}_{[1,1,2]}})\\
=&~h^2_-(D,\mathcal{O}_D)\,.
\end{split}
\end{align}
Because the Newton polytope $\Delta$ defines a big line bundle, the Lefschetz hypersurface theorem implies that $h^{1}(\overline{D},\mathcal{O}_{\overline{D}})=h^{2}(\overline{D},\mathcal{O}_{\overline{D}})=0.$ Hence, we are able to complete the computation of the Hodge vector $h^\bullet(\overline{D},\mathcal{O}_{\overline{D}})$:
\begin{equation}
h^\bullet(\overline{D},\mathcal{O}_{\overline{D}})=(1,0,0,l(\Delta_{\mathcal{L}+K_{\Bbb{P}_{[1,1,2]}}}))\,.
\end{equation}
Computing the Hodge numbers $h^{1,1}(\overline{D})=h^{2,2}(\overline{D})$ is relatively simple. Because the $\Bbb{P}_{[2,3,1]}$ fibration over $\Bbb{P}_{[1,1,2]}$ has six rays, we obtain
\begin{equation}
h^{1,1}(\overline{D})=2\,.
\end{equation}
Finally, by carrying out computations similar to what we carried out for $h^{2,1}(Z_3)$ and $h^{2,1}(B_3)$ in \S\ref{sec:toric CYs} and \S\ref{sec:toric CY orientifolds}, respectively, we are able to compute $h^{2,1}(\overline{D})$:
\begin{equation}
h^{2,1}(\overline{D})=l^*(2\Delta)-5l^*(\Delta)-\sum_{\Delta^{(3)}\subset \Delta}l^*(\Delta^{(3)})+\sum_{\Delta^{(2)}\subset\Delta} l^*(\Delta^{(2)})e^{1,1}(\sigma^{\circ(2)})\,.
\end{equation}
The above formula warrants further explanation. The first three terms count the number of inequivalent monomial deformations as usual. The last term describes the number of complex structure deformation that cannot be captured by the monomial deformations. This arises due to the number of $\Bbb{P}^1$ fibrations over a genus $l^*(\Delta^{(2)})$ curve, where only $1+ l^*(\Delta^{(2)})$ worth of linear combinations are captured by the monomial deformations in $\Delta.$ We find that $\overline{D}$ does not receive corrections from such a term, as long as we do not blow up the orbifold $\Bbb{Z}_2$ singularity. We find that for $a\equiv0\mod 2,$ we can blow up along the orbifold singularity, and increase the dimension of the $H^{2,1}(\overline{D},\Bbb{C}).$  We list the Hodge numbers for a few choices $ 0\leq a \leq 5$ in Table \ref{tab:Hodge-Deligne computation1}.
\begin{table}
\begin{center}
\begin{tabular}{|p{3em}|p{3em}|p{3em}|p{3em}|p{3em}|p{3em}|p{3em}|}
\hline
$a$&0&1&2&3&4&5\\\hline
$h^{1,0}$&1&0&0&0&0&0\\\hline
$h^{2,0}$&0&0&0&0&0&0\\\hline
$h^{3,0}$&0&0&0&0&1&2\\\hline
$h^{1,1}$&2 (1)&2& 2 (1)& 2& 2 (1)&2\\\hline
$h^{2,1}$&1 (1)&8&51 (1)&128&242 (1)&390\\\hline
\end{tabular}
\end{center}
\caption{\label{tab:Hodge-Deligne computation1} The Hodge numbers obtained from the Hodge-Deligne numbers of the strata. We denote the changes in $h^{1,1}$ and $h^{2,1}$ due to the introduction of new toric rays by $(k),$ where $k\in \Bbb{Z}.$}
\end{table}

We next study blowups in the cases $a=3$ and $a=4$ in more detail. One cautionary remark is warranted here: Although it is in principle possible to blow up the $\Bbb{Z}_2$ singularity in $\overline{D}$, if the blowup projects out some monomials in $\Delta$, it will break the CY condition of any elliptic CY fourfold $Y_4$ in which we embed $\overline{D}.$ Thus, the $\Bbb{Z}_2$ singularity in this case corresponds to a true O3-plane, as an O3-plane should be thought of as a terminal $\Bbb{Z}_2$ singularity in $Y_4$ \cite{morrison1984terminal,anno2003four,Garcia-Etxebarria:2015wns}. The terminal nature of the singularity reflects the fact that an O3-plane does not have any moduli associated that can be associated with a (crepant) resolution of the $\mathbb Z_2$ singularity. However, in cases where the blowup does not project out any monomials in $\Delta$, the resolution can be lifted to a blowup of $Y_4$ that resolves the $\Bbb{Z}_2$ singularity in a manner that preserves the CY condition. Hence, in this second case, there is no true O3-plane locus.

In view of the above comments, this blowup analysis should only be interpreted as a means to understand the conditions under which a discrepancy between $\chi_n(\overline{D})$ and $\chi(\overline{D})$ arises. One may worry that the condition that a blowup of $\overline{D}$ does not project out any monomials in $\Delta$ may not be sufficient to imply that such a blowup can be lifted to $Y_4.$ One should bear in mind, however, that blowups are strictly local operations, and if a blowup in $\overline{D}$ does not obstruct any monomials in $\Delta,$ then the same blowup does not obstruct any other monomials in the full Weierstrass model for $Y_4$ regardless of how it is realized in the global Weierstrass model. 

Let us begin with the CY case, $a=4$. To blow up the orbifold $\Bbb{Z}_2$ singularity in the base $\Bbb{P}_{[1,1,2]},$ we need to introduce a new toric ray $\vec v_E=(-1,0).$ Now the goal is to check whether or not one can lift $\vec v_E$ to a four-dimensional vector such that no monomial deformation in $\Delta$ is obstructed under the blowup. The candidate GLSM is
\begin{equation}
\begin{array}{c|c|c|c|c|c|c}
x_1&x_2&x_3&x_E&X&Y&Z\\\hline
2&1&1&0&8&12&0\\\hline
1&0&0&1&b&c&0\\\hline
0&0&0&0&2&3&1
\end{array}
\end{equation}
Not all GLSMs, labeled $(b,c)$, are good candidates, because we want no monomials to be projected out under the blowup. To ensure this, we need to impose the conditions
\begin{align}
2+b&=c\,\\
2c&=3b\,\\
6\times 1&\leq c\,.
\end{align}
The first condition makes sure that the CY condition remains unbroken. The second condition ensures that the form of the Weierstrass model is unchanged, namely
\begin{equation}
Y^2=X^3+f XZ^4+gZ^6\,.
\end{equation}
We can freely impose the above condition because the introduction of the exceptional divisor $\vec v_E$ corresponds to a blowup in the base manifold. The third condition, which is redundant, ensures that no monomial in $g$ is projected out. There is a unique solution to these three conditions, namely 
	\begin{equation}
		(b,c)=(4,6).
	\end{equation} 
This gives us a blow up that resolves the $\Bbb{Z}_2$ singularity in the base. Furthermore, this blowup can be lifted to a blowup of $Y_4,$ which indicates that the $\Bbb{Z}_2$ singularity should not be thought of as an O3-plane.

We now illustrate how to lift the blowup in $\widehat{D}$ to a blowup of $Y_4.$ We consider a simple model of $Y_4,$ such that $Y_4$ is an elliptic fibration over $\Bbb{P}^1\times \Bbb{P}_{[1,1,2]},$ where the GLSM is given by 
\begin{equation}
\begin{array}{c|c|c|c|c|c|c|c}
u_1&u_2&x_1&x_2&x_3&X&Y&Z\\\hline
0&0&2&1&1&8&12&0\\\hline
0&0&0&0&0&2&3&1\\\hline
1&1&0&0&0&4&6&0
\end{array}
\end{equation} 
Then the divisor $\overline{D}$ is in the class $[u_1]=[u_2].$ It should be then noted that the resolution $\pi:Y'_4\rightarrow Y_4,$ such  that the GLSM of $Y'_4$ is
\begin{equation}
\begin{array}{c|c|c|c|c|c|c|c|c}
u_1&u_2&x_1&x_2&x_3&x_E&X&Y&Z\\\hline
0&0&2&1&1&0&8&12&0\\\hline
0&0&1&0&0&1&4&6&0\\\hline
0&0&0&0&0&0&2&3&1\\\hline
1&1&0&0&0&0&4&6&0
\end{array}
\end{equation}
is crepant, meaning that $Y'_4$ is still CY. We emphasize that the blowup of $\overline{D}$ lifts to a blowup of $Y_4$, which implies that the $\Bbb{Z}_2$ singularity did not correspond to an O3-plane because it was not a terminal singularity.

Next, we turn to the case $a=3.$ We follow the same procedure we outlined for $a=4.$ We first find a candidate GLSM,
\begin{equation}
\begin{array}{c|c|c|c|c|c|c}
x_1&x_2&x_3&x_E&X&Y&Z\\\hline
2&1&1&0&6&9&0\\\hline
1&0&0&1&b&c&0\\\hline
0&0&0&0&2&3&1
\end{array}
\end{equation}
or equivalently, in a more convenient form, the GLSM
\begin{equation}
\begin{array}{c|c|c|c|c|c|c}
x_1&x_2&x_3&x_E&X&Y&Z\\\hline
0&1&1&-2&6-2b&9-2c&0\\\hline
1&0&0&1&b&c&0\\\hline
0&0&0&0&2&3&1
\end{array}.
\end{equation}
In order not to project out any monomials, the following conditions must be satisfied
\begin{align}
3b&=2c\\
3&= b\\
9/2&= c\, ,
\end{align}
where the first condition ensures that the canonical form of the Weierstrass form is preserved and the last two conditions make sure that no monomials in $f$ and $g$ are projected out. So far, everything looks good. But, there are two conditions we have not yet spelled out: We also need to make sure that monomials of the form
\begin{equation}
h XYZ\,,~~~~
q YZ^3\,,
\end{equation}
are not projected out. One of the two conditions needed to ensure this is
\begin{equation}
4+2b+2c=18\,,
\end{equation} 
which does not have a solution! As a result, we conclude that there is no blowup in the base that does not project out any monomials in $\Delta.$ Hence, in the case $a=3$, the $\Bbb{Z}_2$ singularity corresponds to a genuine O3-plane.

We now compare the above results to the computation of $\chi_n(\overline{D})$ using the methods described in \S\ref{sec:push forward}. We first note that the intersection ring is
\begin{equation}
J=\frac{1}{2} H^2\,,
\end{equation}
and the first Chern class of $\Bbb{P}_{[1,1,2]}$ is
\begin{equation}
c_1(\Bbb{P}_{[1,1,2]})=4H\,.
\end{equation}
Because there is no rigid O7-plane stack that intersects $\Bbb{P}_{[1,1,2]},$ we have
\begin{align}
\begin{split}
\chi_n(\overline{D})=& 12 (c_1 L-6L^2)\\
=&12 (4 a H^2-6a^2H)\\
=&12 a(2-3a)\,.
\end{split}
\end{align}
From the expression for $\chi_n(\overline{D})$ given above, we can deduce the Hodge number $h^{2,1}_n$ (note the subscript `$n$', which denotes that this is the ``naive'' Hodge number),
\begin{equation}
h_n^{2,1}(\overline{D})=-\frac{1}{2} \chi_n(\overline{D})+\frac{1}{2}\sum_{(p,q)\neq(2,1), (1,2)} (-1)^{p,q}h^{p,q}(\overline{D})\,.
\end{equation}
We list the values of $h_n^{2,1}$ for $a=0,\dots,5$ in Table \ref{tab:h21 comparison1}. We find that when $n$ is even, which corresponds to cases where there exists a blowup in the base that does not project out any monomials in $\Delta,$ the two computations agree: $h^{2,1}=h^{2,1}_n.$ But, when the $\Bbb{Z}_2$ singularity is terminal, and hence there is an O3-plane, we find that
\begin{equation}
h^{2,1}=h^{2,1}_n-1\,.
\end{equation}

\begin{table}
\begin{center}
\begin{tabular}{|p{3em}|p{3em}|p{3em}|p{3em}|p{3em}|p{3em}|p{3em}|}
\hline
$a$&0&1&2&3&4&5\\\hline
$h^{2,1}$&1 (1)&8&51 (1)&128&242 (1)&390\\\hline
$h^{2,1}_n$& 1 (1)& 9&51 (1)&129&242 (1)&391\\\hline
\end{tabular}
\end{center}
\caption{\label{tab:h21 comparison1} $h^{2,1}$ computed using the Hodge-Deligne numbers and the push-forward formula.}
\end{table}

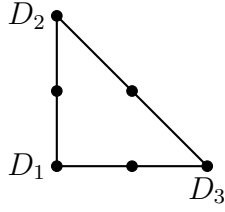
\begin{figure}
\begin{center}
\begin{tikzpicture}
\draw[thick] (0,0) -- (0,2);
\draw[thick] (0,0) -- (2,0);
\draw[thick] (2,0) -- (0,2);
\filldraw[black] (0,0) circle (2pt) node[left]{$D_1$};
\filldraw[black] (0,1) circle (2pt);
\filldraw[black] (0,2) circle (2pt) node[left]{$D_2$};
\filldraw[black] (1,0) circle (2pt);
\filldraw[black] (2,0) circle (2pt) node[below]{$D_3$};
\filldraw[black] (1,1) circle (2pt);
\end{tikzpicture}
\end{center}
\caption{\label{fig:toric fan} A 2D cross-section of the toric fan of $\Bbb{C}^3/\Bbb{Z}_2\times\Bbb{Z}_2.$}
\end{figure}

\begin{figure}
\begin{center}
\begin{tikzpicture}
\draw[thick] (0,0) -- (0,2);
\draw[thick] (0,0) -- (2,0);
\draw[thick] (2,0) -- (0,2);
\draw[thick] (0,1) -- (1,1);
\draw[thick] (1,1) -- (1,0);
\draw[thick] (1,0) -- (0,1);
\filldraw[black] (0,0) circle (2pt) node[left]{$D_1$};
\filldraw[black] (0,1) circle (2pt) node[left]{$E_{12}$};
\filldraw[black] (0,2) circle (2pt) node[left]{$D_2$};
\filldraw[black] (1,0) circle (2pt) node[below]{$E_{31}$};
\filldraw[black] (2,0) circle (2pt) node[below]{$D_3$};
\filldraw[black] (1,1) circle (2pt) node[right]{$E_{23}$};
\end{tikzpicture}
\qquad
\begin{tikzpicture}
\draw[thick] (0,0) -- (0,2);
\draw[thick] (0,0) -- (2,0);
\draw[thick] (2,0) -- (0,2);
\draw[thick] (0,0) -- (1,1);
\draw[thick] (1,0) -- (1,1);
\draw[thick] (0,1) -- (1,1);
\filldraw[black] (0,0) circle (2pt) node[left]{$D_1$};
\filldraw[black] (0,1) circle (2pt) node[left]{$E_{12}$};
\filldraw[black] (0,2) circle (2pt) node[left]{$D_2$};
\filldraw[black] (1,0) circle (2pt) node[below]{$E_{31}$};
\filldraw[black] (2,0) circle (2pt) node[below]{$D_3$};
\filldraw[black] (1,1) circle (2pt) node[right]{$E_{23}$};
\end{tikzpicture}
\end{center}
\caption{\label{fig:toric fan2} The 2D cross-sections of the symmetric and the asymmetric resolutions.}
\end{figure}
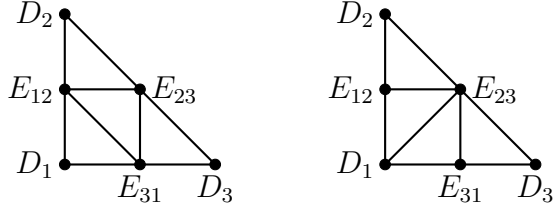

Now, we would like to explain the discrepancy between $h^{2,1}$ and $h^{2,1}_n$ in the presence of an O3-plane. For this purpose, we expect that a local model is sufficient to explain the discrepancy in a large class of geometries. We therefore study a local model of an orientifold of a local CY threefold $\Bbb{C}^3/\Bbb{Z}_2\times \Bbb{Z}_2,$ and its F-theory uplift.\footnote{This local model admits a natural compactification to $T^6/\Bbb{Z}_2\times \Bbb{Z}_2.$ For the detailed study of this model, see \cite{DelaOssa:2001blj,Denef:2005mm}. We thank Jakob Moritz for suggesting this local analysis.} The toric fan is spanned by three rays 
\begin{equation}
\left(\begin{array}{ccc}
\vec v_1&\vec v_2&\vec v_3\\
0&2&0\\
0&0&2\\
1&1&1
\end{array}\right)
\end{equation}
A two-dimensional cross section of the toric fan for $\mathbb C_3 / \mathbb Z_2 \times \mathbb Z_2$ (i.e. the toric diagram) is shown in Figure \ref{fig:toric fan}. We consider two resolutions---see Figure \ref{fig:toric fan2}. Following the terminology of \cite{Denef:2005mm}, we call these resolutions the ``symmetric'' phase and ``asymmetric'' phases, respectively. We use the symbol $Z_3$ to denote the symmetric phase and $\tilde Z_3$ to denote the asymmetric phase. The toric rays for these resolutions are given by the following vectors
\begin{equation}
\left(\begin{array}{cccccc}
\vec v_1&\vec v_2&\vec v_3&\vec v_{12}&\vec v_{23}&\vec v_{31}\\
0&2&0&0&1&1\\
0&0&2&1&1&0\\
1&1&1&1&1&1
\end{array}\right)
\end{equation}
We denote by $x_i$ the homogeneous coordinate whose vanishing locus corresponds to the toric ray $\vec v_i.$ Likewise, we use $x_{ij}$ to denote the homogeneous coordinate associated to the toric ray $\vec v_{ij}.$

In the symmetric phase $Z_3$, the Mori cone is generated by curves $C_1:=E_{12}\cap E_{23},$ $C_2:=E_{23}\cap E_{31},$ and $C_3:= E_{31}\cap E_{12}.$ The intersection numbers between the curves and the divisors are given by
\begin{equation}\label{eqn:int in sym phase}
\begin{array}{c|c|c|c|c|c|c}
&D_1&D_2&D_3&E_{12}&E_{23}&E_{31}\\\hline
C_1&0&1&0&-1&-1&1\\\hline
C_2&0&0&1&1&-1&-1\\\hline
C_3&1&0&0&-1&1&-1
\end{array}
\end{equation}
In the asymmetric phase $\tilde Z_3$, the Mori cone is generated by curves $\tilde{C}_1:=E_{12}\cap E_{23}=C_1+C_3,$ $\tilde{C}_2:=E_{23}\cap E_{31}=C_2+C_3,$ $\tilde{C}_3:=D_1\cap E_{23}=-C_3.$ The intersection numbers in the asymmetric phase are 
\begin{equation}\label{eqn:int in asym phase}
\begin{array}{c|c|c|c|c|c|c}
&D_1&D_2&D_3&E_{12}&E_{23}&E_{31}\\\hline
\tilde{C}_1&1&1&0&-2&0&0\\\hline
\tilde{C}_2&1&0&1&0&0&-2\\\hline
\tilde{C}_3&-1&0&0&1&-1&1
\end{array}
\end{equation}

We next construct an orientifold. Consider the orientifold action $I_1$
\begin{equation}
\mathcal I_1:x_1\mapsto -x_1\,.
\end{equation}
Because $Z_3$ is a projective toric variety equipped with a $\mathbb C^*$-action, $\mathcal I_1$ is equivalent to
\begin{equation}
g \cdot \mathcal I_1 \cdot g^{-1}: x_2\mapsto -x_2\,,
\end{equation}
and
\begin{equation}
g' \cdot \mathcal I_1 \cdot g'^{-1}: x_3\mapsto -x_3\,.
\end{equation}
The above equivalences can easily be seen from the relations (recall the discussion in \S\ref{sec:toric CY orientifolds})
\begin{equation}
v_1+v_2\equiv v_1+v_3\equiv0\mod2\,.
\end{equation}
As a result, we find that there are three O7-plane loci, at $x_1=0,$ $x_2=0,$ and $x_3=0.$ Note that upon compactification, all three O7-planes become rigid. So, we place four D7-branes on top of each O7-plane to make the local computation consistent with the global computation. In the symmetric phase, we in addition find an O3-plane at $x_{12}=x_{23}=x_{31}=0,$ whereas there is no O3-plane in the asymmetric phase. As a result, in the F-theory uplift, those two different phases are related by flips, rather than flops, as flops cannot introduce additional singularities to the geometric background.

We now study how the flip between the symmetric phase and the asymmetric phase changes the Euler characteristic $\chi(D)$ of a divisor $D$, the Euler characteristic of the orientifold fixed loci $\widehat{D}$ (namely, $\chi(\widehat{D}):= \sum_{p,q}(-1)^{p+q}(h_+^{p,q}(D)-h_-^{p,q}(D))$), and the Euler characteristic of the corresponding (F-theory uplift) vertical divisor, $\chi(\overline{D}).$ 

First, note that under a flop $Z_3 \rightarrow \tilde Z_3$ given by shrinking a curve $C \subset Z_3$ and blowing up a curve in $\tilde Z_3$, the intersection numbers and the second Chern class transform as \cite{mcduff1994j,wilson1999flops}
\begin{equation}
D_a \cdot D_b \cdot  D_c\mapsto D_a \cdot D_b \cdot D_c-n_{{C}} w_a w_b w_c\,,~~~~ w_i := D_i C,
\end{equation}
and
\begin{equation}
c_2(Z_3)\mapsto c_2(Z_3)+2n_{{C}} C\,,
\end{equation}
where $D_{i}$ for $i=a,~b,~c,$ are divisors, $C$ is the class of the curve in $Z_3$ that is flopped, and $n_C$ is the genus zero Gopakumar-Vafa (GV) invariant of the curve $C.$ In our local model, we take $C_3 \subset Z_3$ to be the flop curve. In this particular case, the genus zero Gopakumar-Vafa invariant of $C_3$ is equal to 1.

Let us study the topology of $E_{12}$ under the flop. The second Chern class of $E_{12}$ is, by adjunction formula,
\begin{equation}
c_2(E_{12})= c_2(X)+ E_{12}^2\,.
\end{equation}
Therefore, under the flop from the symmetric phase to the asymmetric phase, the Euler characteristic of $E_{12}$ transforms as follows
\begin{align}\label{eqn:euler e12}
\chi(E_{12})\mapsto \chi(E_{12})+2 C_3\cdot E_{12}- (C_3\cdot E_{12})^3
= \chi(E_{12})-1\,.
\end{align}
The transformation \eqref{eqn:euler e12} can be understood as follows. In general, flops can change even the Hodge vector of divisors.\footnote{For a recent study of this phenomenon, see \cite{Gendler:2022qof}.} But, in our local model, the flop does not change the Hodge vectors of all the prime toric divisors. This is rather intuitive to understand: what the flop does to the prime toric divisors can be understood as a blowdown of a nearby $\Bbb{P}^1,$ and the subsequent blowup of another $\Bbb{P}^1$ whose homology class is minus the homology class of the $\mathbb P^1$ that was blown down. Due to the fact that the structure sheaf cohomology is a birational invariant, we conclude that the Hodge vector of the prime toric divisors does not change under a flop. As a result, the flop does not change all Hodge numbers of $E_{12}$, but rather only changes $h^{1,1}(E_{12}).$ Going from the symmetric phase to the asymmetric phase, the number of one-simplices connected to $E_{12}$ decreases by 1. Because the genus zero GV invariant of the curve $C_3$ is 1, this means that we have excised one $\Bbb{P}^1$ from $E_{12},$ leading to a decrease of $h^{1,1}(E_{12})$ by 1. As a result, $h^{1,1}(E_{12})$ also decreases by 1. 

We next turn our attention to the Euler characteristic of fixed loci. In the symmetric phase, the O3-plane at $x_{12}=x_{23}=x_{31}=0,$ is contained in $E_{12}$, but in the asymmetric phase, there is no O3-plane. Hence, the Euler characteristic of the fixed locus corresponding to the O3-plane decreases by 1 under the flop. Next, we compute the Euler characteristic of the O7-plane fixed loci. Since the only O7-plane that is affected by the flop is $D_1,$ we focus on the change in the orientifold fixed locus due to the topology change of $D_1.$ The Euler characteristic of the intersection between $D_1$ and $E_{12}$ is 
\begin{equation}
\chi(D_1\cap E_{12})=\int_{D_1\cap E_{12}} c_1(D_1\cap E_{12})= - D_1\cdot E_{12} \cdot (D_1+E_{12})\,.
\end{equation} 
Under the flop, $\chi(D_1\cap E_{12})$ changes as follows:
\begin{align}
\begin{split}
\chi(D_1\cap E_{12})\mapsto& ~\chi(D_1\cap E_{12})+ (C_3 \cdot D_1)(C_3\cdot E_{12})(C_3\cdot (D_1+E_{12}))\\
=&~\chi(D_1\cap E_{12})\,.
\end{split}
\end{align}
As a result, we obtain $\chi_f(E_{12})\mapsto \chi_f(E_{12})-1.$ Because the equivariant Hodge vector $h^{\bullet}_{\pm}(E_{12},\mathcal{O}_{E_{12}})$ does not change under the flop, we conclude that under this local flop
\begin{equation}
h^{1,1}_+(E_{12})\mapsto h^{1,1}_+(E_{12})+1\,,~h^{1,1}_-(E_{12})\mapsto h^{1,1}_-(E_{12})\,.
\end{equation} 
Lastly, we study $\chi_n(\overline{E}_{12})$:
\begin{equation}
\chi_n(\overline{E}_{12})=12 (c_1(\widehat{E}_{12})\cdot {L}-6{L}^2+6{L}\cdot S-2S^2)\,.
\end{equation}
Because all of the O7-plane loci in the local patch host $SO(8)$ gauge groups, we have $S=2{L},$ where ${L}=D_1+D_2+D_3.$ Note that in the downstairs picture, $2{L}=\widehat{D}_1+\widehat{D}_2+\widehat{D}_3.$ (Because the upstairs picture is more convenient to work with, we continue using the upstairs picture.) Similarly, we have
\begin{equation}
c_1(\widehat{E}_{12})={L} -E_{12}\,.
\end{equation}
Combining these expressions, we find that
\begin{align}
\chi_n(\overline{E}_{12})&=12\int_{\widehat{E}_{12}} (c_1(\widehat{E}_{12}) \cdot {L}-2 {L}^2)\\
&=-12\int_{\widehat{E}_{12}} ({L} \cdot E_{12}+{L}^2)\,.
\end{align}
Under the flop, we find
\begin{equation}
\chi_n(\overline{E}_{12})\mapsto \chi(\overline{E}_{12})\,.
\end{equation}

So, what does this lengthy index computation imply for $h^{2,1}(\overline{E}_{12})$? It should be noted that the local flip/flop analysis is  carefully chosen to ensure that remains $h^{2,1}(\overline{E}_{12})$ invariant. This property is reflected by the fact that the local flop does not change $h^{1,1}_-(E_{12})$ or $h^{p,0}_\pm (E_{12}).$ Furthermore, the number of seven-branes (or more importantly seven-brane moduli) seen by $E_{12}$ is not changed under the flop. Lastly, the elliptic fibration over the flop curve has the topology of four blowups of $\Bbb{F}_2$, whose Hodge diamond is displayed in Figure \ref{cap:hodge bl4F2}. Choosing the flop curve such that the twisting line bundle restricted to the flop curve is big ensures that $h^{2,1}$ of the elliptic fibration over the flop curve is trivial (were this not the case, there could be a non-trivial change in $h^{2,1}(\overline{D})$ due to the flop.) Summarizing, these choices ensure that under the local flop $h^{2,1}(\overline{D})$ cannot change. This implies that $h^{2,1}(\overline{D})$ in the symmetric phase, where $\widehat{D}$ intersects an O3-plane, can be read off by the modified formula
\begin{equation}
\chi_n(\overline{D})+2=\sum_{p,q} (-1)^{p,q}h^{p,q}(\overline{D})\,,
\end{equation}
assuming that $E_{12}$ intersects only one O3-plane in the symmetric phase. In the case that more than O3-plane intersects $E_{12}$ in the symmetric phase, the prescription naturally generalizes to
\begin{equation}
\chi_n(\overline{D})+2n_{O3}(\widehat{D})=\sum_{p,q} (-1)^{p,q}h^{p,q}(\overline{D})\,,
\end{equation}
We emphasize again that in this local analysis, we made sure that the flop does not change $h^{2,1}(\overline{E}_{12}).$ One can perform a similar computation for $E_{23}$ and $E_{31}$ to arrive at the same conclusion. Because the analysis for $E_{23}$ and $E_{31}$ is completely analogous, we leave this an exercise for the interested reader.

From the local analyses involving $\Bbb{C}^3/\Bbb{Z}_2\times \Bbb{Z}_2$ and the elliptic fibration over $\Bbb{P}_{[1,1,2]},$ we thus arrive at the conjecture that in the presence of an O3-plane, $\chi_n(\overline{D})$ does not compute $\sum_{p,q}(-1)^{p,q}h^{p,q}(\overline{D})$ but
\begin{equation}\label{eqn:O3 euler}
\chi_n(\overline{D})=\sum_{p,q} (-1)^{p,q}h^{p,q}(\overline{D})-2n_{O3}(\widehat{D})\,.
\end{equation}
Although we have no general proof of the prescription \eqref{eqn:O3 euler}, we have provided strong evidence that \eqref{eqn:O3 euler} is true. Because the physics in question is local, we expect that the formula \eqref{eqn:O3 euler} should hold more generally. Furthermore, we have found that the prescription \eqref{eqn:O3 euler} matches the results of a direct computation via Hodge-Deligne numbers in a number of explicit examples, which serve as non-trivial cross-checks. 

\begin{figure}
\centering
\begin{tikzcd}[row sep=tiny, column sep=tiny]
&&1&&\\
&0&&0&\\
0&&6&&0\\
&0&&0&\\
&&1&&
\end{tikzcd}
\caption{\label{cap:hodge bl4F2} Hodge diamond of $\Bbb{F}_2$ blown up at four points.}
\end{figure}
\subsection{Twisting line bundle of Iitaka dimension 1}
\label{sec:Iitakadimone}
In this section, we study a local model of a base divisor $\widehat{D} \subset B_3$ over which the Iitaka dimension of the twisting line bundle defining an elliptic fibration $Y_4 \supset B_3$ is equal to 1, in order to clarify some aspects of the results presented in \S\ref{sec:dim1 line bundle}. In this subsection, we simply denote the restriction of the twisting line bundle to $\widehat{D}$ by $\mathcal L$. 

To begin, we consider a divisor $D$ in a CY threefold $Z_3$, which is topologically $\Bbb{F}_2.$ The GLSM defining $\Bbb{F}_2$ is
\begin{equation}
\begin{array}{c|c|c|c}
x_1&x_2&x_3&x_4\\\hline
1&1&0&-2\\\hline
0&0&1&1
\end{array}.
\end{equation}
The toric fan of $\Bbb{F}_2$ is generated by the rays
\begin{equation}
\left(
\begin{array}{cccc}
\vec v_1&\vec v_2&\vec v_3&\vec v_4\\
1&-1&0&0\\
0&-2&1&-1
\end{array}\right)
\end{equation}
We choose an orientifold involution $\mathcal I:X\rightarrow X,$ such that the induced orientifold action on $D$ is 
\begin{equation}
\mathcal I|_D:x_1\mapsto -x_1\,,
\end{equation}
which is equivalent to
\begin{equation}
g\cdot \mathcal I|_D \cdot g^{-1} :x_2\mapsto -x_2\,.
\end{equation}
After the orientifolding, we have $\widehat{D}:=\Bbb{F}_2/\Bbb{Z}_2=\Bbb{F}_1.$ This can be seen as follows. The orientifold action $\mathcal I$ refines the toric fan such that it consists of rays generated by
\begin{equation}\label{eqn:toric ray F1n}
\left(
\begin{array}{cccc}
\vec{\widehat{v}}_1&\vec{\widehat{v}}_2&\vec{\widehat{v}}_3&\vec{\widehat{v}}_4\\
2&-2&0&0\\
0&-2&1&-1
\end{array}\right)
\end{equation}
Because $\vec{\widehat{v}}_1$ and $\vec{\widehat{v}}_2$ are divisible by 2 (and hence not primitive), the identification of the generators of the toric rays suggested by \eqref{eqn:toric ray F1n} is not quite correct. Instead, the correct identification of the generators is
\begin{equation}\label{eqn:toric ray F1}
\left(
\begin{array}{cccc}
\vec{\widehat{v}}_1&\vec{\widehat{v}}_2&\vec{\widehat{v}}_3&\vec{\widehat{v}}_4\\
1&-1&0&0\\
0&-1&1&-1
\end{array}\right)
\end{equation}
As a result, we obtain that $\widehat{D} \equiv \Bbb{F}_1.$ 

To study the elliptic fibration over $\Bbb{F}_1,$ we first compute the twisting line bundle $\mathcal{L}.$ Because the fixed loci of $\mathcal I|_D$ are $x_1=0$ and $x_2=0,$ we obtain
\begin{equation}
{L}=H\,,
\end{equation}
where $H=[\widehat{x}_1]=[\widehat{x}_2]$ is the hyperplane class of $\Bbb{F}_1.$ Because $\Bbb{F}_1$ is projective, to compute the Iitaka dimension of $\mathcal{L},$ we can simply compute the dimension of the Newton polytope of sections of $\mathcal{L}.$ We select a gauge in which
\begin{equation}
{L}=[x_1]+0 [x_2]+0[x_3]+0[x_4]\,,
\end{equation}
which corresponds to the following choice of Newton polytope:
\begin{align}
\begin{split}
\Delta=&\{ \vec{m}\in M| v_1\geq -1\,, -v_1-v_2\geq 0\,,v_2\geq 0\,,-v_2\geq0\}\,,\\
=&\{\vec{m}\in M| -1\leq v_1\leq 0\,, v_2=0\}\,.
\end{split}
\end{align}
Because the Newton polytope is one-dimensional, we conclude that the Iitaka dimension of the twisting line bundle $\mathcal{L}$ is also 1. It is also quite intuitive to understand why the Iitaka dimension must be equal to 1. A section $s_a$ of a line bundle $a H,$ for $a>0$, can be expressed as
\begin{equation}
s_a=\sum_{i=0}^{a}b_i x_1^i x_2^{a-i}\,
\end{equation}
where $b_i\in \Bbb{C}.$ Because the section $s_a$ does not depend on the homogeneous coordinates $x_3$ and $x_4,$ the image of the map $\phi_a$
\begin{equation}
\phi_a:\Bbb{F}_1\rightarrow \mathbb PH^0(\Bbb{F}_1,\mathcal{L}^{\otimes a})\,, 
\end{equation}
has maximal dimension equal to 1. This conclusion is also tied to the fact that the section $s_a$ does not vary along the $\Bbb{P}^1$ fiber of $\Bbb{F}_1.$ 

\begin{figure}
\centering
\begin{tikzcd}[row sep=tiny, column sep=tiny]
&&1&&\\
&0&&0&\\
0&&10&&0\\
&0&&0&\\
&&1&&
\end{tikzcd}
\caption{\label{cap:hodge Enriques}The Hodge diamond of an Enriques surface.}
\end{figure}
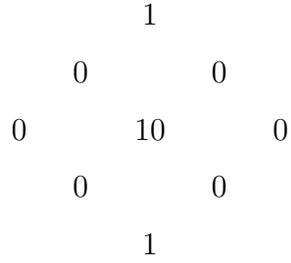

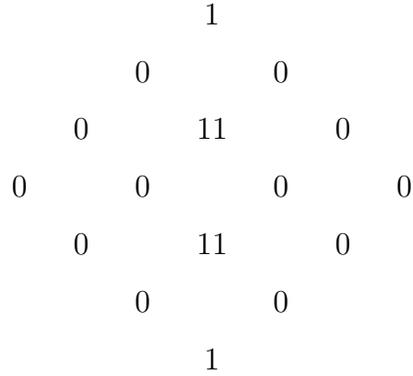
\begin{figure}
\centering
\begin{tikzcd}[row sep=tiny, column sep=tiny]
&&&1&&&\\
&&0&&0&&\\
&0&&11&&0&\\
0&&0&&0&&0\\
&0&&11&&0&\\
&&0&&0&&\\
&&&1&&&
\end{tikzcd}
\caption{\label{cap:hodge overD}The Hodge diamond of $\overline{D}$, in the case that $\overline{D}$ is a $\mathbb P^1$ fibration over an Enriques surface.}
\end{figure}

We next construct a Weierstrass model for the elliptic fibration over $\Bbb{F}_1$ with twisting line bundle ${L}=H$. The GLSM for such a Weierstrass model is given by
\begin{equation}
\begin{array}{c|c|c|c|c|c|c}
\widehat{x}_1&\widehat{x}_2&\widehat{x}_3&\widehat{x}_4&X&Y&Z\\\hline
1&1&0&-1&2&3&0\\\hline
0&0&1&1&0&0&0\\\hline
0&0&0&0&2&3&1
\end{array}
\end{equation}
Because the elliptic fibration does not vary along the $\Bbb{P}^1$ fiber, we can simply swap the order of the fibrations and regard $\overline{D}$ as a $\Bbb{P}^1$ fibration over an elliptic fibration over $\Bbb{P}^1.$ Because an elliptic fibration over $\Bbb{P}^1$ with twisting line bundle ${L}=H$ is an Enriques surface\footnote{For the Hodge numbers of the Enriques surface, see Figure \ref{cap:hodge Enriques}. }, we conclude that $\overline{D}$ is a $\Bbb{P}^1$ fibration over an Enriques surface. We record the Hodge numbers of $\overline{D}$ in Figure \ref{cap:hodge overD}.

As one can check from Figure \ref{cap:hodge Enriques}, applying the Shioda-Tate-Wazir theorem to an elliptic fibration with twisting line bundle equal to dimension 1 must be done with care because the Picard rank of $\overline{D}$ may not necessarily be equal to $h^{1,1}(\overline{D}).$ But, for Enriques surfaces it is known that the Picard rank coincides with $h^{1,1}=10$ \cite{liedtke2016picard}. This is a consequence of $h^{2,0}(\hat{D})=0$ and the Lefschetz theorem. We now illustrate that the Picard rank, hence $h^{1,1}(\overline{D}),$ can be found by tuning the complex structure moduli of the Weierstrass model so that the model exhibits a gauge group of maximal rank. Consider the Weierstrass model
\begin{equation}
Y^2=X^3+f(x_1,x_2) X Z^4+g(x_1,x_2) Z^6\,,
\end{equation}
where $f$ is a polynomial of degree 4 and $g$ is a polynomial of degree 6. The discriminant is a degree 12 polynomial defined as
\begin{equation}
\Delta(x_1,x_2):=4 f(x_1,x_2)^3+27g(x_1,x_2)^2\,.
\end{equation}
One can easily convince oneself that the maximal rank of the gauge group can be achieved either by tuning $E_8$ gauge group on a divisor, or tuning $SO(8)$ gauge group on two disjoint divisors.

\newpage
\section{Examples}\label{Examples}
In this section, we illustrate our algorithm for computing the Hodge numbers of vertical divisors $\overline{D}$ in a few examples of M/F-theory uplifts of type IIB compactifications on O3/O7 orientifolds of CY threefolds. Importantly, we use these examples as cross-checks of the conjectured formula (\ref{O3conjecture}).

\subsection{An orientifold of the quintic threefold}\label{sec:quintic}
In this subsection, we compute the Hodge structure of vertical divisors in an orientifold of the quintic threefold. For a detailed study of the quintic threefold, see \cite{Candelas:1990rm}. The GLSM of $\Bbb{P}^4$ with homogeneous coordinates $x_1,\dots,x_5,$ is
\begin{equation}
\begin{array}{c|c|c|c|c}
x_1&x_2&x_3&x_4&x_5\\\hline
1&1&1&1&1
\end{array}
\end{equation}
Components of the toric vertices $\vec v_i$ that correspond to the vanishing loci of the homogeneous coordinates $x_i$ are
\begin{equation}
\left(
\begin{array}{ccccc}
\vec v_1&\vec v_2&\vec v_3&\vec v_4&\vec v_5\\
1&0&0&0&-1\\
0&1&0&0&-1\\
0&0&1&0&-1\\
0&0&0&1&-1
\end{array}\right)
\end{equation}
The convex hull of the lattice points $\vec v_1,\dots,\vec v_5$ defines a reflexive polytope whose FRST $\mathcal{T}$ is given by
\begin{align}
\mathcal{T}:=&\left\{ \{0,v_1,v_2,v_3,v_4\},~\{0,v_2,v_3,v_4,v_5\},~\{0,v_1,v_2,v_4,v_5\}\right.,\nonumber\\
&\left.~\{0,v_1,v_2,v_3,v_5\},~\{0,v_1,v_2,v_3,v_4\}\right\}\,.
\end{align}
The Stanley-Reisner ideal of this toric variety is
\begin{equation}
SRI(\Bbb{P}^4)=\{x_1x_2x_3x_4x_5\}\,.
\end{equation}

To construct a CY threefold in $\Bbb{P}^4,$ we identify the class of the anti-canonical line bundle,
\begin{equation}
-K_{\Bbb{P}^4}=D_1+D_2+D_3+D_4+D_5\,,
\end{equation}
and the associated Newton polytope
\begin{equation}
\Delta:=\{ \vec{m}\in M| m_1\geq-1,~m_2\geq-1,~m_3\geq-1,~m_4\geq-1,~-m_1-m_2-m_3-m_4\geq-1\}\,.
\end{equation}
The vanishing locus of a section of $-K_{\Bbb{P}^4}$ defines the quintic threefold $Z_3$. 

\begin{figure}
\centering
\begin{tikzcd}[row sep=tiny, column sep=tiny]
&&&1&&&\\
&&0&&0&&\\
&0&&1&&0&\\
1&&101&&101&&1\\
&0&&1&&0&\\
&&0&&0&&\\
&&&1&&&
\end{tikzcd}
\caption{\label{cap:hodge quintic}The Hodge diamond of the quintic threefold}
\end{figure}
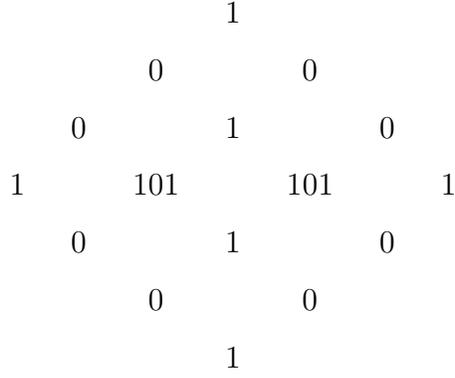

It is relatively simple to compute $h^{1,1}(Z_3).$ Because no zero-dimensional stratum of the quintic is reducible, $h^{1,1}(Z_3)$ of the quintic threefold is the same as $h^{1,1}(\Bbb{P}^4)=1.$ $h^{2,1}(Z_3)$ can be counted by computing the number of inequivalent monomials in $\Delta.$ There are in total 126 monomials in $\Delta,$ and exactly 25 of them are related by $\text{GL}(4,\Bbb{Z})$ transformations; this counting is reproduced by \eqref{eqn:h21 CY3}. We record the Hodge numbers of the quintic threefold in Figure \ref{cap:hodge quintic}.

\begin{figure}
\centering
\begin{tikzcd}[row sep=tiny, column sep=tiny]
&&1&&\\
&0&&0&\\
4&&45&&4\\
&0&&0&\\
&&1&&
\end{tikzcd}
\caption{\label{cap:D1 in quintic}The Hodge diamond of $D_1$ in the quintic threefold.}
\end{figure}
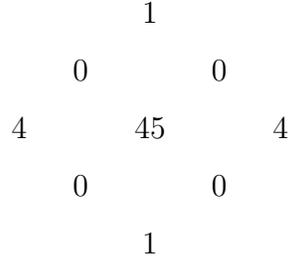

Prime toric divisors of the quintic threefold are all isomorphic to each other due to exchange symmetry. Therefore, we can simply study just one of them, say $D_1.$ A prime toric divisor in the quintic threefold is isomorphic to a degree 5 hypersurface in $\Bbb{P}^3.$ We pick a gauge such that the corresponding Newton polytope is given by
\begin{equation}
\Delta_1^{(3)}:=\{ \vec{m}\in M| m_1=-1,~m_2\geq-1,~m_3\geq-1,~m_4\geq-1,~-m_1-m_2-m_3-m_4\geq-1\}\,.
\end{equation}
Because $\Delta_1^{(3)}$ is three-dimensional, one can easily convince oneself that $D_1$ is a vertex divisor. We find that number of integral points in the strict interior of $\Delta_1^{(3)}$ is 4, which leads to
\begin{equation}
h^{\bullet}(D_1,\mathcal{O}_{D_1})=(1,0,4)\,.
\end{equation}
To compute $h^{1,1}(D_1),$ we shall study $2 \Delta_1^{(3)}$
\begin{equation}
2\Delta_1^{(3)}:=\{ \vec{m}\in M| m_1=-2,~m_2\geq-2,~m_3\geq-2,~m_4\geq-2,~-m_1-m_2-m_3-m_4\geq-2\}\,.
\end{equation}
We find that 
\begin{equation}
l^*(2\Delta_1^{(3)})=84\,.
\end{equation}
Furthermore, we find that every codim 1 face of $\Delta_1^{(3)},$ $\Delta^{(2)}\leq \Delta_1^{(3)}$ has
\begin{equation}
l^*(\Delta^{(2)})=6\,.
\end{equation}
As a result, using \eqref{eqn:h11 vertex}, we obtain
\begin{equation}
h^{1,1}(D_1)= 84-4\times 4-4\times 6+1=45\,.
\end{equation}
For the Hodge diamond of $D_1,$ see Figure \ref{cap:D1 in quintic}. 

Let us next study a $\Bbb{Z}_2$ involution $\mathcal I_{\vec v_1}: x_1\mapsto -x_1.$ Because 
\begin{equation}
\vec v_1+ \vec v_2+ \vec v_3+ \vec v_4+ \vec v_5\equiv0\mod2\,,
\end{equation}
there is a different representation of the involution given by
\begin{equation}
g\cdot \mathcal I_{\vec v_1}\cdot g^{-1} :(x_1,x_2,x_3,x_4,x_5)\mapsto (x_1,-x_2,-x_3,-x_4,-x_5)\,.
\end{equation} 
It follows that there are two fixed loci under the $\Bbb{Z}_2$ involution $\mathcal I_{\vec v_1}$, namely
\begin{equation}
%\mathcal{F}(I_{v_1})=
\{x_1=0\}\cup \{x_2=x_3=x_4=x_5=0\}\,.
\end{equation}
Note that $x_2=x_3=x_4=x_5=0$ is an O3-plane. We can now construct the refinement map $\varphi_{\mathcal I_{\vec v_1}}$,
\begin{equation}
\varphi_{\mathcal I_{\vec v_1}}: \Bbb{P}_{[2,1,1,1,1]}\rightarrow \Bbb{P}^4\,,
\end{equation}
where 
\begin{equation}
\varphi_{\mathcal I_{\vec v_1}}(y_i)=\begin{cases}
x_i^2 &\text{ if } i=1\\
x_i & \text{ otherwise}.
\end{cases}\,.
\end{equation}
We can use the refinement map $\varphi_{\mathcal I_{\vec v_1}}$ to construct the toric rays of $\Bbb{P}_{[2,1,1,1,1]}$:
\begin{equation}
\left(
\begin{array}{ccccc}
\vec{\widehat{v}}_1&\vec{\widehat{v}}_2&\vec{\widehat{v}}_3&\vec{\widehat{v}}_4&\vec{\widehat{v}}_5\\
1&0&0&0&-2\\
0&1&0&0&-1\\
0&0&1&0&-1\\
0&0&0&1&-1
\end{array}
\right)\,,
\end{equation}
and the corresponding toric fan $\varphi_{\mathcal{I}_{\vec v_1}}^{-1}(\mathcal{T}).$

Now, we shall study the topology of the coordinate flip orientifold $B_3=Z_3/\Bbb{Z}_2.$ Let us first compute the equivariant Hodge numbers $h^{p,q}_\pm(Z_3)$ manually, in order to compare with the toric method. Because there are no reducible divisors in the quintic $Z_3,$ the coordinate flip orientifold action does not induce non-trivial $h^{1,1}_-(Z_3).$ Thus, we conclude that $h^{1,1}_-(Z_3)=0$ and $h^{1,1}_+(Z_3)=1.$ Likewise, the coordinate flip orientifold changes the sign of the holomorphic threeform $\Omega_3$ of $Z_3,$ which yields $h^{3,0}_-(Z_3)=1$ and $h^{3,0}_+(Z_3)=0.$ To compute $h^{2,1}_{\pm}(Z_3)$, we first note that the number of monomials in $\Delta$ that are invariant under the orientifold involution is 80, and the number of odd monomials is 46. We then study the root automorphisms \cite{cox1995homogeneous}
\begin{equation}
x_j\mapsto x_j+\sum_{i\neq j}\lambda_i x_i\,,
\end{equation}
and how they transform under the orientifolding. We find that the action of each of the four root automorphisms on $x_1$, namely
\begin{equation}
x_1\mapsto x_1+\sum_{i\neq1}\lambda_i x_i,
\end{equation}
is projected out. On the other hand, three out of four of the root automorphism actions on $x_2,\dots,x_5$ are not projected out. Thus, we conclude that there are in total 12 out of 20 possible root automorphism actions that are even under the orientifold involution. Thus, we find that the number of complex structure moduli preserved after quotienting by the orientifold involution is 
\begin{equation}
h^{2,1}_-(Z_3)= 80-12-5=63\,,
\end{equation}
where the `$5$' subtracted above is due to the toric $\mathbb C^*$ action. As a result, we find that 
\begin{equation}
	h^{2,1}_+(Z_3)=46-8 =38.
\end{equation}

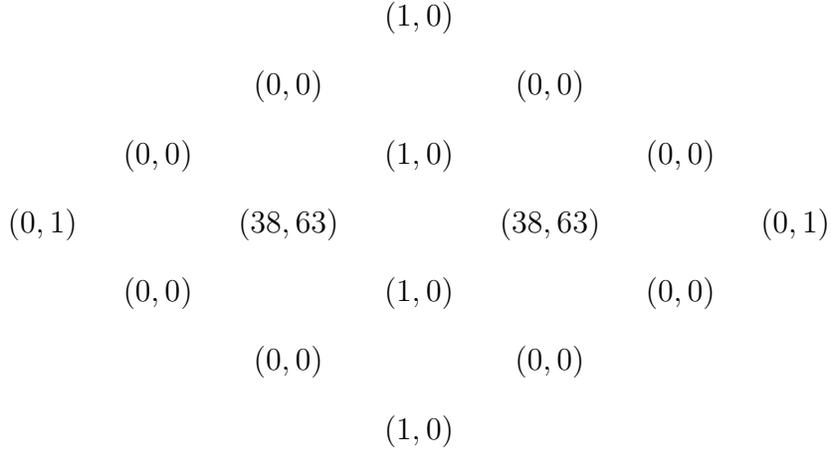
\begin{figure}
\centering
\begin{tikzcd}[row sep=tiny, column sep=tiny]
&&&(1,0)&&&\\
&&(0,0)&&(0,0)&&\\
&(0,0)&&(1,0)&&(0,0)&\\
(0,1)&&(38,63)&&(38,63)&&(0,1)\\
&(0,0)&&(1,0)&&(0,0)&\\
&&(0,0)&&(0,0)&&\\
&&&(1,0)&&&
\end{tikzcd}
\caption{\label{cap:hodge quintic ori}The equivariant Hodge diamond of an orientifold of the quintic threefold. Each entry is $(h^{p,q}_+,h^{p,q}_-).$}
\end{figure}

The coordinate flip orientifold $B_3$ is embedded into $\widehat{V}_4=\Bbb{P}_{[2,1,1,1,1]}$ as a degree 5 hypersurface. Therefore, we have
\begin{equation}
-K_{\widehat{V}_4}-{L}=\widehat{D}_1+\widehat{D_2}+\widehat{D}_3+\widehat{D}_4\,,
\end{equation}
where
\begin{equation}
{L}=[\widehat{D}_5]\,.
\end{equation}
We thus define the Newton polytope for the orientifold $B_3$ to be
\begin{equation}
\widehat{\Delta}:=\{ \vec{m}\in M| m_1\geq-1,~m_2\geq-1,~m_3\geq-1,~m_4\geq-1,~-2m_1-m_2-m_3-m_4\geq0\}\,,
\end{equation}
which implies that the associated polytope $2\widehat{\Delta}$ is
\begin{equation}
2\widehat{\Delta}:=\{ \vec{m}\in M| m_1\geq-2,~m_2\geq-2,~m_3\geq-2,~m_4\geq-2,~-2m_1-m_2-m_3-m_4\geq0\}\,.
\end{equation}
Because there are no reducible divisors in the quintic threefold (since all prime toric divisors are vertex divisors), we simply need to compute $h^{3,0}(B_3)=h^{3,0}_+(Z_3)$ and $h^{2,1}(B_3)=h^{2,1}_+(Z_3)$ of the orientifold $B_3$ via toric methods. First off, $l^*(\widehat{\Delta})$ is determined by the number of points in
\begin{equation}
\widehat{\Delta}^*=\{ \vec{m}\in M| m_1>-1,~m_2>-1,~m_3>-1,~m_4>-1,~-2m_1-m_2-m_3-m_4>0\}\,.
\end{equation}
Because there are no integral points in $\widehat{\Delta}^*,$ we conclude that $h^{3,0}(B_3)=0.$ We find that the number of integral points in
\begin{equation}
(2\widehat{\Delta})^*=\{ \vec{m}\in M| m_1>-2,~m_2>-2,~m_3>-2,~m_4>-2,~-2m_1-m_2-m_3-m_4>0\}\,
\end{equation}
is exactly 46, which matches the number of monomials in $\Delta$ that are odd under the orientifold action. In fact, this correspondence can be made very explicit by associating the following monomial to each point $\vec{m}\in M$:
\begin{equation}
y_1^{3/2 + m_1}y_2^{m_2+1}y_3^{m_3+1}y_4^{m_4+1}y_5^{-2m_1-m_2-m_3-m_4+1}=x_1^{3+2m_1}x_2^{m_2+1}x_3^{m_3+1}x_4^{m_4+1}x_5^{-2m_1-m_2-m_3-m_4+1}\,.
\end{equation}
Similarly, we find that the number of integral-interior points in 
\begin{equation}
\widehat{\Delta}^{(3)}_1:=\{ \vec{m}\in M| m_1=-1,~m_2\geq-1,~m_3\geq-1,~m_4\geq-1,~-2m_1-m_2-m_3-m_4\geq0\}
\end{equation}
is 4, and moreover that the number of integral interior points in
\begin{equation}
\widehat{\Delta}^{(3)}_2:=\{ \vec{m}\in M| m_1\geq-1,~m_2=-1,~m_3\geq-1,~m_4\geq-1,~-2m_1-m_2-m_3-m_4\geq0\}
\end{equation}
is 1, which matches the number of root automorphism group action that are projected out by (the quotient by) the orientifold involution. Therefore, we find that
\begin{align}
\begin{split}
h^{2,1}(B_3)=&~h^{2,1}_+(Z_3)\\
=&~l^*(2\widehat{\Delta})-\sum_{\widehat{\Delta}^{(3)}}l^*(\widehat{\Delta}^{(3)})\\
=&~ 38\,,
\end{split}
\end{align}
which perfectly matches the counting we performed ``manually''. We record the equivariant Hodge numbers of the orientifold in Figure \ref{cap:hodge quintic ori}.

Let us study the topology of prime toric divisors. We first study topology of $\widehat{D}_1.$ In the downstairs picture, the divisor class $\widehat{D}_1$ is twice the O7-plane class $D_1.$ The class of the line bundle, $2D_1,$ has in total 15 sections
\begin{equation}
x_i x_j\,,
\end{equation}
where $i,j=1,\dots,5.$ This implies that the Hodge vector of a divisor in the class $2D_1$ is
\begin{equation}
h^\bullet(2D_1,\mathcal{O}_{2D_1})=(1,0,14)\,.
\end{equation}
The sections of the normal bundle to $2D_1$ are given by
\begin{equation}
x_ix_j\,,
\end{equation}
where $(i,j)\neq(1,1).$ Upon orientifolding, the following sections of the normal bundle to $2D_1$ are projected out:
\begin{equation}
x_1x_j\,,
\end{equation} 
where $j=2,\dots,5.$ As a result, we expect that
\begin{equation}
h^\bullet_+(2D_1,\mathcal{O}_{2D_1})=(1,0,4)\,,
\end{equation}
and
\begin{equation}
h^\bullet_-(2D_1,\mathcal{O}_{2D_1})=(0,0,10)\,.
\end{equation}
We check this conclusion against the polytope data. Recall that $\widehat{D}_1$ is a vertex divisor. It follows that the only non-trivial structure sheaf Hodge number is 
\begin{equation}
h^{2,0}(\widehat{D}_1,\mathcal{O}_{\widehat{D}_1})=l^*(\widehat{\Delta}_1^{(3)})\,.
\end{equation}
As we have studied already, $\widehat{\Delta}_1^{(3)}$ has 4 interior points, which confirms
\begin{equation}
h^\bullet(\widehat{D}_1,\mathcal{O}_{\widehat{D}_1})=h^\bullet_+(2D_1,\mathcal{O}_{2D_1})=(1,0,4)\,.
\end{equation}
We find that
\begin{align}
\begin{split}
h^{1,1}(\widehat{D}_1)=&~l^*(2\widehat{\Delta}_1^{(3)})-4l^*(\widehat{\Delta}_1^{(3)})-\sum_{\Theta^{(2)}\subset \widehat{\Delta}_1^{(3)}} l^*(\Theta^{(2)})+1\,,\\
=&~84-4\times 4-4\times 6+1\\
=&~45\,.
\end{split}
\end{align}
We record the Hodge diamond of $\widehat{D}_1$ in Figure \ref{cap:D1 in quintic ori}. Similarly, we compute the Hodge diamond of $\widehat{D}_i,$ for $i=2,\dots,5$ and record it in Figure \ref{cap:D2 in quintic ori}. Note that because the divisors $D_i,$ for $i=2,\dots,5$ have trivial first Chern class, they are singular K3 surfaces. The fact that $h^{1,1}(D_2)=19$ reflects the fact that $D_2$ is singular, for example.

\begin{figure}
\centering
\begin{tikzcd}[row sep=tiny, column sep=tiny]
&&1&&\\
&0&&0&\\
4&&45&&4\\
&0&&0&\\
&&1&&
\end{tikzcd}
\caption{\label{cap:D1 in quintic ori}The Hodge diamond of $\widehat{D}_1$ in the orientifold of the quintic threefold.}
\end{figure}

\begin{figure}
\centering
\begin{tikzcd}[row sep=tiny, column sep=tiny]
&&1&&\\
&0&&0&\\
1&&19&&1\\
&0&&0&\\
&&1&&
\end{tikzcd}
\caption{\label{cap:D2 in quintic ori}The Hodge diamond of $\widehat{D}_i,$ for $i=2,\dots,5,$ in the orientifold of the quintic threefold.}
\end{figure}

We can now construct the F-theory uplift $Y_4$ of the O3/O7 orientifold of the quintic threefold. We write the GLSM charge matrix of the F-theory uplift as
\begin{equation}\label{eqn:GLSM f-uplift quintic}
\begin{array}{c|c|c|c|c|c|c|c}
\overline{x}_1&\overline{x}_2&\overline{x}_3&\overline{x}_4&\overline{x}_5&X&Y&Z\\\hline
2&1&1&1&1&2&3&0\\\hline
0&0&0&0&0&2&3&1
\end{array}
\end{equation}
Similarly, we choose a gauge so that the toric rays for the GLSM \eqref{eqn:GLSM f-uplift quintic} are
\begin{equation}
\left(
\begin{array}{cccccccc}
\vec{\overline{v}}_1&\vec{\overline{v}}_2&\vec{\overline{v}}_3&\vec{\overline{v}}_4&\vec{\overline{v}}_5&\vec v_X&\vec v_Y&\vec v_Z\\
1&0&0&0&-2&0&0&0\\
0&1&0&0&-1&0&0&0\\
0&0&1&0&-1&0&0&0\\
0&0&0&1&-1&0&0&0\\
0&0&0&0&-2&1&0&-2\\
0&0&0&0&-3&0&1&-3
\end{array}
\right)
\end{equation}
We note that the twisting line bundle for the elliptic fibration is
\begin{equation}
\mathcal{L}=\frac{1}{2}[\overline{x}_1]=[\overline{x}_i]\,,
\end{equation}
where $i=2,\dots,5.$ By applying the Shioda-Tate-Wazir theorem, we find that 
\begin{equation}
h^{1,1}(Y_4)=2\,.
\end{equation}
Using $h^{2,1}(B_3)=38,$ we obtain
\begin{equation}
h^{2,1}(Y_4)=38\,.
\end{equation}
To study the number of seven brane moduli, we examine the Weierstrass model 
\begin{equation}
Y^2=X^3+F XZ+GZ^6\,,
\end{equation}
more closely, where in the above equation $F$ is a section of a line bundle $\mathcal{L}^{\otimes 4}$ and $G$ is a section of a line bundle $\mathcal{L}^{\otimes 6}.$  We find that the number of seven brane moduli is \cite{nefpartition}
\begin{equation}
n_{D7}=160\,.
\end{equation}
The above result completes the Hodge diamond of $Y_4$---see Figure \eqref{cap:Y4 quintic}. The D3-brane tadpole is
\begin{equation}\label{eqn:tadpole Y4 quintic}
\frac{\chi(Y_4)}{24}=49\,.
\end{equation}
Note that the result \eqref{eqn:tadpole Y4 quintic} disagrees with the result obtained in \cite{Collinucci:2008zs}, even after taking the differing conventions into account. Let us explain the discrepancy. The Euler characteristic obtained in \cite{Collinucci:2008zs} was computed using pushforwards, analogous to the methods described in \S\ref{sec:push forward}. However, as was explained in previous sections, in the presence of O3-planes the Euler characteristic obtained via pushforward methods is not equal to $\sum_i (-1)^i b_i,$ and consequently needs to be corrected. In fact, the pushforward formula predicts $h^{3,1}_n(Y_4)=223,$ which is one less than the correct value 224. A strong evidence that our computation is correct comes from the fact that the D3-brane tadpole induced by the seven brane stack studied in \cite{Collinucci:2008zs} via the tachyon condensation is, in our convention,
\begin{equation}
Q^{D7}_{D3}=\frac{195}{4}\,.
\end{equation}
Because the tachyon condensation should be able to capture the correct physical D3-brane tadpole induced by the seven brane stack, as was carefully studied in \cite{Collinucci:2008pf}, we expect that the total D3-brane tadpole is
\begin{equation}
Q^{D7}_{D3}+\frac{n_{O3}}{4}=49\,,
\end{equation}
which agrees with \eqref{eqn:tadpole Y4 quintic}.

\begin{figure}
\centering
\begin{tikzcd}[row sep=tiny, column sep=tiny]
&&&&1&&&\\
&&&0&&0&&\\
&&0&&2&&0&\\
&0&&38&&38&&0&\\
1&&224&&872&&224&&1
\end{tikzcd}
\caption{\label{cap:Y4 quintic}The Hodge diamond of the F-theory uplift of the orientifold of the quintic.}
\end{figure}

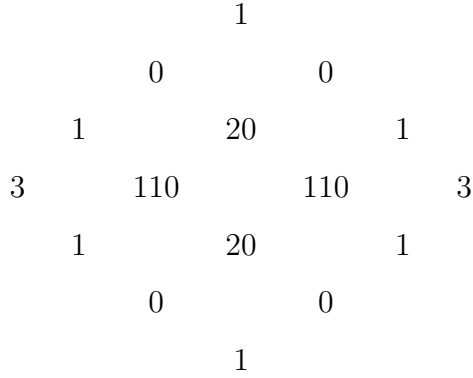
\begin{figure}
\centering
\begin{tikzcd}[row sep=tiny, column sep=tiny]
&&&1&&&\\
&&0&&0&&\\
&1&&20&&1&\\
3&&110&&110&&3\\
&1&&20&&1&\\
&&0&&0&&\\
&&&1&&&
\end{tikzcd}
\caption{\label{cap:hodge D2bar quintic}The Hodge diamond of $\overline{D}_2$ in the F-theory uplift of the quintic.}
\end{figure}

We finally study the vertical prime toric divisors in $Y_4.$ Let us first study $\overline{D}_1.$ We first note 
\begin{equation}
h^\bullet(\overline{D}_1,\mathcal{O}_{\overline{D}_1})=(1,0,4,10)\,,
\end{equation}
and $h^{1,1}(\overline{D}_1)=46,$ where we used Shioda-Tate-Wazir. Because $\overline{D}_1$ does not intersect an O3-plane, we can simply use the pushforward method to correctly compute
\begin{equation}
\chi(\overline{D}_1)=-390\,.
\end{equation}
The above formula implies that $h^{2,1}(\overline{D}_1)=240$, which enables us to complete the Hodge diamond. We record the Hodge diamond of $\overline{D}_1$ in Figure \ref{cap:hodge D1bar quintic}. Next, let us compute the Hodge structure of $\overline{D}_2.$ We note that
\begin{equation}
h^\bullet(\overline{D}_2,\mathcal{O}_{\overline{D}_2})=(1,0,1,3)\,,
\end{equation}
and
\begin{equation}
h^{1,1}(\overline{D}_2)=20\,,
\end{equation}
where we have again used the Shioda-Tate-Wazir theorem. Because $\overline{D}_2$ does in this case intersect an O3-plane, we need to be careful when computing $h^{2,1}(\overline{D}_2).$ We first compute $\chi_n(\overline{D}_2)$:
\begin{equation}
\chi_n(\overline{D}_2)=-180\,.
\end{equation}
Using \eqref{eqn:o3 conjecture}, we relate $\chi_n(\overline{D}_2)$ to the Betti numbers
\begin{equation}
-180=\sum_i (-1)^i b_i(\overline{D}_2)-2\,.
\end{equation}
As a result, we obtain $h^{2,1}(\overline{D}_2)=110.$  We record the Hodge diamond of $\overline{D}_1$ in Figure \ref{cap:hodge D2bar quintic}.

\begin{figure}
\centering
\begin{tikzcd}[row sep=tiny, column sep=tiny]
&&&1&&&\\
&&0&&0&&\\
&4&&46&&4&\\
10&&240&&240&&10\\
&4&&46&&4&\\
&&0&&0&&\\
&&&1&&&
\end{tikzcd}
\caption{\label{cap:hodge D1bar quintic}The Hodge diamond of $\overline{D}_1$ in the F-theory uplift of the quintic.}
\end{figure}

\subsection{An orientifold of $\Bbb{P}_{[1,1,1,6,9]}[18]$}
In this section, we study an orientifold of $Z_3=\Bbb{P}_{[1,1,1,6,9]}[18]$, i.e. a degree $18$ hypersurface of the weighted projective space $\mathbb P_{[1,1,1,6,9]}$. For a more detailed study of this CY threefold, see \cite{Candelas:1994hw}. The CY threefold $\Bbb{P}_{[1,1,1,6,9]}[18]$ can be understood as a generic elliptic fibration over $\Bbb{P}^2.$ The GLSM of $Z_3$ is
\begin{equation}
\begin{array}{c|c|c|c|c|c}
x_1&x_2&x_3&x_4&x_5&x_6\\\hline
1&1&1&6&9&0\\\hline
0&0&0&2&3&1
\end{array}
\end{equation}
We select a gauge such that the corresponding toric rays are generated by the vertices
\begin{equation}
\left(\begin{array}{cccccc}
\vec v_1&\vec v_2&\vec v_3&\vec v_4&\vec v_5&\vec v_6\\
1&0&-1&0&0&0\\
0&1&-1&0&0&0\\
-2&-2&-2&1&0&-2\\
-3&-3&-3&0&1&-3
\end{array}\right)\,.
\end{equation}
We triangulate the reflexive polytope $\Delta^\circ,$ which is defined to be the convex hull of the points $\{ v_1,\dots,v_6\},$ where the FRST $\mathcal{T}$ is defined as follows:
\begin{align}
\mathcal{T}=&\{\{ 0,v_1,v_2,v_4,v_5\},\{0,v_1,v_2,v_4,v_6\},\{0,v_1,v_2,v_5,v_6\}, (v_2\leftrightarrow v_3),(v_1,\leftrightarrow v_3) \}\,.
\end{align}
The $SRI$ of the toric variety $V_4$ corresponding to the triangulation is
\begin{equation}
SRI(V_4)=\{ x_1x_2x_3,x_4x_5x_6\}\,.
\end{equation}
The anti-canonical line bundle $-K_{V_4}$ is
\begin{equation}
-K_{V_4}=\sum_i D_i\,,
\end{equation}
and the associated Newton polytope is
\begin{equation}
\Delta_{-K_{V_4}}:= \{\vec{m}\in M| v_i\cdot \vec{m}\geq -1\,, \forall i\}\,.
\end{equation}
A generic section of $-K_{V_4}$ reads
\begin{equation}
s=-x_5^2+x_4^3+ f_{12}(x_1,x_2,x_3)x_4 x_6^4+g_{18} (x_1,x_2,x_3)x_6^6\,,
\end{equation}
where $f_{12}$ is a degree 12 polynomial in $\{ x_1,x_2,x_3\},$ and $g_{18}$ is a degree 18 polynomial in $\{x_1,x_2,x_3\}.$ The vanishing locus $s=0$ defines the CY threefold $Z_3=\Bbb{P}_{[1,1,1,6,9]}[18].$  We compute the Hodge diamond of $Z_3$ and record it in Figure \ref{cap:hodge 11169}.

\begin{figure}
\centering
\begin{tikzcd}[row sep=tiny, column sep=tiny]
&&&1&&&\\
&&0&&0&&\\
&0&&2&&0&\\
1&&272&&272&&1\\
&0&&2&&0&\\
&&0&&0&&\\
&&&1&&&
\end{tikzcd}
\caption{\label{cap:hodge 11169}The Hodge diamond of $\Bbb{P}_{[1,1,1,6,9]}[18]$}
\end{figure}

We note that divisors $D_1$ through $D_5$ are vertex divisors, and that $D_6$ is the only 2-face divisor. Let us first study topology of the divisor $D_1.$ The Newton polytope for the prime toric divisor $D_1$ is given by\footnote{The superscript `$(3)$' is a reminder that $\Delta_1^{(3)}$ is three-dimensional.}
\begin{equation}
\Delta_1^{(3)}:=\{\vec{m}\in M| v_1\cdot \vec{m}=-1,~ v_i\cdot \vec{m}\geq-1,~\text{ for } i=2,\dots,5 \}\,.
\end{equation}
We find that
\begin{equation}
l^*(\Delta_1^{(3)})=2\,,
\end{equation}
which leads to the Hodge vector
\begin{equation}
h^\bullet(D_1,\mathcal{O}_{D_1})=(1,0,2)\,.
\end{equation}
Similarly, we compute
\begin{equation}
l^*(2\Delta_1^{(3)})=62\,,
\end{equation}
and
\begin{equation}
\sum_{\Theta^{(2)}\subset \Delta_1^{(3)}}l^*(\Theta^{(2)})=26\,,
\end{equation}
which lead to
\begin{equation}
h^{1,1}(D_1)=30\,.
\end{equation}
We record the Hodge diamond of $D_1,~D_2,$ and $D_3$ in Figure \ref{cap:D1 in 11169}. Similarly, we compute the Hodge numbers of $D_4,~D_5,~D_6,$ and record them in Table \ref{cap:D4 in 11169}.

\begin{figure}
\centering
\begin{tikzcd}[row sep=tiny, column sep=tiny]
&&1&&\\
&0&&0&\\
2&&30&&2\\
&0&&0&\\
&&1&&
\end{tikzcd}
\caption{\label{cap:D1 in 11169}The Hodge diamond of $D_i,$ for $i=1,\dots,3,$ in $\Bbb{P}_{[1,1,1,6,9]}[18].$}
\end{figure}

\begin{table}
\centering
\begin{tabular}{|c|c|c|c|c|}
\hline
&$h^{0,0}$&$h^{1,0}$&$h^{2,0}$&$h^{1,1}$\\\hline
$D_4$&1&0&28&218\\\hline
$D_5$&1&0&65&417\\\hline
$D_6$&1&0&0&1\\\hline
\end{tabular}
\caption{\label{cap:D4 in 11169}The Hodge numbers of $D_i,$ for $i=4,\dots,6,$ in $\Bbb{P}_{[1,1,1,6,9]}[18].$}
\end{table}

Let us study the $\Bbb{Z}_2$ involution $\mathcal I_{\vec v_5}:x_5\mapsto -x_5.$ Because $\vec v_5+\vec v_6\equiv \vec 0\mod 2,$ there is a different presentation of the involution, namely
\begin{equation}
g\cdot \mathcal I_{\vec v_5}\cdot g^{-1}:x_6\mapsto -x_6\,.
\end{equation}
As a result, we find two fixed loci
\begin{equation}
%\mathcal{F}(I_{v_5})=
\{ x_5=0\}\cup \{ x_6=0\}\,.
\end{equation}
We construct the refinement map $\varphi_{\mathcal I_{\vec v_5}}$,
\begin{equation}
\varphi_{\mathcal I_{\vec v_1}}: \widehat{V}_4\rightarrow V_4\,,
\end{equation}
where 
\begin{equation}
\varphi_{\mathcal I_{\vec v_5}}(\widehat{x}_i)=\begin{cases}
x_i^2 &\text{ if } i=5,~6\\
x_i & \text{ otherwise}
\end{cases}\,.
\end{equation}
We then use the refinement map $\varphi_{\mathcal I_{\vec v_5}}$ to construct the toric rays of $\widehat{V}_4$,
\begin{equation}
\left(
\begin{array}{cccccc}
\vec{\widehat{v}}_1&\vec{\widehat{v}}_2&\vec{\widehat{v}}_3&\vec{\widehat{v}}_4&\vec{\widehat{v}}_5&\vec{\widehat{v}}_6\\
1&0&-1&0&0&0\\
0&1&-1&0&0&0\\
-2&-2&-2&1&0&-1\\
-3&-3&-3&0&1&-3
\end{array}
\right)\,,
\end{equation}
and the corresponding GLSM
\begin{equation}
\begin{array}{c|c|c|c|c|c}
\widehat{x}_1&\widehat{x}_2&\widehat{x}_3&\widehat{x}_4&\widehat{x}_5&\widehat{x}_6\\\hline
1&1&1&6&18&0\\\hline
0&0&0&2&6&2
\end{array}
\end{equation}
The orientifold is the vanishing locus of a section of the line bundle 
\begin{equation}
-K_{\widehat{V}_4}-{L}=\widehat{D}_1+\widehat{D}_2+\widehat{D}_3+\widehat{D}_4+\frac{1}{2}\widehat{D}_5+\frac{1}{2}\widehat{D}_6\,.
\end{equation}
A comment is in order. Because the line bundle $-K_{\widehat{V}_4}-{L}$ is isomorphic to $\widehat{D}_5,$ the orientifold is in fact isomorphic to a toric threefold, the generalized Hirzebruch threefold with twisting line bundle 6 times the line bundle corresponding to a hyperplane class of the base $\mathbb P^2$. This fact simplifies various computations. Nevertheless, to illustrate the methods we present in this paper, we will be agnostic about the fact that the orientifod $B_3$ is a toric threefold and carry out a systematic computation. For further treatment of the orientifold $B_3$ (regarded as a toric threefold), see e.g. \cite{Kim:2022uni}. 

\begin{figure}
\centering
\begin{tikzcd}[row sep=tiny, column sep=tiny]
&&&1&&&\\
&&0&&0&&\\
&0&&2&&0&\\
0&&0&&0&&0\\
&0&&2&&0&\\
&&0&&0&&\\
&&&1&&&
\end{tikzcd}
\caption{\label{cap:hodge 11169 ori}The Hodge diamond of $B_3:=\Bbb{P}_{[1,1,1,6,9]}[18]/\Bbb{Z}_2.$ Note  $h^{p,q}(B_3)=h^{p,q}_+(Z_3).$}
\end{figure}
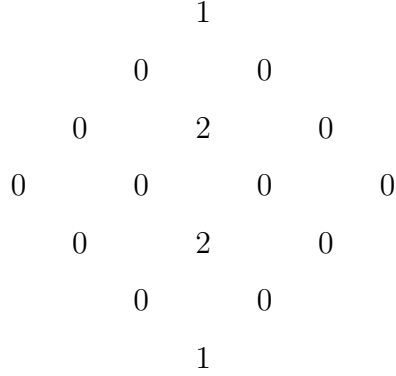

We select a gauge in which the Newton polytope for the line bundle $-K_{\widehat{V}_4}-{L}$ is
\begin{equation}
\widehat{\Delta}:=\{\vec{m}\in M| \widehat{v}_i\cdot\vec{m}\geq -a_i,~\forall \widehat{v}_i \}\,,
\end{equation}
where
\begin{equation}
\vec{a}=(0,0,0,0,1,0)\,.
\end{equation}
We find that the number of interior points in $\widehat{\Delta}$ is zero. We therefore obtain
\begin{equation}
h^\bullet(B_3,\mathcal{O}_{B_3})=(1,0,0,0)\,.
\end{equation}
We define $2\widehat{\Delta}$ to be
\begin{equation}
\widehat{\Delta}:=\{\vec{m}\in M| \widehat{v}_i\cdot\vec{m}\geq -2a_i,~\forall \widehat{v}_i \}\,.
\end{equation}
We compute
\begin{equation}
l^*(2\widehat{\Delta})=65\,,
\end{equation}
and
\begin{equation}
\sum_{\Theta^{(3)}\subset \widehat{\Delta}}l^*(\Theta^{(3)})=65\,.
\end{equation}
As a result, we conclude $h^{2,1}(B_3)=h^{2,1}_+(Z_3)=0.$ We record the Hodge diamond of $B_3$ in Figure \ref{cap:hodge 11169 ori}.

We similarly compute the topology of prime toric divisors. We record the results in Table \ref{cap:D4 in 11169ori}. Let us explain how to understand some of the results in Table \ref{cap:D4 in 11169ori}. Let us first recall that a section of normal bundle to $D_1$ describes the root automorphism
\begin{equation}
x_1\mapsto x_1+a x_2+b x_3\,.
\end{equation}
Because the orientifold involution does not project out any section of the normal bundle to $D_1,$ we obtain $h^{2,0}_-(D_1)=2$ and $h^{2,0}_+(D_1)=0.$ This agrees with the result $h^{2,0}(\widehat{D}_1)=0.$ One can similarly make sense of the result $h^{2,0}(\widehat{D}_4)=0.$ Now let us ``manually'' compute $h^{2,0}(\widehat{D}_5).$ We first note that $\widehat{D}_5$ is in the same class as $2D_5.$  A section of $2D_5$ is a bi-degree $(18,6)$ polynomial. There are in total 376 sections in $2D_5,$ so a naive conclusion is that the number of sections in the normal bundle to $2D_5$ is 375. But, we note that because $2D_5$ is the same as the anti-canonical class in $V_4,$ there is one non-trivial equivalence relation imposed by the vanishing of the defining equation for the CY threefold. This reduces the number of sections, and we obtain
\begin{equation}
h^\bullet(2D_5,\mathcal{O}_{2D_5})=(1,0,374)\,.
\end{equation}
Now, among the sections of $2D_5,$ there are in total 65 monomials that are odd under the orientifold projection and in total 309 monomials that are even. As a result, we obtain
\begin{equation}
h^\bullet_+(2D_5,\mathcal{O}_{2D_5})=(1,0,65)\,,
\end{equation}
and
\begin{equation}
h^\bullet_-(2D_5,\mathcal{O}_{2D_5})=(0,0,309)\,.
\end{equation}
This result reproduces $h^\bullet(\widehat{D}_5,\mathcal{O}_{\widehat{D}_5})=(1,0,65).$

\begin{table}
\centering
\begin{tabular}{|c|c|c|c|c|}
\hline
&$h^{0,0}$&$h^{1,0}$&$h^{2,0}$&$h^{1,1}$\\\hline
&&&&\\[-1em]
$\widehat{D}_1$&1&0&0&2\\\hline
&&&&\\[-1em]
$\widehat{D}_4$&1&0&0&1\\\hline
&&&&\\[-1em]
$\widehat{D}_5$&1&0&65&417\\\hline
&&&&\\[-1em]
$\widehat{D}_6$&1&0&0&1\\\hline
\end{tabular}
\caption{\label{cap:D4 in 11169ori}The Hodge numbers of $\widehat{D}_i,$ in $\Bbb{P}_{[1,1,1,6,9]}[18]/\Bbb{Z}_2.$}
\end{table}

\begin{figure}
\centering
\begin{tikzcd}[row sep=tiny, column sep=tiny]
&&&&1&&&\\
&&&0&&0&&\\
&&0&&7&&0&\\
&0&&0&&0&&0&\\
1&&7341&&29436&&7341&&1
\end{tikzcd}
\caption{\label{cap:Y4 11169}The Hodge diamond of the F-theory uplift of the orientifold of $\Bbb{P}_{[1,1,1,6,9]}[18]$.}
\end{figure}
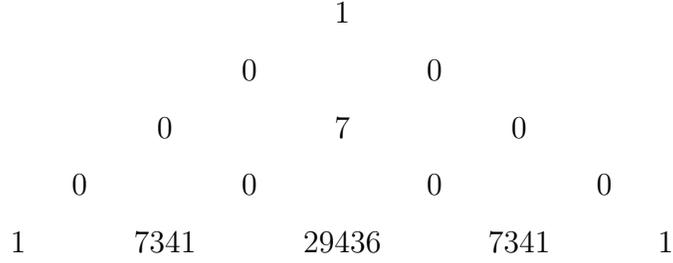

We now construct the F-theory uplift of a type IIB compactification on $B_3.$ The GLSM is 
\begin{equation}
\begin{array}{c|c|c|c|c|c|c|c|c}
\overline{x}_1&\overline{x}_2&\overline{x}_3&\overline{x}_4&\overline{x}_5&\overline{x}_6&X&Y&Z\\\hline
1&1&1&6&18&0&18&27&0\\\hline
0&0&0&2&6&2&8&12&0\\\hline
0&0&0&0&0&0&2&3&1
\end{array}
\end{equation}
and the corresponding toric rays are
\begin{equation}
\left(\begin{array}{ccccccccc}
\vec{\overline{v}}_1&\vec{\overline{v}}_2&\vec{\overline{v}}_3&\vec{\overline{v}}_4&\vec{\overline{v}}_5&\vec{\overline{v}}_6&\vec v_X&\vec v_Y&\vec v_Z\\
1&0&-1&0&0&0&0&0&0\\
0&1&-1&0&0&0&0&0&0\\
-2&-2&-2&1&0&-1&0&0&0\\
-3&-3&-3&0&1&-3&0&0&0\\
-2&-2&-2&-2&0&-2&1&0&-2\\
-3&-3&-3&-3&0&-3&0&1&-3
\end{array}\right)\,.
\end{equation}
Note that $\overline{x}_6 = 0$ hosts a non-Higgsable seven brane stack with $SO(8)$ gauge group. This can be seen from the fact that $D_6$ is a fixed locus of the orientifold involution, hence should host an O7-plane, as well as the fact that $h^{2,0}(D_6)=0$, which implies that the seven branes wrapped on $D_6$ are rigid. As a result, using the Shioda-Tate-Wazir theorem, we conclude that
\begin{equation}
h^{1,1}(Y_4)=7\,.
\end{equation}
Because $h^{2,1}(B_3)=0,$ we conclude that $h^{2,1}(Y_4)=0.$ We find the number of seven brane moduli to be 
\begin{equation}
n_{D7}=7068\,.
\end{equation}
As a result, we obtain
\begin{equation}
h^{3,1}(Y_4)=7341\,.
\end{equation}
We record the Hodge diamond of $Y_4$ in Figure \ref{cap:Y4 11169}.

We are finally ready to compute the Hodge numbers of vertical prime toric divisors. We record their Hodge numbers in Table \ref{cap:D4 in F 11169}.

\begin{table}
\centering
\begin{tabular}{|c|c|c|c|c|c|c|}
\hline
&$h^{0,0}$&$h^{1,0}$&$h^{2,0}$&$h^{3,0}$&$h^{1,1}$&$h^{2,1}$\\\hline
&&&&&&\\[-1em]
$\overline{D}_1$&1&0&0&2&7&414\\\hline
&&&&&&\\[-1em]
$\overline{D}_2$&1&0&0&2&7&414\\\hline
&&&&&&\\[-1em]
$\overline{D}_3$&1&0&0&2&7&414\\\hline
&&&&&&\\[-1em]
$\overline{D}_4$&1&0&0&28&2&2729\\\hline
&&&&&&\\[-1em]
$\overline{D}_5$&1&0&65&309&418&10381\\\hline
&&&&&&\\[-1em]
$\overline{D}_6$&1&0&0&0&2&0\\\hline
\end{tabular}
\caption{\label{cap:D4 in F 11169}The Hodge numbers of $\overline{D}_i,$ for $i=1,\dots,6,$ in the F-theory uplift of the orientifold of $\Bbb{P}_{[1,1,1,6,9]}[18].$}
\end{table}

\subsection{An orientifold with twisting line bundle of Iitaka dimension 1}
\begin{figure}
\centering
\begin{tikzcd}[row sep=tiny, column sep=tiny]
&&&1&&&\\
&&0&&0&&\\
&0&&3&&0&\\
1&&75&&75&&1\\
&0&&3&&0&\\
&&0&&0&&\\
&&&1&&&
\end{tikzcd}
\caption{\label{cap:hodge P2P112}The Hodge diamond of $Z_3\subset \Bbb{P}^2\times \Bbb{P}_{[2,1,1]}.$ }
\end{figure}
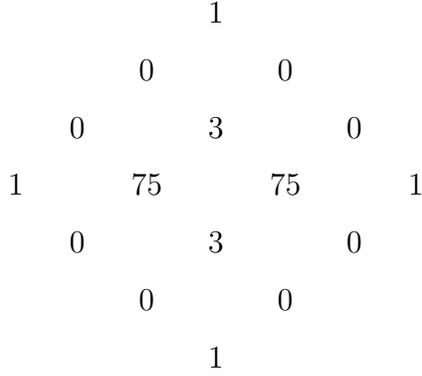

In this section, we study a CY hypersurface $Z_3$ in $V_4:=\Bbb{P}^2\times \Bbb{P}_{[2,1,1]}$ and the F-theory uplift of the orientifold $B_3:=Z_3/\Bbb{Z}_2.$ The GLSM of the geometry in question is 
\begin{equation}
\begin{array}{c|c|c|c|c|c|c}
x_1&x_2&x_3&x_4&x_5&x_6&x_7\\\hline
1&1&1&0&0&0&0\\\hline
0&0&0&2&1&1&0\\\hline
0&0&0&1&0&0&1
\end{array}
\end{equation}
and the toric rays are given as
\begin{equation}
\left(
\begin{array}{ccccccc}
\vec v_1&\vec v_2&\vec v_3&\vec v_4&\vec v_5&\vec v_6&\vec v_7\\
1&0&-1&0&0&0&0\\
0&1&-1&0&0&0&0\\
0&0&0&1&-1&-1&-1\\
0&0&0&0&1&-1&0
\end{array}
\right)\,.
\end{equation}
The geometry can be understood as a blow up of $\Bbb{P}^2 \times \Bbb{P}_{[2,1,1]}$. This results in the Stanley-Reisner ideal
\begin{equation}
SRI(V_4)=\{ x_1x_2x_3,x_4x_7,x_5x_6\}\,.
\end{equation}
Again, we define the CY hypersurface $Z_3$ to be the vanishing locus of the anti-canonical class. We record the Hodge numbers of $Z_3$ in Figure \ref{cap:hodge P2P112}. We compute the Hodge numbers of prime toric divisors in $Z_3,$ and record them in Table \ref{cap:D4 in P2P112}.

\begin{table}
\centering
\begin{tabular}{|c|c|c|c|c|}
\hline
&$h^{0,0}$&$h^{1,0}$&$h^{2,0}$&$h^{1,1}$\\\hline
$D_1$&1&0&2&30\\\hline
$D_2$&1&0&2&30\\\hline
$D_3$&1&0&2&30\\\hline
$D_4$&1&0&3&40\\\hline
$D_5$&1&0&1&20\\\hline
$D_6$&1&0&1&20\\\hline
$D_7$&1&1&0&2\\\hline
\end{tabular}
\caption{\label{cap:D4 in P2P112}The Hodge numbers of $D_i$  in $Z_3\subset \Bbb{P}^2\times\Bbb{P}_{[2,1,1]}.$}
\end{table}

To construct an orientifold of $Z_3,$ we consider an involution $\mathcal I_{\vec v_4}$ 
\begin{equation}
\mathcal I_{\vec v_4}: x_4\mapsto -x_4\,.
\end{equation}
It is easy to check that there is a different (but equivalent) presentation of the involution, namely
\begin{equation}
g\cdot \mathcal I_{\vec v_4}\cdot g^{-1} : x_7\mapsto -x_7\,.
\end{equation}
The refined ``image'' GLSM under the refinement map $\varphi_{\mathcal I_{\vec v_4}}$ is
\begin{equation}
\begin{array}{c|c|c|c|c|c|c}
\widehat{x}_1&\widehat{x}_2&\widehat{x}_3&\widehat{x}_4&\widehat{x}_5&\widehat{x}_6&\widehat{x}_7\\\hline
1&1&1&0&0&0&0\\\hline
0&0&0&4&1&1&0\\\hline
0&0&0&1&0&0&1
\end{array}
\end{equation}
and the generators of the toric rays are
\begin{equation}
\left(\begin{array}{ccccccc}
\vec{\widehat{v}}_1&\vec{\widehat{v}}_2&\vec{\widehat{v}}_3&\vec{\widehat{v}}_4&\vec{\widehat{v}}_5&\vec{\widehat{v}}_6&\vec{\widehat{v}}_7\\
1&0&-1&0&0&0&0\\
0&1&-1&0&0&0&0\\
0&0&0&1&-2&-2&-1\\
0&0&0&0&1&-1&0
\end{array}
\right)\,.
\end{equation}
The orientifold class $B_3$ is 
\begin{equation}
-K_{\widehat{V}_4}-{L}=\widehat{D}_1+\widehat{D}_2+\widehat{D}_3+\widehat{D}_5+\widehat{D}_6+\frac{1}{2}\left(\widehat{D}_4+\widehat{D}_7\right)\,,
\end{equation}
where ${L}$ is the class
\begin{equation}
{L}= \frac{1}{2}\left(\widehat{D}_4+\widehat{D}_7\right)=\widehat{D}_5+\widehat{D}_6+\widehat{D}_7\,.
\end{equation}
Using \eqref{eqn:h21p}, we compute
\begin{equation}
h^{2,1}(B_3)=28\,
\end{equation}
and we obtain 
	\begin{equation} 
	h^{1,1}(B_3)=3.
\end{equation} 
We record the Hodge numbers of $B_3$ in Figure \ref{cap:hodge P2P211ori}. Similarly, we compute the Hodge numbers of prime toric divisors in $B_3$ and record the results in Table \ref{cap:D4 in P2P112ori}.

\begin{figure}
\centering
\begin{tikzcd}[row sep=tiny, column sep=tiny]
&&&1&&&\\
&&0&&0&&\\
&0&&3&&0&\\
0&&28&&28&&0\\
&0&&3&&0&\\
&&0&&0&&\\
&&&1&&&
\end{tikzcd}
\caption{\label{cap:hodge P2P211ori}The Hodge diamond of the Calabi-Yau orientifold in $\Bbb{P}^2\times\Bbb{P}_{[4,1,1]}.$ Note  $h^{p,q}(B_3)=h^{p,q}_+(Z_3).$}
\end{figure}

\begin{table}
\centering
\begin{tabular}{|c|c|c|c|c|}
\hline
&$h^{0,0}$&$h^{1,0}$&$h^{2,0}$&$h^{1,1}$\\\hline
&&&&\\[-1em]
$\widehat{D}_1$&1&0&0&14\\\hline
&&&&\\[-1em]
$\widehat{D}_2$&1&0&0&14\\\hline
&&&&\\[-1em]
$\widehat{D}_3$&1&0&0&14\\\hline
&&&&\\[-1em]
$\widehat{D}_4$&1&0&3&40\\\hline
&&&&\\[-1em]
$\widehat{D}_5$&1&0&0&10\\\hline
&&&&\\[-1em]
$\widehat{D}_6$&1&0&0&10\\\hline
&&&&\\[-1em]
$\widehat{D}_7$&1&1&0&2\\\hline
\end{tabular}
\caption{\label{cap:D4 in P2P112ori}The Hodge numbers of $D_i$  in $B_3\subset \Bbb{P}^2\times\Bbb{P}_{[4,1,1]}.$}
\end{table}

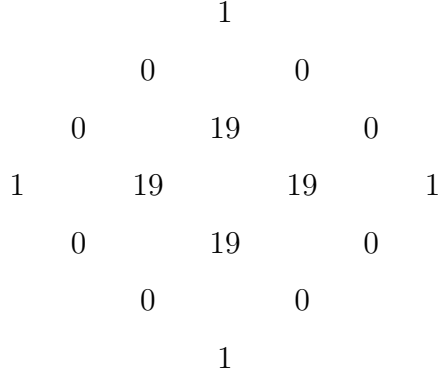
\begin{figure}
\centering
\begin{tikzcd}[row sep=tiny, column sep=tiny]
&&&1&&&\\
&&0&&0&&\\
&0&&19&&0&\\
1&&19&&19&&1\\
&0&&19&&0&\\
&&0&&0&&\\
&&&1&&&
\end{tikzcd}
\caption{\label{enrique three}The Hodge diamond of the Enriques threefold.}
\end{figure}

We claim that $\mathcal{L}$ resctricted to the prime toric divisors $\widehat{D}_i$ for $i=4,5,6$ has Iitaka dimension 1. To explain this claim in detail, we study the divisor $\widehat{D}_5.$ The divisor $\widehat{D}_5 =[\widehat{x}_5]$ can be understood as a hypersurface in a three-dimensional toric variety $V_3$ defined by the following GLSM,
\begin{equation}
\begin{array}{c|c|c|c|c}
\widehat{x}_1&\widehat{x}_2&\widehat{x}_3&\widehat{x}_4&\widehat{x}_7\\\hline
1&1&1&0&0\\\hline
0&0&0&1&1
\end{array}
\end{equation}
where $\widehat{x}_5 = 0$ is the vanishing locus of a bi-degree (3,1) polynomial in $V_3.$ As a result, $[\widehat{x}_5]$ is a genus one fibration over $\Bbb{P}^1,$ which makes it an Enriques surface. Now, the first Chern class of the twisting line bundle $\mathcal{L}$ restricted to $[\widehat{x}_5]$ is isomorphic to $[\widehat{x}_4].$ As a result, the sections of $\mathcal{L}$ only vary along the base of the genus one fibration. This is enough to prove that the Iitaka dimension of $\mathcal{L}|_{\widehat{D}_5}$ is 1. Now let us construct an elliptic fibration over $\widehat{D}_5$ with twisting line bundle $\mathcal{L}.$ The corresponding GLSM is 
\begin{equation}
\begin{array}{c|c|c|c|c|c|c|c}
\overline{x}_1&\overline{x}_2&\overline{x}_3&\overline{x}_4&\overline{x}_7&X&Y&Z\\\hline
1&1&1&0&0&0&0&0\\\hline
0&0&0&1&1&2&3&0\\\hline
0&0&0&0&0&2&3&1
\end{array}
\end{equation}
where the uplifted divisor $\overline{D}_5$ is a complete intersection of the form 
\begin{equation}
\overline{D}_5:= \left(\overline{D}_1+\overline{D}_2+\overline{D}_3+\overline{D}_4\right)\cap \left(\overline{D}_7+\overline{D}_X+\overline{D}_Y+\overline{D}_Z\right)\,.
\end{equation}
As a result, one can easily check that $\overline{D}_5$ is a CY threefold with a genus one fibration and an elliptic fibration over $\Bbb{P}^1.$ In fact, $\overline{D}_5$ is the Enriques threefold, whose Hodge numbers are recorded in Figure \ref{enrique three}. As is expected, $h^{1,1}(\overline{D}_5)$ is different from the results that would be obtained by a naive application of the Shioda-Tate-Wazir theorem. The Shioda-Tate-Wazir theorem tells us that
\begin{equation}
\text{rank} (\text{Pic}(\overline{D}_5)) = 11+1+4\,.
\end{equation}
This discrepancy is due to the fact that for the Enriques threefold, $h^{1,1}$ need not be the same as the Picard rank. In fact, $h^{1,1}$ can be obtained by counting the Picard rank at a point in the moduli space where the maximal gauge rank is engineered, which correctly indicates
\begin{equation}
\text{rank} (\text{Pic}(\overline{D}_5))=11+1+8=19\,.
\end{equation}
The above Picard rank agrees with $h^{1,1}(\overline{D}_5).$ Combining $h^{1,1}(\overline{D}_5)$ with $\chi(\overline{D}_5)$ computed using pushforward techniques, we can compute $h^{2,1}(\overline{D}_5)=19,$ which agrees with the expected value of $h^{2,1}$ of the Enriques threefold. 

We remark that one can similarly show that $\mathcal{L}|_{\overline{D}_4}$ also has Iitaka dimension 1. The GLSM for the ambient space $V_5$ of the divisor $\overline{D}_4$ is given as
\begin{equation}\label{eqn:GLSM D4 P2P211}
\begin{array}{c|c|c|c|c|c|c|c}
\overline{x}_1&\overline{x}_2&\overline{x}_3&\overline{x}_5&\overline{x}_6&X&Y&Z\\\hline
1&1&1&0&0&0&0&0\\\hline
0&0&0&1&1&4&6&0\\\hline
0&0&0&0&0&2&3&1
\end{array}
\end{equation}
where $\overline{D}_4$ is defined as a complete intersection of multi-degree (3,4,0) and (0,12,6) polynomials. Because the Iitaka dimension is one and at the same time $\overline{D}_4$ is not rigid, a study of the divisor $\overline{D}_4$ requires a lot of care. Let us recall the GLSM of the CY threefold
\begin{equation}
\begin{array}{c|c|c|c|c|c|c}
x_1&x_2&x_3&x_4&x_5&x_6&x_7\\\hline
1&1&1&0&0&0&0\\\hline
0&0&0&2&1&1&0\\\hline
0&0&0&1&0&0&1
\end{array}
\end{equation}
and first study the topology of $\varphi_{\mathcal I_4}(\widehat{D}_4)=2D_4.$ The linear system $|2D_4|$ is obtained by collecting monomials of multi-degree $(0,4,2),$ of the form
\begin{equation}
x_4^2 +x_4 x_7 f_2(x_5,x_6)+x_7^2f_4(x_5,x_6)\,,
\end{equation}
where $f_n$ is a degree $n$ polynomial in $x_5,x_6.$ As a result, we find that there are in total 9 monomials in the linear system $|2D_4|.$ This implies 
\begin{equation}
h^{2,0}(2D_4)=8\,.
\end{equation}
Because $2D_4$ is connected, we obtain $h^{0,0}(2D_4)=1.$ To compute $h^{1,0}(2D_4),$ we can use the Hirzebruch-Riemann-Roch theorem
\begin{equation}
\chi(2D_4,\mathcal{O}_{2D_4})=\int_{2D_4}\frac{c_1^2+c_2}{12}=8\,.
\end{equation}
As a result, we obtain 
\begin{equation}
h^{1,0}(2D_4)=1\,.
\end{equation}
Comparing this result to 
\begin{equation}\label{eqn:d4P2P211 1}
h^{\bullet}(\widehat{D}_4,\mathcal{O}_{\widehat{D}_4})=(1,0,3)\,,
\end{equation}
we obtain
\begin{equation}\label{eqn:d4P2P211 2}
h^\bullet_-(2D_4,\mathcal{O}_{2D_4})=(0,1,5)\,.
\end{equation}
We can combine \eqref{eqn:d4P2P211 1} and \eqref{eqn:d4P2P211 2} to compute
\begin{equation}
h^\bullet(\overline{D}_4,\mathcal{O}_{\overline{D}_4})=(1,0,4,5)\,.
\end{equation}
To compute $h^{1,1}(\overline{D}),$ we use the following trick. From the structure of the GLSM \eqref{eqn:GLSM D4 P2P211}, we know that $\overline{D}_4$ has two fibration structures, one of which is characterized by a genus one fiber, while the other consists of an  elliptic fiber. Quite crucially, these two fibrations are independent of one another. This implies that we can count the contributions to $h^{1,1}$ from each fiber to compute $h^{1,1}(\overline{D}).$ This is enough to conclude that
\begin{equation}
h^{1,1}(\overline{D}_4)=3+18+38=59\,.
\end{equation}
To complete the computation of the Hodge numbers of $\overline{D}_4,$ we can compute 
\begin{equation}
\chi(\overline{D}_4)=0\,,
\end{equation}
either by using pushforward techniques or by using the toric description \eqref{eqn:GLSM D4 P2P211}. We finally complete the computation of the Hodge numbers of $\overline{D}_4$ by computing
\begin{equation}
h^{2,1}(\overline{D}_4)=63\,.
\end{equation}
We record the Hodge numbers of the vertical divisors in the F-theory uplift in Table \ref{cap:D4 in P2P112f}.

\begin{table}
\centering
\begin{tabular}{|c|c|c|c|c|c|c|}
\hline
&$h^{0,0}$&$h^{1,0}$&$h^{2,0}$&$h^{3,0}$&$h^{1,1}$&$h^{2,1}$\\\hline
&&&&&&\\[-1em]
$\overline{D}_1$&1&0&0&2&19&102\\\hline
&&&&&&\\[-1em]
$\overline{D}_2$&1&0&0&2&19&102\\\hline
&&&&&&\\[-1em]
$\overline{D}_3$&1&0&0&2&19&102\\\hline
&&&&&&\\[-1em]
$\overline{D}_4$&1&0&4&5&59&63\\\hline
&&&&&&\\[-1em]
$\overline{D}_5$&1&0&0&1&19&19\\\hline
&&&&&&\\[-1em]
$\overline{D}_6$&1&0&0&1&19&19\\\hline
&&&&&&\\[-1em]
$\overline{D}_7$&1&1&0&0&3&1\\\hline
\end{tabular}
\caption{\label{cap:D4 in P2P112f}The Hodge numbers of $\overline{D}_i$  in the F-theory uplift of $B_3\subset \Bbb{P}^2\times\Bbb{P}_{[4,1,1]}.$}
\end{table}
\subsection{An orientifold of the mirror of $\Bbb{P}_{[1,1,1,6,9]}[18]$}
\label{mirrorexample}
In this section, we study the F-theory uplift of an orientifold of the mirror of $\Bbb{P}_{[1,1,1,6,9]}[18].$ Because the toric data of this CY manifold is quite complicated, it would not be efficient to record all of the data in the body of this paper. Much of the toric data is contained in ancillary files accompanying this arXiv upload. To analyze this example, we used the software package `CYtools' \cite{cytools}.

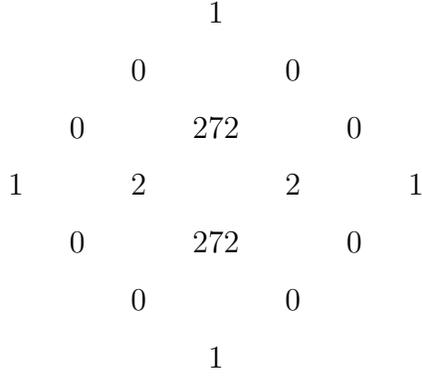
\begin{figure}
\centering
\begin{tikzcd}[row sep=tiny, column sep=tiny]
&&&1&&&\\
&&0&&0&&\\
&0&&272&&0&\\
1&&2&&2&&1\\
&0&&272&&0&\\
&&0&&0&&\\
&&&1&&&
\end{tikzcd}
\caption{\label{mirror P11169}The Hodge diamond of the mirror of $\Bbb{P}_{[1,1,1,6,9]}[18]$.}
\end{figure}

The mirror of $\Bbb{P}_{[1,1,1,6,9]}[18]$ is an anti-canonical hypersurface in a toric fourfold $V_4.$ The toric variety $V_4$ is defined by an FRST $\mathcal{T}$\footnote{The simplices of the FRST $\mathcal{T}$ are recorded in the file `simplices.dat'.} of a reflexive polytope $\Delta^\circ,$ which is itself taken to be the convex hull of the following points
\begin{equation}
\left(
\begin{array}{ccccc}
\vec v_1&\vec v_2&\vec v_3&\vec v_4&\vec v_5\\
 -15 & 0 & 1 & 3 & 3 \\
 -14 & 1 & 0 & 4 & 4 \\
 -12 & 0 & 0 & 6 & 6 \\
 -18 & 0 & 0 & 0 & 18 \\
\end{array}
\right)\,.
\end{equation}
Note that $\vec v_i,$ for $i=1,\dots,5,$ are in fact vertices of $\Delta^\circ.$ There are in total 375 points in $\partial\Delta^\circ,$ and 276 of them are not in strict interior of facets of $\Delta^\circ.$ As a result, only 276 of prime toric divisors intersect the CY threefold $Z_3$. We record the Hodge diamond of $Z_3$ in Figure \ref{mirror P11169}. We record the Hodge numbers of prime toric divisors in \eqref{tab:hodge mirror P11169} and \eqref{tab:hodge mirror P11169 2}. Note that $D_i$ denotes the prime toric divisor that corresponds to $\vec v_i$, which can be found in the ancillary file `points.dat'.

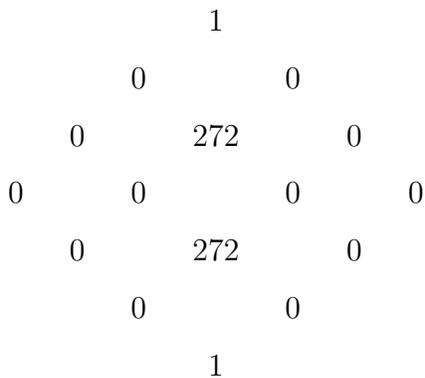
\begin{figure}
\centering
\begin{tikzcd}[row sep=tiny, column sep=tiny]
&&&1&&&\\
&&0&&0&&\\
&0&&272&&0&\\
0&&0&&0&&0\\
&0&&272&&0&\\
&&0&&0&&\\
&&&1&&&
\end{tikzcd}
\caption{\label{mirror P11169ori}The Hodge diamond of the orientifold of the mirror of $\Bbb{P}_{[1,1,1,6,9]}[18]$.}
\end{figure}

Let us study the orientifold involution
\begin{equation}
\mathcal I_{\vec v_3}:x_3\mapsto -x_3\,.
\end{equation}
The resulting refinement map $\varphi_{\mathcal I_{\vec p}}^{-1}$ is defined as follows:
\begin{equation}
\varphi_{\mathcal I_{\vec v_3}}^{-1}: \vec v=(v_1,v_2,v_3,v_4)\mapsto\vec{\widehat{v}}= 
\begin{cases}
(2v_1,v_2,v_3,v_4)& \text{ if } \vec v_3+\vec v\not\equiv \vec 0\mod 2\\
\frac{1}{2}(2v_1,v_2,v_3,v_4)& \text{ if } \vec v_3+\vec v\equiv \vec 0\mod2
\end{cases}\,,
\end{equation}
for $\vec v\in \partial\Delta^\circ.$ We find that there are in total 65 O7-planes. Indices for the O7-planes are given by
\begin{align}
\{&1,3,4,5,8,9,12,13,17,18,22,23,26,27,28,31,32,35,36,39,40,41,42,50,52,54,\nonumber\\
&56,58,60,62,64,70,79,81,94,96,98,117,119,121,123,143,144,148,150,152,154\nonumber\\
&156,179,180,184,186,188,190,192,194,221,223,228,230,232,234,236,238,240\}\,.
\end{align}
Note that all O7-planes but $D_3$ host non-Higgsable $SO(8)$ stacks. We find that there are in total 103 O3-planes. We record the O3-planes in the ancillary file `o3planes.dat'. We compute the Hodge numbers of the orientifold $B_3=Z_3/\Bbb{Z}_2$ and record them in Figure \ref{mirror P11169ori}. We find that, except for $\widehat{D}_6,$ all the prime toric divisors in the orientifold satisfy
\begin{equation}\label{eqn:hodge divs P11169ori}
h^{p,q}(\widehat{D}_i)=h^{p,q}(D_i)\,.
\end{equation}
Note that the right-hand side of \eqref{eqn:hodge divs P11169ori} is $D_i$, not $\varphi_{\mathcal I_{\vec v_3}}(\widehat{D}_i).$ For $\widehat{D}_6,$ we obtain
\begin{equation}
h^\bullet(\widehat{D}_6,\mathcal{O}_{\widehat{D}})=(1,0,0)\,,
\end{equation}
and
\begin{equation}
h^{1,1}(\widehat{D}_6)=19\,.
\end{equation}

\begin{figure}
\centering
\begin{tikzcd}[row sep=tiny, column sep=tiny]
&&&&1&&&\\
&&&0&&0&&\\
&&0&&529&&0&\\
&0&&8&&8&&0&\\
1&&23&&2236&&23&&1
\end{tikzcd}
\caption{\label{cap:Y4 11169mirror}The Hodge diamond of the F-theory uplift of the orientifold of the mirror of $\Bbb{P}_{[1,1,1,6,9]}[18]$.}
\end{figure}
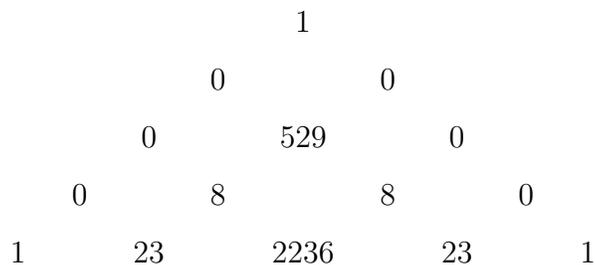

We construct the F-theory uplift $Y_4$ as a generic elliptic fibration over $B_3.$ The nef partition\footnote{For the definition of a nef partition, see \cite{Batyrev:1994pg}.} data is recorded in the ancillary file `nef.dat', and the six-dimensional polytope data is recorded in the ancillary file `6dpoints.dat'. We record the Hodge numbers of the elliptic fourfold $Y_4$ in Figure \ref{cap:Y4 11169mirror}. Note that the D3-brane tadpole computed from the weakly coupled type IIB description is
\begin{equation}
Q_{D3}=138\,,
\end{equation}
which matches
\begin{equation}
\frac{\chi(Y_4)}{24}=138\,,
\end{equation}
as it should. We record the Hodge numbers of vertical prime toric divisors in \eqref{eqn:D4 mirror P11169 1}, \eqref{eqn:D4 mirror P11169 2}, and \eqref{eqn:D4 mirror P11169 3}. We note that all but two vertical prime toric divisors are rigid and have trivial intermediate Jacobian.

\section{Discussion}\label{sec:discussion}
We have presented an algorithm for analyzing the Hodge structure of vertical divisors $\overline{D}$ in elliptic Calabi-Yau fourfolds that admit a global Sen limit. Understanding the Hodge structure of vertical divisors $\overline {D}$ is crucial for analyzing the intermediate Jacobian of $\overline{D}$, which is the relevant mathematical structure that determines the moduli dependence of the one-loop Pfaffian appearing in non-perturbative contributions to the 4D $\mathcal N=1$ superpotential that can be generated by Euclidean M5-branes wrapping $\overline{D}$. Euclidean M5 branes in M/F-theory compactifications on elliptic Calabi-Yau fourfolds that admit a global Sen limit are particularly interesting because these string constructions admit a weakly-coupled description in type IIB string theory and thus represent one of the brightest lampposts in the 4D $\mathcal N=1$ string landscape. Furthermore, elliptic Calabi-Yau fourfolds admitting a global Sen limit also often admit dual descriptions in heterotic string theory, which makes them a unique playground for exploring non-perturbative physics relevant for understanding the string landscape in multiple branches of string theory.

As mentioned at the beginning of this paper, several recent papers (see, e.g., \cite{Dudas:2019pls,Demirtas:2019sip,Demirtas:2020ffz,Alvarez-Garcia:2020pxd}) studying flux vacua with small flux superpotential have provided compelling evidence that light complex structure moduli cannot be avoided in generic KKLT-like constructions; in particular, this suggests that the one-loop Pfaffian of Euclidean D3-instanton contributions to the superpotential cannot generically be ignored. These papers are thus strong motivation to develop robust and efficient methods for analyzing the geometric properties of vertical divisors of elliptic Calabi-Yau fourfolds that can be wrapped by EM5 instantons in the dual M-theory picture, leading to a non-perturbative superpotential. Our algorithm addresses this need by providing a combinatorial procedure for computing Hodge numbers and related topological data in large classes of global Sen limit constructions, which bypasses the need for cumbersome ``manual'' computations---in particular, our algorithm only requires the topological data of a Calabi-Yau orientifold defining the type IIB compactification as input, and does not require a explicit construction of the full elliptic Calabi-Yau fourfold that defines the F-theory uplift. 

In order to devise our algorithm, we first prove combinatorial formulae for the equivariant Hodge numbers of coordinate flip O3/O7 Calabi-Yau orientifolds and their prime toric divisors, and then we use (local) pushforward techniques to study the Hodge structure of vertical divisors corresponding to the F-theory uplifts of prime toric divisors in the orientifold. A key part of our algorithm involves computing the Euler characteristic of these vertical divisors in order to recover the Hodge number $h^{2,1}(\overline{D})$, which in certain cases depends on detailed information about the F-theory uplift that is difficult to compute more directly. This strategy provides sufficient information to compute the Hodge structure of an EM5-instanton wrapping a vertical prime toric divisor. 

Although the procedure described above works without complication for suitable vertical divisors (e.g., smooth vertical divisors), because orientifold Calabi-Yau threefolds can contain O3 planes, vertical divisors more generally may contain pointlike terminal $\mathbb Z_2$ singularities whose physics cannot easily be studied using classic geometric formulae. In particular, the formula for the Euler characteristic of such a vertical divisor, given by the integral of its third Chern class, is demonstrably incorrect. To address this issue, we conjecture a combinatorial formula for the Euler characteristic for vertical divisors that contain point-like terminal $\mathbb Z_2$ orbifold singularities. The formula we propose consists of a correction term to a more familiar expression for the Euler characteristic, where the correction term depends on the number of O3-planes intersecting the vertical divisor. We have provided local geometric arguments for why our conjectured formula is correct, as well as numerous cross-checks in examples for which the Hodge structure can be computed ``manually'' in terms of the Hodge-Deligne numbers of a toric realization of an elliptic Calabi-Yau fourfold. 

There are a number of interesting future research directions closely related to the results of this paper that await exploration. We list a few below, for concreteness:
\begin{itemize}
	\item In this work, although we devised a method to compute the dimension of the intermediate Jacobian $\mathcal J(\overline{D}):=H^3(\overline{D},\Bbb{R})/H^3(\overline{D},\Bbb{Z}),$ we did not study how the complex structure of $\mathcal J(\overline{D})$ is explicitly related to the complex structure of the elliptic Calabi-Yau fourfold $Y_4$ containing the vertical divisor $\overline{D}$. To our best knowledge, this problem has not been solved in general, and has only been solved in the very special case that $\mathcal J(\overline{D})$ only receives contributions from $H^{1,1}_-(D)$ in the global Sen limit \cite{Kerstan:2012cy}. To fully determine the complex structure moduli-dependence of the one-loop Pfaffian in the non-perturbative superpotential, it is necessary to understand how the complex structure moduli of $\mathcal J(\overline{D})$ encode the complex structure moduli of the Calabi-Yau fourfold $Y_4$. It is therefore extremely important to make progress on understanding the precise map between the complex structure of the elliptic Calabi-Yau fourfold and the complex structure of the intermediate Jacobian. 
	\item One of the main complications we have addressed in this paper is the presence of terminal $\mathbb Z_2$ singularities in elliptic Calabi-Yau fourfold backgrounds. By focusing exclusively on elliptic Calabi-Yau fourfolds that admit a global Sen limit, we have severely restricted the types of singularities that can arise, terminal or otherwise. However, in more general F-theory compactifications, it is possible for more general terminal singularities to occur, for example, terminal $\mathbb Z_k$ orbifold singularities with $ k >2$. It would be very interesting to study the M5-brane partition function in the presence of more general terminal singularities, such as pointlike $\Bbb{Q}$ factorial $\Bbb{Z}_{k>2}$ singularities, or perhaps to be able to use our conjectured formula for the topological data of elliptic Calabi-Yau fourfolds in the presence of such singularities to better understand other physical properties of their corresponding vacua. An interesting, but still rather mysterious class of singularities corresponds to the cases $k=3,4,6$, which are expected to correspond to non-perturbative O3-planes \cite{Garcia-Etxebarria:2015wns}. 
	\item In this work, we studied properties of M/F-theory compactifications admitting a weakly-coupled type IIB string theory description. It would be illuminating to relate the data characterizing these M/F-theory vacua to geometric data characterizing dual heterotic string vacua, specifically by identifying the cohomological data of the spectral cover construction of heterotic string vacua \cite{Friedman:1997yq} with the cohomological data readily available in F-theory. In addition to potentially sharpening our understanding of F-theory/heterotic string theory duality, the complementary perspective provided by the heterotic string description may produce new insights into the physics of these vacua that are not obvious in the type IIB/F/M-theory duality frames.
	\item The central result of this paper is a combinatorial algorithm for computing topological data of elliptic Calabi-Yau fourfolds that admit a global Sen limit, which, as we have emphasized, does not require an explicit construction of the global fourfold. Part of the motivation to produce such an algorithm was the observation that not all triangulations of orientifold Calabi-Yau threefolds can be lifted to a resolution of a fourfold corresponding to a convex triangulation of the ambient toric space. Rather, in some cases, the F-theory uplift is forced to be a complete intersection inside a vex triangulation \cite{Berglund:2016yqo,Berglund:2016nvh,vex} of a toric polytope. Although our algorithm sidesteps the difficulties of analyzing vex triangulations (due to a lack of mathematical tools), there is nevertheless strong motivation to generalize conventional techniques for analyzing convex triangulations to the case of vex triangulations, as this would considerably enlarge the set of possible geometries that can be analyzed using combinatorial methods. 
\end{itemize}

\section*{Acknowledgements}
We thank Jakob Moritz, Liam McAllister, Washington Taylor for numerous discussions and comments. We thank Andreas Schachner for catching typos in the manuscript. The work of MK was supported by the Pappalardo Fellowship. PJ was supported by the DOE grant DE-SC00012567.

\newgeometry{top=25mm, bottom=25mm}     
\appendix
\label{app:mirrorexample}
\begin{adjustwidth}{-20pt}{-20pt}

\section{Hodge numbers of prime toric divisors in the mirror of $\Bbb{P}_{[1,1,1,6,9]}[18]$}
In this Appendix, we record the Hodge numbers for divisors $D_i$ in the mirror of $Z_3= \Bbb{P}_{[1,1,1,6,9]}[18]$, along with vertical divisors $\overline{D}$ in the elliptic Calabi-Yau fourfold uplift of the orientfiold of the mirror of $Z_3$. This example is discussed in \S\ref{mirrorexample}. The Hodge numbers are as follows:

\begin{center}
\begin{equation}\label{tab:hodge mirror P11169}
\scalebox{.8}{$
\begin{array}{c|c|c|c|c}
&h^{0,0}&h^{1,0}&h^{2,0}&h^{1,1}\\
 D_1 & 1 & 0 & 0 & 1 \\
 D_2 & 1 & 0 & 1 & 20 \\
 D_3 & 1 & 0 & 2 & 30 \\
 D_4 & 1 & 0 & 0 & 5 \\
 D_5 & 1 & 0 & 0 & 1 \\
 D_6 & 1 & 0 & 0 & 2 \\
 D_7 & 1 & 0 & 0 & 4 \\
 D_8 & 1 & 0 & 0 & 3 \\
 D_9 & 1 & 0 & 0 & 5 \\
 D_{10} & 1 & 0 & 0 & 4 \\
 D_{11} & 1 & 0 & 0 & 5 \\
 D_{12} & 1 & 0 & 0 & 4 \\
 D_{13} & 1 & 0 & 0 & 6 \\
 D_{14} & 1 & 0 & 0 & 11 \\
 D_{15} & 1 & 0 & 0 & 5 \\
 D_{16} & 1 & 0 & 0 & 6 \\
 D_{17} & 1 & 0 & 0 & 5 \\
 D_{18} & 1 & 0 & 0 & 7 \\
 D_{19} & 1 & 0 & 0 & 5 \\
 D_{20} & 1 & 0 & 0 & 7 \\
 D_{21} & 1 & 0 & 0 & 18 \\
 D_{22} & 1 & 0 & 0 & 6 \\
 D_{23} & 1 & 0 & 0 & 7 \\
 D_{24} & 1 & 0 & 0 & 6 \\
 D_{25} & 1 & 0 & 0 & 7 \\
 D_{26} & 1 & 0 & 0 & 12 \\
 D_{27} & 1 & 0 & 0 & 6 \\
 D_{28} & 1 & 0 & 0 & 7 \\
 D_{29} & 1 & 0 & 0 & 6 \\
 D_{30} & 1 & 0 & 0 & 7 \\
 D_{31} & 1 & 0 & 0 & 6 \\
 D_{32} & 1 & 0 & 0 & 7 \\
 D_{33} & 1 & 0 & 0 & 6 \\
 D_{34} & 1 & 0 & 0 & 7 \\
 D_{35} & 1 & 0 & 0 & 6 \\
 D_{36} & 1 & 0 & 0 & 7 \\

\end{array}$}
\quad
\scalebox{.8}{$
\begin{array}{c|c|c|c|c}
&h^{0,0}&h^{1,0}&h^{2,0}&h^{1,1}\\
 D_{37} & 1 & 0 & 0 & 6 \\
 D_{38} & 1 & 0 & 0 & 6 \\
 D_{39} & 1 & 0 & 0 & 5 \\
 D_{40} & 1 & 0 & 0 & 13 \\
 D_{41} & 1 & 0 & 0 & 6 \\
 D_{42} & 1 & 0 & 0 & 5 \\
 D_{43} & 1 & 0 & 0 & 6 \\
 D_{44} & 1 & 0 & 0 & 18 \\
 D_{45} & 1 & 0 & 0 & 5 \\
 D_{46} & 1 & 0 & 0 & 12 \\
 D_{47} & 1 & 0 & 0 & 4 \\
 D_{48} & 1 & 0 & 0 & 5 \\
 D_{49} & 1 & 0 & 0 & 5 \\
 D_{50} & 1 & 0 & 0 & 5 \\
 D_{51} & 1 & 0 & 0 & 5 \\
 D_{52} & 1 & 0 & 0 & 5 \\
 D_{53} & 1 & 0 & 0 & 5 \\
 D_{54} & 1 & 0 & 0 & 5 \\
 D_{55} & 1 & 0 & 0 & 5 \\
 D_{56} & 1 & 0 & 0 & 5 \\
 D_{57} & 1 & 0 & 0 & 4 \\
 D_{58} & 1 & 0 & 0 & 4 \\
 D_{59} & 1 & 0 & 0 & 4 \\
 D_{60} & 1 & 0 & 0 & 3 \\
 D_{61} & 1 & 0 & 0 & 3 \\
 D_{62} & 1 & 0 & 0 & 3 \\
 D_{63} & 1 & 0 & 0 & 3 \\
 D_{64} & 1 & 0 & 0 & 2 \\
 D_{65} & 1 & 0 & 0 & 2 \\
 D_{66} & 1 & 0 & 0 & 2 \\
 D_{67} & 1 & 0 & 0 & 2 \\
 D_{68} & 1 & 0 & 0 & 3 \\
 D_{69} & 1 & 0 & 0 & 3 \\
 D_{70} & 1 & 0 & 0 & 3 \\
 D_{71} & 1 & 0 & 0 & 3 \\
 D_{72} & 1 & 0 & 0 & 3 \\

\end{array}
$}
\quad
\scalebox{.8}{$
\begin{array}{c|c|c|c|c}
&h^{0,0}&h^{1,0}&h^{2,0}&h^{1,1}\\
 D_{73} & 1 & 0 & 0 & 4 \\
 D_{74} & 1 & 0 & 0 & 3 \\
 D_{75} & 1 & 0 & 0 & 3 \\
 D_{76} & 1 & 0 & 0 & 3 \\
 D_{77} & 1 & 0 & 0 & 3 \\
 D_{78} & 1 & 0 & 0 & 3 \\
 D_{79} & 1 & 0 & 0 & 4 \\
 D_{80} & 1 & 0 & 0 & 4 \\
 D_{81} & 1 & 0 & 0 & 3 \\
 D_{82} & 1 & 0 & 0 & 3 \\
 D_{83} & 1 & 0 & 0 & 3 \\
 D_{84} & 1 & 0 & 0 & 3 \\
 D_{85} & 1 & 0 & 0 & 3 \\
 D_{86} & 1 & 0 & 0 & 4 \\
 D_{87} & 1 & 0 & 0 & 4 \\
 D_{88} & 1 & 0 & 0 & 4 \\
 D_{89} & 1 & 0 & 0 & 3 \\
 D_{90} & 1 & 0 & 0 & 3 \\
 D_{91} & 1 & 0 & 0 & 3 \\
 D_{92} & 1 & 0 & 0 & 3 \\
 D_{93} & 1 & 0 & 0 & 3 \\
 D_{94} & 1 & 0 & 0 & 4 \\
 D_{95} & 1 & 0 & 0 & 4 \\
 D_{96} & 1 & 0 & 0 & 4 \\
 D_{97} & 1 & 0 & 0 & 4 \\
 D_{98} & 1 & 0 & 0 & 3 \\
 D_{99} & 1 & 0 & 0 & 3 \\
 D_{100} & 1 & 0 & 0 & 3 \\
 D_{101} & 1 & 0 & 0 & 3 \\
 D_{102} & 1 & 0 & 0 & 3 \\
 D_{103} & 1 & 0 & 0 & 4 \\
 D_{104} & 1 & 0 & 0 & 3 \\
 D_{105} & 1 & 0 & 0 & 4 \\
 D_{106} & 1 & 0 & 0 & 4 \\
 D_{107} & 1 & 0 & 0 & 4 \\
 D_{108} & 1 & 0 & 0 & 4 \\

\end{array}
$}
\quad
\scalebox{.8}{$
\begin{array}{c|c|c|c|c}
&h^{0,0}&h^{1,0}&h^{2,0}&h^{1,1}\\
 D_{109} & 1 & 0 & 0 & 4 \\
 D_{110} & 1 & 0 & 0 & 3 \\
 D_{111} & 1 & 0 & 0 & 3 \\
 D_{112} & 1 & 0 & 0 & 3 \\
 D_{113} & 1 & 0 & 0 & 3 \\
 D_{114} & 1 & 0 & 0 & 4 \\
 D_{115} & 1 & 0 & 0 & 4 \\
 D_{116} & 1 & 0 & 0 & 3 \\
 D_{117} & 1 & 0 & 0 & 4 \\
 D_{118} & 1 & 0 & 0 & 4 \\
 D_{119} & 1 & 0 & 0 & 4 \\
 D_{120} & 1 & 0 & 0 & 4 \\
 D_{121} & 1 & 0 & 0 & 4 \\
 D_{122} & 1 & 0 & 0 & 4 \\
 D_{123} & 1 & 0 & 0 & 3 \\
 D_{124} & 1 & 0 & 0 & 3 \\
 D_{125} & 1 & 0 & 0 & 3 \\
 D_{126} & 1 & 0 & 0 & 3 \\
 D_{127} & 1 & 0 & 0 & 3 \\
 D_{128} & 1 & 0 & 0 & 3 \\
 D_{129} & 1 & 0 & 0 & 4 \\
 D_{130} & 1 & 0 & 0 & 4 \\
 D_{131} & 1 & 0 & 0 & 3 \\
 D_{132} & 1 & 0 & 0 & 4 \\
 D_{133} & 1 & 0 & 0 & 4 \\
 D_{134} & 1 & 0 & 0 & 4 \\
 D_{135} & 1 & 0 & 0 & 4 \\
 D_{136} & 1 & 0 & 0 & 4 \\
 D_{137} & 1 & 0 & 0 & 4 \\
 D_{138} & 1 & 0 & 0 & 4 \\
 D_{139} & 1 & 0 & 0 & 3 \\
 D_{140} & 1 & 0 & 0 & 3 \\
 D_{141} & 1 & 0 & 0 & 3 \\
 D_{142} & 1 & 0 & 0 & 3 \\
 D_{143} & 1 & 0 & 0 & 3 \\
 D_{144} & 1 & 0 & 0 & 3 \\

\end{array}
$}
\end{equation}
\end{center}
\begin{center}
\begin{equation}\label{tab:hodge mirror P11169 2}
\scalebox{.8}{$
\begin{array}{c|c|c|c|c}
&h^{0,0}&h^{1,0}&h^{2,0}&h^{1,1}\\
 D_{145} & 1 & 0 & 0 & 4 \\
 D_{146} & 1 & 0 & 0 & 4 \\
 D_{147} & 1 & 0 & 0 & 3 \\
 D_{148} & 1 & 0 & 0 & 4 \\
 D_{149} & 1 & 0 & 0 & 4 \\
 D_{150} & 1 & 0 & 0 & 4 \\
 D_{151} & 1 & 0 & 0 & 4 \\
 D_{152} & 1 & 0 & 0 & 4 \\
 D_{153} & 1 & 0 & 0 & 4 \\
 D_{154} & 1 & 0 & 0 & 4 \\
 D_{155} & 1 & 0 & 0 & 4 \\
 D_{156} & 1 & 0 & 0 & 3 \\
 D_{157} & 1 & 0 & 0 & 3 \\
 D_{158} & 1 & 0 & 0 & 2 \\
 D_{159} & 1 & 0 & 0 & 3 \\
 D_{160} & 1 & 0 & 0 & 3 \\
 D_{161} & 1 & 0 & 0 & 3 \\
 D_{162} & 1 & 0 & 0 & 3 \\
 D_{163} & 1 & 0 & 0 & 4 \\
 D_{164} & 1 & 0 & 0 & 4 \\
 D_{165} & 1 & 0 & 0 & 3 \\
 D_{166} & 1 & 0 & 0 & 4 \\
 D_{167} & 1 & 0 & 0 & 4 \\
 D_{168} & 1 & 0 & 0 & 4 \\
 D_{169} & 1 & 0 & 0 & 4 \\
 D_{170} & 1 & 0 & 0 & 4 \\
 D_{171} & 1 & 0 & 0 & 4 \\
 D_{172} & 1 & 0 & 0 & 4 \\
 D_{173} & 1 & 0 & 0 & 4 \\
 D_{174} & 1 & 0 & 0 & 4 \\
 D_{175} & 1 & 0 & 0 & 3 \\
 D_{176} & 1 & 0 & 0 & 3 \\
 D_{177} & 1 & 0 & 0 & 3 \\
 D_{178} & 1 & 0 & 0 & 3 \\
 D_{179} & 1 & 0 & 0 & 3 \\
 D_{180} & 1 & 0 & 0 & 3 \\

\end{array}
\quad
\begin{array}{c|c|c|c|c}
&h^{0,0}&h^{1,0}&h^{2,0}&h^{1,1}\\
 D_{181} & 1 & 0 & 0 & 4 \\
 D_{182} & 1 & 0 & 0 & 4 \\
 D_{183} & 1 & 0 & 0 & 3 \\
 D_{184} & 1 & 0 & 0 & 4 \\
 D_{185} & 1 & 0 & 0 & 4 \\
 D_{186} & 1 & 0 & 0 & 4 \\
 D_{187} & 1 & 0 & 0 & 4 \\
 D_{188} & 1 & 0 & 0 & 4 \\
 D_{189} & 1 & 0 & 0 & 4 \\
 D_{190} & 1 & 0 & 0 & 4 \\
 D_{191} & 1 & 0 & 0 & 4 \\
 D_{192} & 1 & 0 & 0 & 4 \\
 D_{193} & 1 & 0 & 0 & 4 \\
 D_{194} & 1 & 0 & 0 & 3 \\
 D_{195} & 1 & 0 & 0 & 3 \\
 D_{196} & 1 & 0 & 0 & 3 \\
 D_{197} & 1 & 0 & 0 & 3 \\
 D_{198} & 1 & 0 & 0 & 3 \\
 D_{199} & 1 & 0 & 0 & 3 \\
 D_{200} & 1 & 0 & 0 & 4 \\
 D_{201} & 1 & 0 & 0 & 3 \\
 D_{202} & 1 & 0 & 0 & 3 \\
 D_{203} & 1 & 0 & 0 & 4 \\
 D_{204} & 1 & 0 & 0 & 4 \\
 D_{205} & 1 & 0 & 0 & 4 \\
 D_{206} & 1 & 0 & 0 & 4 \\
 D_{207} & 1 & 0 & 0 & 4 \\
 D_{208} & 1 & 0 & 0 & 4 \\
 D_{209} & 1 & 0 & 0 & 4 \\
 D_{210} & 1 & 0 & 0 & 4 \\
 D_{211} & 1 & 0 & 0 & 4 \\
 D_{212} & 1 & 0 & 0 & 4 \\
 D_{213} & 1 & 0 & 0 & 4 \\
 D_{214} & 1 & 0 & 0 & 3 \\
 D_{215} & 1 & 0 & 0 & 3 \\
 D_{216} & 1 & 0 & 0 & 2 \\

\end{array}
$}
\quad
\scalebox{.8}{$
\begin{array}{c|c|c|c|c}
&h^{0,0}&h^{1,0}&h^{2,0}&h^{1,1}\\
 D_{217} & 1 & 0 & 0 & 2 \\
 D_{218} & 1 & 0 & 0 & 3 \\
 D_{219} & 1 & 0 & 0 & 3 \\
 D_{220} & 1 & 0 & 0 & 3 \\
 D_{221} & 1 & 0 & 0 & 3 \\
 D_{222} & 1 & 0 & 0 & 3 \\
 D_{223} & 1 & 0 & 0 & 3 \\
 D_{224} & 1 & 0 & 0 & 3 \\
 D_{225} & 1 & 0 & 0 & 3 \\
 D_{226} & 1 & 0 & 0 & 3 \\
 D_{227} & 1 & 0 & 0 & 3 \\
 D_{228} & 1 & 0 & 0 & 4 \\
 D_{229} & 1 & 0 & 0 & 4 \\
 D_{230} & 1 & 0 & 0 & 4 \\
 D_{231} & 1 & 0 & 0 & 4 \\
 D_{232} & 1 & 0 & 0 & 4 \\
 D_{233} & 1 & 0 & 0 & 4 \\
 D_{234} & 1 & 0 & 0 & 4 \\
 D_{235} & 1 & 0 & 0 & 4 \\
 D_{236} & 1 & 0 & 0 & 4 \\
 D_{237} & 1 & 0 & 0 & 4 \\
 D_{238} & 1 & 0 & 0 & 4 \\
 D_{239} & 1 & 0 & 0 & 4 \\
 D_{240} & 1 & 0 & 0 & 3 \\
 D_{241} & 1 & 0 & 0 & 3 \\
 D_{242} & 1 & 0 & 0 & 3 \\
 D_{243} & 1 & 0 & 0 & 3 \\
 D_{244} & 1 & 0 & 0 & 3 \\
 D_{245} & 1 & 0 & 0 & 3 \\
 D_{246} & 1 & 0 & 0 & 3 \\
 D_{247} & 1 & 0 & 0 & 3 \\
 D_{248} & 1 & 0 & 0 & 3 \\
 D_{249} & 1 & 0 & 0 & 3 \\
 D_{250} & 1 & 0 & 0 & 4 \\
 D_{251} & 1 & 0 & 0 & 4 \\
 D_{252} & 1 & 0 & 0 & 4 \\

\end{array}
$}
\quad
\scalebox{.8}{$
\begin{array}{c|c|c|c|c}
&h^{0,0}&h^{1,0}&h^{2,0}&h^{1,1}\\
 D_{253} & 1 & 0 & 0 & 4 \\
 D_{254} & 1 & 0 & 0 & 4 \\
 D_{255} & 1 & 0 & 0 & 3 \\
 D_{256} & 1 & 0 & 0 & 3 \\
 D_{257} & 1 & 0 & 0 & 3 \\
 D_{258} & 1 & 0 & 0 & 3 \\
 D_{259} & 1 & 0 & 0 & 3 \\
 D_{260} & 1 & 0 & 0 & 3 \\
 D_{261} & 1 & 0 & 0 & 3 \\
 D_{262} & 1 & 0 & 0 & 4 \\
 D_{263} & 1 & 0 & 0 & 4 \\
 D_{264} & 1 & 0 & 0 & 4 \\
 D_{265} & 1 & 0 & 0 & 4 \\
 D_{266} & 1 & 0 & 0 & 4 \\
 D_{266} & 1 & 0 & 0 & 4 \\
 D_{267} & 1 & 0 & 0 & 4 \\
 D_{268} & 1 & 0 & 0 & 4 \\
 D_{269} & 1 & 0 & 0 & 4 \\
 D_{270} & 1 & 0 & 0 & 4 \\
 D_{271} & 1 & 0 & 0 & 4 \\
 D_{272} & 1 & 0 & 0 & 4 \\
 D_{273} & 1 & 0 & 0 & 4 \\
 D_{274} & 1 & 0 & 0 & 4 \\
 D_{275} & 1 & 0 & 0 & 3 \\
 D_{276} & 1 & 0 & 0 & 3 \\
\end{array}
$}
\end{equation}
\end{center}

\begin{center}
\begin{equation}\label{eqn:D4 mirror P11169 1}
\scalebox{.8}{$
\begin{array}{c|c|c|c|c|c|c}
&h^{0,0}&h^{1,0}&h^{2,0}&h^{3,0}&h^{1,1}&h^{2,1}\\
 \overline{D}_1 & 1 & 0 & 0 & 0 & 2 & 0 \\
 \overline{D}_2 & 1 & 0 & 0 & 1 & 28 & 7 \\
 \overline{D}_3 & 1 & 0 & 2 & 3  & 31 & 33  \\
 \overline{D}_4 & 1 & 0 & 0 & 0 & 6 & 0 \\
 \overline{D}_5 & 1 & 0 & 0 & 0 & 2 & 0 \\
 \overline{D}_6 & 1 & 0 & 0 & 0 & 11 & 0 \\
 \overline{D}_7 & 1 & 0 & 0 & 0 & 17 & 0 \\
 \overline{D}_8 & 1 & 0 & 0 & 0 & 4 & 0 \\
 \overline{D}_9 & 1 & 0 & 0 & 0 & 6 & 0 \\
 \overline{D}_{10} & 1 & 0 & 0 & 0 & 17 & 0 \\
 \overline{D}_{11} & 1 & 0 & 0 & 0 & 18 & 0 \\
 \overline{D}_{12} & 1 & 0 & 0 & 0 & 5 & 0 \\
 \overline{D}_{13} & 1 & 0 & 0 & 0 & 7 & 0 \\
 \overline{D}_{14} & 1 & 0 & 0 & 0 & 36 & 0 \\
 \overline{D}_{15} & 1 & 0 & 0 & 0 & 18 & 0 \\
 \overline{D}_{16} & 1 & 0 & 0 & 0 & 19 & 0 \\
 \overline{D}_{17} & 1 & 0 & 0 & 0 & 6 & 0 \\
 \overline{D}_{18} & 1 & 0 & 0 & 0 & 8 & 0 \\
 \overline{D}_{19} & 1 & 0 & 0 & 0 & 18 & 0 \\
 \overline{D}_{20} & 1 & 0 & 0 & 0 & 20 & 0 \\
 \overline{D}_{21} & 1 & 0 & 0 & 0 & 59 & 0 \\
 \overline{D}_{22} & 1 & 0 & 0 & 0 & 7 & 0 \\
 \overline{D}_{23} & 1 & 0 & 0 & 0 & 8 & 0 \\
 \overline{D}_{24} & 1 & 0 & 0 & 0 & 19 & 0 \\
 \overline{D}_{25} & 1 & 0 & 0 & 0 & 20 & 0 \\
 \overline{D}_{26} & 1 & 0 & 0 & 0 & 13 & 0 \\
 \overline{D}_{27} & 1 & 0 & 0 & 0 & 7 & 0 \\
 \overline{D}_{28} & 1 & 0 & 0 & 0 & 8 & 0 \\
 \overline{D}_{29} & 1 & 0 & 0 & 0 & 19 & 0 \\
 \overline{D}_{30} & 1 & 0 & 0 & 0 & 20 & 0 \\
 \overline{D}_{31} & 1 & 0 & 0 & 0 & 7 & 0 \\
 \overline{D}_{32} & 1 & 0 & 0 & 0 & 8 & 0 \\
 \overline{D}_{33} & 1 & 0 & 0 & 0 & 19 & 0 \\
 \overline{D}_{34} & 1 & 0 & 0 & 0 & 20 & 0 \\
 \overline{D}_{35} & 1 & 0 & 0 & 0 & 7 & 0 \\
 \overline{D}_{36} & 1 & 0 & 0 & 0 & 8 & 0 \\
 \overline{D}_{37} & 1 & 0 & 0 & 0 & 19 & 0 \\
 \overline{D}_{38} & 1 & 0 & 0 & 0 & 19 & 0 \\
\end{array}
$}
\quad
\scalebox{.8}{$
\begin{array}{c|c|c|c|c|c|c}
&h^{0,0}&h^{1,0}&h^{2,0}&h^{3,0}&h^{1,1}&h^{2,1}\\
 \overline{D}_{39} & 1 & 0 & 0 & 0 & 6 & 0 \\
 \overline{D}_{40} & 1 & 0 & 0 & 0 & 14 & 0 \\
 \overline{D}_{41} & 1 & 0 & 0 & 0 & 7 & 0 \\
 \overline{D}_{42} & 1 & 0 & 0 & 0 & 6 & 0 \\
 \overline{D}_{43} & 1 & 0 & 0 & 0 & 23 & 0 \\
 \overline{D}_{44} & 1 & 0 & 0 & 0 & 59 & 0 \\
 \overline{D}_{45} & 1 & 0 & 0 & 0 & 18 & 0 \\
 \overline{D}_{46} & 1 & 0 & 0 & 0 & 37 & 0 \\
 \overline{D}_{47} & 1 & 0 & 0 & 0 & 17 & 0 \\
 \overline{D}_{48} & 1 & 0 & 0 & 0 & 18 & 0 \\
 \overline{D}_{49} & 1 & 0 & 0 & 0 & 14 & 0 \\
 \overline{D}_{50} & 1 & 0 & 0 & 0 & 6 & 0 \\
 \overline{D}_{51} & 1 & 0 & 0 & 0 & 14 & 0 \\
 \overline{D}_{52} & 1 & 0 & 0 & 0 & 6 & 0 \\
 \overline{D}_{53} & 1 & 0 & 0 & 0 & 14 & 0 \\
 \overline{D}_{54} & 1 & 0 & 0 & 0 & 6 & 0 \\
 \overline{D}_{55} & 1 & 0 & 0 & 0 & 14 & 0 \\
 \overline{D}_{56} & 1 & 0 & 0 & 0 & 6 & 0 \\
 \overline{D}_{57} & 1 & 0 & 0 & 0 & 13 & 0 \\
 \overline{D}_{58} & 1 & 0 & 0 & 0 & 5 & 0 \\
 \overline{D}_{59} & 1 & 0 & 0 & 0 & 13 & 0 \\
 \overline{D}_{60} & 1 & 0 & 0 & 0 & 4 & 0 \\
 \overline{D}_{61} & 1 & 0 & 0 & 0 & 12 & 0 \\
 \overline{D}_{62} & 1 & 0 & 0 & 0 & 4 & 0 \\
 \overline{D}_{63} & 1 & 0 & 0 & 0 & 12 & 0 \\
 \overline{D}_{64} & 1 & 0 & 0 & 0 & 3 & 0 \\
 \overline{D}_{65} & 1 & 0 & 0 & 0 & 11 & 0 \\
 \overline{D}_{66} & 1 & 0 & 0 & 0 & 11 & 0 \\
 \overline{D}_{67} & 1 & 0 & 0 & 0 & 11 & 0 \\
 \overline{D}_{68} & 1 & 0 & 0 & 0 & 12 & 0 \\
 \overline{D}_{69} & 1 & 0 & 0 & 0 & 12 & 0 \\
 \overline{D}_{70} & 1 & 0 & 0 & 0 & 4 & 0 \\
 \overline{D}_{71} & 1 & 0 & 0 & 0 & 12 & 0 \\
 \overline{D}_{72} & 1 & 0 & 0 & 0 & 12 & 0 \\
 \overline{D}_{73} & 1 & 0 & 0 & 0 & 13 & 0 \\
 \overline{D}_{74} & 1 & 0 & 0 & 0 & 12 & 0 \\
 \overline{D}_{75} & 1 & 0 & 0 & 0 & 12 & 0 \\
 \overline{D}_{76} & 1 & 0 & 0 & 0 & 12 & 0 \\
\end{array}
$}
\quad
\scalebox{.8}{$
\begin{array}{c|c|c|c|c|c|c}
&h^{0,0}&h^{1,0}&h^{2,0}&h^{3,0}&h^{1,1}&h^{2,1}\\
 \overline{D}_{77} & 1 & 0 & 0 & 0 & 12 & 0 \\
 \overline{D}_{78} & 1 & 0 & 0 & 0 & 12 & 0 \\
 \overline{D}_{79} & 1 & 0 & 0 & 0 & 5 & 0 \\
 \overline{D}_{80} & 1 & 0 & 0 & 0 & 13 & 0 \\
 \overline{D}_{81} & 1 & 0 & 0 & 0 & 4 & 0 \\
 \overline{D}_{82} & 1 & 0 & 0 & 0 & 12 & 0 \\
 \overline{D}_{83} & 1 & 0 & 0 & 0 & 12 & 0 \\
 \overline{D}_{84} & 1 & 0 & 0 & 0 & 12 & 0 \\
 \overline{D}_{85} & 1 & 0 & 0 & 0 & 12 & 0 \\
 \overline{D}_{86} & 1 & 0 & 0 & 0 & 13 & 0 \\
 \overline{D}_{87} & 1 & 0 & 0 & 0 & 13 & 0 \\
 \overline{D}_{88} & 1 & 0 & 0 & 0 & 13 & 0 \\
 \overline{D}_{89} & 1 & 0 & 0 & 0 & 12 & 0 \\
 \overline{D}_{90} & 1 & 0 & 0 & 0 & 12 & 0 \\
 \overline{D}_{91} & 1 & 0 & 0 & 0 & 12 & 0 \\
 \overline{D}_{92} & 1 & 0 & 0 & 0 & 12 & 0 \\
 \overline{D}_{93} & 1 & 0 & 0 & 0 & 12 & 0 \\
 \overline{D}_{94} & 1 & 0 & 0 & 0 & 5 & 0 \\
 \overline{D}_{95} & 1 & 0 & 0 & 0 & 13 & 0 \\
 \overline{D}_{96} & 1 & 0 & 0 & 0 & 5 & 0 \\
 \overline{D}_{97} & 1 & 0 & 0 & 0 & 13 & 0 \\
 \overline{D}_{98} & 1 & 0 & 0 & 0 & 4 & 0 \\
 \overline{D}_{99} & 1 & 0 & 0 & 0 & 12 & 0 \\
 \overline{D}_{100} & 1 & 0 & 0 & 0 & 12 & 0 \\
 \overline{D}_{101} & 1 & 0 & 0 & 0 & 12 & 0 \\
 \overline{D}_{102} & 1 & 0 & 0 & 0 & 12 & 0 \\
 \overline{D}_{103} & 1 & 0 & 0 & 0 & 13 & 0 \\
 \overline{D}_{104} & 1 & 0 & 0 & 0 & 12 & 0 \\
 \overline{D}_{105} & 1 & 0 & 0 & 0 & 13 & 0 \\
 \overline{D}_{106} & 1 & 0 & 0 & 0 & 13 & 0 \\
 \overline{D}_{107} & 1 & 0 & 0 & 0 & 13 & 0 \\
 \overline{D}_{108} & 1 & 0 & 0 & 0 & 13 & 0 \\
 \overline{D}_{109} & 1 & 0 & 0 & 0 & 13 & 0 \\
 \overline{D}_{110} & 1 & 0 & 0 & 0 & 12 & 0 \\
 \overline{D}_{111} & 1 & 0 & 0 & 0 & 12 & 0 \\
 \overline{D}_{112} & 1 & 0 & 0 & 0 & 12 & 0 \\
 \overline{D}_{113} & 1 & 0 & 0 & 0 & 12 & 0 \\
 \overline{D}_{114} & 1 & 0 & 0 & 0 & 13 & 0 \\
\end{array}
$}
\end{equation}
\end{center}

\begin{center}
\begin{equation}\label{eqn:D4 mirror P11169 2}
\scalebox{.8}{$
\begin{array}{c|c|c|c|c|c|c}
&h^{0,0}&h^{1,0}&h^{2,0}&h^{3,0}&h^{1,1}&h^{2,1}\\
 \overline{D}_{115} & 1 & 0 & 0 & 0 & 13 & 0 \\
 \overline{D}_{116} & 1 & 0 & 0 & 0 & 12 & 0 \\
 \overline{D}_{117} & 1 & 0 & 0 & 0 & 5 & 0 \\
 \overline{D}_{118} & 1 & 0 & 0 & 0 & 13 & 0 \\
 \overline{D}_{119} & 1 & 0 & 0 & 0 & 5 & 0 \\
 \overline{D}_{120} & 1 & 0 & 0 & 0 & 13 & 0 \\
 \overline{D}_{121} & 1 & 0 & 0 & 0 & 5 & 0 \\
 \overline{D}_{122} & 1 & 0 & 0 & 0 & 13 & 0 \\
 \overline{D}_{123} & 1 & 0 & 0 & 0 & 4 & 0 \\
 \overline{D}_{124} & 1 & 0 & 0 & 0 & 12 & 0 \\
 \overline{D}_{125} & 1 & 0 & 0 & 0 & 12 & 0 \\
 \overline{D}_{126} & 1 & 0 & 0 & 0 & 12 & 0 \\
 \overline{D}_{127} & 1 & 0 & 0 & 0 & 12 & 0 \\
 \overline{D}_{128} & 1 & 0 & 0 & 0 & 12 & 0 \\
 \overline{D}_{129} & 1 & 0 & 0 & 0 & 13 & 0 \\
 \overline{D}_{130} & 1 & 0 & 0 & 0 & 13 & 0 \\
 \overline{D}_{131} & 1 & 0 & 0 & 0 & 12 & 0 \\
 \overline{D}_{132} & 1 & 0 & 0 & 0 & 13 & 0 \\
 \overline{D}_{133} & 1 & 0 & 0 & 0 & 13 & 0 \\
 \overline{D}_{134} & 1 & 0 & 0 & 0 & 13 & 0 \\
 \overline{D}_{135} & 1 & 0 & 0 & 0 & 13 & 0 \\
 \overline{D}_{136} & 1 & 0 & 0 & 0 & 13 & 0 \\
 \overline{D}_{137} & 1 & 0 & 0 & 0 & 13 & 0 \\
 \overline{D}_{138} & 1 & 0 & 0 & 0 & 13 & 0 \\
 \overline{D}_{139} & 1 & 0 & 0 & 0 & 12 & 0 \\
 \overline{D}_{140} & 1 & 0 & 0 & 0 & 12 & 0 \\
 \overline{D}_{141} & 1 & 0 & 0 & 0 & 12 & 0 \\
 \overline{D}_{142} & 1 & 0 & 0 & 0 & 12 & 0 \\
 \overline{D}_{143} & 1 & 0 & 0 & 0 & 4 & 0 \\
 \overline{D}_{144} & 1 & 0 & 0 & 0 & 4 & 0 \\
 \overline{D}_{145} & 1 & 0 & 0 & 0 & 13 & 0 \\
 \overline{D}_{146} & 1 & 0 & 0 & 0 & 13 & 0 \\
 \overline{D}_{147} & 1 & 0 & 0 & 0 & 12 & 0 \\
 \overline{D}_{148} & 1 & 0 & 0 & 0 & 5 & 0 \\
 \overline{D}_{149} & 1 & 0 & 0 & 0 & 13 & 0 \\
 \overline{D}_{150} & 1 & 0 & 0 & 0 & 5 & 0 \\
 \overline{D}_{151} & 1 & 0 & 0 & 0 & 13 & 0 \\
 \overline{D}_{152} & 1 & 0 & 0 & 0 & 5 & 0 \\
\end{array}
$}
\quad
\scalebox{.8}{$
\begin{array}{c|c|c|c|c|c|c}
&h^{0,0}&h^{1,0}&h^{2,0}&h^{3,0}&h^{1,1}&h^{2,1}\\
 \overline{D}_{153} & 1 & 0 & 0 & 0 & 13 & 0 \\
 \overline{D}_{154} & 1 & 0 & 0 & 0 & 5 & 0 \\
 \overline{D}_{155} & 1 & 0 & 0 & 0 & 13 & 0 \\
 \overline{D}_{156} & 1 & 0 & 0 & 0 & 4 & 0 \\
 \overline{D}_{157} & 1 & 0 & 0 & 0 & 12 & 0 \\
 \overline{D}_{158} & 1 & 0 & 0 & 0 & 11 & 0 \\
 \overline{D}_{159} & 1 & 0 & 0 & 0 & 12 & 0 \\
 \overline{D}_{160} & 1 & 0 & 0 & 0 & 12 & 0 \\
 \overline{D}_{161} & 1 & 0 & 0 & 0 & 12 & 0 \\
 \overline{D}_{162} & 1 & 0 & 0 & 0 & 12 & 0 \\
 \overline{D}_{163} & 1 & 0 & 0 & 0 & 13 & 0 \\
 \overline{D}_{164} & 1 & 0 & 0 & 0 & 13 & 0 \\
 \overline{D}_{165} & 1 & 0 & 0 & 0 & 12 & 0 \\
 \overline{D}_{166} & 1 & 0 & 0 & 0 & 13 & 0 \\
 \overline{D}_{167} & 1 & 0 & 0 & 0 & 13 & 0 \\
 \overline{D}_{168} & 1 & 0 & 0 & 0 & 13 & 0 \\
 \overline{D}_{169} & 1 & 0 & 0 & 0 & 13 & 0 \\
 \overline{D}_{170} & 1 & 0 & 0 & 0 & 13 & 0 \\
 \overline{D}_{171} & 1 & 0 & 0 & 0 & 13 & 0 \\
 \overline{D}_{172} & 1 & 0 & 0 & 0 & 13 & 0 \\
 \overline{D}_{173} & 1 & 0 & 0 & 0 & 13 & 0 \\
 \overline{D}_{174} & 1 & 0 & 0 & 0 & 13 & 0 \\
 \overline{D}_{175} & 1 & 0 & 0 & 0 & 12 & 0 \\
 \overline{D}_{176} & 1 & 0 & 0 & 0 & 12 & 0 \\
 \overline{D}_{177} & 1 & 0 & 0 & 0 & 12 & 0 \\
 \overline{D}_{178} & 1 & 0 & 0 & 0 & 12 & 0 \\
 \overline{D}_{179} & 1 & 0 & 0 & 0 & 4 & 0 \\
 \overline{D}_{180} & 1 & 0 & 0 & 0 & 4 & 0 \\
 \overline{D}_{181} & 1 & 0 & 0 & 0 & 13 & 0 \\
 \overline{D}_{182} & 1 & 0 & 0 & 0 & 13 & 0 \\
 \overline{D}_{183} & 1 & 0 & 0 & 0 & 12 & 0 \\
 \overline{D}_{184} & 1 & 0 & 0 & 0 & 5 & 0 \\
 \overline{D}_{185} & 1 & 0 & 0 & 0 & 13 & 0 \\
 \overline{D}_{186} & 1 & 0 & 0 & 0 & 5 & 0 \\
 \overline{D}_{187} & 1 & 0 & 0 & 0 & 13 & 0 \\
 \overline{D}_{188} & 1 & 0 & 0 & 0 & 5 & 0 \\
 \overline{D}_{189} & 1 & 0 & 0 & 0 & 13 & 0 \\
 \overline{D}_{190} & 1 & 0 & 0 & 0 & 5 & 0 \\
\end{array}
$}
\quad
\scalebox{.8}{$
\begin{array}{c|c|c|c|c|c|c}
&h^{0,0}&h^{1,0}&h^{2,0}&h^{3,0}&h^{1,1}&h^{2,1}\\
 \overline{D}_{191} & 1 & 0 & 0 & 0 & 13 & 0 \\
 \overline{D}_{192} & 1 & 0 & 0 & 0 & 5 & 0 \\
 \overline{D}_{193} & 1 & 0 & 0 & 0 & 13 & 0 \\
 \overline{D}_{194} & 1 & 0 & 0 & 0 & 4 & 0 \\
 \overline{D}_{195} & 1 & 0 & 0 & 0 & 12 & 0 \\
 \overline{D}_{196} & 1 & 0 & 0 & 0 & 12 & 0 \\
 \overline{D}_{197} & 1 & 0 & 0 & 0 & 12 & 0 \\
 \overline{D}_{198} & 1 & 0 & 0 & 0 & 12 & 0 \\
 \overline{D}_{199} & 1 & 0 & 0 & 0 & 12 & 0 \\
 \overline{D}_{200} & 1 & 0 & 0 & 0 & 13 & 0 \\
 \overline{D}_{201} & 1 & 0 & 0 & 0 & 12 & 0 \\
 \overline{D}_{202} & 1 & 0 & 0 & 0 & 12 & 0 \\
 \overline{D}_{203} & 1 & 0 & 0 & 0 & 13 & 0 \\
 \overline{D}_{204} & 1 & 0 & 0 & 0 & 13 & 0 \\
 \overline{D}_{205} & 1 & 0 & 0 & 0 & 13 & 0 \\
 \overline{D}_{206} & 1 & 0 & 0 & 0 & 13 & 0 \\
 \overline{D}_{207} & 1 & 0 & 0 & 0 & 13 & 0 \\
 \overline{D}_{208} & 1 & 0 & 0 & 0 & 13 & 0 \\
 \overline{D}_{209} & 1 & 0 & 0 & 0 & 13 & 0 \\
 \overline{D}_{210} & 1 & 0 & 0 & 0 & 13 & 0 \\
 \overline{D}_{211} & 1 & 0 & 0 & 0 & 13 & 0 \\
 \overline{D}_{212} & 1 & 0 & 0 & 0 & 13 & 0 \\
 \overline{D}_{213} & 1 & 0 & 0 & 0 & 13 & 0 \\
 \overline{D}_{214} & 1 & 0 & 0 & 0 & 12 & 0 \\
 \overline{D}_{215} & 1 & 0 & 0 & 0 & 12 & 0 \\
 \overline{D}_{216} & 1 & 0 & 0 & 0 & 11 & 0 \\
 \overline{D}_{217} & 1 & 0 & 0 & 0 & 11 & 0 \\
 \overline{D}_{218} & 1 & 0 & 0 & 0 & 12 & 0 \\
 \overline{D}_{219} & 1 & 0 & 0 & 0 & 12 & 0 \\
 \overline{D}_{220} & 1 & 0 & 0 & 0 & 12 & 0 \\
 \overline{D}_{221} & 1 & 0 & 0 & 0 & 4 & 0 \\
 \overline{D}_{222} & 1 & 0 & 0 & 0 & 12 & 0 \\
 \overline{D}_{223} & 1 & 0 & 0 & 0 & 4 & 0 \\
 \overline{D}_{224} & 1 & 0 & 0 & 0 & 12 & 0 \\
 \overline{D}_{225} & 1 & 0 & 0 & 0 & 12 & 0 \\
 \overline{D}_{226} & 1 & 0 & 0 & 0 & 12 & 0 \\
 \overline{D}_{227} & 1 & 0 & 0 & 0 & 12 & 0 \\
 \overline{D}_{228} & 1 & 0 & 0 & 0 & 5 & 0 \\
\end{array}
$}
\end{equation}
\end{center}

\begin{center}
\begin{equation}\label{eqn:D4 mirror P11169 3}
\scalebox{.8}{$
\begin{array}{c|c|c|c|c|c|c}
&h^{0,0}&h^{1,0}&h^{2,0}&h^{3,0}&h^{1,1}&h^{2,1}\\
 \overline{D}_{229} & 1 & 0 & 0 & 0 & 13 & 0 \\
 \overline{D}_{230} & 1 & 0 & 0 & 0 & 5 & 0 \\
 \overline{D}_{231} & 1 & 0 & 0 & 0 & 13 & 0 \\
 \overline{D}_{232} & 1 & 0 & 0 & 0 & 5 & 0 \\
 \overline{D}_{233} & 1 & 0 & 0 & 0 & 13 & 0 \\
 \overline{D}_{234} & 1 & 0 & 0 & 0 & 5 & 0 \\
 \overline{D}_{235} & 1 & 0 & 0 & 0 & 13 & 0 \\
 \overline{D}_{236} & 1 & 0 & 0 & 0 & 5 & 0 \\
 \overline{D}_{237} & 1 & 0 & 0 & 0 & 13 & 0 \\
 \overline{D}_{238} & 1 & 0 & 0 & 0 & 5 & 0 \\
 \overline{D}_{239} & 1 & 0 & 0 & 0 & 13 & 0 \\
 \overline{D}_{240} & 1 & 0 & 0 & 0 & 4 & 0 \\
 \overline{D}_{241} & 1 & 0 & 0 & 0 & 12 & 0 \\
 \overline{D}_{242} & 1 & 0 & 0 & 0 & 12 & 0 \\
 \overline{D}_{243} & 1 & 0 & 0 & 0 & 12 & 0 \\
 \overline{D}_{244} & 1 & 0 & 0 & 0 & 12 & 0 \\
 \overline{D}_{245} & 1 & 0 & 0 & 0 & 12 & 0 \\
 \overline{D}_{246} & 1 & 0 & 0 & 0 & 12 & 0 \\
 \overline{D}_{247} & 1 & 0 & 0 & 0 & 12 & 0 \\
 \overline{D}_{248} & 1 & 0 & 0 & 0 & 12 & 0 \\
 \overline{D}_{249} & 1 & 0 & 0 & 0 & 12 & 0 \\
 \overline{D}_{250} & 1 & 0 & 0 & 0 & 13 & 0 \\
 \overline{D}_{251} & 1 & 0 & 0 & 0 & 13 & 0 \\
 \overline{D}_{252} & 1 & 0 & 0 & 0 & 13 & 0 \\
 \overline{D}_{253} & 1 & 0 & 0 & 0 & 13 & 0 \\
 \overline{D}_{254} & 1 & 0 & 0 & 0 & 13 & 0 \\
 \overline{D}_{255} & 1 & 0 & 0 & 0 & 12 & 0 \\
 \overline{D}_{256} & 1 & 0 & 0 & 0 & 12 & 0 \\
 \overline{D}_{257} & 1 & 0 & 0 & 0 & 12 & 0 \\
 \overline{D}_{258} & 1 & 0 & 0 & 0 & 12 & 0 \\
 \overline{D}_{259} & 1 & 0 & 0 & 0 & 12 & 0 \\
 \overline{D}_{260} & 1 & 0 & 0 & 0 & 12 & 0 \\
 \overline{D}_{261} & 1 & 0 & 0 & 0 & 12 & 0 \\
 \overline{D}_{262} & 1 & 0 & 0 & 0 & 13 & 0 \\
 \overline{D}_{263} & 1 & 0 & 0 & 0 & 13 & 0 \\
 \overline{D}_{264} & 1 & 0 & 0 & 0 & 13 & 0 \\
 \overline{D}_{265} & 1 & 0 & 0 & 0 & 13 & 0 \\
 \overline{D}_{266} & 1 & 0 & 0 & 0 & 13 & 0 \\
\end{array}
$}
\quad
\scalebox{.8}{$
\begin{array}{c|c|c|c|c|c|c}
&h^{0,0}&h^{1,0}&h^{2,0}&h^{3,0}&h^{1,1}&h^{2,1}\\
 \overline{D}_{267} & 1 & 0 & 0 & 0 & 13 & 0 \\
 \overline{D}_{268} & 1 & 0 & 0 & 0 & 13 & 0 \\
 \overline{D}_{269} & 1 & 0 & 0 & 0 & 13 & 0 \\
 \overline{D}_{270} & 1 & 0 & 0 & 0 & 13 & 0 \\
 \overline{D}_{271} & 1 & 0 & 0 & 0 & 13 & 0 \\
 \overline{D}_{272} & 1 & 0 & 0 & 0 & 13 & 0 \\
 \overline{D}_{273} & 1 & 0 & 0 & 0 & 13 & 0 \\
 \overline{D}_{274} & 1 & 0 & 0 & 0 & 13 & 0 \\
 \overline{D}_{275} & 1 & 0 & 0 & 0 & 12 & 0 \\
 \overline{D}_{276} & 1 & 0 & 0 & 0 & 12 & 0 \\
\end{array}
$}
\end{equation}
\end{center}
\end{adjustwidth}
\restoregeometry

\section{Notation}
\label{app:notation}
In this Appendix, we collect and define notation commonly used throughout the draft. The symbols are arranged in (roughly) alphabetical order.
	\begin{itemize}
			\item{} $\boxed{B_3}$: Threefold base of an elliptic CY fourfold, $\pi_E : Y_4 \rightarrow B_3$. In this paper, we focus on the case that $B_3$ is an orientifold of a CY threefold, i.e. $B_3 = Z_3 /\mathbb Z_2$. 
			\item{} $\boxed{c_n(X)}$: $n$th Chern class of the tangent bundle of a (smooth) projective variety $X$.
			\item{} $\boxed{\chi_n(\overline{D})}$: ``Naive'' Euler characteristic of a vertical divisor $\overline{D} \subset Y_4$. When $\overline{D}$ is smooth, then $\chi_n(\overline{D})$ is equal to the usual topological Euler characteristic $\chi(\overline{D}) = \sum_{p,q}(-1)^{p+q} h^{p,q}(\overline{D})$. When $\overline{D}$ contains terminal point-like $\mathbb Z_2$ orbifold singularities, we conjecture that $\chi_n(\overline{D})$ differs from $\chi(\overline{D})$ by twice the number of $\mathbb Z_2$ orbifold points. 
			\item{} $\boxed{\chi(X,\mathcal F)}$: Holomorphic Euler characteristic of the sheaf $\mathcal F$ defined on an algebraic variety $X$. We use the standard definition $\chi(X,\mathcal F) := \sum_{k} (-1)^k h^{k}(X,\mathcal F)$ where $h^{k}(X,\mathcal F)$ is the dimension of the $k$th sheaf cohomology group of $H^{k}(X,\mathcal F)$. 
			\item{} $\boxed{\overline{D}}$: Vertical divisor of an elliptic CY fourfold $\pi_E : Y_4 \rightarrow B_3$. By definition, we have $\overline{D}= \pi_E^{-1}(\widehat{D})$ for some divisor $\widehat{D} \subset B_3$. In the special case that $\widehat{D}$ corresponds to an $SO(8)$ seven brane stack, $\pi_E^{-1}(\widehat{D})$ is reducible and we take $\overline{D} \subset \pi_E^{-1}(\widehat{D})$ to be an irreducible component. 
			\item{} $\boxed{\overline{D}^{(0)}}$: Singular divisor of a singular elliptic CY fourfold $Y_4'$ admitting a global Sen limit. By construction, such elliptic CY fourfolds contain $SO(8)$ Kodaira singularities. We assume the existence of a crepant (partial) resolution $Y_4 \rightarrow Y_4'$ that resolves the $SO(8)$ singularities such that the blowdown of $\overline{D}$ is given by $Y_4 \supset \overline{D} \rightarrow \overline{D}^{(0)} \subset Y_4'$. 
			\item{} $\boxed{\widehat{D}}$: Divisor of the orientifold base $B_3 = Z_3/\mathbb Z_2$. By definition, the image $\varphi_{\mathcal I_{\vec p}}(\widehat{D})$ of $\widehat{D}$ under the refinement map $\varphi_{\mathcal I_{\vec p}}$, is a divisor in $Z_3$. 
			\item{} $\boxed{D_\Delta}$: Divisor defined by a Newton polytope $\Delta$; see (\ref{Newtonpolytopedivisor}). 
			\item{} $\boxed{\dim \, \mathcal L}$: Iitaka dimension of a line bundle $\mathcal L$; see the paragraph above (\ref{Iitakadims}) for a definition. When $\mathcal L $ is a line bundle over a toric 2-fold, then $\dim\, \mathcal L$ is equal to the dimension $ \dim \,\Delta_L$ of the Newton polytope $\Delta_L$ encoding the global sections of $\mathcal L$. In the case that $\mathcal L$ is a line bundle defined over an algebraic variety $X$, we use $\mathcal L|_{D}$ to denote the line bundle given by the restriction of $\mathcal L$ to the divisor $D \subset X$. 
			\item{} $\boxed{\Delta}$: Newton polytope associated to a hypersurface $Z \subset \mathbb P_\Sigma$. Note that when $\Delta$ is regarded as corresponding to some choice of divisor whose class is $L = \sum_{\vec v} a_{\vec v } D_{\vec v}$, we sometimes write $\Delta_L$. In such cases, by definition the divisor $D_\Delta$ defined in (\ref{Newtonpolytopedivisor}) is equal to $L$.
			\item{} $\boxed{\Delta^\circ}$: Polar dual of the Newton polytope $\Delta$; see (\ref{polardual}). We use reflexive pairs of polytopes $\Delta, \Delta^\circ$ to construct Calabi-Yau toric hypersurfaces. In \S\ref{sec:toric CYs}, we introduce the notation $e,f$ to refer to (resp.) edges and 2-faces of $\Delta^\circ$. The vectors $\vec v_e, \vec v_f$ then refer to points interior to $e, f$, respectively. Moreover, we use $t_i$ to denote an $i$-dimensional simplex in $\mathcal T \cap \partial \Delta^\circ$, where $\mathcal T$ is an fine regular star trianguation.  
			\item{} $\boxed{\widehat{\Delta}}$: Newton polytope for the embedded image of a $\mathbb Z_2$-symmetric CY threefold $Z_3$ in $\widehat{V}_4$. A similar notation holds for the image of the Newton polytope of a hypersurface in a toric stratum, namely $\widehat{\Delta}^{(n)}$. 
			\item{} $\boxed{\Delta_{\pm{}}}$: Subset of monomials $x^{\vec m}= \prod_{\vec v \in \partial \Delta^\circ} x_{\vec v}^{\vec m \cdot \vec v + 1}$, which define a basis for a general section of the anticanonical class $-K_{V_4}$ of a toric fourfold $V_4$, that are even ($+$)/odd ($-$) under the orientifold involution $\mathcal I_{\vec p}$. Note here that $\Delta^\circ$ is part of a reflexive pair $\Delta, \Delta^\circ$. 
			\item{} $\boxed{e}$: Edge of the polar dual $\Delta^\circ$ of a reflexive Newton polytope $\Delta$. First used in \S\ref{sec:CY3hyptopology}.
			\item{} $\boxed{\widehat{e}}$: Codimension-one subpolytope in $\widehat{v}$ given by the preimage of the $e$ under the refinement $\varphi_{\mathcal I_{\vec p}}$; see (\ref{ehat}). 
			\item{} $\boxed{e(X)}$: Character $\sum_k (-1)^k \dim(H^k_c)$; see (\ref{echar}).
			\item{} $\boxed{e^{p,q}(X)}$: Hodge-Deligne number $e^{p,q}(X)$ $=$ $\sum_k (-1)^{k} h^{p,q}(H_c^k(X))$. Note that $h^{p,q}(H_c^k(X))$ vanishes for $p+q > k$.
			\item{} $\boxed{e(X;x,\bar x)}$: Characteristic polynomial $e(X;x,\bar x) = \sum_{p,q} e^{p,q}(X) x^p \bar x^q$; see (\ref{HDcharpoly}) and discussion below.
			\item{} $\boxed{\mathbb F_n}$: Hirzebruch surface, i.e. $\mathbb P( \mathcal O \oplus \mathcal O(n) ) \rightarrow C$, $ C \equiv \mathbb P^1$. 
			\item{} $\boxed{f}$: Face of the polar dual $\Delta^\circ$ of a reflexive Newton polytope $\Delta$. First used in \S\ref{sec:CY3hyptopology}.
			\item{} $\boxed{\widehat{f}}$: Preimage of the refinement map, i.e. $\varphi_{\mathcal I_{\vec p}}^{-1}(f),$ where $f$ is a face of $\Delta^\circ.$
			\item{} $\boxed{\varphi_n(\Theta)}$: A weighted and signed sum over $l*(j\Theta^{(k)})$, where $j \Theta$ denotes the polytope obtained by scaling a face $\Theta$ by a factor $j \in \mathbb Z_{>0}$; see (\ref{phifn}).
			\item{} $\boxed{G}$: Abelian group defining a $\mathbb C^*$-action on the homogeneous coordinates of a projective variety $\mathbb P_\Sigma$.
			\item{} $\boxed{g}$: Subgroup of the abelian group $G$, which defines the $\mathbb C^*$ action $x_{\vec v} \rightarrow \lambda^{a_{\vec v}} x_{\vec v}$, where $\lambda \in \mathbb C^*$. See (\ref{morphism}). 
			\item{} $\boxed{H^k(X,R)}$: $k$th cohomology group with coefficients in the ring $R$. 
			\item{} $\boxed{H^k(X,\mathcal F)}$: $k$th sheaf cohomology group. 
			\item{} $\boxed{H^k_c(X)}$: $k$th cohomology group with compact support associated to an algebraic variety $X$. 
			\item{} $\boxed{H_{\pm{}}^k(X)}$: Subgroup of the $k$th cohomology group $H^k(X)$ of an algebraic variety $X$ that is even/odd under the action of the orientifold involution. 
			\item{} $\boxed{h_{\pm{}}^k(X)}$: Dimension of $H_{\pm{}}^k(X)$. An important case, when $X$ is the ambient fourfold $V_4$ corresponding to the image of $\widehat{V}_4$ under the refinement map $\varphi_{\mathcal I_{\vec p}}$, or a divisor therein. 
			\item{} $\boxed{h^{\bullet,0}(X)}$: Hodge vector $(h^{0,0}(X), h^{1,0}(X),\cdots)$, where the $k$th component $h^{k,0}(X)$ is the dimension of the $k$th structure sheaf cohomology group $H^{k}(X,\mathcal O_X)$.
			\item{} $\boxed{\mathcal I_{\vec p}}$: Orientifold involution defined by a choice of vertex $\vec p \in \partial \Delta^\circ$, where $\Delta^\circ$ is part of a reflexive pair $\Delta, \Delta^\circ$ defining a CY threefold toric hypersurface $Z_3 \subset V_4$; see (\ref{pinvolution}) and the surrounding discussion.
			\item{} $\boxed{\widehat{I}_{\vec p}}$: Set of homogeneous coordinates $x_i$ whose vanishing loci correspond to divisors $D_{\vec v_i}$ that are fixed under the orientifold involution $ \mathcal I_{\vec p} $ conjugated by some subgroup $g \subset G$; see (\ref{hatI}). 
			\item{} $\boxed{K_X}$: Canonical class of the algebraic variety $X$, i.e. the first Chern class of the canonical sheaf. 
			\item{} $\boxed{\mathcal L}$: Line bundle. 
			\item{} $\boxed{l(\Theta)}$: Number of lattice points contained in the face $\Theta$.
			\item{} $\boxed{l^*(\Theta)}$: Number of lattice points contained in the relative interior of the face $\Theta$. 
			\item{} $\boxed{l^n(\Theta)}$: Number of points contained in the $n$-skeleton of a face $\Theta$.
			\item{} $\boxed{M}$: Given a lattice $N$, $M$ is the dual lattice, i.e. $M = \text{Hom}(N,\mathbb Z)$.
			\item{} $\boxed{N}$: Lattice. Typically, we regard each cone $\sigma \in \Sigma$ as being spanned by a finite number of rays, which are themselves the spans of primitive lattice vectors $\vec v \in N$. 
			\item{} $\boxed{\mathcal O_{X}(D)}$: Line bundle over a projective variety $X$ that corresponds to the divisor $D \subset X$. By abuse of notation, we denote the Chow ring representative of the divisor by $D = c_1(\mathcal O_X(D))$. We sometimes write $\mathcal O(D)$ when the space over which the line bundle is defined is clear from the context. 
				\item{} $\boxed{\mathbb P^n}$: Complex $n$-dimensional projective space. 
				\item{} $\boxed{\mathbb P_{[w_0,\dots,w_n]}}$: Complex $n$-dimensional weighted projective space with weights $w_{i=0,\dots,n}$.
				\item{} $\boxed{\mathbb P_{[w_0,\dots,w_n]}[d]}$: Degree-$d$ hypersurface in the weighted projective space $\mathbb P_{[w_0,\dots,w_n]}$. 
					\item{} $\boxed{pt}$: A point.
			\item{} $\boxed{\pi}$: Blowdown map, i.e. birational map contracting the exceptional divisors of a blowup, e.g. $\pi : \mathbb P_{\Sigma(\mathcal T)} \rightarrow \mathbb P_\Sigma$. 
			\item{} $\boxed{\pi_E}$: Canonical projection to the base $B_d$ of an elliptic fibration $E \hookrightarrow Y_{d+1} \rightarrow B_{d}$. See \S\ref{sec:Comments on the F-theory uplift} for relevant discussion.
		\item{} $\boxed{\Theta^{(k)}}$: A $k$-dimensional face of a Newton polytope $\Delta$. 
		\item{} $\boxed{\Sigma}$: Fan, i.e. a set of strongly complex rational polyhedral cones $\sigma$. 
		\item{} $\boxed{\Sigma(n)}$: Subset of $\Sigma$ consisting of $n$-dimensional cones $\sigma^{(n)} \in \Sigma$; see (\ref{normalfan}) and the discussion below.
		\item{} $\boxed{\Sigma(\Delta)}$: Normal fan defined with respect to a Newton polytope $\Delta$; see (\ref{normalfan}).
		\item{} $\boxed{\Sigma(\Delta^\circ)}$: Fan over the faces of $\Delta^{\circ}$, or equivalently, normal fan over $\Sigma(\Delta)$. 
		\item{} $\boxed{\sigma^{(k)}}$:  A cone $\sigma$ of real dimension $k$. 
		\item{} $\boxed{\sigma^\vee}$: Dual cone, i.e. the subset of lattice vectors $\vec m \in M$ that satisfy $\vec m \cdot \vec v$ for all $\vec v \in \sigma$. 
			\item{} $\boxed{\sigma(\Theta^{(k)})}$: Cone in $N$ dual to $\sigma^\vee(\Theta^{(k)})$. For the special case of maximal-dimensional cones dual to vertices $\vec m = \Theta^{(0)}$, we write $\sigma(\vec m)$. 
		\item{} $\boxed{\sigma^\vee(\Theta^{(k)})}$: Cone in $M$ defined with respect to a $k$-dimensional face $\Theta^{(k)}$ over $\Delta$, see (\ref{normalcone}). 
		\item{} $\boxed{\widehat{\sigma}^{(n)}}$: Preimage of a cone $\sigma^{(n)} \in \Sigma(\mathcal T)$ where $\Sigma(\mathcal T)$ defines a toric fourfold $V_4$, i.e. $\widehat{\sigma}^{(n)} = \varphi_{\mathcal I_{\vec p}}^{-1}(\sigma^{(n)})$. 
		\item{} \boxed{{SRI}}: Stanley-Reisner ideal, i.e. the set of all subsets $I \subset \{ \vec v\}$ for which the one-dimensional cones $\sigma \in \Sigma$ generated by $\vec v \in I$ do not share a common higher-dimensional cone in a toric fan $\Sigma$.
		\item{} \boxed{\mathcal{T}}: Fine regular star triangulation; see \S\ref{sec:resolution} for further discussion. 
		\item{} $\boxed{T_{\sigma^{(n)}}}$: $(d-n)$-diensional algebraic torus $(\mathbb C^*)^{d-n}$ corresponding to an $n$-dimensional cone $\sigma^{(n)}\in \Sigma(n)$.  
		\item{} $\boxed{\mathcal V_\Delta}$: Set of vertices $\vec v \in \Sigma(\mathcal T)$ for some choice of FRST $\mathcal T$ defining a refinement $\Sigma(\mathcal T) \rightarrow \Sigma(\Delta)$. See \S\ref{sec:resolution} for a discussion of refinements of toric fans $\Sigma$. 
		\item{} $\boxed{t_i}$: $i$-dimensional simplex in $\mathcal T \cap \partial \Delta^\circ$, where $\mathcal T$ is an FRST and $\Delta^\circ$ is the polar dual of a reflexive Newton polytope $\Delta$. First introduced in \S\ref{sec:CY3hyptopology}.
		\item{} $\boxed{V_d}$: Toric variety of complex dimension $d$. We sometimes write $V$ when the dimension is not important.
		\item{} $\boxed{\widehat{V}_4}$: A toric fourfold that is the quotient of the orientifold involution $\mathcal I_{\vec p} : V_4 \rightarrow V_4$ defined by the action $x_{\vec p} \mapsto -x_{\vec p}$ for a subset of homogeneous coordinates $x_{\vec p}$ corresponding to points $\vec p \in \partial \Delta^\circ$. Note that $\widehat{V}_4$ can be regarded as the preimage of the refinement map $\varphi_{I_{\vec p}} : \widehat{V}_4 \rightarrow V_4$ defined in (\ref{refinementmap}). See \S\ref{sec:orientifoldinvolution} for further discussion.
		\item{} $\boxed{\widehat{v}}$: Shorthand notation for the subpolytope of the orientifold Newton polytope $\widehat{\Delta}$ for which no corresponding monomial contains a factor of the homogeneous coordinate $\widehat{x}_{\vec{\widehat{v}}}$; see (\ref{qhat}). 
		\item{} $\boxed{\Omega_V}$: Cotangent bundle of a toric variety $V$.
		\item{} $\boxed{x_{\vec v}}$: Homogeneous coordinate whose vanishing locus is the divisor corresponding to the ray $\sigma_{\vec v}$ in a toric fan $\Sigma$.
		\item{} $\boxed{[x]}$: Divisor class of the vanishing locus $x=0$.
		\item{} $\boxed{Y_4}$: Elliptic Calabi-Yau fourfold, $\pi_E : Y_4 \rightarrow B_3$, where $\pi_E^{-1}(pt) \equiv T^2$ for a point $pt \subset B_3$. 
			\item{} $\boxed{Z}$: Hypersurface of a projective toric variety $\mathbb P_{\Sigma}$.
		\item{} $\boxed{Z_{\sigma^{(n)}}}$: $(d-n-1)$-dimensional stratum $Z \cap T_{\sigma^{(n)}}$ associated to a hypersurface $Z$ of a projective toric variety containing strata $T_{\sigma^{(n)}}$.  
	\end{itemize}

\bibliographystyle{JHEP}
\bibliography{refs}

\providecommand{\href}[2]{#2}\begingroup\raggedright\begin{thebibliography}{100}

\bibitem{Kachru:2003aw}
S.~Kachru, R.~Kallosh, A.~D. Linde and S.~P. Trivedi, \emph{{De Sitter vacua in
  string theory}},
  \href{http://dx.doi.org/10.1103/PhysRevD.68.046005}{\emph{Phys. Rev. D} {\bf
  68} (2003) 046005}, [\href{http://arxiv.org/abs/hep-th/0301240}{{\tt
  hep-th/0301240}}].

\bibitem{Balasubramanian:2005zx}
V.~Balasubramanian, P.~Berglund, J.~P. Conlon and F.~Quevedo,
  \emph{{Systematics of moduli stabilisation in Calabi-Yau flux
  compactifications}},
  \href{http://dx.doi.org/10.1088/1126-6708/2005/03/007}{\emph{JHEP} {\bf 03}
  (2005) 007}, [\href{http://arxiv.org/abs/hep-th/0502058}{{\tt
  hep-th/0502058}}].

\bibitem{Demirtas:2021ote}
M.~Demirtas, M.~Kim, L.~McAllister, J.~Moritz and A.~Rios-Tascon,
  \emph{{Exponentially Small Cosmological Constant in String Theory}},
  \href{http://dx.doi.org/10.1103/PhysRevLett.128.011602}{\emph{Phys. Rev.
  Lett.} {\bf 128} (2022) 011602}, [\href{http://arxiv.org/abs/2107.09065}{{\tt
  2107.09065}}].

\bibitem{Demirtas:2021nlu}
M.~Demirtas, M.~Kim, L.~McAllister, J.~Moritz and A.~Rios-Tascon, \emph{{Small
  cosmological constants in string theory}},
  \href{http://dx.doi.org/10.1007/JHEP12(2021)136}{\emph{JHEP} {\bf 12} (2021)
  136}, [\href{http://arxiv.org/abs/2107.09064}{{\tt 2107.09064}}].

\bibitem{Dudas:2019pls}
E.~Dudas and S.~L\"ust, \emph{{An update on moduli stabilization with antibrane
  uplift}}, \href{http://dx.doi.org/10.1007/JHEP03(2021)107}{\emph{JHEP} {\bf
  03} (2021) 107}, [\href{http://arxiv.org/abs/1912.09948}{{\tt 1912.09948}}].

\bibitem{Demirtas:2019sip}
M.~Demirtas, M.~Kim, L.~Mcallister and J.~Moritz, \emph{{Vacua with Small Flux
  Superpotential}},
  \href{http://dx.doi.org/10.1103/PhysRevLett.124.211603}{\emph{Phys. Rev.
  Lett.} {\bf 124} (2020) 211603}, [\href{http://arxiv.org/abs/1912.10047}{{\tt
  1912.10047}}].

\bibitem{Demirtas:2020ffz}
M.~Demirtas, M.~Kim, L.~McAllister and J.~Moritz, \emph{{Conifold Vacua with
  Small Flux Superpotential}},
  \href{http://dx.doi.org/10.1002/prop.202000085}{\emph{Fortsch. Phys.} {\bf
  68} (2020) 2000085}, [\href{http://arxiv.org/abs/2009.03312}{{\tt
  2009.03312}}].

\bibitem{Alvarez-Garcia:2020pxd}
R.~\'Alvarez-Garc\'\i{}a, R.~Blumenhagen, M.~Brinkmann and L.~Schlechter,
  \emph{{Small Flux Superpotentials for Type IIB Flux Vacua Close to a
  Conifold}}, \href{http://dx.doi.org/10.1002/prop.202000088}{\emph{Fortsch.
  Phys.} {\bf 68} (2020) 2000088}, [\href{http://arxiv.org/abs/2009.03325}{{\tt
  2009.03325}}].

\bibitem{Bena:2018fqc}
I.~Bena, E.~Dudas, M.~Gra\~na and S.~L\"ust, \emph{{Uplifting Runaways}},
  \href{http://dx.doi.org/10.1002/prop.201800100}{\emph{Fortsch. Phys.} {\bf
  67} (2019) 1800100}, [\href{http://arxiv.org/abs/1809.06861}{{\tt
  1809.06861}}].

\bibitem{Bena:2019sxm}
I.~Bena, A.~Buchel and S.~L\"ust, \emph{{Throat destabilization (for profit and
  for fun)}},  \href{http://arxiv.org/abs/1910.08094}{{\tt 1910.08094}}.

\bibitem{Lust:2022xoq}
S.~L\"ust and L.~Randall, \emph{{Effective Theory of Warped Compactifications
  and the Implications for KKLT}},
  \href{http://dx.doi.org/10.1002/prop.202200103}{\emph{Fortsch. Phys.} {\bf
  70} (2022) 2200103}, [\href{http://arxiv.org/abs/2206.04708}{{\tt
  2206.04708}}].

\bibitem{Honma:2021klo}
Y.~Honma and H.~Otsuka, \emph{{Small flux superpotential in F-theory
  compactifications}},
  \href{http://dx.doi.org/10.1103/PhysRevD.103.126022}{\emph{Phys. Rev. D} {\bf
  103} (2021) 126022}, [\href{http://arxiv.org/abs/2103.03003}{{\tt
  2103.03003}}].

\bibitem{Marchesano:2021gyv}
F.~Marchesano, D.~Prieto and M.~Wiesner, \emph{{F-theory flux vacua at large
  complex structure}},
  \href{http://dx.doi.org/10.1007/JHEP08(2021)077}{\emph{JHEP} {\bf 08} (2021)
  077}, [\href{http://arxiv.org/abs/2105.09326}{{\tt 2105.09326}}].

\bibitem{Broeckel:2021uty}
I.~Broeckel, M.~Cicoli, A.~Maharana, K.~Singh and K.~Sinha, \emph{{On the
  Search for Low $W_0$}},  \href{http://arxiv.org/abs/2108.04266}{{\tt
  2108.04266}}.

\bibitem{Bastian:2021hpc}
B.~Bastian, T.~W. Grimm and D.~van~de Heisteeg, \emph{{Engineering Small Flux
  Superpotentials and Mass Hierarchies}},
  \href{http://arxiv.org/abs/2108.11962}{{\tt 2108.11962}}.

\bibitem{Grimm:2021ckh}
T.~W. Grimm, E.~Plauschinn and D.~van~de Heisteeg, \emph{{Moduli stabilization
  in asymptotic flux compactifications}},
  \href{http://dx.doi.org/10.1007/JHEP03(2022)117}{\emph{JHEP} {\bf 03} (2022)
  117}, [\href{http://arxiv.org/abs/2110.05511}{{\tt 2110.05511}}].

\bibitem{Carta:2021kpk}
F.~Carta, A.~Mininno and P.~Shukla, \emph{{Systematics of perturbatively flat
  flux vacua}}, \href{http://dx.doi.org/10.1007/JHEP02(2022)205}{\emph{JHEP}
  {\bf 02} (2022) 205}, [\href{http://arxiv.org/abs/2112.13863}{{\tt
  2112.13863}}].

\bibitem{Carta:2022oex}
F.~Carta, A.~Mininno and P.~Shukla, \emph{{Systematics of perturbatively flat
  flux vacua for CICYs}},
  \href{http://dx.doi.org/10.1007/JHEP08(2022)297}{\emph{JHEP} {\bf 08} (2022)
  297}, [\href{http://arxiv.org/abs/2201.10581}{{\tt 2201.10581}}].

\bibitem{Cicoli:2022vny}
M.~Cicoli, M.~Licheri, R.~Mahanta and A.~Maharana, \emph{{Flux Vacua with
  Approximate Flat Directions}},  \href{http://arxiv.org/abs/2209.02720}{{\tt
  2209.02720}}.

\bibitem{Witten:1996bn}
E.~Witten, \emph{{Nonperturbative superpotentials in string theory}},
  \href{http://dx.doi.org/10.1016/0550-3213(96)00283-0}{\emph{Nucl. Phys. B}
  {\bf 474} (1996) 343--360}, [\href{http://arxiv.org/abs/hep-th/9604030}{{\tt
  hep-th/9604030}}].

\bibitem{Bianchi:2011qh}
M.~Bianchi, A.~Collinucci and L.~Martucci, \emph{{Magnetized E3-brane
  instantons in F-theory}},
  \href{http://dx.doi.org/10.1007/JHEP12(2011)045}{\emph{JHEP} {\bf 12} (2011)
  045}, [\href{http://arxiv.org/abs/1107.3732}{{\tt 1107.3732}}].

\bibitem{Palti:2020qlc}
E.~Palti, C.~Vafa and T.~Weigand, \emph{{Supersymmetric Protection and the
  Swampland}}, \href{http://dx.doi.org/10.1007/JHEP06(2020)168}{\emph{JHEP}
  {\bf 06} (2020) 168}, [\href{http://arxiv.org/abs/2003.10452}{{\tt
  2003.10452}}].

\bibitem{Gendler:2022qof}
N.~Gendler, M.~Kim, L.~McAllister, J.~Moritz and M.~Stillman,
  \emph{{Superpotentials from Singular Divisors}},
  \href{http://arxiv.org/abs/2204.06566}{{\tt 2204.06566}}.

\bibitem{Katz:1996fh}
S.~H. Katz, A.~Klemm and C.~Vafa, \emph{{Geometric engineering of quantum field
  theories}},
  \href{http://dx.doi.org/10.1016/S0550-3213(97)00282-4}{\emph{Nucl. Phys. B}
  {\bf 497} (1997) 173--195}, [\href{http://arxiv.org/abs/hep-th/9609239}{{\tt
  hep-th/9609239}}].

\bibitem{Katz:1996th}
S.~H. Katz and C.~Vafa, \emph{{Geometric engineering of N=1 quantum field
  theories}},
  \href{http://dx.doi.org/10.1016/S0550-3213(97)00283-6}{\emph{Nucl. Phys. B}
  {\bf 497} (1997) 196--204}, [\href{http://arxiv.org/abs/hep-th/9611090}{{\tt
  hep-th/9611090}}].

\bibitem{Diaconescu:1998ua}
D.-E. Diaconescu and S.~Gukov, \emph{{Three-dimensional N=2 gauge theories and
  degenerations of Calabi-Yau four folds}},
  \href{http://dx.doi.org/10.1016/S0550-3213(98)00597-5}{\emph{Nucl. Phys. B}
  {\bf 535} (1998) 171--196}, [\href{http://arxiv.org/abs/hep-th/9804059}{{\tt
  hep-th/9804059}}].

\bibitem{Denef:2008wq}
F.~Denef, \emph{{Les Houches Lectures on Constructing String Vacua}},
  {\emph{Les Houches} {\bf 87} (2008) 483--610},
  [\href{http://arxiv.org/abs/0803.1194}{{\tt 0803.1194}}].

\bibitem{Witten:1999eg}
E.~Witten, \emph{{World sheet corrections via D instantons}},
  \href{http://dx.doi.org/10.1088/1126-6708/2000/02/030}{\emph{JHEP} {\bf 02}
  (2000) 030}, [\href{http://arxiv.org/abs/hep-th/9907041}{{\tt
  hep-th/9907041}}].

\bibitem{Garcia-Etxebarria:2007fvo}
I.~Garcia-Etxebarria and A.~M. Uranga, \emph{{Non-perturbative superpotentials
  across lines of marginal stability}},
  \href{http://dx.doi.org/10.1088/1126-6708/2008/01/033}{\emph{JHEP} {\bf 01}
  (2008) 033}, [\href{http://arxiv.org/abs/0711.1430}{{\tt 0711.1430}}].

\bibitem{Blumenhagen:2008ji}
R.~Blumenhagen and M.~Schmidt-Sommerfeld, \emph{{Power Towers of String
  Instantons for N=1 Vacua}},
  \href{http://dx.doi.org/10.1088/1126-6708/2008/07/027}{\emph{JHEP} {\bf 07}
  (2008) 027}, [\href{http://arxiv.org/abs/0803.1562}{{\tt 0803.1562}}].

\bibitem{Blumenhagen:2012kz}
R.~Blumenhagen, X.~Gao, T.~Rahn and P.~Shukla, \emph{{A Note on Poly-Instanton
  Effects in Type IIB Orientifolds on Calabi-Yau Threefolds}},
  \href{http://dx.doi.org/10.1007/JHEP06(2012)162}{\emph{JHEP} {\bf 06} (2012)
  162}, [\href{http://arxiv.org/abs/1205.2485}{{\tt 1205.2485}}].

\bibitem{Alexandrov:2022mmy}
S.~Alexandrov, A.~H. F\i{}rat, M.~Kim, A.~Sen and B.~Stefa\'nski,
  \emph{{D-instanton induced superpotential}},
  \href{http://dx.doi.org/10.1007/JHEP07(2022)090}{\emph{JHEP} {\bf 07} (2022)
  090}, [\href{http://arxiv.org/abs/2204.02981}{{\tt 2204.02981}}].

\bibitem{Ganor:1996pe}
O.~J. Ganor, \emph{{A Note on zeros of superpotentials in F theory}},
  \href{http://dx.doi.org/10.1016/S0550-3213(97)00311-8}{\emph{Nucl. Phys. B}
  {\bf 499} (1997) 55--66}, [\href{http://arxiv.org/abs/hep-th/9612077}{{\tt
  hep-th/9612077}}].

\bibitem{Harvey:1999as}
J.~A. Harvey and G.~W. Moore, \emph{{Superpotentials and membrane instantons}},
   \href{http://arxiv.org/abs/hep-th/9907026}{{\tt hep-th/9907026}}.

\bibitem{Buchbinder:2002ic}
E.~I. Buchbinder, R.~Donagi and B.~A. Ovrut, \emph{{Superpotentials for vector
  bundle moduli}},
  \href{http://dx.doi.org/10.1016/S0550-3213(02)01093-3}{\emph{Nucl. Phys. B}
  {\bf 653} (2003) 400--420}, [\href{http://arxiv.org/abs/hep-th/0205190}{{\tt
  hep-th/0205190}}].

\bibitem{Buchbinder:2002pr}
E.~I. Buchbinder, R.~Donagi and B.~A. Ovrut, \emph{{Vector bundle moduli
  superpotentials in heterotic superstrings and M theory}},
  \href{http://dx.doi.org/10.1088/1126-6708/2002/07/066}{\emph{JHEP} {\bf 07}
  (2002) 066}, [\href{http://arxiv.org/abs/hep-th/0206203}{{\tt
  hep-th/0206203}}].

\bibitem{Beasley:2003fx}
C.~Beasley and E.~Witten, \emph{{Residues and world sheet instantons}},
  \href{http://dx.doi.org/10.1088/1126-6708/2003/10/065}{\emph{JHEP} {\bf 10}
  (2003) 065}, [\href{http://arxiv.org/abs/hep-th/0304115}{{\tt
  hep-th/0304115}}].

\bibitem{Berg:2004ek}
M.~Berg, M.~Haack and B.~Kors, \emph{{Loop corrections to volume moduli and
  inflation in string theory}},
  \href{http://dx.doi.org/10.1103/PhysRevD.71.026005}{\emph{Phys. Rev. D} {\bf
  71} (2005) 026005}, [\href{http://arxiv.org/abs/hep-th/0404087}{{\tt
  hep-th/0404087}}].

\bibitem{Kallosh:2005gs}
R.~Kallosh, A.-K. Kashani-Poor and A.~Tomasiello, \emph{{Counting fermionic
  zero modes on M5 with fluxes}},
  \href{http://dx.doi.org/10.1088/1126-6708/2005/06/069}{\emph{JHEP} {\bf 06}
  (2005) 069}, [\href{http://arxiv.org/abs/hep-th/0503138}{{\tt
  hep-th/0503138}}].

\bibitem{Lust:2005cu}
D.~Lust, S.~Reffert, W.~Schulgin and P.~K. Tripathy, \emph{{Fermion zero modes
  in the presence of fluxes and a non-perturbative superpotential}},
  \href{http://dx.doi.org/10.1088/1126-6708/2006/08/071}{\emph{JHEP} {\bf 08}
  (2006) 071}, [\href{http://arxiv.org/abs/hep-th/0509082}{{\tt
  hep-th/0509082}}].

\bibitem{Ibanez:2006da}
L.~E. Ibanez and A.~M. Uranga, \emph{{Neutrino Majorana Masses from String
  Theory Instanton Effects}},
  \href{http://dx.doi.org/10.1088/1126-6708/2007/03/052}{\emph{JHEP} {\bf 03}
  (2007) 052}, [\href{http://arxiv.org/abs/hep-th/0609213}{{\tt
  hep-th/0609213}}].

\bibitem{Akerblom:2006hx}
N.~Akerblom, R.~Blumenhagen, D.~Lust, E.~Plauschinn and M.~Schmidt-Sommerfeld,
  \emph{{Non-perturbative SQCD Superpotentials from String Instantons}},
  \href{http://dx.doi.org/10.1088/1126-6708/2007/04/076}{\emph{JHEP} {\bf 04}
  (2007) 076}, [\href{http://arxiv.org/abs/hep-th/0612132}{{\tt
  hep-th/0612132}}].

\bibitem{Blumenhagen:2006xt}
R.~Blumenhagen, M.~Cvetic and T.~Weigand, \emph{{Spacetime instanton
  corrections in 4D string vacua: The Seesaw mechanism for D-Brane models}},
  \href{http://dx.doi.org/10.1016/j.nuclphysb.2007.02.016}{\emph{Nucl. Phys. B}
  {\bf 771} (2007) 113--142}, [\href{http://arxiv.org/abs/hep-th/0609191}{{\tt
  hep-th/0609191}}].

\bibitem{Baumann:2006th}
D.~Baumann, A.~Dymarsky, I.~R. Klebanov, J.~M. Maldacena, L.~P. McAllister and
  A.~Murugan, \emph{{On D3-brane Potentials in Compactifications with Fluxes
  and Wrapped D-branes}},
  \href{http://dx.doi.org/10.1088/1126-6708/2006/11/031}{\emph{JHEP} {\bf 11}
  (2006) 031}, [\href{http://arxiv.org/abs/hep-th/0607050}{{\tt
  hep-th/0607050}}].

\bibitem{Braun:2007xh}
V.~Braun, M.~Kreuzer, B.~A. Ovrut and E.~Scheidegger, \emph{{Worldsheet
  instantons and torsion curves, part A: Direct computation}},
  \href{http://dx.doi.org/10.1088/1126-6708/2007/10/022}{\emph{JHEP} {\bf 10}
  (2007) 022}, [\href{http://arxiv.org/abs/hep-th/0703182}{{\tt
  hep-th/0703182}}].

\bibitem{Braun:2007vy}
V.~Braun, M.~Kreuzer, B.~A. Ovrut and E.~Scheidegger, \emph{{Worldsheet
  Instantons and Torsion Curves, Part B: Mirror Symmetry}},
  \href{http://dx.doi.org/10.1088/1126-6708/2007/10/023}{\emph{JHEP} {\bf 10}
  (2007) 023}, [\href{http://arxiv.org/abs/0704.0449}{{\tt 0704.0449}}].

\bibitem{Koerber:2007xk}
P.~Koerber and L.~Martucci, \emph{{From ten to four and back again: How to
  generalize the geometry}},
  \href{http://dx.doi.org/10.1088/1126-6708/2007/08/059}{\emph{JHEP} {\bf 08}
  (2007) 059}, [\href{http://arxiv.org/abs/0707.1038}{{\tt 0707.1038}}].

\bibitem{Blumenhagen:2007zk}
R.~Blumenhagen, M.~Cvetic, D.~Lust, R.~Richter and T.~Weigand,
  \emph{{Non-perturbative Yukawa Couplings from String Instantons}},
  \href{http://dx.doi.org/10.1103/PhysRevLett.100.061602}{\emph{Phys. Rev.
  Lett.} {\bf 100} (2008) 061602}, [\href{http://arxiv.org/abs/0707.1871}{{\tt
  0707.1871}}].

\bibitem{Beasley:2008dc}
C.~Beasley, J.~J. Heckman and C.~Vafa, \emph{{GUTs and Exceptional Branes in
  F-theory - I}},
  \href{http://dx.doi.org/10.1088/1126-6708/2009/01/058}{\emph{JHEP} {\bf 01}
  (2009) 058}, [\href{http://arxiv.org/abs/0802.3391}{{\tt 0802.3391}}].

\bibitem{Blumenhagen:2010ja}
R.~Blumenhagen, A.~Collinucci and B.~Jurke, \emph{{On Instanton Effects in
  F-theory}}, \href{http://dx.doi.org/10.1007/JHEP08(2010)079}{\emph{JHEP} {\bf
  08} (2010) 079}, [\href{http://arxiv.org/abs/1002.1894}{{\tt 1002.1894}}].

\bibitem{Baumann:2010sx}
D.~Baumann, A.~Dymarsky, S.~Kachru, I.~R. Klebanov and L.~McAllister,
  \emph{{D3-brane Potentials from Fluxes in AdS/CFT}},
  \href{http://dx.doi.org/10.1007/JHEP06(2010)072}{\emph{JHEP} {\bf 06} (2010)
  072}, [\href{http://arxiv.org/abs/1001.5028}{{\tt 1001.5028}}].

\bibitem{Donagi:2010pd}
R.~Donagi and M.~Wijnholt, \emph{{MSW Instantons}},
  \href{http://dx.doi.org/10.1007/JHEP06(2013)050}{\emph{JHEP} {\bf 06} (2013)
  050}, [\href{http://arxiv.org/abs/1005.5391}{{\tt 1005.5391}}].

\bibitem{Marsano:2011xfj}
J.~Marsano, N.~Saulina and S.~Sch\"afer-Nameki, \emph{{G-flux, M5 instantons,
  and U(1) symmetries in F-theory}},
  \href{http://dx.doi.org/10.1103/PhysRevD.87.066007}{\emph{Phys. Rev. D} {\bf
  87} (2013) 066007}, [\href{http://arxiv.org/abs/1107.1718}{{\tt 1107.1718}}].

\bibitem{Grimm:2011dj}
T.~W. Grimm, M.~Kerstan, E.~Palti and T.~Weigand, \emph{{On Fluxed Instantons
  and Moduli Stabilisation in IIB Orientifolds and F-theory}},
  \href{http://dx.doi.org/10.1103/PhysRevD.84.066001}{\emph{Phys. Rev. D} {\bf
  84} (2011) 066001}, [\href{http://arxiv.org/abs/1105.3193}{{\tt 1105.3193}}].

\bibitem{Kerstan:2012cy}
M.~Kerstan and T.~Weigand, \emph{{Fluxed M5-instantons in F-theory}},
  \href{http://dx.doi.org/10.1016/j.nuclphysb.2012.07.008}{\emph{Nucl. Phys. B}
  {\bf 864} (2012) 597--639}, [\href{http://arxiv.org/abs/1205.4720}{{\tt
  1205.4720}}].

\bibitem{Cvetic:2012ts}
M.~Cvetic, R.~Donagi, J.~Halverson and J.~Marsano, \emph{{On Seven-Brane
  Dependent Instanton Prefactors in F-theory}},
  \href{http://dx.doi.org/10.1007/JHEP11(2012)004}{\emph{JHEP} {\bf 11} (2012)
  004}, [\href{http://arxiv.org/abs/1209.4906}{{\tt 1209.4906}}].

\bibitem{Anderson:2015yzz}
L.~B. Anderson, F.~Apruzzi, X.~Gao, J.~Gray and S.-J. Lee, \emph{{Instanton
  superpotentials, Calabi-Yau geometry, and fibrations}},
  \href{http://dx.doi.org/10.1103/PhysRevD.93.086001}{\emph{Phys. Rev. D} {\bf
  93} (2016) 086001}, [\href{http://arxiv.org/abs/1511.05188}{{\tt
  1511.05188}}].

\bibitem{Hamada:2018qef}
Y.~Hamada, A.~Hebecker, G.~Shiu and P.~Soler, \emph{{On brane gaugino
  condensates in 10d}},
  \href{http://dx.doi.org/10.1007/JHEP04(2019)008}{\emph{JHEP} {\bf 04} (2019)
  008}, [\href{http://arxiv.org/abs/1812.06097}{{\tt 1812.06097}}].

\bibitem{Kim:2018vgz}
M.~Kim and L.~McAllister, \emph{{Monodromy Charge in D7-brane Inflation}},
  \href{http://dx.doi.org/10.1007/JHEP10(2020)060}{\emph{JHEP} {\bf 10} (2020)
  060}, [\href{http://arxiv.org/abs/1812.03532}{{\tt 1812.03532}}].

\bibitem{Kachru:2019dvo}
S.~Kachru, M.~Kim, L.~Mcallister and M.~Zimet, \emph{{de Sitter vacua from ten
  dimensions}}, \href{http://dx.doi.org/10.1007/JHEP12(2021)111}{\emph{JHEP}
  {\bf 12} (2021) 111}, [\href{http://arxiv.org/abs/1908.04788}{{\tt
  1908.04788}}].

\bibitem{Bena:2019mte}
I.~Bena, M.~Gra\~na, N.~Kovensky and A.~Retolaza, \emph{{K\"ahler moduli
  stabilization from ten dimensions}},
  \href{http://dx.doi.org/10.1007/JHEP10(2019)200}{\emph{JHEP} {\bf 10} (2019)
  200}, [\href{http://arxiv.org/abs/1908.01785}{{\tt 1908.01785}}].

\bibitem{Hamada:2019ack}
Y.~Hamada, A.~Hebecker, G.~Shiu and P.~Soler, \emph{{Understanding KKLT from a
  10d perspective}},
  \href{http://dx.doi.org/10.1007/JHEP06(2019)019}{\emph{JHEP} {\bf 06} (2019)
  019}, [\href{http://arxiv.org/abs/1902.01410}{{\tt 1902.01410}}].

\bibitem{Gautason:2019jwq}
F.~F. Gautason, V.~Van~Hemelryck, T.~Van~Riet and G.~Venken, \emph{{A 10d view
  on the KKLT AdS vacuum and uplifting}},
  \href{http://dx.doi.org/10.1007/JHEP06(2020)074}{\emph{JHEP} {\bf 06} (2020)
  074}, [\href{http://arxiv.org/abs/1902.01415}{{\tt 1902.01415}}].

\bibitem{Carta:2019rhx}
F.~Carta, J.~Moritz and A.~Westphal, \emph{{Gaugino condensation and small
  uplifts in KKLT}},
  \href{http://dx.doi.org/10.1007/JHEP08(2019)141}{\emph{JHEP} {\bf 08} (2019)
  141}, [\href{http://arxiv.org/abs/1902.01412}{{\tt 1902.01412}}].

\bibitem{Kim:2022uni}
M.~Kim, \emph{{On D3-brane Superpotential}},
  \href{http://arxiv.org/abs/2207.01440}{{\tt 2207.01440}}.

\bibitem{Witten:1996hc}
E.~Witten, \emph{{Five-brane effective action in M theory}},
  \href{http://dx.doi.org/10.1016/S0393-0440(97)80160-X}{\emph{J. Geom. Phys.}
  {\bf 22} (1997) 103--133}, [\href{http://arxiv.org/abs/hep-th/9610234}{{\tt
  hep-th/9610234}}].

\bibitem{Belov:2006jd}
D.~Belov and G.~W. Moore, \emph{{Holographic Action for the Self-Dual Field}},
  \href{http://arxiv.org/abs/hep-th/0605038}{{\tt hep-th/0605038}}.

\bibitem{Denef:2005mm}
F.~Denef, M.~R. Douglas, B.~Florea, A.~Grassi and S.~Kachru, \emph{{Fixing all
  moduli in a simple f-theory compactification}},
  \href{http://dx.doi.org/10.4310/ATMP.2005.v9.n6.a1}{\emph{Adv. Theor. Math.
  Phys.} {\bf 9} (2005) 861--929},
  [\href{http://arxiv.org/abs/hep-th/0503124}{{\tt hep-th/0503124}}].

\bibitem{Kim:2021lwo}
M.~Kim, \emph{{A note on h$^{2,1}$ of divisors in CY fourfolds. Part I}},
  \href{http://dx.doi.org/10.1007/JHEP03(2022)168}{\emph{JHEP} {\bf 03} (2022)
  168}, [\href{http://arxiv.org/abs/2107.09779}{{\tt 2107.09779}}].

\bibitem{Sen:1996vd}
A.~Sen, \emph{{F theory and orientifolds}},
  \href{http://dx.doi.org/10.1016/0550-3213(96)00347-1}{\emph{Nucl. Phys. B}
  {\bf 475} (1996) 562--578}, [\href{http://arxiv.org/abs/hep-th/9605150}{{\tt
  hep-th/9605150}}].

\bibitem{Sen:1997gv}
A.~Sen, \emph{{Orientifold limit of F theory vacua}},
  \href{http://dx.doi.org/10.1103/PhysRevD.55.R7345}{\emph{Phys. Rev. D} {\bf
  55} (1997) R7345--R7349}, [\href{http://arxiv.org/abs/hep-th/9702165}{{\tt
  hep-th/9702165}}].

\bibitem{Gao:2013pra}
X.~Gao and P.~Shukla, \emph{{On Classifying the Divisor Involutions in
  Calabi-Yau Threefolds}},
  \href{http://dx.doi.org/10.1007/JHEP11(2013)170}{\emph{JHEP} {\bf 11} (2013)
  170}, [\href{http://arxiv.org/abs/1307.1139}{{\tt 1307.1139}}].

\bibitem{Carta:2020ohw}
F.~Carta, J.~Moritz and A.~Westphal, \emph{{A landscape of orientifold vacua}},
  \href{http://dx.doi.org/10.1007/JHEP05(2020)107}{\emph{JHEP} {\bf 05} (2020)
  107}, [\href{http://arxiv.org/abs/2003.04902}{{\tt 2003.04902}}].

\bibitem{Altman:2021pyc}
R.~Altman, J.~Carifio, X.~Gao and B.~D. Nelson, \emph{{Orientifold Calabi-Yau
  threefolds with divisor involutions and string landscape}},
  \href{http://dx.doi.org/10.1007/JHEP03(2022)087}{\emph{JHEP} {\bf 03} (2022)
  087}, [\href{http://arxiv.org/abs/2111.03078}{{\tt 2111.03078}}].

\bibitem{Crino:2022zjk}
C.~Crin\`o, F.~Quevedo, A.~Schachner and R.~Valandro, \emph{{A database of
  Calabi-Yau orientifolds and the size of D3-tadpoles}},
  \href{http://dx.doi.org/10.1007/JHEP08(2022)050}{\emph{JHEP} {\bf 08} (2022)
  050}, [\href{http://arxiv.org/abs/2204.13115}{{\tt 2204.13115}}].

\bibitem{Gao:2021xbs}
X.~Gao and H.~Zou, \emph{{Applying machine learning to the Calabi-Yau
  orientifolds with string vacua}},
  \href{http://dx.doi.org/10.1103/PhysRevD.105.046017}{\emph{Phys. Rev. D} {\bf
  105} (2022) 046017}, [\href{http://arxiv.org/abs/2112.04950}{{\tt
  2112.04950}}].

\bibitem{Esole:2017kyr}
M.~Esole, P.~Jefferson and M.~J. Kang, \emph{{Euler Characteristics of Crepant
  Resolutions of Weierstrass Models}},
  \href{http://dx.doi.org/10.1007/s00220-019-03517-1}{\emph{Commun. Math.
  Phys.} {\bf 371} (2019) 99--144},
  [\href{http://arxiv.org/abs/1703.00905}{{\tt 1703.00905}}].

\bibitem{Berglund:2016yqo}
P.~Berglund and T.~H\"ubsch, \emph{{On Calabi\textendash{}Yau generalized
  complete intersections from Hirzebruch varieties and novel $K3$-fibrations}},
  \href{http://dx.doi.org/10.4310/ATMP.2018.v22.n2.a1}{\emph{Adv. Theor. Math.
  Phys.} {\bf 22} (2018) 261--303},
  [\href{http://arxiv.org/abs/1606.07420}{{\tt 1606.07420}}].

\bibitem{Berglund:2016nvh}
P.~Berglund and T.~Hubsch, \emph{{A Generalized Construction of Calabi-Yau
  Models and Mirror Symmetry}},
  \href{http://dx.doi.org/10.21468/SciPostPhys.4.2.009}{\emph{SciPost Phys.}
  {\bf 4} (2018) 009}, [\href{http://arxiv.org/abs/1611.10300}{{\tt
  1611.10300}}].

\bibitem{vex}
P.~Berglund, Y.-C. Huang, W.~Taylor and Y.~Wang, \emph{Unpublished work}.

\bibitem{Batyrev:1994ju}
V.~V. Batyrev and D.~I. Dais, \emph{{Strong McKay correspondence, string
  theoretic Hodge numbers and mirror symmetry}},
  \href{http://arxiv.org/abs/alg-geom/9410001}{{\tt alg-geom/9410001}}.

\bibitem{Batyrev:1994pg}
V.~V. Batyrev and L.~A. Borisov, \emph{{On Calabi-Yau complete intersections in
  toric varieties}},  \href{http://arxiv.org/abs/alg-geom/9412017}{{\tt
  alg-geom/9412017}}.

\bibitem{Batyrev:1995ca}
V.~V. Batyrev and L.~A. Borisov, \emph{{Mirror duality and string theoretic
  Hodge numbers}}, \href{http://dx.doi.org/10.1007/s002220050093}{\emph{Invent.
  Math.} {\bf 126} (1996) 183},
  [\href{http://arxiv.org/abs/alg-geom/9509009}{{\tt alg-geom/9509009}}].

\bibitem{Kreuzer:2001fu}
M.~Kreuzer, E.~Riegler and D.~A. Sahakyan, \emph{{Toric complete intersections
  and weighted projective space}},
  \href{http://dx.doi.org/10.1016/S0393-0440(02)00124-9}{\emph{J. Geom. Phys.}
  {\bf 46} (2003) 159--173}, [\href{http://arxiv.org/abs/math/0103214}{{\tt
  math/0103214}}].

\bibitem{nefpartition}
M.~Kim, \emph{Work in progress}.

\bibitem{batyrev1993dual}
V.~V. Batyrev, \emph{Dual polyhedra and mirror symmetry for calabi-yau
  hypersurfaces in toric varieties}, {\emph{arXiv preprint alg-geom/9310003}
  (1993) }.

\bibitem{morrison1984terminal}
D.~R. Morrison and G.~Stevens, \emph{Terminal quotient singularities in
  dimensions three and four}, {\emph{Proceedings of the American Mathematical
  Society} {\bf 90} (1984) 15--20}.

\bibitem{Collinucci:2008zs}
A.~Collinucci, \emph{{New F-theory lifts}},
  \href{http://dx.doi.org/10.1088/1126-6708/2009/08/076}{\emph{JHEP} {\bf 08}
  (2009) 076}, [\href{http://arxiv.org/abs/0812.0175}{{\tt 0812.0175}}].

\bibitem{Collinucci:2009uh}
A.~Collinucci, \emph{{New F-theory lifts. II. Permutation orientifolds and
  enhanced singularities}},
  \href{http://dx.doi.org/10.1007/JHEP04(2010)076}{\emph{JHEP} {\bf 04} (2010)
  076}, [\href{http://arxiv.org/abs/0906.0003}{{\tt 0906.0003}}].

\bibitem{Garcia-Etxebarria:2015wns}
I.~n. Garc\'\i{}a-Etxebarria and D.~Regalado, \emph{{$ \mathcal{N}=3 $ four
  dimensional field theories}},
  \href{http://dx.doi.org/10.1007/JHEP03(2016)083}{\emph{JHEP} {\bf 03} (2016)
  083}, [\href{http://arxiv.org/abs/1512.06434}{{\tt 1512.06434}}].

\bibitem{stats}
P.~Jefferson and M.~Kim, \emph{Work in progress}.

\bibitem{reid1983decomposition}
M.~Reid, \emph{Decomposition of toric morphisms},  in \emph{Arithmetic and
  geometry}, pp.~395--418.
\newblock Springer, 1983.

\bibitem{cox2011toric}
D.~A. Cox, J.~B. Little and H.~K. Schenck, \emph{Toric varieties}, vol.~124.
\newblock American Mathematical Soc., 2011.

\bibitem{Danilov_1987}
V.~I. Danilov and A.~G. Khovanski{\u{\i}}, \emph{{NEWTON} {POLYHEDRA} {AND}
  {AN} {ALGORITHM} {FOR} {COMPUTING} {HODGE}{\textendash}{DELIGNE} {NUMBERS}},
  \href{http://dx.doi.org/10.1070/im1987v029n02abeh000970}{\emph{Mathematics of
  the {USSR}-Izvestiya} {\bf 29} (apr, 1987) 279--298}.

\bibitem{Braun:2014xka}
A.~P. Braun and T.~Watari, \emph{{The Vertical, the Horizontal and the Rest:
  anatomy of the middle cohomology of Calabi-Yau fourfolds and F-theory
  applications}}, \href{http://dx.doi.org/10.1007/JHEP01(2015)047}{\emph{JHEP}
  {\bf 01} (2015) 047}, [\href{http://arxiv.org/abs/1408.6167}{{\tt
  1408.6167}}].

\bibitem{Braun:2017nhi}
A.~P. Braun, C.~Long, L.~McAllister, M.~Stillman and B.~Sung, \emph{{The Hodge
  Numbers of Divisors of Calabi-Yau Threefold Hypersurfaces}},
  \href{http://arxiv.org/abs/1712.04946}{{\tt 1712.04946}}.

\bibitem{Fulton}
W.~Fulton, \emph{Introduction to toric varieties.}
\newblock Princeton University Press, Princeton, 1993.

\bibitem{danilov1978geometry}
V.~I. Danilov, \emph{The geometry of toric varieties}, {\emph{Uspekhi
  Matematicheskikh Nauk} {\bf 33} (1978) 85--134}.

\bibitem{khovanskii1977newton}
A.~G. Khovanskii, \emph{Newton polyhedra and toroidal varieties},
  {\emph{Functional analysis and its applications} {\bf 11} (1977) 289--296}.

\bibitem{cox1995homogeneous}
D.~A. Cox, \emph{The homogeneous coordinate ring of a toric variety},
  {\emph{arXiv preprint alg-geom/9210008} (1995) }.

\bibitem{orientifoldingks}
J.~Moritz, \emph{Work in progress}.

\bibitem{Witten:1993yc}
E.~Witten, \emph{{Phases of N=2 theories in two-dimensions}},
  \href{http://dx.doi.org/10.1016/0550-3213(93)90033-L}{\emph{Nucl. Phys. B}
  {\bf 403} (1993) 159--222}, [\href{http://arxiv.org/abs/hep-th/9301042}{{\tt
  hep-th/9301042}}].

\bibitem{hori2003mirror}
K.~Hori, S.~Katz, C.~Vafa, R.~Thomas, A.~M. Society, C.~M. Institute et~al.,
  \emph{Mirror Symmetry}.
\newblock Clay mathematics monographs. American Mathematical Society, 2003.

\bibitem{Gimon:1996rq}
E.~G. Gimon and J.~Polchinski, \emph{{Consistency conditions for orientifolds
  and D-manifolds}},
  \href{http://dx.doi.org/10.1103/PhysRevD.54.1667}{\emph{Phys. Rev. D} {\bf
  54} (1996) 1667--1676}, [\href{http://arxiv.org/abs/hep-th/9601038}{{\tt
  hep-th/9601038}}].

\bibitem{Candelas:1996su}
P.~Candelas and A.~Font, \emph{{Duality between the webs of heterotic and type
  II vacua}},
  \href{http://dx.doi.org/10.1016/S0550-3213(96)00410-5}{\emph{Nucl. Phys. B}
  {\bf 511} (1998) 295--325}, [\href{http://arxiv.org/abs/hep-th/9603170}{{\tt
  hep-th/9603170}}].

\bibitem{Kreuzer:1997zg}
M.~Kreuzer and H.~Skarke, \emph{{Calabi-Yau four folds and toric fibrations}},
  \href{http://dx.doi.org/10.1016/S0393-0440(97)00059-4}{\emph{J. Geom. Phys.}
  {\bf 26} (1998) 272--290}, [\href{http://arxiv.org/abs/hep-th/9701175}{{\tt
  hep-th/9701175}}].

\bibitem{Candelas:1997eh}
P.~Candelas, E.~Perevalov and G.~Rajesh, \emph{{Toric geometry and enhanced
  gauge symmetry of F theory / heterotic vacua}},
  \href{http://dx.doi.org/10.1016/S0550-3213(97)00563-4}{\emph{Nucl. Phys. B}
  {\bf 507} (1997) 445--474}, [\href{http://arxiv.org/abs/hep-th/9704097}{{\tt
  hep-th/9704097}}].

\bibitem{Perevalov:1997vw}
E.~Perevalov and H.~Skarke, \emph{{Enhanced gauged symmetry in type II and F
  theory compactifications: Dynkin diagrams from polyhedra}},
  \href{http://dx.doi.org/10.1016/S0550-3213(97)00477-X}{\emph{Nucl. Phys. B}
  {\bf 505} (1997) 679--700}, [\href{http://arxiv.org/abs/hep-th/9704129}{{\tt
  hep-th/9704129}}].

\bibitem{Skarke:1998yk}
H.~Skarke, \emph{{String dualities and toric geometry: An Introduction}},
  \href{http://dx.doi.org/10.1016/S0960-0779(98)00161-1}{\emph{Chaos Solitons
  Fractals} {\bf 10} (1999) 543},
  [\href{http://arxiv.org/abs/hep-th/9806059}{{\tt hep-th/9806059}}].

\bibitem{Huang:2018gpl}
Y.-C. Huang and W.~Taylor, \emph{{Comparing elliptic and toric hypersurface
  Calabi-Yau threefolds at large Hodge numbers}},
  \href{http://dx.doi.org/10.1007/JHEP02(2019)087}{\emph{JHEP} {\bf 02} (2019)
  087}, [\href{http://arxiv.org/abs/1805.05907}{{\tt 1805.05907}}].

\bibitem{Huang:2018vup}
Y.-C. Huang and W.~Taylor, \emph{{Mirror symmetry and elliptic Calabi-Yau
  manifolds}}, \href{http://dx.doi.org/10.1007/JHEP04(2019)083}{\emph{JHEP}
  {\bf 04} (2019) 083}, [\href{http://arxiv.org/abs/1811.04947}{{\tt
  1811.04947}}].

\bibitem{Morrison:2012np}
D.~R. Morrison and W.~Taylor, \emph{{Classifying bases for 6D F-theory
  models}}, \href{http://dx.doi.org/10.2478/s11534-012-0065-4}{\emph{Central
  Eur. J. Phys.} {\bf 10} (2012) 1072--1088},
  [\href{http://arxiv.org/abs/1201.1943}{{\tt 1201.1943}}].

\bibitem{Morrison:2012js}
D.~R. Morrison and W.~Taylor, \emph{{Toric bases for 6D F-theory models}},
  \href{http://dx.doi.org/10.1002/prop.201200086}{\emph{Fortsch. Phys.} {\bf
  60} (2012) 1187--1216}, [\href{http://arxiv.org/abs/1204.0283}{{\tt
  1204.0283}}].

\bibitem{Morrison:2014lca}
D.~R. Morrison and W.~Taylor, \emph{{Non-Higgsable clusters for 4D F-theory
  models}}, \href{http://dx.doi.org/10.1007/JHEP05(2015)080}{\emph{JHEP} {\bf
  05} (2015) 080}, [\href{http://arxiv.org/abs/1412.6112}{{\tt 1412.6112}}].

\bibitem{Halverson:2015jua}
J.~Halverson and W.~Taylor, \emph{{$ {\mathrm{\mathbb{P}}}^1 $-bundle bases and
  the prevalence of non-Higgsable structure in 4D F-theory models}},
  \href{http://dx.doi.org/10.1007/JHEP09(2015)086}{\emph{JHEP} {\bf 09} (2015)
  086}, [\href{http://arxiv.org/abs/1506.03204}{{\tt 1506.03204}}].

\bibitem{lazarsfeld2017positivity}
R.~K. Lazarsfeld, \emph{Positivity in algebraic geometry I: Classical setting:
  line bundles and linear series}, vol.~48.
\newblock Springer, 2017.

\bibitem{Grassi:1997mr}
A.~Grassi, \emph{{Divisors on elliptic Calabi-Yau four folds and the
  superpotential in F theory. 1.}},
  \href{http://dx.doi.org/10.1016/S0393-0440(98)00004-7}{\emph{J. Geom. Phys.}
  {\bf 28} (1998) 289--319}, [\href{http://arxiv.org/abs/alg-geom/9704008}{{\tt
  alg-geom/9704008}}].

\bibitem{Polchinski:1995mt}
J.~Polchinski, \emph{{Dirichlet Branes and Ramond-Ramond charges}},
  \href{http://dx.doi.org/10.1103/PhysRevLett.75.4724}{\emph{Phys. Rev. Lett.}
  {\bf 75} (1995) 4724--4727}, [\href{http://arxiv.org/abs/hep-th/9510017}{{\tt
  hep-th/9510017}}].

\bibitem{Polchinski:1998rq}
J.~Polchinski, \emph{{String theory. Vol. 1: An introduction to the bosonic
  string}}.
\newblock Cambridge Monographs on Mathematical Physics. Cambridge University
  Press, 12, 2007,
  \href{http://dx.doi.org/10.1017/CBO9780511816079}{10.1017/CBO9780511816079}.

\bibitem{Polchinski:1998rr}
J.~Polchinski, \emph{{String theory. Vol. 2: Superstring theory and beyond}}.
\newblock Cambridge Monographs on Mathematical Physics. Cambridge University
  Press, 12, 2007,
  \href{http://dx.doi.org/10.1017/CBO9780511618123}{10.1017/CBO9780511618123}.

\bibitem{Esole:2017qeh}
M.~Esole, R.~Jagadeesan and M.~J. Kang, \emph{{The Geometry of G$_2$, Spin(7),
  and Spin(8)-models}},  \href{http://arxiv.org/abs/1709.04913}{{\tt
  1709.04913}}.

\bibitem{shioda1972elliptic}
T.~Shioda, \emph{On elliptic modular surfaces}, {\emph{Journal of the
  Mathematical Society of Japan} {\bf 24} (1972) 20--59}.

\bibitem{tate1965algebraic}
J.~Tate, \emph{Algebraic cycles and poles of zeta functions},
  {\emph{Arithmetical algebraic geometry} (1965) 93--110}.

\bibitem{tate1965conjectures}
J.~Tate, \emph{On the conjectures of birch and swinnerton-dyer and a geometric
  analog}, {\emph{S{\'e}minaire Bourbaki} {\bf 9} (1965) 415--440}.

\bibitem{wazir2004arithmetic}
R.~Wazir, \emph{Arithmetic on elliptic threefolds}, {\emph{Compositio
  Mathematica} {\bf 140} (2004) 567--580}.

\bibitem{Weigand:2018rez}
T.~Weigand, \emph{{F-theory}}, {\emph{PoS} {\bf TASI2017} (2018) 016},
  [\href{http://arxiv.org/abs/1806.01854}{{\tt 1806.01854}}].

\bibitem{Jefferson:2022xft}
P.~Jefferson and A.~P. Turner, \emph{{Generating functions for intersection
  products of divisors in resolved F-theory models}},
  \href{http://arxiv.org/abs/2206.11527}{{\tt 2206.11527}}.

\bibitem{Esole:2018tuz}
M.~Esole and M.~J. Kang, \emph{{Characteristic numbers of crepant resolutions
  of Weierstrass models}},  \href{http://arxiv.org/abs/1807.08755}{{\tt
  1807.08755}}.

\bibitem{Esole:2018bmf}
M.~Esole and M.~J. Kang, \emph{{Characteristic numbers of elliptic fibrations
  with non-trivial Mordell-Weil groups}},
  \href{http://arxiv.org/abs/1808.07054}{{\tt 1808.07054}}.

\bibitem{Esole:2019asj}
M.~Esole, R.~Jagadeesan and M.~J. Kang, \emph{{48 Crepant Paths to
  $\text{SU}(2)\!\times\!\text{SU}(3)$}},
  \href{http://arxiv.org/abs/1905.05174}{{\tt 1905.05174}}.

\bibitem{Esole:2019hgr}
M.~Esole and P.~Jefferson, \emph{{The Geometry of SO(3), SO(5), and SO(6)
  models}},  \href{http://arxiv.org/abs/1905.12620}{{\tt 1905.12620}}.

\bibitem{Esole:2019ocl}
M.~Esole and P.~Jefferson, \emph{{USp(4)-models}},
  \href{http://arxiv.org/abs/1910.09536}{{\tt 1910.09536}}.

\bibitem{Esole:2020alo}
M.~Esole and S.~Pasterski, \emph{{Flops and Fibral Geometry of E$_7$-models}},
  \href{http://arxiv.org/abs/2004.06104}{{\tt 2004.06104}}.

\bibitem{anno2003four}
R.~E. Anno, \emph{Four-dimensional terminal gorenstein quotient singularities},
  {\emph{Mathematical Notes} {\bf 73} (2003) 769--776}.

\bibitem{DelaOssa:2001blj}
X.~De~la Ossa, B.~Florea and H.~Skarke, \emph{{D-branes on noncompact
  Calabi-Yau manifolds: K theory and monodromy}},
  \href{http://dx.doi.org/10.1016/S0550-3213(02)00762-9}{\emph{Nucl. Phys. B}
  {\bf 644} (2002) 170--200}, [\href{http://arxiv.org/abs/hep-th/0104254}{{\tt
  hep-th/0104254}}].

\bibitem{mcduff1994j}
D.~W. McDuff, D.~Salamon et~al., \emph{$ J $-Holomorphic Curves and Quantum
  Cohomology}.
\newblock No.~6. American Mathematical Soc., 1994.

\bibitem{wilson1999flops}
P.~Wilson, \emph{Flops, type iii contractions and gromov-witten invariants on
  calabi-yau threefolds}, {\emph{New trends in algebraic geometry} {\bf 264}
  (1999) 465}.

\bibitem{liedtke2016picard}
C.~Liedtke, \emph{The picard rank of an enriques surface}, {\emph{arXiv
  preprint arXiv:1606.01771} (2016) }.

\bibitem{Candelas:1990rm}
P.~Candelas, X.~C. De~La~Ossa, P.~S. Green and L.~Parkes, \emph{{A Pair of
  Calabi-Yau manifolds as an exactly soluble superconformal theory}},
  \href{http://dx.doi.org/10.1016/0550-3213(91)90292-6}{\emph{Nucl. Phys. B}
  {\bf 359} (1991) 21--74}.

\bibitem{Collinucci:2008pf}
A.~Collinucci, F.~Denef and M.~Esole, \emph{{D-brane Deconstructions in IIB
  Orientifolds}},
  \href{http://dx.doi.org/10.1088/1126-6708/2009/02/005}{\emph{JHEP} {\bf 02}
  (2009) 005}, [\href{http://arxiv.org/abs/0805.1573}{{\tt 0805.1573}}].

\bibitem{Candelas:1994hw}
P.~Candelas, A.~Font, S.~H. Katz and D.~R. Morrison, \emph{{Mirror symmetry for
  two parameter models. 2.}},
  \href{http://dx.doi.org/10.1016/0550-3213(94)90155-4}{\emph{Nucl. Phys. B}
  {\bf 429} (1994) 626--674}, [\href{http://arxiv.org/abs/hep-th/9403187}{{\tt
  hep-th/9403187}}].

\bibitem{cytools}
M.~Demirtas, L.~McAllister and A.~Rios-Tascon, \emph{{{\tt{CYTools}}: A
  Software Package for Analyzing Calabi-Yau Manifolds, to appear}}.

\bibitem{Friedman:1997yq}
R.~Friedman, J.~Morgan and E.~Witten, \emph{{Vector bundles and F theory}},
  \href{http://dx.doi.org/10.1007/s002200050154}{\emph{Commun. Math. Phys.}
  {\bf 187} (1997) 679--743}, [\href{http://arxiv.org/abs/hep-th/9701162}{{\tt
  hep-th/9701162}}].

\end{thebibliography}\endgroup
\end{document}